%% file: SMEFiT-topHiggs.tex
\documentclass[11pt,a4paper]{article}
\usepackage{placeins}
\usepackage{graphicx}
\usepackage{xcolor}
\usepackage{float}
\usepackage{afterpage}
\usepackage{amssymb,amsmath,bm}
\usepackage{multirow,booktabs,ulem}
\usepackage{cite}
\usepackage[colorlinks=true, linkcolor=black!50!blue, urlcolor=blue, citecolor=blue, anchorcolor=blue]{hyperref}
\usepackage[font=small,labelfont=bf,labelsep=period]{caption}
\usepackage[a4paper,top=3cm,bottom=2.5cm,left=2.5cm,right=2.5cm,bindingoffset=0mm]{geometry}
\usepackage{tabularx}
\newcolumntype{C}[1]{>{\centering\arraybackslash}p{#1}}

\usepackage{scalerel}

\def\vecsign#1{\rule[1.388\LMex]{\dimexpr#1-2.5pt}{.36\LMpt}%
  \kern-6.0\LMpt\mathchar"017E}

\usepackage{stackengine,amsmath}
\stackMath

\newcommand{\be}{\begin{equation}}
\newcommand{\ee}{\end{equation}}
\newcommand{\bea}{\begin{eqnarray}}
\newcommand{\eea}{\end{eqnarray}}
\newcommand{\bi}{\begin{itemize}}
\newcommand{\ei}{\end{itemize}}
\newcommand{\ben}{\begin{enumerate}}
\newcommand{\een}{\end{enumerate}}
\newcommand{\la}{\left\langle}
\newcommand{\ra}{\right\rangle}
\newcommand{\lc}{\left[}
\newcommand{\rc}{\right]}
\newcommand{\lp}{\left(}
\newcommand{\rp}{\right)}

\def\frac#1#2{{{#1}\over {#2}}}
\def\gsim{\gtrsim}
\def\lsim{\lesssim}
\newcommand{\mrexp}{\mathrm{exp}}

\newcommand{\art}{\mathrm{art}}
\newcommand{\rep}{\mathrm{rep}}

\newcommand{\draft}[1]{}

\newcommand{\sss}{\scriptscriptstyle}
\newcommand{\OO}{\ensuremath{\mathcal{O}}}
\newcommand{\Op}[1]{\OO_{\sss #1}}
\newcommand{\pdp}{\ensuremath{\varphi^\dagger\varphi}}
\def\lra#1{\overset{\text{\scriptsize$\leftrightarrow$}}{#1}}

\def\beq{\begin{equation}}
\def\eeq{\end{equation}}

\def\({\left(}
\def\){\right)}
\def\[{\left[}
\def\]{\right]}
\let\originalleft\left
\let\originalright\right
\renewcommand{\left}{\mathopen{}\mathclose\bgroup\originalleft}
\renewcommand{\right}{\aftergroup\egroup\originalright}

\numberwithin{equation}{section}
\numberwithin{figure}{section}
\numberwithin{table}{section}

\let\oldsubsection\subsection
\renewcommand\subsection[2][\subsectiontoc]{%
  \def\subsectiontoc{#2}%
  \oldsubsection[#1]{\boldmath #2}%
}

\let\oldsubsubsection\subsubsection
\renewcommand\subsubsection[2][\subsubsectiontoc]{%
  \def\subsubsectiontoc{#2}%
  \oldsubsubsection[#1]{\boldmath #2}%
}

\usepackage{xcolor}
\input{Macros.tex}

\begin{document}
\newgeometry{top=1.0cm,bottom=1.5cm,left=2.5cm,right=2.5cm,bindingoffset=0mm}
\begin{titlepage}
\thispagestyle{empty}
\noindent
\begin{flushleft}
\begin{figure}[h]
  \includegraphics[width=0.32\textwidth]{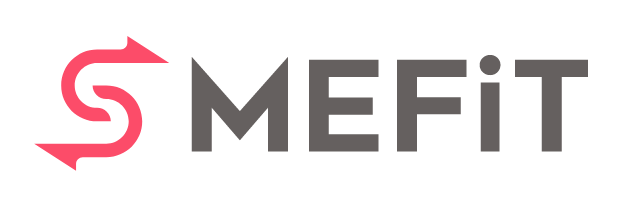}
\end{figure}
\end{flushleft}
\vspace{-3cm}
\begin{flushright}
OUTP-20-05P \\
Nikhef-2020-020 \\
 CP3-21-12 \\
 MCNET-21-07\\
 MAN/HEP/2021/004
\end{flushright}
\vspace{0.7cm}
\begin{center}
  {\LARGE \bf\boldmath Combined SMEFT interpretation of Higgs, diboson,  \\[0.3cm]
  and top quark data from the LHC   }\vspace{1.4cm}

  {\bf  The SMEFiT Collaboration:} \\[0.1cm]
  
  Jacob~J.~Ethier,$^{1,2}$
  Giacomo~Magni,$^{1,2}$
Fabio~Maltoni,$^{3,4}$
Luca~Mantani,$^{3}$\\[0.1cm]
Emanuele~R.~Nocera,$^{2,5}$
Juan~Rojo,$^{1,2}$,
Emma~Slade,$^6$ Eleni~Vryonidou,$^7$ and
Cen~Zhang$^{8,9}$

\vspace{0.7cm}
{\it \small
~$^1$ Department of Physics and Astronomy, Vrije Universiteit Amsterdam,\\ NL-1081 HV Amsterdam, The Netherlands\\[0.1cm]
 ~$^2$  Nikhef Theory Group, Science Park 105, 1098 XG Amsterdam, The Netherlands\\[0.1cm]
~$^3$ Centre for Cosmology, Particle Physics and Phenomenology (CP3),\\
  Universit\'e Catholique de Louvain, B-1348 Louvain-la-Neuve, Belgium\\[0.1cm]
~$^4$
Dipartimento di Fisica e Astronomia, Universit\`a di Bologna \\and INFN, Sezione di Bologna, 
via Irnerio 46, 40126 Bologna, Italy\\[0.1cm]
~$^5$
The Higgs Centre for Theoretical Physics, The University of Edinburgh,\\
JCMB, KB, Mayfield Rd, Edinburgh EH9 3JZ, Scotland\\[0.1cm]
~$^6$ Rudolf Peierls Centre for Theoretical Physics, University of Oxford, \\
  Clarendon Laboratory, Parks Road, Oxford OX1 3PU, United Kingdom\\[0.1cm]
~$^7$ Department of Physics and Astronomy, University of Manchester, \\Oxford Road, Manchester M13 9PL, United Kingdom\\[0.05cm]
~$^8$ Institute of High Energy Physics, and School of Physical Sciences,\\
University of Chinese Academy of Sciences, Beijing 100049, China\\[0.05cm]
~$^9$ Center for High Energy Physics, Peking University, Beijing 100871, China
}

\vspace{0.7cm}

{\bf \large Abstract}

\end{center}

We present a global interpretation of Higgs, diboson, and top quark
production and decay measurements from the LHC in the framework of the
Standard Model Effective Field Theory (SMEFT) at dimension six.
We constrain simultaneously 36 independent directions in its parameter space,
and compare the outcome of the global analysis  with that from
individual and two-parameter fits.
Our results are obtained by means of state-of-the-art theoretical
calculations for the SM and the EFT
cross-sections, and account for both linear and
quadratic corrections in the $1/\Lambda^2$
expansion.
We demonstrate how the inclusion of NLO QCD
and $\mathcal{O}\lp \Lambda^{-4}\rp$ effects
is instrumental to accurately map the  posterior distributions associated
to the fitted Wilson coefficients.
We assess the interplay and complementarity
 between  the top quark, Higgs, and diboson measurements,
 deploy a variety of statistical estimators to quantify the 
 impact of each dataset in the parameter space, and carry out
 fits in  BSM-inspired scenarios such as the top-philic model.
 Our results represent a stepping stone in the ongoing program of model-independent searches
  at the LHC from precision measurements, and pave the way
  towards yet more global SMEFT interpretations extended to other high-$p_T$ processes
   as well as to low-energy observables.

   \vspace{0.4cm}

   \begin{center}
\bf \small This paper is dedicated to the memory of our beloved friend and collaborator Cen Zhang.
   \end{center}  

\end{titlepage}

\restoregeometry

\tableofcontents

\input{sec-introduction.tex}
\input{sec-smeftth.tex}

\input{sec-dataset.tex}
\input{sec-settings.tex}

\input{sec-results.tex}

\input{sec-summary.tex}

\appendix
\input{app-comparisondata.tex}
\input{app-Higgs-Signal-Strenghts.tex}

\input{app-covariance-mat.tex}

\input{app-delivery.tex}

\FloatBarrier

\phantomsection
\addcontentsline{toc}{section}{References}


\providecommand{\href}[2]{#2}\begingroup\raggedright\endgroup

\end{document}

%% file: Macros.tex
\usepackage[utf8]{inputenc}
\usepackage[T1]{fontenc}
\usepackage[numbers, elide, sort&compress]{natbib}
\usepackage{lmodern, amsmath, amssymb, xcolor, a4wide, graphicx, datetime, rotating, adjustbox, array}

\makeatletter\g@addto@macro\bfseries{\boldmath}\makeatother

\def\equationautorefname~#1\null{Eq.\,(#1)\null}
\newcommand{\fullref}[2]{\hyperref[#2]{#1\,\ref*{#2}}}
\makeatletter

\newcommand{\rcite}[1]{\hyper@@link[cite]{}{cite.#1}{Ref.\,\cite*{#1}}}
\makeatother
\newcommand*{\eqsref}[2]{\hyperref[#1]{Eqs.\,(\ref*{#1}--}\hyperref[#2]{\ref*{#2})}}

\allowdisplaybreaks[4]

\makeatletter
\newcommand{\parenbar}{\mathpalette\p@renb@r}
\def\p@renb@r#1#2{%
	\vbox{%
		\ifx#1\scriptscriptstyle\dimen@.7em\dimen@ii.20em\else%
		\ifx#1\scriptstyle	\dimen@.8em\dimen@ii.25em\else%
					\dimen@.9em\dimen@ii.35em\fi\fi%
		\offinterlineskip%
		\ialign{%
			\hfill##\hfill\cr
			\vbox{\hrule width\dimen@ii height .35pt}\cr
			\noalign{\vskip-.35ex}%
			\hbox to\dimen@{$\mathchar300\hfil\mathchar301$}\cr
			\noalign{\vskip-.35ex}%
			$#1#2$\cr%
		}%
	}%
}%
\makeatother

\newcommand*{\tmp}[4]{\ensuremath{%
	{#4%
	\ifx\empty#3\empty\ifx\empty#1\empty\else^{#1}\fi\else^{#1(#3)}\fi%
	\ifx\empty#2\empty\else_{#2}\fi}%
}}

\newcommand*{\ccc}[4][]{\tmp{#2}{#3}{#4}{#1{c}}}

\newcommand*{\qq }[4][]{\tmp{#2}{#3}{#4}{#1{\mathcal{O}}}}

\newcommand{\FDF}{(\varphi^\dagger i\!\!\overleftrightarrow{D}_\mu\varphi)}
\newcommand{\FDFI}{(\varphi^\dagger i\!\!\overleftrightarrow{D}^I_\mu\varphi)}

\newcommand{\rien}[1]{}

\newcounter{affiliation}

%% file: sec-introduction.tex
\section{Introduction}

A powerful, model-independent framework to constrain, identify,  and parametrise potential
deviations with respect to the predictions of the Standard
Model (SM) is
provided by the Standard Model Effective Field
Theory (SMEFT)~\cite{Weinberg:1979sa,Buchmuller:1985jz,Grzadkowski:2010es}, see also~\cite{Brivio:2017vri}
for a review.
A particularly attractive feature of the SMEFT is its capability to systematically correlate
deviations from the SM between different processes, for example between  Higgs 
and top quark cross-sections, or between high-$p_T$ and flavor observables.

A direct consequence of this model independence  is the high dimensionality
of the parameter space spanned by the relevant higher-dimensional EFT operators.
Indeed, the number of Wilson coefficients
constrained in  typical SMEFT analyses can vary between
just a few up to the several tens or even hundreds, depending
on the specific  assumptions adopted concerning
the flavour, family (non-)universality of the couplings, and CP-symmetry
structure (among others) of the UV-complete theory.
For this reason, the full exploitation of the SMEFT potential
for indirect New Physics searches from precision measurements requires
combining the information provided by the broadest possible dataset.

The phenomenology of the SMEFT has attracted significant
 attention, with most analyses focusing on specific sectors
of the parameter space and groups of processes.
Some of these recent studies have targeted the top quark
properties~\cite{Buckley:2016cfg,Buckley:2015lku,Hartland:2019bjb,Brivio:2019ius},
the Higgs and electroweak gauge sector~\cite{Biekotter:2018rhp,Ellis:2018gqa,Almeida:2018cld},
single and double gauge boson
production~\cite{Baglio:2020ibv,Alioli:2018ljm,Ethier:2021ydt,Greljo:2017vvb},
vector-boson scattering~\cite{Gomez-Ambrosio:2018pnl,Ethier:2021ydt,Dedes:2020xmo},
and flavour and low-energy
observables~\cite{Aebischer:2018iyb,Falkowski:2019xoe,Falkowski:2017pss}, among
several others.
Furthermore, analyses that combine the constraints of different
groups of processes in the EFT parameter space, such as the Higgs and electroweak sector with
the top quark one~\cite{Ellis:2020unq} or 
top quark data with $B$-meson observables~\cite{Bissmann:2020mfi,Bruggisser:2021duo},
have also been presented.
These and related studies demonstrate that a
global interpretation of the SMEFT is unavoidable and makes possible
benefiting from hitherto unexpected connections, such as the  correlation
of the LHCb flavour anomalies~\cite{Pich:2019pzg,Aaij:2021vac}
at the $B$-meson scale with the high-$p_T$ tails
at the LHC~\cite{Greljo:2017vvb,Fuentes-Martin:2020lea}.

With the ultimate motivation of performing a truly global EFT interpretation
of particle physics data, the {\tt SMEFiT} fitting framework was developed in~\cite{Hartland:2019bjb}
and applied to the analysis of the top quark properties at the LHC as a proof-of-concept.
This novel EFT fitting methodology, inspired by techniques deployed by the NNPDF
Collaboration to
determine the proton's parton distribution
functions (PDFs)~\cite{Ball:2008by,Ball:2010de,Ball:2014uwa,Rojo:2018qdd,Forte:2020yip},
made possible constraining the Wilson coefficients
associated to 34 independent dimension-six
operators that modify the production cross-sections of top
quarks.
Our results improved over
existing bounds~\cite{AguilarSaavedra:2018nen} for the wide majority of
directions in the SMEFT parameter space and in several
cases the associated Wilson coefficients were constrained for the first time.
Subsequently, {\tt SMEFiT} was extended with the Bayesian reweighting method~\cite{vanBeek:2019evb}
developed for PDFs~\cite{Ball:2011gg,Ball:2010gb} which  allows one
constraining the EFT parameter space {\it a posteriori}
with novel measurements without requiring a dedicated fit.
 {\tt SMEFiT} has also been recently applied for the first SMEFT interpretation of vector boson
 scattering data~\cite{Ethier:2021ydt} from the full Run II dataset.

In this work, we complement and
extend the {\tt SMEFiT} analysis framework of~\cite{Hartland:2019bjb} in several directions.
First and foremost, we extend the dimension-six EFT operator basis
in order to simultaneously describe top-quark
measurements together with Higgs boson production and decay cross-sections,
as well as with weak gauge boson pair production from LEP and the LHC.
Specifically, we consider Higgs signal strengths, differential distributions,
and simplified template cross-section (STXS) measurements from 
ATLAS and CMS taken at Runs I and II.
Furthermore, we account for the most recent top-quark observables
from the Run II dataset, such as updated measurements of four-top,
top quark pair in association with a $Z$ boson, and differential
single-top and top quark pair production.
We also include the differential distributions
in gauge boson pair production from LEP
and the LHC, which constrain complementary directions in the EFT space.
In addition, we account in an indirect manner for the information
provided by electroweak precision observables (EWPO) from
LEP~\cite{ALEPH:2005ab}
by means of imposing restrictions on specific combinations
of the EFT coefficients.

A second improvement as compared to~\cite{Hartland:2019bjb}
concerns the fitting methodology.
On the one hand, the  Monte Carlo replica fitting method has been
upgraded by means of more efficient optimizers and the imposition of
post-fit quality selection criteria for the replicas.
On the other hand, we have implemented a novel, independent
approach to constrain the  parameter space based on
Nested Sampling (NS) by means of the MultiNest algorithm~\cite{Feroz:2013hea}.
As opposed to the replica fitting method, which is an
optimisation problem,
NS aims to reconstruct the posterior probability distribution given
the model and the data by means of Bayesian inference.
We have cross-validated the performance
of  the two methods and demonstrated that they
lead to equivalent results.
The availability of two orthogonal fitting strategies
strengthens the robustness of {\tt SMEFiT} and facilitates the
combined interpretation of data from different processes.

From the combination of the  improved fitting framework
and the extensive input dataset, we 
derive individual, two-dimensional, and global (marginalised) bounds for  36
independent directions (and 14 dependent ones) in the EFT parameter space.
The EFT cross-sections used in this analysis account for
either only the linear
or for both linear and quadratic effects, $\mathcal{O}\lp \Lambda^{-2}\rp$ and
$\mathcal{O}\lp \Lambda^{-4}\rp$ respectively, and
include NLO QCD corrections whenever available.
We demonstrate in detail how the inclusion of NLO QCD
and $\mathcal{O}\lp \Lambda^{-4}\rp$ corrections
in the EFT calculations
is instrumental in order to accurately pin down the  posterior distributions associated
to the fitted Wilson coefficients.

By means of information geometry and principal component analysis techniques,
we quantify the sensitivity of each
of the input datasets to the various Wilson coefficients.
We validate these statistical diagnosis tools by means of a series of  fits restricted
to subsets of processes, such as Higgs-only and top-only EFT analyses.
Specifically, we quantify the interplay between the top-quark and Higgs measurements
in the determination of EFT degrees of freedom sensitive to both processes,
such as the modifications of the top Yukawa coupling.
Furthermore, we explore how the EFT fit results are modified when additional, UV-inspired theory restrictions
are imposed in the parameter space, and present results
for the case of a top-philic model.

The paper is organised as follows.
First of all, Sect.~\ref{sec:smefttheory} discusses the operator basis,
flavour assumptions, the fitted degrees of freedom,
and the top-philic scenario.
Then Sect.~\ref{sec:settings_expdata} describes the top-quark, Higgs,
and diboson datasets that are used as input to the analysis together
with the corresponding SM and EFT calculations.
The methodological improvements in {\tt SMEFiT}, together with the description of  the fit settings,
are presented in Sect.~\ref{sec:fitsettings}.
The main results of this work, namely the combined SMEFT
interpretation of top-quark, Higgs, and diboson measurements at the LHC,
are presented and discussed in Sect.~\ref{sec:results}.
Finally, in Sect.~\ref{sec:summary} we summarise and discuss future steps
in this project.

Supplementary information is provided in three appendices.
In App.~\ref{sect:app_comparison_data} we present
the comparison between the SM and SMEFT theory predictions
with the experimental datasets used as input to
the fit;
in App.~\ref{sec:signalstrenghts} we describe the implementation
of  the Higgs signal strength measurements;
{  in App.~\ref{sec:fullcovmat} we present the 
  correlation matrices for the complete set of operators
considered in the analysis; and}
then
in App.~\ref{sec:delivery} we discuss how
the results of this work are rendered publicly available
and provide usage instruction.

%% file: sec-smeftth.tex
\section{EFT description of the top, Higgs, and electroweak sectors}
\label{sec:smefttheory}

In this section we collect the definitions and conventions that will
be used to construct the dimension-six operators and the associated degrees of freedom (DoFs)
relevant for the theoretical description of the processes considered
in this analysis.
These are operators that modify the production and decay of Higgs bosons
and top quarks at hadron colliders, precision electroweak measurements from LEP/SLC,
and gauge-boson pair production cross-sections both at LEP2 and at the LHC.

First of all,
we provide explicit definitions for the operators
and for the physical EFT coefficients adopted in this work,
as well as the corresponding notational conventions.
Following the recommendation of the LHC Top Quark Working Group
\cite{AguilarSaavedra:2018nen} as well as the strategy of our previous work
\cite{Hartland:2019bjb}, in the top-quark sector we fit specific degrees
of freedom closely related to the experimental measurements, instead of directly using the
Warsaw-basis operator coefficients.
Our degrees of freedom are therefore linear combinations of the
Warsaw-basis operator coefficients, which appear in the interference with SM
amplitudes, and represent interactions of physical fields after 
electroweak symmetry breaking.
These combinations are then aligned with
physically relevant directions of the parameter space, and thus have
a more transparent physical interpretation. They also represent the
maximal information that can be extracted from measuring a certain process.

We will then discuss how the constraints provided by the electroweak precision
observables (EWPOs) from LEP/SLC can be approximately accounted for by means of a series of
restrictions on the EFT parameter space.
We also discuss theoretical constraints on the operator coefficients following
a more restrictive assumption about the UV-complete theory, namely the
so-called top-philic scenario.
Finally, we discuss several theoretical relations that must be satisfied
by the EFT cross-sections following the requirement that physical cross-sections are positive-definite
quantities.

\subsection{Operator basis and degrees of freedom}
\label{sec:operatorbasis}

\paragraph{Conventions.}
Let us start by summarizing the notation and conventions that are adopted
in this work concerning the relevant dimension-six SMEFT operators.
Here we follow the notation of the Warsaw basis presented in~\cite{Grzadkowski:2010es}.
In this notation, flavour indices are labelled by $i,j,k$ and $l$; left-handed
quark and lepton fermion SU(2)$_L$ doublets are denoted by $q_i$, $\ell_i$;
the right-handed quark singlets by $u_i$, $d_i$, while
the right-handed lepton singlets are denoted by $e$, $\mu$, $\tau$ without using
flavor index. Given the special role of the top-quark in this work, we 
use $Q$ and $t$ to denote the left-handed top-bottom doublet and the right-handed
top singlet, instead of using $q_3$ and $u_3$.
The Higgs doublet is denoted by $\varphi$;
the antisymmetric SU(2) tensor by $\varepsilon\equiv i\tau^2$;
$\tilde{\varphi}=\varepsilon\varphi^*$; and we define
\be
\FDF\equiv \varphi^\dagger(iD_\mu \varphi) - (iD_\mu\varphi^\dagger) \varphi \,, \qquad 
\FDFI\equiv \varphi^\dagger\tau^I(iD_\mu \varphi) - (iD_\mu\varphi^\dagger) \tau^I\varphi \, ,
\ee where $\tau^I$ are the Pauli matrices.
In the following,
$G^A_{\mu\nu}$, $W^I_{\mu\nu}$, and $B_{\mu\nu}$ stand for the SU(3) strong and SU(2)$_L$ and U(1)$_Y$ weak gauge
field strengths respectively, and the covariant derivatives include all the relevant
interaction terms.
For instance, the gluon field strength tensor is given by
\be
G_{\mu\nu}^A = \partial_{\mu} G_{\nu}^A - \partial_{\nu} G_{\mu}^A + g_s f^{ABC}G_{\mu}^B G_{\nu}^C \, ,
\ee
where $G_{\mu}^A$ is the gluon field, $A, B, C$ are color indices in the adjoint
representation, $g_s$ is the strong coupling and $f^{ABC}$ are the structure
constants of SU(3).
Similar definitions hold for the electroweak $W_I^{\mu\nu}$ and $B^{\mu\nu}$
field strength tensors, for instance one has
\be
W_{\mu\nu}^I = \partial_{\mu} W_{\nu}^I - \partial_{\nu}W_{\mu}^I + g_w \epsilon_{JK}^I W_{\mu}^J W_{\mu}^K \, ,
\ee
where $g_w$ is the SU(2)$_L$ coupling constant.

\paragraph{Flavour assumptions.}
The number of independent dimension-six operators can be unfeasibly large, if all three
generations of the SM fermions are taken into account: there are 2499 in total
\cite{Alonso:2013hga}, with 572 four-fermion operators that are in principle relevant for top-quark
physics~\cite{AguilarSaavedra:2010zi}.
In this analysis, we follow closely the strategy which we adopted in our previous top-quark
sector study~\cite{Hartland:2019bjb} and that has been documented in the LHC Top
Quark Working Group note~\cite{AguilarSaavedra:2018nen}: we implement the
Minimal Flavour Violation (MFV) hypothesis~\cite{DAmbrosio:2002vsn} in the
quark sector as the baseline scenario.
A slight difference is that instead of a $U(2)_q\times U(2)_u \times U(2)_d$
flavour symmetry among the first two generations,
we now impose the $U(2)_q\times U(2)_u \times U(3)_d$ symmetry,
under the assumption that the Yukawa couplings are nonzero only for the top quark.
This flavour assumption is consistent with the {\tt SMEFT@NLO} model~\cite{Degrande:2020evl},
the implementation of automated one-loop calculation in the SMEFT which we will use to the provide
theoretical inputs for our global fit, as discussed in the next section.

As a result of the different flavour assumption,
the EFT parameter space is further
reduced compared to~\cite{Hartland:2019bjb}.
In particular, the coefficients of operators with right-handed bottom quarks
are either set to zero or set equal to the corresponding down-quark ones.
Furthermore, we then slightly relax our assumptions by keeping the bottom and
charm quark Yukawa operators in our fit, to account for the current LHC
sensitivity to these parameters.
All other light quark Yukawa operators are set to zero,
since we do not expect to have any sensitivity on their coefficients.

Concerning the leptonic sector, the adopted flavour symmetry is $(U(1)_\ell \times
U(1)_e)^3$, also following~\cite{AguilarSaavedra:2018nen}.
This assumption sets all the lepton masses as well as their Yukawa couplings to
zero in the SM, while leaving independent parameters for each lepton-antilepton
pair of a given generation.
This is then relaxed by including the $\tau$ Yukawa operator, to account 
for the expected LHC sensitivity arising from dedicated measurements.
In practice, the lepton flavor assumptions do not have implications for the EFT fit
given the constraints from $Z$-pole measurements at LEP and SLC, see the discussion below.

\paragraph{Purely bosonic operators.}
Table~\ref{tab:oper_bos} reports the purely bosonic dimension-six operators that
modify the production and decay of Higgs bosons as well as the interactions
of the electroweak gauge bosons.
For each operator, we indicate its definition in terms of the SM fields
and the notation that we will use both
for the operators and for the Wilson coefficients.
These operators modify several important Higgs boson production and decay processes
that are (or will become) accessible at the LHC, as well as the production
of gauge boson pairs both in electron-positron and in proton-proton collisions.

\input{tables/table-bosonicoperators.tex}

One can comment on some interesting features of the operators defined
in Table~\ref{tab:oper_bos}.
To begin with, the operators $O_{\varphi WB}$ and $O_{\varphi D}$ are the ones
often identified as the $S$ and $T$ oblique parameters, though this
identification is basis-dependent and is not strictly correct in the Warsaw
basis.
Together with several of the two-fermion operators listed in Table~\ref{tab:oper_ferm_bos}, they are severely
constrained by the $Z$-pole and $W$-pole measurements available from LEP and
SLC, but with 2 linear combinations left unconstrained. These two combinations
in turn modify the electroweak triple gauge boson (TGC) couplings and the Higgs-electroweak
interactions.  They are thus constrained mainly by the diboson measurements
at the LEP2 and the LHC, as well as the Higgs measurements at the LHC. We will
discuss this property in more detail in the following section.
The operator $O_{W}$ generates a TGC coupling modification which is purely transversal
and is hence constrained only by diboson data.

The rest of the bosonic operators listed in Table~\ref{tab:oper_bos} modify only the Higgs boson couplings, and represent 
degrees of freedom that are accessible only with Higgs data.
First, the operators $\mathcal{O}_{\varphi W}$ and $\mathcal{O}_{\varphi B}$
modify the interaction between Higgs bosons and electroweak gauge bosons. 
At the LHC, they can be probed for example by means of the Higgs decays into
weak vector bosons, $h\to ZZ^*$ and $h \to W^+W^-$, as well as in the
vector-boson-fusion (VBF) process and in associated production with vector bosons, $hW$
and $hZ$.
In addition, the $\mathcal{O}_{\varphi G}$ operator is similar but 
introduces a direct coupling between the Higgs boson and gluons.
It therefore enters the Higgs total width and branching ratios,
the production cross section in gluon fusion channel, 
as well as the associated production channel $t\bar{t}h$.
Finally, the $O_{\varphi d}$ operator generates a wavefunction correction to the
Higgs boson, which rescales all the Higgs boson couplings in a universal manner.

{  In principle, one could also include in
  Table~\ref{tab:oper_bos} the triple-gluon
  operator $\mathcal{O}_{G}$, which contributes to $t\bar{t}(V)$ and Higgs+jets production.
  However this operator is already tightly constrained by
  multi-jet production measurements at the LHC~\cite{Hirschi:2018etq},
  as discussed also in~\cite{Hartland:2019bjb}.
  It is found that the bounds on the coefficient $c_G$ obtained
  from multijet data are very stringent and
  beyond the sensitivity achievable via either top quark or Higgs production measurements
  at the LHC.
  For this reason, $\mathcal{O}_{G}$ is not considered in the present analysis.
}
  
\paragraph{Two-fermion operators.}
%
\input{tables/table-2foperators.tex}
%
Table~\ref{tab:oper_ferm_bos} collects, using the same format
as in Table~\ref{tab:oper_bos}, the relevant Warsaw-basis operators
that contain two fermion fields, either quarks or leptons,
plus a single four-lepton operator.
From top to bottom, we list the two-fermion operators involving 3rd generation quarks,
those involving 1st and 2nd generation quarks, and
operators containing two leptonic fields (of any generation).
We also include in this list the four-lepton operator $\mathcal{O}_{\ell\ell}$.

The operators that involve a top-quark field, either $Q$ (left-handed doublet) or $t$
(right-handed singlet),
are crucial for the interpretation of LHC top-quark measurements.
Interestingly, all of them involve at least one Higgs-boson field, which
introduces an interplay between the top and Higgs sectors of the SMEFT.
For example, the chromo-magnetic dipole operator $O_{tG}$ and the dimension-six Yukawa
operator $O_{t\varphi}$ are constrained by both top quark measurements, such
as $t\bar t{h}$ associated production, as well as Higgs measurements, such
as Higgs production through gluon fusion.
Furthermore, the electroweak-dipole operators, $O_{tW}$ and $O_{tB}$, as well as the
current operators, 
$O_{\varphi Q}^{(3)}$ and $O_{\varphi t}$, can be constrained by the associated production
of single top-quarks and Higgs bosons,
as well as by the loop-induced Higgs decays into a $Z\gamma$ final state.

In Table~\ref{tab:oper_ferm_bos} we also list operators that contain light quark
(1st and 2nd generation)
and leptonic fields (of any generation).
The light quark operators enter the Higgs production through the $Vh$ and
VBF channels, as well as the diboson processes.
These operators also modify the Higgs boson width and branching ratios.
For example, the Higgs decay width to $q\bar{q}\ell^+\ell^-$ becomes modified by operators that
induce an effective $Zhq\bar{q}$ vertex, such as $\mathcal{O}_{\varphi u}$.
The leptonic operators are relevant for the same reason,
once we account for the leptonic decays of the Higgs and gauge bosons.
In addition, indirect contributions arise from  the $\mathcal{O}_{\varphi
\ell_1}^{(3)}$, $\mathcal{O}_{\varphi \ell_2}^{(3)}$, and $\mathcal{O}_{\ell\ell}$ operators, which
 modify the measurement of the Fermi constant, $G_F$,
and this affects the extracted SM parameters. They therefore introduce a universal
contribution to all electroweak interactions, and are relevant for $Vh$, VBF, and for the
diboson channels.

\begin{table}[htbp]
  \begin{center}
    \renewcommand{\arraystretch}{1.49}
    \begin{tabular}{ll}
      \toprule
      DoF $\qquad$  & Definition \\
                \midrule
$c_{\varphi Q}^{(-)}$ &  $c_{\varphi Q}^{(1)} - c_{\varphi Q}^{(3)}$ \\
\midrule
$c_{tZ}$ &   $-\sin\theta_W c_{tB} + \cos\theta_W c_{tW} $\\
\midrule
$c_{\varphi q}^{(-)}$ & $ c_{\varphi q}^{(1)} - c_{\varphi q}^{(3)}$ \\
  \bottomrule
\end{tabular}
    \caption{Additional degrees of freedom defined from linear combinations of
      the two-fermion operators listed in Table~\ref{tab:oper_ferm_bos}.
The first two DoFs modify the $t\bar{t}Z$ couplings,
while the third combination is introduced for consistency with the first one.
These are the DoFs that enter at the fit level, replacing those
marked with  (*) in Table~\ref{tab:oper_ferm_bos}.
\label{tab:oper_ferm_bos2}}
\end{center}
\end{table}

We point out that
most of the operator coefficients defined in Table~\ref{tab:oper_ferm_bos} correspond
directly to degrees of freedom used in
the fit, except for three of them, which are
indicated with a (*) in the second column.
Instead, following Ref.~\cite{AguilarSaavedra:2018nen},
three additional degrees of freedom are defined from the linear
combinations indicated in Table~\ref{tab:oper_ferm_bos2}.
These are the DoFs that enter at the fit level, replacing those
marked with  a (*) in Table~\ref{tab:oper_ferm_bos}.

Finally, we note that, as mentioned above, here flavour universality in the
leptonic sector is not imposed, and thus the coefficients of the operators
involving bilinears in the electron, muon, and tau lepton fields are in
principle independent.  In total we have 23 independent fit parameters, defined
from two-fermion operators, plus in addition the four-lepton
operator $c_{\ell\ell}$.
However, in practice, this flexibility will not be relevant for the present
fit due to the constraints from the EWPOs, to be discussed next.

\paragraph{The role of electroweak precision observables.}
At this point, one should note that a subset
of the dimension-six operators defined in
Tables~\ref{tab:oper_bos} and~\ref{tab:oper_ferm_bos} are already well
constrained by the electroweak precision observables (EWPO)~\cite{Han:2004az}
measured at the $Z$-pole~\cite{ALEPH:2005ab} and  the $W$-pole at the LEP and
SLC electron-position colliders.
Given in particular the high accuracy of these LEP measurements, these constraints are known
to dominate in many cases when compared to those provided by the LHC
cross-sections.  
Specifically, the operators sensitive to the EWPO are the following (with definitions
presented in Tables~\ref{tab:oper_bos} and \ref{tab:oper_ferm_bos})
\be
\mathcal{O}_{\varphi WB},
\mathcal{O}_{\varphi D},
\mathcal{O}_{\varphi q}^{\sss(1)},
\mathcal{O}_{\varphi q}^{\sss(3)}, \mathcal{O}_{\varphi u i}, \mathcal{O}_{\varphi d i},
\mathcal{O}_{\varphi \ell_i}^{\sss(3)}, \mathcal{O}_{\varphi \ell_i}^{\sss(1)},
\mathcal{O}_{\varphi e/\mu/\tau}, \mathcal{O}_{\ell\ell} \, . \label{eq:LEPconstrainedDoFs}
\ee
Note that, with $i=1,2,3$, these add up to 16 operators, rather than the 10 which would
correspond to the flavour universal configuration in the leptonic sector.

Fourteen linear combinations of the coefficients associated to these 16 operators  
are constrained by the LEP EWPOs~\cite{Falkowski:2014tna},
leaving therefore only two linear combinations unconstrained. 
These two remaining unconstrained directions  can be
determined  from the information contained in
diboson production cross-sections~\cite{Grojean:2006nn,Alonso:2013hga,Brivio:2017bnu}
as well as by the Higgs production and decay measurements.
{ 
For completeness, the 14 linear combinations of bosonic and two-fermion Wilson coefficients 
which are constrained by the EWPOs measured at LEP are the following~\cite{Brivio:2017bnu}:
\begin{align}
\nonumber
\delta g_{V}^{l_i}&=\delta \bar{g}_{Z} \bar{g}_{V}^{l_i}+Q^{l_i} \delta s_{\theta}^{2}+\Delta_{V}^{l_i} = 0\, , \quad i=1,2,3\, , \\\nonumber
\delta g_{A}^{l_i}&=\delta \bar{g}_{Z} \bar{g}_{A}^{l_i}+\Delta_{A}^{l_i} = 0\, , \quad i=1,2,3\, , \\\nonumber
\delta g_{V}^{u}&=\delta \bar{g}_{Z} \bar{g}_{V}^{u}+Q^{u} \delta s_{\theta}^{2}+\Delta_{V}^{u} = 0\, , \\\nonumber
\delta g_{A}^{u}&=\delta \bar{g}_{Z} \bar{g}_{A}^{u}+\Delta_{A}^{u} = 0\, ,\label{LEPconstraints} \\
\delta g_{V}^{d}&=\delta \bar{g}_{Z} \bar{g}_{V}^{d}+Q^{d} \delta s_{\theta}^{2}+\Delta_{V}^{d} = 0\, , \\\nonumber
\delta g_{A}^{d}&=\delta \bar{g}_{Z} \bar{g}_{A}^{d}+\Delta_{A}^{d} = 0\, , \\\nonumber
\delta g_V^{W, l_i}&= \frac{c_{ll} + 2 c_{\varphi \ell_i}^{(3)} - c_{\varphi \ell_1}^{(3)} - c_{\varphi \ell_2}^{(3)}}{4 \sqrt{2} G_F} = 0 \, , \quad i=1,2,3\, , \\\nonumber
\delta g_V^{W, q}&= \frac{ c_{ll} + c_{\varphi q}^{(3)} - c_{\varphi \ell_1}^{(3)} - c_{\varphi \ell_2}^{(3)}}{4 \sqrt{2} G_F} = 0 \, ,
\end{align}
where $g_1$ and $g_w$ are the corresponding electroweak couplings, $\bar{g}_{V}^{f}=T_{3} / 2-Q^{f} \bar{s}_{\theta}^{2}$, $\bar{g}_{A}^{f}=T_{3} / 2$ and
\begin{align*}
\Delta_{V}^{\ell_i} &=-\frac{1}{4 \sqrt{2} \hat{G}_{F}}\left(c_{\varphi \ell_i}^{(1)}+c_{\varphi \ell_i}^{(3)}+c_{\varphi e_i}\right) & \Delta_{A}^{\ell_i}&=-\frac{1}{4 \sqrt{2} \hat{G}_{F}}\left(c_{\varphi \ell_i}^{(1)}+c_{\varphi \ell_i}^{(3)}-c_{\varphi e_i}\right) \\
\Delta_{V}^{u} &=-\frac{1}{4 \sqrt{2} \hat{G}_{F}}\left(c_{\varphi q}^{(1)}-c_{\varphi q}^{(3)}+c_{\varphi u i}\right) & \Delta_{A}^{u} &=-\frac{1}{4 \sqrt{2} \hat{G}_{F}}\left(c_{\varphi q}^{(1)}-c_{\varphi q}^{(3)}-c_{\varphi u i}\right) \\
\Delta_{V}^{d} &=-\frac{1}{4 \sqrt{2} \hat{G}_{F}}\left(c_{\varphi q}^{(1)}+c_{\varphi q}^{(3)}+c_{\varphi d i}\right) & \Delta_{A}^{d} &=-\frac{1}{4 \sqrt{2} \hat{G}_{F}}\left(c_{\varphi q}^{(1)}+c_{\varphi q}^{(3)}-c_{\varphi d i}\right) \\
\delta g_Z &=-\frac{1}{4 \sqrt{2} \hat{G}_{F}}\left(c_{\varphi D}+2 c_{\varphi \ell_1}^{(3)} + 2 c_{\varphi \ell_2}^{(3)}-2 c_{l l}\right) & \delta s_{\theta}^{2}&= \frac{\hat{m}_{W}^{2}}{2 \sqrt{2} \hat{G}_{F} \hat{m}_{Z}^{2}}\left(c_{\varphi D} + \sqrt{\frac{\hat{m}_{Z}^{2}}{\hat{m}_{W}^{2}} - 1} \, c_{\varphi W B}  \right) \, ,\\ 
\end{align*}
where we have used the notation of Ref.~\cite{Brivio:2017bnu}. We note that the modifications of the
$W$ and $Z$ couplings in Eq.~(\ref{LEPconstraints}) are given in the $\lp m_W,m_Z, G_F\rp$ scheme.
Similar 
expressions in the $\lp a_{\rm ew},m_Z, G_F\rp$ scheme can be found in 
 App.~A of~\cite{Falkowski:2001958}.
}
 In a flavour universal scenario, the 
 $Z$- and $W$-pole observables used to 
 constrain these couplings
 are those listed in Table~1 of~\cite{Falkowski:2014tna}. These constrain 8 out of 10 linear combinations of 
 the Warsaw operators which are present in the flavour universal scenario.
 Our
 assumption in this work is stronger, as we assume that there are enough observables to constrain all but 2 degrees of
 freedom of Eq.~(\ref{eq:LEPconstrainedDoFs}). In order to achieve this, one
 would need to go beyond the standard EWPO observables e.g.
beyond those of Table 1 of~\cite{Falkowski:2014tna},
in particular by  including more data which will allow one to constrain more
degrees of freedom which appear in our non-flavour universal scenario.
For example, we distinguish
between leptons of different generations, and therefore a setup like the one
of~\cite{Efrati:2015eaa} would be more appropriate.
Our assumption is that these observables are precise enough to constrain all but two linear combinations. This is supported for example by Ref.~\cite{Efrati:2015eaa} which suggests that even with less
restrictive flavour assumptions,
the constraints on these Wilson coefficients remain relatively stringent.

While in this work we do not explicitly include any EWPO data in the present
fit, we still need to account for the information that they provide on the
SMEFT parameter space.
{  As motivated above,}
this is achieved by assuming that the EWPOs are precise enough to allow us to
set the 14 linear combinations of Eq.~\eqref{LEPconstraints} to zero in our
fit.
{  In other words, one can derive
  constraints in the SMEFT parameter space, Eq.~(\ref{eq:2independents}) below,
  that emulate the information contained in the EWPOs
  by setting all quantities in Eq.~(\ref{LEPconstraints}) to zero.}
The remaining two degrees of freedom can be parametrized by, say,
$ c_{\varphi WB}$ and $c_{\varphi D}$, if the following replacements are made
\begin{flalign}
	\left(
\begin{array}{c}
c_{\varphi \ell_i}^{\sss(3)} \\
 c_{\varphi \ell_i}^{\sss(1)} \\
 c_{\varphi e/\mu/\tau} \\
 c_{\varphi q}^{(-)} \\
 c_{\varphi q}^{(3)} \\
 c_{\varphi u} \\
 c_{\varphi d} \\
 c_{\ell\ell} \\
\end{array}
\right)
= 
\left(
\begin{array}{cc}
 -\frac{1}{t_W} & -\frac{1}{4 t_W^2} \\
 0 & -\frac{1}{4} \\
 0 & -\frac{1}{2} \\
 \frac{1}{t_W} & \frac{1}{4 s_W^2}-\frac{1}{6} \\
 -\frac{1}{t_W} & -\frac{1}{4 t_W^2} \\
 0 & \frac{1}{3} \\
 0 & -\frac{1}{6} \\
 0 & 0 \\
\end{array}
\right)
\left(
\begin{array}{c}
	c_{\varphi WB}\\ c_{\varphi D}
\end{array}
\right) \, .
\label{eq:2independents}
\end{flalign}
These relations will emulate the impact of LEP EWPOs in the fit, and allow us to
produce a consistent fit without explicitly including the EWPOs.

{ 
  We note that there is one additional combination of Wilson coefficients that
  could be constrained by the EWPOs, namely
  \be
  \label{eq:ZbbLEP}
  c_{\varphi Q}^{(1)} +  c_{\varphi Q}^{(3)} =  c_{\varphi Q}^{(-)}
  + 2c_{\varphi Q}^{(3)} \, ,
  \ee
  which modifies the left-handed coupling of the $Z$ boson
  to bottom quarks.
  However, in this work we prefer to constrain this
  combination directly from top quark production
  measurements rather than from the EWPOs.
  Therefore we have kept both $c_{\varphi Q}^{(-)}$
   and $c_{\varphi Q}^{(3)}$ in our fit to be able to assess 
   how well top measurements can constrain these two degrees of freedom. 
 As a cross-check,
  we have verified that the global fit results for other operators are essentially
  unchanged if Eq.~(\ref{eq:ZbbLEP}) is assumed to be constrained
  by LEP rather than by top quark data in the fit, whilst bounds on $c_{\varphi Q}^{(-)}$
  would improve by about a factor of two had we applied this constraint in the fit. 
}

Thanks to these 14 constraints, the 7 and
24 operators listed in Tables~\ref{tab:oper_bos} and~\ref{tab:oper_ferm_bos} respectively
are then reduced to 17 independent degrees of freedom to be constrained by the
LHC experimental data and the LEP diboson cross-sections.
This allows us to set bounds on all operator coefficients
listed in Tables~\ref{tab:oper_bos} and~\ref{tab:oper_ferm_bos}.
Of course, the bounds on the 16
operators of Eq.~(\ref{eq:LEPconstrainedDoFs})
will be highly correlated as indicated by Eq.~(\ref{eq:2independents}).
When presenting results for the independent DoFs, for example when evaluating
the Fisher Information matrix or the principal components, we will select
$c_{\varphi W B}$ and $c_{\varphi D}$, with the understanding that the
replacements of Eq.~(\ref{eq:2independents}) have been made.
Note that it has been argued that the diboson channels at the LHC can in
principle compete with EWPO \cite{Zhang:2016zsp,Grojean:2018dqj}, which
indicates that in an accurate fit one should always include the full set of
EWPO constraints explicitly, as has been done, for example, in the combined
Higgs/electroweak fits of~\cite{Ellis:2018gqa,Ellis:2020unq}.
We  however leave this option to future work.

\paragraph{Four-fermion top quark operators.}
We finally discuss the four-quark operators which involve the top quark 
fields and thus modify the production of top quarks at hadron colliders.
The dimension-six four-fermion operators sensitive to top quarks can be classified into
two categories: operators composed by four heavy quark fields (top and/or bottom quarks) and 
operators composed by two light and two heavy quark fields.
The physical degrees of freedom corresponding to four-heavy and
two-light-two-heavy interactions that we use in the present analysis are
constructed in terms of suitable linear combinations of the four fermion
coefficients in the Warsaw basis, whose corresponding operators are defined as
\begin{align}
	\qq{1}{qq}{ijkl}
	&= (\bar q_i \gamma^\mu q_j)(\bar q_k\gamma_\mu q_l)
	 \nonumber
	,\\
	\qq{3}{qq}{ijkl}
	&= (\bar q_i \gamma^\mu \tau^I q_j)(\bar q_k\gamma_\mu \tau^I q_l)
 \nonumber
	,\\
	\qq{1}{qu}{ijkl}
	&= (\bar q_i \gamma^\mu q_j)(\bar u_k\gamma_\mu u_l)
         \nonumber
	,\\
	\qq{8}{qu}{ijkl}
	&= (\bar q_i \gamma^\mu T^A q_j)(\bar u_k\gamma_\mu T^A u_l)
         \nonumber
	,\\
	\qq{1}{qd}{ijkl}
	&= (\bar q_i \gamma^\mu q_j)(\bar d_k\gamma_\mu d_l)
         \nonumber
	,\\
	\qq{8}{qd}{ijkl}
	&= (\bar q_i \gamma^\mu T^A q_j)(\bar d_k\gamma_\mu T^A d_l)
        \label{eq:FourQuarkOp} 
	,\\
	\qq{}{uu}{ijkl}
	&=(\bar u_i\gamma^\mu u_j)(\bar u_k\gamma_\mu u_l)
         \nonumber
	,\\
	\qq{1}{ud}{ijkl}
	&=(\bar u_i\gamma^\mu u_j)(\bar d_k\gamma_\mu d_l)
         \nonumber
	,\\
	\qq{8}{ud}{ijkl}
	&=(\bar u_i\gamma^\mu T^A u_j)(\bar d_k\gamma_\mu T^A d_l)
         \nonumber \, ,
\end{align}
where recall that $i,j,k,l$ are fermion generation indices.
In Table~\ref{eq:summaryOperatorsTop} we provide the definition of all degrees of freedom
that enter the fit
in terms of the coefficients of Warsaw basis operators of Eq.~(\ref{eq:FourQuarkOp}).
Within our flavour assumptions, the coefficients associated to different values of
the generation indices $i$ ($i=1,2$) or $j$ ($j=1,2,3$) will be the same.

\input{tables/table-fourfermion.tex}


Comparing with our previous EFT analysis of the top quark sector,
in this work due to the different flavor assumptions
several degrees of freedom that were used there as  independent fit parameters are now absent.
{  The reason
  is that here we assume  $U (2)_q \times U (2)_u \times U (3)_d $
  as compared to  $U (2)_q \times U (2)_u \times U (2)_d$ in~\cite{Hartland:2019bjb}.
  The difference is that right-handed bottom  quarks are now treated on the same footing as the right-handed down-type quarks of the first two generations.
  Furthermore, this
  flavour assumption forbids quark
  bilinears such as the chirality-flipping $\bar{Q}b$ and the right-handed charged current $\bar{t}b$.
  These modified flavor assumptions have two main consequences.
}
First of all, the coefficients $c_{QtQb}^1$ and $c_{QtQb}^8$ are set to zero.
In addition, four-heavy operators that involve
right-handed bottom quarks are not free parameters anymore.
The correspondence between these four-heavy degrees of freedom
from~\cite{Hartland:2019bjb} and those of the present work is
\be
c_{Qb}^1=c_{Qd}^1 \,,
\quad c_{Qb}^8=c_{Qd}^8\,,
\quad c_{tb}^1=c_{td}^1 \,,
\quad c_{tb}^8=c_{td}^8 \, .
\ee
Furthermore, we do not have $c_{Qb,tb}^{1,8}$ in the present fit anymore.
These considerations explain why the 11 four-heavy operators of our previous study
are now reduced to the 5 listed in Table~\ref{eq:summaryOperatorsTop}.

All in all, in total we end up with 5 degrees of
freedom involving four heavy quark fields and 14 involving two light and two
heavy quark fields, for a total of 19 independent parameters at the fit level
associated to four-quark operators.
The more stringent flavour assumptions restricting the four-heavy operators
imply that the constraints that we will obtain in the present fit for the four-fermion
operators will be superior, thanks to these new constraints as well as the
addition of the latest top production measurements from Run II of the LHC.

{  One should also mention that the flavour assumptions
  adopted in this work allow in principle for the presence of
  additional four-fermion operators that do not involve
  the top quark.
  One example would be operators containing four bottom quarks.
  These operators are however not directly constrained by any of the measurements
  that we consider in this work, and hence we do not take them into account.
  Future work with an extended dataset e.g. with LHC dijet and multijet measurements~\cite{Domenech:2012ai,Biekotter:2018rhp}, Drell-Yan production~\cite{Dawson:2018dxp},
  and low-energy measurements~\cite{Falkowski:2017pss} will allow directly constraining such
  light four-fermion operators.
  }

\paragraph{Overview of the degrees of freedom.}
We summarise in Table~\ref{tab:operatorbasis}
the degrees of freedom  considered in the present work.
These are associated either to the Wilson coefficients of Warsaw-basis operators
or to linear combinations of those.
We categorize the DoFs into five disjoint classes, from top to bottom: four-quark (two-light-two-heavy), four-quark (four-heavy), four-lepton, two-fermion, and purely bosonic DoFs.
We end up with 50 EFT coefficients that enter the theory predictions associated to the processes
input to the fit, of which 36 are independent.
The 16 DoFs displayed in the last columns are subject to the 14 constraints from the EWPOs
listed in Eq.~(\ref{eq:2independents}),
leaving only 2 independent combinations to be constrained by the fit.
When presenting results for the independent DoFs, for example when evaluating the
Fisher Information matrix, we will select $c_{\varphi W B}$ and $c_{\varphi D}$,
for illustration purposes.
Then in Table~\ref{tab:notation_coeffs} we indicate the
notation that will be used to indicate the EFT coefficients
listed in  Table~\ref{tab:operatorbasis} in the
subsequent sections, as well as in the released output files
with the results of the global analysis,
where again
only two of the 16 EFT coefficients labelled in blue are independent fit parameters.

\input{tables/table-operatorbasis.tex}

\input{tables/notation-coeffs.tex}

\subsection{The top-philic scenario}
\label{sec:topphilic}

The four-fermion operators defined in the previous section and listed in
Table~\ref{eq:summaryOperatorsTop}
correspond to a specific set of assumptions concerning
the flavour structure of the UV-completion of the Standard Model.
However, there exist well-motivated BSM scenarios that suggest further
restrictions in the SMEFT parameter space spanned
by these four-fermion operators.
Therefore, phenomenological explorations of the SMEFT would benefit from comparing
results obtained in different scenarios concerning the possible UV completion,
from more restrictive to more general.

With this motivation, we have implemented a new
feature in the {\tt SMEFiT} analysis framework
which allows one to implement arbitrary restrictions
in the EFT parameter space, for example those motivated by specific BSM scenarios 
or existing constraints such as those from EWPO, as discussed in the previous section.
As a proof of concept, here we will present results
for the top-philic scenario introduced in~\cite{AguilarSaavedra:2018nen}.
This  scenario is not constructed by imposing a specific flavour symmetry, but
rather by assuming that new physics couples predominantly to the third-generation left-handed doublet,
the third-generation right-handed up-type quark singlet, the gauge bosons, and the Higgs boson.
In other words, that new physics interacts mostly with the top and bottom quarks as well as with
the bosonic sector.
The top-philic scenario satisfies the flavour assumptions that we are imposing in this work,
but is based on a more restrictive theoretical assumption.

The restrictions in the EFT parameter space introduced by the top-philic assumption
lead to a number of relations between the DoFs listed in
Table~\ref{tab:operatorbasis}.
These relations are the following:
\bea
c_{QDW} &=& c_{Qq}^{3,1}  \, , \nonumber \\
c_{QDB} &=& 6c_{Qq}^{1,1} = \frac{3}{2}c_{Qu}^1 = -3c_{Qd}^1 \, , \nonumber  \\
c_{tDB} &=& 6c_{tq}^1 = \frac{3}{2}c_{tu}^1 = -3c_{td}^1 \, ,  \label{eq:topphilic}\\
c_{QDG} &=& c_{Qq}^{1,8} = c_{Qu}^{8} = c_{Qd}^8 \, , \nonumber \\
c_{tDG} &=& c_{tq}^8 = c_{tu}^8 = c_{td}^8 \, ,\nonumber \\
c_{Qq}^{3,8} &=& 0 \,,\nonumber
\eea
which can be implemented as an additional restriction at the fitting level.
Therefore, we now have 9 equations that relate a subset of the 14
two-heavy-two-light degrees of freedom listed in Table~\ref{tab:operatorbasis} among them,
which leave 5 independent two-heavy-two-light  degrees of freedom.
The number of operators coupling the top quark with gauge bosons,
as well as that of the four-heavy operators, is not modified.
By comparing with Table~\ref{tab:operatorbasis}, we see that in the top-philic
scenario the EFT fit will constrain 41 DoFs, of which 27 are independent.

In principle, the top-philic assumption also implies non-trivial correlations between 
the light-fermion couplings to the gauge and Higgs bosons. However, following our
strategy to include the EWPO, most of them are already set to zero, while the two remaining
degrees of freedom are not affected.
The same assumptions also imply that the light fermion
Yukawa operator coefficients are proportional to the SM Yukawa couplings. 
As will be shown in Sect.~\ref{sec:results}, imposing the additional relations of the
top-philic scenario leads to more stringent bounds on all the
relevant Wilson coefficients,
due to the fact that the same amount of experimental information is now
used to constrain a significantly more limited parameter space.

\subsection{Cross-section positivity}

The constraint that physical cross-sections are (semi-)positive definite
quantities can also be accounted for in global SMEFT analyses. 
This positivity requirement has different implications depending on whether
the EFT expansion is considered up to either the linear or quadratic level. 

The expansion up to linear terms, $\mathcal{O}(\Lambda^{-2})$, does not automatically
lead to positive-definite cross sections, as in this case the new physics terms are
generated by interference with the SM amplitudes, and their sign and size directly
depend on the Wilson coefficients $c_i$. Imposing the positivity of the cross sections
will therefore set (possibly one-sided) bounds on the Wilson coefficients. 
This can be easily implemented in the fitting procedure if helpful.
In fact, we do not find the need to do so, since the fitted experimental data
already leads to positive-definite cross-sections.

The expansion up to quadratic terms, $\mathcal{O}(\Lambda^{-4})$,
i.e. specifically those coming from squaring the 
linearly expanded amplitudes, obviously automatically leads to positive definite cross sections. No constraints
on the Wilson coefficients can therefore be obtained or need to be imposed.
On the other hand, verifying
the positivity of the cross section at the quadratic level provides a sanity check that the theoretical calculation  of the various contributions is correctly performed,
also taking into account the MC generation uncertainties.
The conditions that have to be  met are simple to obtain. 
Consider the  SMEFT Lagrangian
\begin{equation}
\mathcal{L} =   \mathcal{L}_{\rm SM}  + \sum^{ n_{\rm op}}_{i=1} \frac{c_i}{\Lambda^2}\mathcal{O}_i \,,
\end{equation}
where $\mathcal{O}_i$ stand for dimension-six operators and $c_i$ are the
corresponding Wilson coefficients, which we assume to be real. 
Any observable calculated using this Lagrangian can be written as a quadratic form
\begin{eqnarray}
\Sigma&=&c_0^2 \Sigma_{00} \nonumber \\
 &+&c_0 c_1 \Sigma_{01} + c_1 c_0 \Sigma_{10}  +  c_0 c_2 \Sigma_{02} +\dots \nonumber \\
 &+&c_1^2 \Sigma_{11} + c_1 c_2 \Sigma_{12} +  c_1 c_3 \Sigma_{13} + \ldots \nonumber \\
&=& {\boldsymbol c^T} \cdot  \boldsymbol{\Sigma} \cdot {\boldsymbol c}. 
\end{eqnarray}
The first line corresponds to the SM contribution, where $c_0$ is an auxiliary coefficient that can be set to unity at the end.  The second line corresponds to the linear ${\cal O}(\Lambda^{-2})$ EFT contributions, while the third line to the ${\cal O}(\Lambda^{-4})$ contributions.
$\boldsymbol{\Sigma}$ is by construction a symmetric matrix.\footnote{Note that with respect to
  the convention where $\sigma= \sigma_{\rm SM} + \sum^{n_{\rm op}}_{i=1} c_i \sigma_i  + \sum^{n_{\rm op}}_{i<j} c_i c_j \sigma_{ij}$, one has to account for factors of 2, {\it e.g.}, $\sigma_i=2 \Sigma_{i0}$ and $\sigma_{ij}=2 \Sigma_{ij}$ for $i\neq  j$.}

Given that a physical cross-section must be either positive or null,
the matrix $\boldsymbol{\Sigma}$ must be semi-positive-definite.
We can therefore use the Sylvester criterion that states that a symmetric matrix is semi-positive-definite if and only if all principal minors are greater or equal to zero. As a simple example, the constraints coming from the $2\times 2$ minors are:
\begin{equation}
  \label{eq:sylvester}
  \lp \Sigma_{ii}\Sigma_{jj} - \Sigma_{ij}^2 \rp   \ge 0 \, , \qquad i,j=0,\ldots\, n_{\rm op}\,.
\end{equation}
We have verified that the Sylvester criterion, and Eq.~(\ref{eq:sylvester}) in particular,
are satisfied by the EFT calculations used as input to the present analysis.

%% file: tables/table-bosonicoperators.tex
\begin{table}[t] 
  \begin{center}
    \renewcommand{\arraystretch}{1.90}
        \begin{tabular}{lll}
          \toprule
          Operator $\qquad$ & Coefficient $\qquad\qquad\qquad$ & Definition \\
        \midrule
        $\Op{\varphi G}$ & $c_{\varphi G}$  & $\left(\pdp\right)G^{\mu\nu}_{\sss A}\,
        G_{\mu\nu}^{\sss A}$  \\ \hline
        $\Op{\varphi B}$ & $c_{\varphi B}$ & $\left(\pdp\right)B^{\mu\nu}\,B_{\mu\nu}$\\ \hline
        $\Op{\varphi W}$ &$c_{\varphi W}$ & $\left(\pdp\right)W^{\mu\nu}_{\sss I}\,
        W_{\mu\nu}^{\sss I}$ \\ \hline
        $\Op{\varphi W B}$ &$c_{\varphi W B}$ & $(\varphi^\dagger \tau_{\sss I}\varphi)\,B^{\mu\nu}W_{\mu\nu}^{\sss I}\,$ \\ \hline
        $\Op{\varphi d}$ & $c_{\varphi d}$ & $\partial_\mu(\pdp)\partial^\mu(\pdp)$ \\ \hline
        $\Op{\varphi D}$ & $c_{\varphi D}$ & $(\varphi^\dagger D^\mu\varphi)^\dagger(\varphi^\dagger D_\mu\varphi)$ \\ \hline
         $\mathcal{O}_{W}$&   $c_{WWW}$ & $\epsilon_{IJK}W_{\mu\nu}^I W^{J,\nu\rho} W^{K,\mu}_\rho$ \\
       \bottomrule
        \end{tabular}
        \caption{Purely bosonic dimension-six operators that
          modify the production and decay of Higgs bosons and
          the interactions of the electroweak gauge bosons.
          For each operator, we indicate its definition in terms of the SM
          fields,
          and the notational conventions that will be used
          both for the operator and for the Wilson coefficient.
          The operators $O_{\varphi WB}$ and $O_{\varphi D}$
          are severely
          constrained by the EWPOs together with several of
          the two-fermion operators from Table~\ref{tab:oper_ferm_bos}.
           \label{tab:oper_bos}}
\end{center}
\end{table}

%% file: tables/table-2foperators.tex
\begin{table}[htbp]
  \begin{center}
    \renewcommand{\arraystretch}{1.45}
    \begin{tabular}{lll}
      \toprule
      Operator $\qquad$ & Coefficient & Definition \\
                \midrule \midrule
		&3rd generation quarks&\\
                \midrule \midrule
    $\Op{\varphi Q}^{(1)}$ & $c_{\varphi Q}^{(1)}$~(*) & $i\big(\varphi^\dagger\lra{D}_\mu\,\varphi\big)
 \big(\bar{Q}\,\gamma^\mu\,Q\big)$ \\\hline
    $\Op{\varphi Q}^{(3)}$ & $c_{\varphi Q}^{(3)}$  & $i\big(\varphi^\dagger\lra{D}_\mu\,\tau_{\sss I}\varphi\big)
 \big(\bar{Q}\,\gamma^\mu\,\tau^{\sss I}Q\big)$ \\ \hline
    $\Op{\varphi t}$ & $c_{\varphi t}$& $i\big(\varphi^\dagger\,\lra{D}_\mu\,\,\varphi\big)
 \big(\bar{t}\,\gamma^\mu\,t\big)$ \\ \hline
      $\Op{tW}$ & $c_{tW}$ & $i\big(\bar{Q}\tau^{\mu\nu}\,\tau_{\sss I}\,t\big)\,
 \tilde{\varphi}\,W^I_{\mu\nu}
 + \text{h.c.}$ \\  \hline
 $\Op{tB}$ & $c_{tB}$~(*) &
 $i\big(\bar{Q}\tau^{\mu\nu}\,t\big)
 \,\tilde{\varphi}\,B_{\mu\nu}
 + \text{h.c.}$ \\\hline
    $\Op{t G}$ & $c_{tG}$ & $ig{\sss S}\,\big(\bar{Q}\tau^{\mu\nu}\,T_{\sss A}\,t\big)\,
 \tilde{\varphi}\,G^A_{\mu\nu}
 + \text{h.c.}$ \\  \hline
    $\Op{t \varphi}$ & $c_{t\varphi}$ & $\left(\pdp\right)
 \bar{Q}\,t\,\tilde{\varphi} + \text{h.c.}$  \\\hline
    $\Op{b \varphi}$ & $c_{b\varphi}$ & $\left(\pdp\right)
 \bar{Q}\,b\,\varphi + \text{h.c.}$ \\
                \midrule \midrule
		&1st, 2nd generation quarks&\\
                \midrule \midrule
    $\Op{\varphi q}^{(1)}$ & $c_{\varphi q}^{(1)}$~(*) & $\sum\limits_{\sss i=1,2} i\big(\varphi^\dagger\lra{D}_\mu\,\varphi\big)
 \big(\bar{q}_i\,\gamma^\mu\,q_i\big)$ \\\hline
    $\Op{\varphi q}^{(3)}$ & $c_{\varphi q}^{(3)}$ & $\sum\limits_{\sss i=1,2} i\big(\varphi^\dagger\lra{D}_\mu\,\tau_{\sss I}\varphi\big)
 \big(\bar{q}_i\,\gamma^\mu\,\tau^{\sss I}q_i\big)$ \\  \hline
  ${\Op{\varphi u i}}$ &
      ${{c_{\varphi u i}}}$ & $\sum\limits_{\sss i=1,2,3} i\big(\varphi^\dagger\,\lra{D}_\mu\,\,\varphi\big)
 \big(\bar{u}_i\,\gamma^\mu\,u_i\big)$\\ \hline
 ${\Op{\varphi d i}}$ &
      ${{c_{\varphi d i}}}$ & $\sum\limits_{\sss i=1,2,3} i\big(\varphi^\dagger\,\lra{D}_\mu\,\,\varphi\big)
 \big(\bar{d}_i\,\gamma^\mu\,d_i\big)$\\ \hline
    $\Op{c \varphi}$ & $c_{c \varphi}$ & $\left(\pdp\right)
 \bar{q}_2\,c\,\tilde\varphi + \text{h.c.}$ \\
                \midrule \midrule
		&two-leptons&\\
                \midrule \midrule
    $\Op{\varphi \ell_i}^{(1)}$ & $c_{\varphi \ell_i}^{(1)}$ & $ i\big(\varphi^\dagger\lra{D}_\mu\,\varphi\big)
   \big(\bar{\ell}_i\,\gamma^\mu\,\ell_i\big)$ \\\hline 
    $\Op{\varphi \ell_i}^{(3)}$ & $c_{\varphi \ell_i}^{(3)}$ & $ i\big(\varphi^\dagger\lra{D}_\mu\,\tau_{\sss I}\varphi\big)
 \big(\bar{\ell}_i\,\gamma^\mu\,\tau^{\sss I}\ell_i\big)$ \\  \hline
    $\Op{\varphi e}$ & $c_{\varphi e}$ & $ i\big(\varphi^\dagger\lra{D}_\mu\,\varphi\big)
 \big(\bar{e}\,\gamma^\mu\,e\big)$ \\\hline
    $\Op{\varphi \mu}$ & $c_{\varphi \mu}$ & $ i\big(\varphi^\dagger\lra{D}_\mu\,\varphi\big)
 \big(\bar{\mu}\,\gamma^\mu\,\mu\big)$ \\  \hline
    $\Op{\varphi \tau}$ & $c_{\varphi \tau}$ & $i\big(\varphi^\dagger\lra{D}_\mu\,\varphi\big)
 \big(\bar{\tau}\,\gamma^\mu\,\tau\big)$ \\  \hline
    $\Op{\tau \varphi}$ & $c_{\tau \varphi}$ & $\left(\pdp\right)
 \bar{\ell_3}\,\tau\,{\varphi} + \text{h.c.}$ \\
                \midrule \midrule
		&four-lepton &\\
                \midrule \midrule
 $\Op{\ell\ell}$ & $c_{\ell\ell}$ & $\left(\bar \ell_1\gamma_\mu \ell_2\right) \left(\bar \ell_2\gamma^\mu \ell_1\right)$ \\
 \hline
  \bottomrule
\end{tabular}
\caption{Same as Table~\ref{tab:oper_bos}
  for the operators containing two fermion fields, either
  quarks or leptons, as well as the four-lepton operator $\Op{\ell\ell}$.
  The flavor index $i$ runs from 1 to 3.
  The coefficients indicated with (*) in the second column do not correspond to physical degrees of freedom
  in the fit, but are rather replaced by  $c_{\varphi q_i}^{(-)}$, $c_{\varphi Q_i}^{(-)}$, and
  $c_{tZ}$ defined in Table~\ref{tab:oper_ferm_bos2}.
\label{tab:oper_ferm_bos}}
\end{center}
\end{table}

%% file: tables/table-fourfermion.tex
\begin{table}[htbp] 
  \begin{center}
    \renewcommand{\arraystretch}{1.53}
        \begin{tabular}{ll}
          \toprule
          DoF $\qquad$ &  Definition (in  Warsaw basis notation) \\
          \midrule
      $c_{QQ}^1$    &   $2\ccc{1}{qq}{3333}-\frac{2}{3}\ccc{3}{qq}{3333}$ \\ \hline
    $c_{QQ}^8$       &         $8\ccc{3}{qq}{3333}$\\  \hline
 $c_{Qt}^1$         &         $\ccc{1}{qu}{3333}$\\   \hline
 $c_{Qt}^8$         &         $\ccc{8}{qu}{3333}$\\   \hline
  $c_{tt}^1$         &     $\ccc{}{uu}{3333}$  \\    \hline
            \midrule      
  $c_{Qq}^{1,8}$       &  	 $\ccc{1}{qq}{i33i}+3\ccc{3}{qq}{i33i}$     \\   \hline
  $c_{Qq}^{1,1}$         &   $\ccc{1}{qq}{ii33}+\frac{1}{6}\ccc{1}{qq}{i33i}+\frac{1}{2}\ccc{3}{qq}{i33i} $   \\    \hline
   $c_{Qq}^{3,8}$         &   $\ccc{1}{qq}{i33i}-\ccc{3}{qq}{i33i} $   \\   \hline
  $c_{Qq}^{3,1}$          & 	$\ccc{3}{qq}{ii33}+\frac{1}{6}(\ccc{1}{qq}{i33i}-\ccc{3}{qq}{i33i}) $   \\     \hline
   $c_{tq}^{8}$         &  $ \ccc{8}{qu}{ii33}   $ \\    \hline
   $c_{tq}^{1}$       &   $  \ccc{1}{qu}{ii33} $\\    \hline
   $c_{tu}^{8}$      &   $2\ccc{}{uu}{i33i}$  \\     \hline
    $c_{tu}^{1}$        &   $ \ccc{}{uu}{ii33} +\frac{1}{3} \ccc{}{uu}{i33i} $ \\   \hline
    $c_{Qu}^{8}$         &  $  \ccc{8}{qu}{33ii}$\\     \hline
    $c_{Qu}^{1}$     &  $  \ccc{1}{qu}{33ii}$  \\     \hline
    $c_{td}^{8}$        &   $\ccc{8}{ud}{33jj}$ \\    \hline
    $c_{td}^{1}$          &  $ \ccc{1}{ud}{33jj}$ \\     \hline
    $c_{Qd}^{8}$        &   $ \ccc{8}{qd}{33jj}$ \\     \hline
    $c_{Qd}^{1}$         &   $ \ccc{1}{qd}{33jj}$\\
         \bottomrule
  \end{tabular}
  \caption{\small Definition of the four-fermion degrees of freedom that enter into
    the fit in terms of the coefficients of Warsaw basis operators of Eq.~(\ref{eq:FourQuarkOp}).
    These DoFs are classified into four-heavy (upper) and two-light-two-heavy
    (bottom part) operators. The flavor index $i$ is either 1 or 2, 
    and $j$ is either 1, 2 or 3: with our flavor assumptions,  these coefficients will be the same
    regardless of the specific values that $i$ and $j$ take.
\label{eq:summaryOperatorsTop}}
  \end{center}
\end{table}

%% file: tables/table-operatorbasis.tex
\begin{table}[htbp] 
  \begin{center}
    \renewcommand{\arraystretch}{1.70}
        \begin{tabular}{cccc}
          \toprule
  Class  &  $N_{\rm dof}$ &  Independent DOFs  & DoF in EWPOs\\
          \midrule
          \multirow{4}{*}{four-quark}    &  \multirow{5}{*}{14}
          & $c_{Qq}^{1,8}$, $c_{Qq}^{1,1}$, $c_{Qq}^{3,8}$,   \\
          \multirow{4}{*}{(two-light-two-heavy)}   &
          &  $c_{Qq}^{3,1}$,  $c_{tq}^{8}$,  $c_{tq}^{1}$,  \\
             &    & $c_{tu}^{8}$, $c_{tu}^{1}$, $c_{Qu}^{8}$,\\
            &    & $c_{Qu}^{1}$, $c_{td}^{8}$, $c_{td}^{1}$,   \\
          &    &  $c_{Qd}^{8}$, $c_{Qd}^{1}$   \\
          \midrule
                    \multirow{1}{*}{four-quark}      &  \multirow{2}{*}{5}
                    & $c_{QQ}^1$, $c_{QQ}^8$, $c_{Qt}^1$, &   \\
          \multirow{1}{*}{(four-heavy)}      &   & $c_{Qt}^8$, $c_{tt}^1$ &  \\
\midrule
                    \multirow{1}{*}{four-lepton}      &  \multirow{1}{*}{1}
                    &   &  $c_{\ell\ell}$ \\
	  \midrule
      \multirow{4}{*}{two-fermion}     &  \multirow{5}{*}{23} &  $c_{t\varphi}$, $c_{tG}$,   $c_{b\varphi}$,    &  
                     $c_{\varphi \ell_1}^{(1)}$, $c_{\varphi \ell_1}^{(3)}$, $c_{\varphi \ell_2}^{(1)}$ \\
       \multirow{4}{*}{(+ bosonic fields)}      &    & $c_{c\varphi}$, $c_{\tau\varphi}$,   $c_{tW}$,   &
       $c_{\varphi \ell_2}^{(3)}$, $c_{\varphi \ell_3}^{(1)}$, $c_{\varphi \ell_3}^{(3)}$,\\
            &    &  $c_{tZ}$,  $c_{\varphi Q}^{(3)}$, $c_{\varphi Q}^{(-)}$,     &
            $c_{\varphi e}$, $c_{\varphi \mu}$, $c_{\varphi \tau}$,  \\
       &    &  $c_{\varphi t}$     & $c_{\varphi q}^{(3)}$, $c_{\varphi q}^{(-)}$, \\
       &    &       &   ${{c_{\varphi u i}}}$,
        ${{c_{\varphi d i}}}$ \\
       \midrule
      \multirow{2}{*}{Purely bosonic}     &  \multirow{2}{*}{7} &
      $c_{\varphi G}$, $c_{\varphi B}$, $c_{\varphi W}$,   &  $c_{\varphi W B}$, $c_{\varphi D}$   \\
               &   &  $c_{\varphi d}$,  $c_{WWW}$   &  \\
            \midrule
          Total  & 50 (36 independent)   & 34   & 16 (2 independent)   \\
         \bottomrule
  \end{tabular}
  \caption{\small \label{tab:operatorbasis} Summary of the degrees of freedom
  considered in the present work.
   We categorize these DoFs into five disjoint classes: four-quark (two-light-two-heavy),
   four-quark (four-heavy), four-lepton, two-fermion, and purely bosonic DoFs.
   The 16 DoFs displayed in the last columns are subject to 14 constraints from the EWPOs,
   leaving only 2 independent combinations to be constrained by the fit.
 }
  \end{center}
\end{table}

%% file: tables/notation-coeffs.tex
\begin{table}[htbp] 
  \begin{center}
    \renewcommand{\arraystretch}{1.80}
        \begin{tabular}{ccc}
          \toprule
  Class  &   $\qquad\qquad$ DoF$\qquad\qquad$   & $\qquad \qquad$ Notation $\qquad\qquad$ \\
          \midrule
          \multirow{4}{*}{four-quark}     & $c_{Qq}^{1,8}$, $c_{Qq}^{1,1}$, $c_{Qq}^{3,8}$,
                                            & {\tt c81qq}, {\tt c11qq}, {\tt c83qq},   \\
          \multirow{4}{*}{(two-light-two-heavy)}     & $c_{Qq}^{3,1}$,  $c_{tq}^{8}$,  $c_{tq}^{1}$,
                                                     &  {\tt c13qq}, {\tt c8qt}, {\tt c1qt},  \\
          &  $c_{tu}^{8}$, $c_{tu}^{1}$, $c_{Qu}^{8}$,
          &{\tt c8ut}, {\tt c1ut}, {\tt c8qu},\\
          &  $c_{Qu}^{1}$, $c_{td}^{8}$, $c_{td}^{1}$,
          &{\tt c1qu}, {\tt c8dt}, {\tt c1dt},   \\
              & $c_{Qd}^{8}$, $c_{Qd}^{1}$   & {\tt c8qd}, {\tt c1qd}   \\
          \midrule
          \multirow{1}{*}{four-quark}      & $c_{QQ}^1$, $c_{QQ}^8$, $c_{Qt}^1$,
                                              & {\tt cQQ1}, {\tt  cQQ8}, {\tt cQt1},   \\
          \multirow{1}{*}{(four-heavy)}      &  $c_{Qt}^8$, $c_{tt}^1$
                                            & {\tt cQt8}, {\tt ctt1} \\
          \midrule
	  four-lepton  & $c_{\ell\ell}$ & {\tt \color{blue} cll}\\
	  \midrule
          \multirow{7}{*}{two-fermion}      & $c_{t\varphi}$, $c_{tG}$,   $c_{b\varphi}$,
                                            &  {\tt ctp}, {\tt ctG}, {\tt cbp}, \\
          \multirow{7}{*}{(+ bosonic fields)}&       $c_{c\varphi}$, $c_{\tau\varphi}$,   $c_{tW}$,
                                             &        {\tt ccp}, {\tt ctap}, {\tt  ctW},       \\
          &    $c_{tZ}$,  $c_{\varphi Q}^{(3)}$, $c_{\varphi Q}^{(-)}$,
          &    {\tt ctZ}, {\tt c3pQ3}, {\tt cpQM},            \\
          &     $c_{\varphi t}$, $c_{\varphi \ell_1}^{(1)}$, $c_{\varphi \ell_1}^{(3)}$,
          &     {\tt cpt}, {\tt \color{blue} cpl1},  {\tt \color{blue} c3pl1},          \\
          &   $c_{\varphi \ell_2}^{(1)}$, $c_{\varphi \ell_2}^{(3)}$, $c_{\varphi \ell_3}^{(1)}$,
          &  {\tt \color{blue} cpl2}, {\tt \color{blue} c3pl2},  {\tt \color{blue} cpl3},         \\
          &   $c_{\varphi \ell_3}^{(3)}$, $c_{\varphi e}$, $c_{\varphi \mu}$,
          &  {\tt \color{blue} c3pl3}, {\tt \color{blue} cpe},   {\tt \color{blue} cpmu},    \\
          &  $c_{\varphi \tau}$, $c_{\varphi q}^{(3)}$, $c_{\varphi q}^{(-)}$, 
          &     {\tt \color{blue}cpta},{\tt \color{blue} c3pq}, {\tt \color{blue}cpqMi},          \\
          &     ${ {c_{\varphi u i}}}$,
        ${ {c_{\varphi d i}}}$
          &     {\tt \color{blue}cpui}, {\tt \color{blue} cpdi}           \\
       \midrule
       \multirow{3}{*}{Purely bosonic}      & $c_{\varphi G}$, $c_{\varphi B}$, $c_{\varphi W}$,
                                            &   {\tt cpG}, {\tt cpB}, {\tt cpW}, \\
       &   $c_{\varphi d}$, $c_{\varphi W B}$, $c_{\varphi D}$,
       &   {\tt cpd}, {\tt \color{blue} cpWB}, {\tt \color{blue} cpD}, \\    
       &  $c_{WWW}$
       &    {\tt   cWWW}  \\
         \bottomrule
  \end{tabular}
        \caption{The notation that will be used to indicate the EFT coefficients
          listed in  Table~\ref{tab:operatorbasis} in the
          subsequent sections, as well as in the released output files
          with the results of the global analysis.
Only two of the 16 EFT coefficients labelled in blue are independent fit parameters.
\label{tab:notation_coeffs}}
\end{center}
\end{table}

%% file: sec-dataset.tex
\section{Experimental data and theoretical calculations}
\label{sec:settings_expdata}

In this section we present the experimental measurements and the theoretical
computations used to constrain the SMEFT operators
introduced in
Sect.~\ref{sec:smefttheory}.
We focus in turn on each of the three
groups of LHC processes that we consider in the current analysis:
top quark, Higgs boson,
and gauge boson pair production.

\subsection{Top-quark production data}
\label{sec:exptop}

The top-quark production measurements included in this analysis belong to four
different categories: inclusive top-quark pair production, top-quark pair
production in association with vector bosons or heavy quarks, inclusive single
top-quark production, and single top-quark production in association with
vector bosons.
In the following we present the datasets that belong to each of
these categories. Top-quark pair production in association
with a Higgs boson is discussed in Sect.~\ref{sec:HiggsProductionDecay}.

\paragraph{Inclusive top-quark pair production.}
The experimental measurements of inclusive top-quark pair production
included in this analysis are summarised in Table~\ref{eq:input_datasets}. 
For each of them, we indicate the dataset label, the center of mass energy
$\sqrt{s}$, the integrated luminosity $\mathcal{L}$, the final state or the
specific production mechanism, the physical observable, the number of data 
points $n_{\rm dat}$, and the publication reference. Measurements indicated with
a {\bf (*)} were not included in our earlier analysis~\cite{Hartland:2019bjb}.

The bulk of the measurements correspond to datasets already included
in~\cite{Hartland:2019bjb}: at 8~TeV,
the ATLAS top-quark pair invariant mass distribution~\cite{Aad:2015mbv} and
the CMS top-quark pair normalized invariant rapidity
distribution~\cite{Khachatryan:2015oqa}, both in the lepton+jets final state,
the CMS top-quark pair normalized invariant mass and rapidity two-dimensional
distribution in the dilepton final state~\cite{Sirunyan:2017azo},
and the ATLAS and CMS $W$ helicity fractions~\cite{Aaboud:2016hsq,
  Khachatryan:2016fky}; at 13~TeV, the CMS top-quark pair invariant mass
distributions in the lepton+jets and dilepton final states based on integrated
luminosities of up to $\mathcal{L}=35.8$~fb$^{-1}$~\cite{Khachatryan:2016mnb,
  Sirunyan:2018wem,Sirunyan:2017mzl}.
In addition to these, we now  consider further top-quark pair
invariant mass distributions: at 8~TeV, the ATLAS measurement in the dilepton
final state~\cite{Aaboud:2016iot}; at 13~TeV, and the ATLAS and CMS
measurements, respectively in the lepton+jets and dilepton final states, 
corresponding to an integrated luminosity of
$\mathcal{L}=35.8$~fb$^{-1}$~\cite{Aad:2019ntk,Sirunyan:2018ucr}.
We also include top-quark pair charge asymmetry measurements:
the ATLAS and CMS combined dataset at 8~TeV~\cite{Sirunyan:2017lvd},
and the ATLAS dataset at 13~TeV~\cite{ATLAS:2019czt}.

\input{tables/table-input-dataset-top1.tex}


Although several distributions differential in various kinematic variables are
available for the measurements presented in~\cite{Khachatryan:2015oqa,
  Sirunyan:2017azo,Khachatryan:2016mnb,Sirunyan:2018wem,
  Sirunyan:2017mzl,Aaboud:2016iot,Aad:2019ntk,Sirunyan:2018ucr}, only one of
them can typically be included in the fit at a time.
The reason is that
experimental correlations between pairs of distributions are unknown:
including more than one distribution at a time will therefore result in
a double counting.
An exception to this state of affairs is represented
by the ATLAS measurement of~\cite{Aad:2015mbv}, which is provided with the
correlations among differential distributions.
Unfortunately, they significantly deteriorate the fit quality when
an analysis of all the available distributions is attempted, a fact that
questions their reliability (see also~\cite{Amoroso:2020lgh,Bailey:2019yze}).
We therefore include only one distribution also in this case.
In general, we
include the invariant mass distribution $m_{t\bar{t}}$, whose high-energy
tail 
is known to be particularly
sensitive to deviations from the SM expectations.
For~\cite{Khachatryan:2015oqa}
we include instead the invariant rapidity distribution as in our earlier
analysis~\cite{Hartland:2019bjb}, due to difficulties in achieving an
acceptable fit quality to $m_{t\bar{t}}$.

The additional top-quark pair measurements considered
in this work do not expand the kinematic coverage
in the EFT parameter space in comparison to those already included
in~\cite{Hartland:2019bjb}.
Nevertheless, they provide additional weight for
the inclusive top-quark pair differential distributions in the global fit,
which are known to provide the dominant constraints on several of the EFT
coefficients.
All in all, we end up with $n_{\rm dat}=94$ data points in this category.

Additional sensitivity to EFT effects could be achieved
by means of LHC Run-II measurements with an extended coverage in the
invariant mass or transverse momentum.
However, differential distributions based on luminosities larger than
$\mathcal{L}\simeq 36$~fb$^{-1}$ are not available yet: the statistical
precision of the data, and consequently their constraining power, remain
therefore limited.
For instance, the ATLAS fully hadronic final state
measurement~\cite{Aad:2020nsf} is available, but it exhibits
larger uncertainties than in the cleaner lepton+jets and dilepton final states.
Furthermore, some measurements are not reconstructed at the parton level,
as required in our analysis.
This is the case of the ATLAS and CMS measurements
at high top-quark
transverse momentum~\cite{Aad:2020nsf,Sirunyan:2020vwa}, that are
based on reconstructing boosted topologies, and of the dilepton distributions
from ATLAS~\cite{Aad:2019hzw}, that are restricted to the particle level.

Concerning theoretical calculations, the SM cross-sections are evaluated at NLO
using {\tt MadGraph5\_aMC@NLO}~\cite{Alwall:2014hca} and supplemented with
NNLO $K$-factors~\cite{Czakon:2016dgf,Czakon:2016olj}.
The input PDF set is NNPDF3.1NNLO no-top~\cite{Ball:2017nwa},
to avoid possible contamination between PDF and EFT
effects.\footnote{See~\cite{Greljo:2021kvv,Carrazza:2019sec} for a detailed
  discussion of the interplay between PDF and EFT fits.}
The EFT cross-sections are evaluated with
{\tt MadGraph5\_aMC@NLO}~\cite{Alwall:2014hca}
combined with the {\tt SMEFT@NLO} model~\cite{Degrande:2020evl}.
Unless otherwise specified, the same EFT settings will be used also for the
other processes considered in this analysis.
Specifically, NLO QCD effects to
the EFT corrections are accounted systematically whenever available.

\paragraph{Associated top-quark pair production.}
Table~\ref{eq:input_datasets2} lists, in the same format as
Table~\ref{eq:input_datasets}, the experimental measurements for top
quark pair production in association with heavy quarks or weak vector bosons.
The dataset considered in~\cite{Hartland:2019bjb} consisted of
the CMS measurements of total cross-sections for $t\bar{t}t\bar{t}$
and $b\bar{b}b\bar{b}$ at 13~TeV~\cite{Sirunyan:2017snr,Sirunyan:2017roi},
and in the ATLAS and CMS measurements of inclusive $tW$ and $tZ$ production at
8~TeV and 13~TeV~\cite{Khachatryan:2015sha,Sirunyan:2017uzs,Aad:2015eua,
  Aaboud:2016xve}.
In the present analysis, we augment this dataset with the most updated
measurements of total cross-sections for $t\bar{t}t\bar{t}$ and
$t\bar{t}b\bar{b}$ production at 13~TeV: for $t\bar{t}b\bar{b}$, with the ATLAS
and CMS measurements based on
$\mathcal{L}=137$~fb$^{-1}$~\cite{Sirunyan:2019wxt,Aad:2020klt};
for $\sigma_{\rm tot}(t\bar{t}b\bar{b})$, with the ATLAS and CMS measurements
based on $\mathcal{L}=36$~fb$^{-1}$~\cite{Aaboud:2018eki,Sirunyan:2019jud}.
These measurements are comparatively more precise than the measurements
already included in~\cite{Hartland:2019bjb} thanks to the increased luminosity.

Concerning top-quark pair production in association with an electroweak gauge
boson, we include here the ATLAS total cross-section measurements of
$t\bar{t}W$ and $t\bar{t}Z$ based on
$\mathcal{L}=36$~fb$^{-1}$~\cite{Aaboud:2019njj},
as well as the CMS differential
measurements of $d\sigma(t\bar{t}Z)/dp_T^Z$ based on
$\mathcal{L}=78$~fb$^{-1}$~\cite{CMS:2019too}, which
is the first differential measurement of $t\bar{t}V$ associated production
presented at the LHC. We do not include the still preliminary
ATLAS measurement of $\sigma_{\rm tot}(t\bar{t}Z)$ based on
$\mathcal{L}=139$~fb$^{-1}$~\cite{ATLAS-CONF-2020-028}.
The $t\bar{t}V$ measurements are
especially useful to constrain EFT effects that modify the electroweak
couplings of the top-quark.
In total, we include $n_{\rm dat}=20$
data points in the category of $t\bar{t}$ associated production with heavy quark pairs
or weak vector bosons.

\input{tables/table-input-dataset-top2.tex}


Theoretical predictions are computed at NLO both in the SM and in the EFT.
We use {\tt MCFM} for the SM cross-sections and {\tt SMEFT@NLO} for the EFT corrections,
with NLO QCD effects accounted for exactly for the 2-fermion operators. 
The exception is the $p_T^Z$ distribution in $t\bar{t}Z$ events, for which
{\tt MadGraph5\_aMC@NLO} is used instead to evaluate the SM cross-section
at NLO.

\paragraph{Inclusive single top-quark production.}
We now move to consider the inclusive production of single top-quarks,
both in the $t$-channel and in the $s$-channel ($tW$ associated production is
discussed separately below).
Table~\ref{eq:input_datasets3} displays the information on
the experimental data for these processes that is being considered in the
present analysis.
The dataset in this category that was already included in our
previous analysis~\cite{Hartland:2019bjb} consisted, at 8 TeV, of the
$t$-channel total cross-sections and in the top-quark rapidity differential
distributions from CMS~\cite{Khachatryan:2014iya,CMS-PAS-TOP-14-004}
and from ATLAS~\cite{Aaboud:2017pdi}, and in the $s$-channel total
cross-sections from ATLAS~\cite{Aad:2015upn} and CMS~\cite{Khachatryan:2016ewo};
at 13 TeV, in  the $t$-channel total cross-sections and top-quark rapidity
differential distributions from ATLAS~\cite{Aaboud:2016ymp} and
CMS~\cite{CMS:2016xnv,Sirunyan:2016cdg}.

Here we augment this dataset with one additional measurement, namely the CMS
top-quark rapidity differential cross-section for $t$-channel single top-quark
production at 13 TeV based on
$\mathcal{L}=35.9$~fb$^{-1}$~\cite{Sirunyan:2019hqb}.
As customary, we consider the distribution reconstructed at parton level for
consistency with the theoretical predictions.
No differential measurements of single top-quark production
based on the Run II dataset have been presented by ATLAS so far.
Furthermore, while the ATLAS and CMS combination of total cross-sections for
single top-quark production at 7 TeV and 8 TeV has been presented
in~\cite{Aaboud:2019pkc}, here we include instead the original individual
measurements. We end up with $n_{\rm dat}=27$ data points in this category.

\input{tables/table-input-dataset-top3.tex}

The calculation of the SM and EFT cross-sections has been carried out with the
same settings as for inclusive $t\bar{t}$ production. Note that for single top
we work with a 5-flavour number scheme (5FNS) where the bottom quark is
considered as massless, and thus enters the initial state of the reaction,
see~\cite{Nocera:2019wyk} for details.
The NNLO QCD $K$-factors in the 5FNS are
obtained from the calculation of~\cite{Berger:2016oht}.

\paragraph{Associated single top-quark production with weak bosons.}
Finally, in Table~\ref{eq:input_datasets4} we consider the experimental
measurements on the associated production of single top-quarks together with
a weak gauge boson $V$.
The dataset in this category that was already part of our
original analysis~\cite{Hartland:2019bjb} consisted of the total inclusive
cross-sections for $tW$ production by ATLAS and CMS at
8~TeV~\cite{Aad:2015eto,Chatrchyan:2014tua}
and at 13~TeV~\cite{Aaboud:2016lpj,Sirunyan:2018lcp},
as well as in the ATLAS and CMS measurements of the $tZ$ total
cross-sections at 13~TeV~\cite{Sirunyan:2017nbr,Aaboud:2017ylb},
in the latter case restricted to the fiducial region
in the $Wb\ell^+\ell^-q$ final state.

\input{tables/table-input-dataset-top4.tex}

In addition to these datasets, we include here several new measurements
of $tW$ and $tZ$ production.
First of all, we include a new total cross-section measurement of
$tW$ production by ATLAS at 8 TeV~\cite{Aad:2020zhd}.
This measurement is carried out in the single lepton channel, and thus does
not overlap with~\cite{Aad:2015eto}, which instead was obtained
in the two leptons with one central $b$-jet channel.
Then we include the ATLAS measurement of the fiducial cross-section
for $tZ$ production~\cite{Aad:2020wog} using the $t \ell^+\ell^- q$ final state
(in the tri-lepton channel) based on the full Run II luminosity of
$\mathcal{L}=139$ fb$^{-1}$. In this analysis, the cross-section measurement
differs from the background-only hypothesis
(dominated by $t\bar{t}Z$ and dibosons) by more than five sigma and thus
corresponds to an observation of this process.
We also consider the corresponding measurement from CMS,
where the observation of $tZ$ associated production
is reported by reconstructing the $t\ell^+\ell^-q$
final state~\cite{Sirunyan:2018zgs}
based on a luminosity of $\mathcal{L}=77.4$ fb$^{-1}$.
No differential distributions for $tZ$ have been reported so far.
The settings of the theoretical calculations for these
$n_{\rm dat}=9$ data points are the same as of~\cite{Hartland:2019bjb}.

In addition to these measurements, both ATLAS and CMS have
measured differential distributions in $tW$ production
at 13~TeV based on a luminosity of
$\mathcal{L}=35.9$~fb$^{-1}$~\cite{Aaboud:2017qyi,CMS-PAS-TOP-19-003}.
However, these measurements are reported
at the particle rather than
at the parton level, and therefore they are not suitable for inclusion in the
present analysis, which is restricted to top-quark level observables.
We also note that CMS has reported on the EFT interpretation of the associated
production of top-quarks, including with vector bosons, in an analysis
based on a luminosity of $\mathcal{L}=41.5$ fb$^{-1}$~\cite{CMS-PAS-TOP-19-001}.
\\
[-0.3cm]

Combining the four categories discussed above, the present analysis contains
$n_{\rm dat}=150$ top-quark cross-sections, to be compared with
$n_{\rm dat}=103$ in~\cite{Hartland:2019bjb}.
In Sect.~\ref{sec:dataset_dependence} we will quantify the
impact of the new top-quark
measurements by comparing two fits, one based on the dataset
of~\cite{Hartland:2019bjb} and one based on the extended top-quark dataset
included here.

\subsection{Higgs production and decay}
\label{sec:HiggsProductionDecay}

We now turn to the Higgs boson production and decay measurements.
We consider first inclusive cross-section measurements, presented as signal
strengths normalised to the SM predictions, and then differential
distributions and STXS measurements.

\paragraph{Signal strengths.} 
First of all, we consider the inclusive Higgs boson production signal strengths
$\mu_i^f$ measured by ATLAS and CMS from LHC Run I and Run II.
These  signal strengths are defined for each combination
of production and decay channels in terms of cross-section $\sigma_i$
and the branching fraction $B_f$ as
\be
\mu_i^f \equiv \frac{\sigma_i \times B_f}{\lp \sigma_i\rp_{\rm SM} \times
  \lp B_f \rp_{\rm SM}} = \mu_i \cdot \mu^f = \lp \frac{\sigma_i }{\lp \sigma_i\rp_{\rm SM} } \rp
\lp  \frac{  B_f}{ \lp B_f \rp_{\rm SM}}\rp \, ,
\ee
that is, as the ratio of the experimentally measured production cross-sections
in specific decay channels to the corresponding (state-of-the-art)
SM predictions.
These inclusive signal strengths can also be expressed as
\be
\label{eq:signal_strength_def}
\mu_i^f = \lp \frac{\sigma_i }{\lp \sigma_i\rp_{\rm SM} } \rp
\lp  \frac{  \Gamma(h \to f) }{  \Gamma(h \to f)\big|_{\rm SM} }\rp
\lp  \frac{  \Gamma(h \to {\rm all}) }{  \Gamma(h \to {\rm all})\big|_{\rm SM} }\rp^{-1} \, ,
\ee
in terms of the partial and total decay widths. The measurements of
signal strengths that we consider in the present analysis
are collected in Table~\ref{eq:input_datasets_higgsSS}.
In contrast to the differential distributions and STXS discussed below,
these signal strengths are typically extrapolated to the full phase space and
do not include selection or acceptance cuts.

\input{tables/table-input-dataset-HiggsSS.tex}


For the LHC Run I, we take into account the inclusive
measurements of  Higgs boson production and decay rates from the ATLAS and CMS
combination based on the full 7 and 8~TeV datasets~\cite{Khachatryan:2016vau}.
Specifically, we include the 20 measurements presented in Table~8
of~\cite{Khachatryan:2016vau}. These measurements correspond to five different
production channels ($gg{\rm F}$, VBF, $Wh$, $Zh$, $tth$) for five final
states ($\gamma\gamma$, $ZZ$, $WW$, $\tau\tau$, $b\bar{b}$), 
excluding those combinations that are either not measured with
a meaningful precision or not measured at all.
We account for the experimental correlations between the measured
signal strengths using the information provided in~\cite{Khachatryan:2016vau}.
In addition to these ATLAS+CMS combination results from Run I, we also include
two more signal strengths measurements from Run I, namely the ATLAS
constraints on the $Z\gamma$ and $\mu\mu$ decays from~\cite{Aad:2015gba}.

For the LHC Run II, we consider the ATLAS measurement of signal strengths
corresponding to an integrated luminosity of
$\mathcal{L}=80$~fb$^{-1}$~\cite{Aad:2019mbh}, and the CMS measurement
corresponding to an integrated luminosity of
$\mathcal{L}=35.9$~fb$^{-1}$~\cite{Sirunyan:2018koj}.
As in the case of the Run I signal strengths, we keep into account
correlations between the various production and final state combinations.
The ATLAS combination contains 16 signal strengths
for the $gg$F, VBF, $Vh$ and $t\bar{t}h$ production channels
and the $\gamma\gamma$, $ZZ$, $WW$, $\tau\tau$ and $b\bar{b}$
final states. As in the case of Run I, measurements are sometimes not available
for all final states for a given production channel, for example the
$h\to b\bar{b}$ decay is not available for $gg$F while $\tau\tau$ is not
provided in the case of $Vh$ associate production.
The CMS analysis contains 24 signal strengths measurements
in the $gg$F, VBF, $Wh$, $Zh$, and $t\bar{t}h$ production channels
for the same final states as in the ATLAS case. Results for the
$WW$, $ZZ$ and$\gamma\gamma$ final states are available for all production
channels, while for the other final states, $\mu\mu$, $\tau\tau$,
and $b\bar{b}$, signal strength measurements
are only available for specific production channels.
In total, we have $n_{\rm dat}=62$ measurements
of Higgs inclusive signal strengths from Runs I and II.

Concerning the theoretical calculations corresponding to these
measurements, the SM production cross-sections and decay branching fractions
are obtained from the associated experimental publications.
In turn,
these represent the
most updated available predictions, and are provided in the LHC Higgs
Cross-Section Working Group (HXSWG)
reports~\cite{Heinemeyer:2013tqa,deFlorian:2016spz,Dittmaier:2012vm}.
As in the case of top-quark production processes,
EFT calculations are obtained at NLO QCD
using {\tt MadGraph5\_aMC@NLO}~\cite{Alwall:2014hca} with the
{\tt SMEFT@NLO} model.
Additional details about the implementation of EFT
corrections to the Higgs signal strengths are provided
in App.~\ref{sec:signalstrenghts}.

\paragraph{Differential distributions and STXS.}
Table~\ref{eq:input_datasets_higgs} summarizes the experimental measurements of
differential distributions and STXS for Higgs boson production and decay at the
LHC considered in the present analysis.
Whenever one has a potential double counting between a
signal strength measurement and the corresponding differential distribution or
STXS measurement, we always select the latter, which provides more information
on the EFT parameter space due to its enhanced
kinematical sensitivity.

\input{tables/table-input-dataset-Higgs.tex}


To being with, we consider the ATLAS and CMS differential distributions
in the Higgs boson kinematic variables obtained from the combination of the
$h\to \gamma\gamma$, $h\to ZZ$, and (in the CMS case)
$h \to b\bar{b}$ final states at 13~TeV based
on $\mathcal{L}=36$~fb$^{-1}$~\cite{Aaboud:2018ezd,Sirunyan:2018sgc}.
Specifically, we consider the differential distributions in the Higgs boson
transverse momentum $p_T^h$.
These distributions are particularly
sensitive probes of new physics, for instance through new particles circulating
in the gluon-fusion loop.

We also include the ATLAS measurement of the associated production of Higgs
bosons, $Vh$, in the $h\to b\bar{b}$ final state at
13~TeV~\cite{Aaboud:2019nan}.
These measurements, performed in kinematic
fiducial volumes defined in the simplified template cross-section framework,
correspond to an integrated luminosity of $\mathcal{L}=79.8$ fb$^{-1}$.
Specifically, here we include the data corresponding to the 5-POI
(parameters of interest) category, which refers to three cross-sections
for $Zh$ production in the bins $75 < p_T^Z < 150$ GeV,
$150 < p_T^Z < 250$ GeV, and $p_T^Z > 250$ GeV, and two cross-sections
for $Wh$ production, one for $150 < p_T^W < 250$ GeV and the other for
$p_T^W > 250$ GeV. Gauge bosons are reconstructed by means of their leptonic
decays.

Then we also include selected differential measurements presented in the ATLAS
Run II Higgs combination paper~\cite{Aad:2019mbh}.
Specifically, we include the measurements of Higgs production
in gluon fusion, $gg \to h$, in different bins of
$p_T^h$ and in the number of jets in the event.
These measurements are presented as $\sigma_i \times B_{ZZ}/B_{ZZ}^{\rm (SM)}$,
since the $ZZ$ branching fraction is used to normalise the data.
We include the 0-jet cross-section, the 1-jet cross-section
for $p_T^h < 60$ GeV, $60 \le p_T^h \le 120$ GeV, and
$120 \le p_T^h \le 200$ GeV, and the $\le 1$ jet and $\le 2$ jet cross-sections
for $p_T^h \ge 200$ GeV and $p_T^h<200$ GeV respectively.

Furthermore, we consider the differential Higgs boson production measurements
presented by CMS at 13~TeV based on an integrated luminosity of
$\mathcal{L}=77.4$~fb$^{-1}$ and corresponding to the final state
$\gamma\gamma$~\cite{CMS:1900lgv}.
The STXS measurements associated to different final-state topologies
and kinematic values such as $p_T^h$ are presented.
These inclusive measurements are dominated by the gluon-fusion production
channel. Note that the CMS diphoton measurement of~\cite{CMS:1900lgv}
supersedes~\cite{Aaboud:2018xdt}, which was based on the 2016 dataset only.

Whenever available, the
information on the experimental correlated systematic uncertainties is included.
As mentioned above, the SM theoretical predictions are taken
from the HXSWG reports~\cite{deFlorian:2016spz,
  Heinemeyer:2013tqa,Dittmaier:2012vm}.
In total, we include
$n_{\rm dat}=35$ measurements of differential cross-sections and STXS
on Higgs production and decay from the LHC Run II.

We note that additional Higgs production
and decay measurements have been recently presented by ATLAS and CMS based
on the full Run II luminosity of $\mathcal{L}=139$ fb$^{-1}$.
Two examples of these are the CMS measurement of the $p_T^h$ distribution in the
$h\to WW$ fully leptonic final state~\cite{Sirunyan:2020tzo} and the updated
ATLAS measurement of $Vh$ associated production in the $b\bar{b}$ final
state~\cite{Aad:2020jym}.
These measurements are however not expected to modify
significantly the results of the present analysis,
since the constraints they provide on the EFT parameter space are already
covered by other measurements, and  their inclusion is left for future work.

\subsection{Diboson production from LEP and the LHC}

We complement the constraints provided by the Higgs data with those
provided by diboson production cross-sections measured by LEP and the LHC.
The dataset is summarised in Table~\ref{eq:input_datasets_diboson}.
To begin with, we consider the LEP-2
legacy measurements of $WW$ production~\cite{Schael:2013ita}.
Specifically, we include the cross-sections differential
in $\cos\theta_W$ in four different bins in the center of
mass energy, from $\sqrt{s}=182$ GeV up to $\sqrt{s}=206$ GeV.
Here $\theta_{W}$ is defined as the polar angle of the produced
$W^-$ boson with respect to the incoming electron beam direction.
Each set of bins with a different center-of-mass energy
correspond to a different integrated luminosity, ranging
between $\mathcal{L}=163.9$ pb$^{-1}$ and 630.5 pb$^{-1}$.
For each value of $\sqrt{s}$, there are 10 bins in $\cos\theta_W$,
adding up to a total of $n_{\rm dat}=40$ data points.
The theoretical calculations of the SM predictions,
which include higher-order electroweak but not NLO QCD
corrections, are also taken from~\cite{Schael:2013ita}.
For this process, the squared terms in the EFT
proportional to
$c_ic_j/\Lambda^{-4}$ are small and will be neglected.

\input{tables/table-input-dataset-diboson.tex}


Concerning the LHC datasets, we include measurements of the differential
distributions for $W^{\pm}Z$ production at 13~TeV from
ATLAS~\cite{ATLAS-CONF-2018-034} and CMS~\cite{Sirunyan:2019bez} based on a
luminosity of $\mathcal{L}=36.1$ fb$^{-1}$. In both cases, the two gauge bosons
are reconstructed by means of the fully leptonic final state, whereby events
of the type  $WZ \to \ell^+ \ell^- \ell^{(')\pm}$ are selected.
The different leptonic final states are then combined into an inclusive
measurement. For the ATLAS measurement three fiducial distributions are
presented, respectively differential in $p_T^W$, $p_T^Z$ and $m_T^{WZ}$.
As indicated in Table~\ref{eq:input_datasets_diboson},
in this analysis, our baseline choice will be to include the $m_{T}^{WZ}$
distribution, which extends up to transverse masses of $m_{T}^{WZ}=600$ GeV.
In the case of the corresponding CMS measurement, normalised
differential distributions in $p_T^Z$, $m_{WZ}$,
$p_T^{W}$, and $p_T^{\rm jet,lead}$ are available.
Here the baseline will be the $p^Z_{T}$ distribution.

In addition to these measurements, we also consider
the differential distributions for $WW $production from ATLAS at 13~TeV
based on a luminosity of $\mathcal{L}=36.1$ fb$^{-1}$~\cite{Aaboud:2019nkz}.
Events are selected by requiring one electron
and one muon in the final state, corresponding to the decay
mode $WW \to e^\pm \nu \mu^\pm \nu$.
Several differential distributions in the fiducial
region are provided, including  $m_{e\mu}$,
$p_T^{e\mu}$ and $|y_{e\mu}|$.
Here our baseline choice will be the $m_{e\mu}$ distribution,
the invariant mass of the dilepton system, which reaches values
of up to $m_{e\mu}\simeq 1$~TeV.
The total number of  data points in the
LHC diboson category is $n_{\rm dat}=30$.

Other diboson measurements from the LHC have been presented but their EFT
interpretation is left for future work.
For instance, the data for the CMS differential distributions of $WW$ production
at 13 TeV based on $\mathcal{L}=36.1$ fb$^{-1}$~\cite{CMS-PAS-SMP-18-004} is
still preliminary.
ATLAS has presented recent measurements
of the differential cross-sections in four-lepton events in 13~TeV
based on $\mathcal{L}=139$~fb$^{-1}$~\cite{ATLAS:2020xtq}, though here the
measured
distributions receive contributions from single $Z$ and Higgs boson production,
in addition to those from $ZZ$ production.

The theoretical predictions for the SM cross-sections of these LHC diboson
processes are accurate to NNLO QCD and were computed with
{\tt MATRIX}~\cite{Grazzini:2017mhc}.
The EFT contributions for this process include NLO QCD corrections
and take into account the constraints from Eq.~(\ref{LEPconstraints})
to express the calculation in terms of only three Wilson
coefficients, one being the triple-gauge operator $c_{WWW}$
and the other two the purely bosonic coefficients
$c_{\varphi D}$ and $c_{\varphi W B}$.
This choice is ultimately arbitrary and has no physical implications;
any other two coefficients
out of Eq.~(\ref{eq:LEPconstrainedDoFs}) would lead to
the same results. Its only motivation is to facilitate the event generation of
the diboson processes.

\subsection{Dataset and theory overview and EFT sensitivity}

We conclude this section by presenting an overview of the datasets
considered (and of the corresponding theoretical
calculations), summarizing their dependence
on the EFT coefficients defined in Sect.~\ref{sec:smefttheory},
and
quantifying the sensitivity that each process
has on these coefficients by means of information geometry.

\paragraph{Dataset overview.}
In Table~\ref{eq:table_dataset_overview} we summarise the number of data points in our baseline dataset
for each of the data categories and processes considered in this analysis, as well
as the total per category and the overall total.
We include 150, 97, and 70 cross-sections from top-quark production, Higgs boson production
and decay, and diboson production
cross-sections from LEP and the LHC respectively in the baseline dataset,
for a total of 317 cross-section measurements.

\input{tables/table-dataset-overview.tex}


\paragraph{Overview of theoretical calculations.}
Table~\ref{eq:table-processes-theory} displays a
summary of the theoretical calculations used for the 
description various datasets included in the 
present analysis. We indicate, for both the SM and the SMEFT contributions 
to the cross-sections, the perturbative accuracy and the codes used to 
produce the corresponding theoretical predictions.

\input{tables/table-processes-theory.tex}

{ 
\paragraph{Electroweak scheme.}
The theoretical calculations presented
in this work are based on
the $(G_F, m_Z, m_W)$ electroweak scheme,
which is the default in the
{\tt SMEFTatNLO} model.
The corresponding values of the SM parameters are
set to be the following:
  \begin{equation}
\begin{aligned}
     m_\mathrm{W}  &= 80.352~{\rm GeV} \text{,} &
\Gamma_\mathrm{W}  &= 2.084~{\rm GeV} \text{,} &
     m_t           &= 172.5~{\rm GeV} \text{,} \\
m_\mathrm{Z}       &= 91.1535~{\rm GeV} \text{,} &
\Gamma_\mathrm{Z}  &= 2.4943~{\rm GeV} \text{,} &
\Gamma_t           &= 1.37758~{\rm GeV} \text{,} \\
m_\mathrm{H}       &= 125.0~{\rm GeV} \text{,} &
\Gamma_\mathrm{H}  &= 4.07468 \times 10^{-3}~{\rm GeV} \text{,} &
G_\mu              &= 1.166378 \times 10^{-5}~{\rm GeV}^{-2} \text{.}
\end{aligned}
\end{equation}
}

\paragraph{Dependence on the EFT coefficients.}
In order to interpret the results of the global EFT analyses
which will be presented in Sect.~\ref{sec:results},
it is useful to collect the dependence
of the various datasets described
in this section with respect to the degrees
of freedom defined in Sect.~\ref{sec:smefttheory}.
Table~\ref{table:operatorprocess} indicates
which EFT coefficients contribute to the theoretical description of each of the
processes considered in this analysis.
Recall that the 16 coefficients listed in Eq.~(\ref{eq:LEPconstrainedDoFs}) are related
among them by the EWPO relations, and that only two of them are independent.

In Table~\ref{table:operatorprocess} we display
from top to bottom the coefficients associated to
the  two-light-two-heavy, four-heavy, four-lepton, two-fermion plus bosonic,
and purely bosonic dimension-six operators.
The Higgs measurements are separated between the Run I and Run II datasets,
and in the latter case also between signal strengths and differential
distributions and STXS.
A check mark outside (inside) brackets indicates that a given
process constrains the corresponding coefficients
starting at $\mathcal{O}( \Lambda^{-2})$ 
($\mathcal{O}( \Lambda^{-4})$) at LO.
Entries labelled with (b) indicate that the sensitivity
to the associated coefficients enters via bottom-initiated
processes, which arise due to contributions from the $b$-PDF
in the 5FNS adopted here.

\input{tables/table-operatorprocess.tex}

Several observations can be drawn from this table.
First of all, we observe that the four-heavy coefficients are constrained only by the $t\bar{t}Q\bar{Q}$ production
data, either $t\bar{t}t\bar{t}$ or $t\bar{t}b\bar{b}$.
Such measurements also depend on the 2-light-2-heavy operators, as well as on $c_{tG}$, although in practice
this correlation is  small.
Furthermore, the four-heavy coefficients are essentially left undetermined at
$\mathcal{O}\lp \Lambda^{-2}\rp$, and can only be meaningfully constrained only
the quadratic corrections are accounted for.
One can also note how the two-light-two-heavy operators are constrained by top-quark pair production
(inclusive and in association with vector bosons) as well as by the $t\bar{t}h$ production measurements.
As will be shown below, by far the dominant constraints on these coefficients
arise from the differential
distributions in inclusive top
quark pair production.

Concerning the two-fermion operators, most of them are constrained both by top and by Higgs production
process.
Recall that the top and Higgs sectors are connected, among others,
by means of the gluon-fusion production process (with its virtual
top-quark loop) as well as by $t\bar{t}h$ associated production.
In particular, we note that $c_{t\varphi}$, which modifies the top
Yukawa coupling,
is constrained by these Higgs production measurements.
The purely bosonic operators exhibit sensitivity only to Higgs and diboson processes, since
these do not affect the properties of top quarks.
The diboson data is uniquely sensitive to the triple-gauge coefficient
$c_{WWW}$, which modifies the triple (and quartic) electroweak gauge couplings,
as well as to $c_{\varphi D}$ and $c_{\varphi WB}$, which  are also constrained by Higgs data.

\paragraph{The Fisher matrix and information geometry.}
\label{sec:information}
The information presented in Table~\ref{table:operatorprocess} does not allow one to compare
the sensitivity brought in by different datasets on a given EFT coefficient.
To achieve this, here we adopt the ideas underlying information geometry~\cite{Brehmer:2017lrt}
and define the Fisher information  matrix $I_{ij}$ as
\be
\label{eq:FisherDef}
I_{ij}\lp {\boldsymbol c} \rp = -{\rm E}\lc \frac{\partial^2 \ln f \lp {\boldsymbol \sigma}_{\rm exp}|
{\boldsymbol c} \rp}{\partial c_i \partial c_j} \rc \, , \qquad i,j=1,\ldots,n_{\rm op} \, ,
\ee
where ${\rm E}\lc~\rc$ indicates the expectation value and
$ f \lp {\boldsymbol \sigma}_{\rm exp}|{\boldsymbol c} \rp$ indicates the relation
between a set of experimental measurements and the assumed true values of
the EFT coefficients
$ {\boldsymbol c}$.
The covariance matrix in the EFT parameter space, $C_{ij} \lp {\boldsymbol c} \rp$,
is then bounded by the Fisher information matrix:
\be
C_{ij} \ge \lp I^{-1}\rp_{ij} \, ,
\ee
which is known as the Cramer-Rao bound.
The diagonal entries of the Cramer-Rao bound are
$C_{ii} =\lp \delta c_i\rp^2 \ge \lp I^{-1}\rp_{ii}$ and indicate that the smallest possible
uncertainty achievable on the coefficient $c_i$  given the input
data is $\delta{c_i}|^{\rm (best)}=\sqrt{ (I^{-1})_{ii}}$.

For a given set of EFT
coefficients, comparing the values of $I_{ij}$ between different datasets highlights
those which provide the highest information.
The larger the entries of the Fisher matrix, the better (smaller uncertainty) that
these  coefficients can be constrained by the data considered.
The Fisher information matrix can also be understood as a metric in
model space.
If one has two sets of coefficients $\boldsymbol{c}_a$ and $\boldsymbol{c}_b$,
corresponding to two different points in the EFT parameter space,
then the local distance between them is defined as
\be
d_{\rm loc}\lp \boldsymbol{c}_a,\boldsymbol{c}_b \rp =  \lc \sum_{i,j}
(\boldsymbol{c}_a-\boldsymbol{c}_b)_i I_{ij}(\boldsymbol{c}_a)
(\boldsymbol{c}_a-\boldsymbol{c}_b)_j\rc^{1/2} \, ,
\ee
a feature which provides a robust method to quantify how (di)similar
are two points in this model space.

If one has $n_{\rm dat}$ experimental measurements
$\sigma_m^{\rm (exp)}$ whose theoretical predictions depend on $n_{\rm op}$ coefficients
${\boldsymbol c}$,
assuming that these measurements are Gaussianly distributed, one has
\be
\label{eq:fisherf}
f \lp {\boldsymbol \sigma}_{\rm exp})|
{\boldsymbol c} \rp = \prod_{m=1}^{n_{\rm dat}}\frac{1}{\sqrt{2\pi \delta_{{\rm exp},m}^2}}
\exp \lp -\frac{ \lp \sigma_m^{\rm (exp)}-\sigma_m^{\rm (th)}({\boldsymbol c})\rp^2  }{ 2\delta_{{\rm exp},m}^2}\rp \, ,
\ee
where $\delta_{{\rm exp},m}$ stands for the total experimental uncertainty associated to this
cross-section measurement.
Here we neglect for simplicity the point-by-point correlations, the extension
to the full correlation covariance matrix is straightforward.
The theoretical predictions that enter Eq.~(\ref{eq:fisherf})  include the  SM contribution
as well as the terms linear and quadratic on the Wilson coefficients,
\be
\label{eq:quadraticTHform}
\sigma_m^{\rm (th)}({\boldsymbol c})= \sigma_m^{\rm (sm)} + \sum_{i=1}^{n_{\rm op}}c_i\sigma^{(\rm eft)}_{m,i} +
\sum_{i<j}^{n_{\rm op}}c_ic_j \sigma^{(\rm eft)}_{m,ij} \, ,
\ee
where we assume $\Lambda=1$ TeV, so that one can write
\be
-\ln f \lp {\boldsymbol \sigma}_{\rm exp}|
{\boldsymbol c} \rp = \sum_{m=1}^{n_{\rm dat}} \frac{1}{2\delta_{{\rm exp},m}^2} \lp
\lp \sigma_m^{\rm (exp)} - \sigma_m^{\rm (sm)}\rp - \sum_{i=1}^{n_{\rm op}}c_i\sigma^{(\rm eft)}_{m,i} -
\sum_{i<j}^{n_{\rm op}}c_ic_j \sigma^{(\rm eft)}_{m,ij} \rp^2 + A \, ,
\ee
where $A$ is a constant that does not depend on the value of the coefficients,
and thus the Fisher information matrix can be evaluated using Eq.~(\ref{eq:FisherDef}) to yield
\be
\label{eq:fisherinformation}
I_{ij} = {\rm E}\Bigg[ \sum_{m=1}^{n_{\rm dat}} \frac{1}{\delta_{{\rm exp},m}^2}  \Bigg( \sigma_{m,ij}\lp
  \sigma_m^{\rm (th)}-\sigma_m^{\rm (exp)} \rp  
  + \lp  \sigma^{\rm (eft)}_{m,i} + \sum_{l=1}^{n_{\rm op}}
  c_l \sigma^{\rm (eft)}_{m,il} \rp
  \lp \sigma^{\rm (eft)}_{m,j}+ \sum_{l'=1}^{n_{\rm op}} 
 c_{l'}  \sigma_{m,jl'} \rp \Bigg)\Bigg] \, ,
\ee
where this expectation value can be evaluated  by averaging over the
$N_{\rm rep}$ replicas (or $N_{\rm spl}$ samples) that provide a sampling of the probability density in the space
of coefficients within our approach, see also Sect.~\ref{sec:fitsettings}.
 
The Fisher information matrix becomes specially simple if we restrict
 ourselves to the linear approximation~\cite{Ellis:2018gqa}, i.e. $\sigma_{m,ij}=0$, since in this case
\be
\label{eq:fisherinformation2}
I_{ij} = \sum_{m=1}^{n_{\rm dat}} \frac{\sigma^{\rm (eft)}_{m,i}\sigma^{\rm (eft)}_{m,j}}{\delta_{{\rm exp},m}^2} \, ,
\ee
which is independent of the values of the coefficients $\boldsymbol{c}$
and therefore of the actual fit results.
The diagonal entries of the Fisher matrix $I_{ii}$ are then given by the
sum over a given dataset (or group of processes) of the square of
the linear
EFT cross-sections over the total experimental uncertainty.
In the specific case of one-parameter fits, the Cramer-Rao bound reads
\be
\label{cramerraobound}
\delta c_i  \ge \lp \sum_{m=1}^{n_{\rm dat}} \frac{\sigma^{\rm (eft)}_{m,i}\sigma^{\rm (eft)}_{m,j}}{\delta_{{\rm exp},m}^2}   \rp ^{-1/2} \,,\qquad  i=1,\ldots, n_{\rm op}\, ,
\ee
which, provided that the sum is over the global dataset, can be used to cross-check
the results of individual (one-parameter) fits.

One should emphasize that the absolute size of the entries of the Fisher matrix does not
contain physical information: one is always allowed to redefine the overall normalisation
of an operator such that $c_i\sigma^{(\rm eft)}_{m,i} = c_i'\sigma'^{(\rm eft)}_{m,i}$, with
$c_i' = B_i c_i$ and $\sigma^{(\rm eft)}_{m,i} =\sigma'^{(\rm eft)}_{m,i}/B_i$
with $B_i$ being arbitrary constants.
However, for a given operator the relative value of $I_{ii}$ between two groups of processes is independent
of this choice of normalisation and
thus conveys meaningful information.
For this reason, in the following we present results for the Fisher information matrix normalised
such that 
the sum of the diagonal entries associated to a given EFT coefficient adds up to a fixed reference value
which is taken to be 100.

\begin{figure}[htbp]
  \begin{center}
    \includegraphics[width=0.99\linewidth]{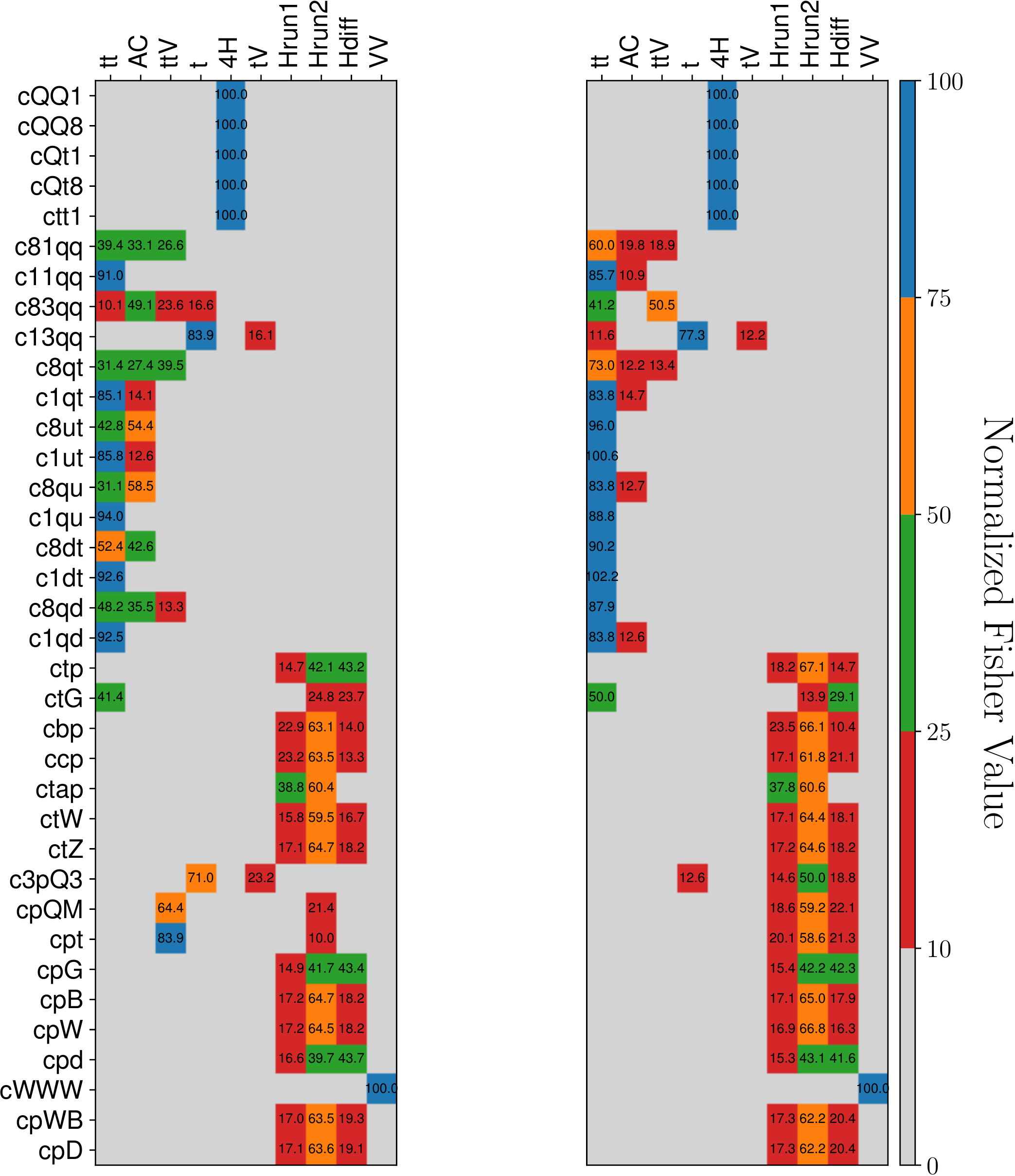}
    \caption{\small The values of the diagonal entries of the Fisher
      information matrix, Eq.~(\ref{eq:fisherinformation}), evaluated for
      the same groups of processes as in Table~\ref{table:operatorprocess} (except
      with the charge asymmetry $A_C$ data considered separately).
      The normalisation here is such that the sum of the entries associated to each EFT
      coefficient adds up to 100.
      We show results for the Fisher information matrix
      both at the linear level, Eq.~(\ref{eq:fisherinformation2}),
      and with the quadratic corrections included,  left and right panels respectively.
      For entries in the heat map larger than 10, we also indicate the corresponding
      numerical values.
      {  The quadratic Fisher information matrix (right panel)
        is evaluated
        using the best-fit values of the corresponding global baseline
      fit, to be presented in Sect.~5.}
     \label{fig:FisherMatrix} }
  \end{center}
\end{figure}

Fig.~\ref{fig:FisherMatrix} displays the values of the diagonal entries of the Fisher
information matrix, Eq.~(\ref{eq:fisherinformation}), evaluated for
the same groups of processes as in Table~\ref{table:operatorprocess}.
The normalisation is such that the sum of the entries associated to each coefficients
adds up to 100.
We show results for the Fisher information both at the linear level, Eq.~(\ref{eq:fisherinformation2}),
and with the quadratic corrections included, Eq.~(\ref{eq:fisherinformation}),
in the left and right panels respectively.
The entries in blue indicate those groups of processes
which provide more than 75\% of the information on the corresponding
EFT coefficient.
Entries in grey indicate relative contributions
of less than 10\%.
As mentioned above, the sum of the entries over columns does not contain a physical
      interpretation.
  
The information contained in Fig.~\ref{fig:FisherMatrix} is consistent with that of
Table~\ref{table:operatorprocess}, but now we can identify, for each coefficient, which datasets
provide the dominant constraints.
For instance, one observes that the two-light-two-heavy operators are overwhelmingly constrained
by inclusive top quark pair production data, except for $c_{Qq}^{3,1}$ for
which single top is the most important set of processes.
At the linear level, the information on the two-light-two-heavy coefficients provided
by the differential distributions and by the charge asymmetry $A_C$ data is comparable,
while the latter is less important in the quadratic fits.
In the case of the two-fermion operators, the leading constraints typically arise from Higgs data, in particular
from the Run II signal strengths measurements, and then to a lesser extent from the Run I data
and the Run II differential distributions.
Two  exceptions are $c_{\varphi t}$, which at the linear level (but not at the quadratic one)
is dominated by $t\bar{t}V$, and the chromo-magnetic operator $c_{tG}$, for which inclusive
$t\bar{t}$ production is most important.
Also for the purely bosonic operators the Higgs data provides most of the information,
except for $c_{WWW}$, as expected since this operator is only accessible in
diboson processes.
Furthermore, one observes that the $\mathcal{O}\lp \Lambda^{-4}\rp$ corrections induce in most
cases a moderate change in the Fisher information
matrix, but in others they can significantly alter
the balance between processes.
As a representative example, the two-fermion operators
such as $c_{\varphi t}$ and $c_{\varphi Q}^{(3)}$
become dominated by the Higgs data only once quadratic corrections are accounted for.

Another relevant application of the Fisher information matrix is the determination
of optimal directions in the EFT parameter space by means of principal component
analysis (PCA), and in particular the assessment of whether or not the coefficients basis adopted
for the fit contains flat directions.
We will discuss this related application in the Sect.~\ref{sec:pca}.

%% file: tables/table-input-dataset-top1.tex
\begin{table}[t]
  \centering
  \scriptsize
   \renewcommand{\arraystretch}{1.80}
  \begin{tabular}{c|c|c|c|c|c}
 Dataset   &  $\sqrt{s}$, $\mathcal{L}$  & Info  &  Observables  & $n_{\rm dat}$ & Ref   \\
    \toprule
      {\tt ATLAS\_tt\_8TeV\_ljets}
      & { \bf 8 TeV, 20.3~fb$^{-1}$}
      & lepton+jets
      & $d\sigma/dm_{t\bar{t}}$
      & 7
      & \cite{Aad:2015mbv} \\
    \midrule
      {\tt CMS\_tt\_8TeV\_ljets}
      & {\bf 8 TeV, 20.3~fb$^{-1}$}
      & lepton+jets
      & $1/\sigma d\sigma/dy_{t\bar{t}}$
      & 10
      & \cite{Khachatryan:2015oqa} \\
    \midrule
      {\tt CMS\_tt2D\_8TeV\_dilep}
      & {\bf 8 TeV, 20.3~fb$^{-1}$}
      & dileptons
      & $1/\sigma d^2\sigma/dy_{t\bar{t}}dm_{t\bar{t}}$
      & 16
      & \cite{Sirunyan:2017azo} \\
    \midrule
      {\tt ATLAS\_tt\_8TeV\_dilep} {\bf(*)}
      & {\bf 8 TeV, 20.3~fb$^{-1}$}
      & dileptons
      & $d\sigma/dm_{t\bar{t}}$
      & 6
      & \cite{Aaboud:2016iot}  \\
    \midrule
    \midrule
      {\tt CMS\_tt\_13TeV\_ljets\_2015 }
      & {\bf 13 TeV, 2.3~fb$^{-1}$}
      & lepton+jets
      & $d\sigma/dm_{t\bar{t}}$
      & 8
      & \cite{Khachatryan:2016mnb}  \\
    \midrule
      {\tt CMS\_tt\_13TeV\_dilep\_2015 }
      & {\bf 13 TeV, 2.1~fb$^{-1}$}
      & dileptons
      & $d\sigma/dm_{t\bar{t}}$
      & 6
      & \cite{Sirunyan:2017mzl}  \\
    \midrule
      {\tt CMS\_tt\_13TeV\_ljets\_2016 }
      & {\bf 13 TeV, 35.8~fb$^{-1}$}
      & lepton+jets
      & $d\sigma/dm_{t\bar{t}}$
      & 10
      & \cite{Sirunyan:2018wem}  \\
    \midrule
      {\tt CMS\_tt\_13TeV\_dilep\_2016 } {\bf(*)}
      & {\bf 13 TeV, 35.8~fb$^{-1}$}
      & dileptons
      & $d\sigma/dm_{t\bar{t}}$
      & 7
      & \cite{Sirunyan:2018ucr}  \\
    \midrule
      {\tt ATLAS\_tt\_13TeV\_ljets\_2016 } {\bf (*)}
      & {\bf 13 TeV, 35.8~fb$^{-1}$}
      & lepton+jets
      & $d\sigma/dm_{t\bar{t}}$
      & 9
      & \cite{Aad:2019ntk}  \\
   \midrule
\midrule
 {\tt ATLAS\_WhelF\_8TeV}  & {\bf 8 TeV, 20.3~fb$^{-1}$}  & $W$ hel. fract &
$F_0, F_L, F_R$  &  3  &  \cite{Aaboud:2016hsq}  \\
\midrule
{\tt CMS\_WhelF\_8TeV}  & {\bf 8 TeV, 20.3~fb$^{-1}$}  & $W$ hel. fract &
$F_0, F_L, F_R$  &  3  &  \cite{Khachatryan:2016fky}  \\
\midrule
\midrule
    {\tt ATLAS\_CMS\_tt\_AC\_8TeV}  {\bf(*)} & {\bf 8 TeV, 20.3~fb$^{-1}$} & charge asymmetry &
    $ A_C$ & 6 & \cite{Sirunyan:2017lvd}\\
\midrule   
    {\tt ATLAS\_tt\_AC\_13TeV
}  {\bf(*)} & {\bf 13 TeV, 139~fb$^{-1}$} & charge asymmetry &
    $ A_C$ & 5 & \cite{ATLAS:2019czt}\\
\bottomrule
  \end{tabular}
  \caption{\small The experimental measurements of inclusive top-quark pair 
    production at the LHC considered in the present analysis.
    For each dataset we indicate the label, 
    the center of mass energy $\sqrt{s}$, the
    integrated luminosity $\mathcal{L}$,
    the final state or the specific 
    production mechanism, the physical observable, the number of data 
    points $n_{\rm dat}$, and the publication reference.
    Measurements indicated with {\bf (*)} were not included 
    in~\cite{Hartland:2019bjb}.
    We also include in this category the $W$ helicity fractions
    from top quark decay and the charge asymmetries.
\label{eq:input_datasets}
}
\end{table}

%% file: tables/table-input-dataset-top2.tex
\begin{table}[t]
  \centering
  \scriptsize
   \renewcommand{\arraystretch}{1.40}
  \begin{tabular}{c|c|c|c|c|c}
 Dataset   &  $\sqrt{s}, \mathcal{L}$ & Info  &  Observables  & $N_{\rm dat}$ & Ref   \\
    \toprule
 {\tt CMS\_ttbb\_13TeV}  & {\bf 13 TeV, 2.3~fb$^{-1}$}  & total xsec & $\sigma_{\rm tot}(t\bar{t}b\bar{b})$  &  1  &  \cite{Sirunyan:2017snr}  \\
\midrule
{\tt CMS\_ttbb\_13TeV\_2016}  {\bf (*)}  & {\bf 13 TeV, 35.9~fb$^{-1}$}  & total xsec & $\sigma_{\rm tot}(t\bar{t}b\bar{b})$  &  1  &  \cite{Sirunyan:2019jud}  \\
\midrule
{\tt ATLAS\_ttbb\_13TeV\_2016}  {\bf (*)}  & {\bf 13 TeV, 35.9~fb$^{-1}$}  & total xsec & $\sigma_{\rm tot}(t\bar{t}b\bar{b})$  &  1  &  \cite{Aaboud:2018eki}  \\
\midrule
 {\tt CMS\_tttt\_13TeV}  & {\bf 13 TeV, 35.9~fb$^{-1}$}  & total xsec & $\sigma_{\rm tot}(t\bar{t}t\bar{t})$  &  1  &  \cite{Sirunyan:2017roi}  \\
\midrule
 {\tt CMS\_tttt\_13TeV\_run2} {\bf (*)} & {\bf 13 TeV, 137~fb$^{-1}$}  & total xsec & $\sigma_{\rm tot}(t\bar{t}t\bar{t})$  &  1  &  \cite{Sirunyan:2019wxt}  \\
 \midrule
 {\tt ATLAS\_tttt\_13TeV\_run2} {\bf (*)} & {\bf 13 TeV, 137~fb$^{-1}$}  & total xsec & $\sigma_{\rm tot}(t\bar{t}t\bar{t})$  &  1  &  \cite{Aad:2020klt}  \\
\midrule
\midrule
  {\tt CMS\_ttZ\_8TeV}  & {\bf 8 TeV, 19.5~fb$^{-1}$}  & total xsec & $\sigma_{\rm tot}(t\bar{t}Z)$  &  1  &  \cite{Khachatryan:2015sha}  \\ \midrule
   {\tt CMS\_ttZ\_13TeV}  & {\bf 13 TeV, 35.9~fb$^{-1}$ }  & total xsec & $\sigma_{\rm tot}(t\bar{t}Z)$  &  1  &  \cite{Sirunyan:2017uzs}  \\
   \midrule
    {\tt CMS\_ttZ\_ptZ\_13TeV}  {\bf (*)} & {\bf 13 TeV, 77.5~fb$^{-1}$ }  & total xsec & $d\sigma(t\bar{t}Z)/dp_T^Z $  &  4  &  \cite{CMS:2019too}  \\
   \midrule
  {\tt ATLAS\_ttZ\_8TeV}  & {\bf 8 TeV, 20.3~fb$^{-1}$}  & total xsec & $\sigma_{\rm tot}(t\bar{t}Z)$  &  1  &  \cite{Aad:2015eua}  \\
    \midrule
        {\tt ATLAS\_ttZ\_13TeV}  & {\bf 13 TeV, 3.2~fb$^{-1}$}  & total xsec & $\sigma_{\rm tot}(t\bar{t}Z)$  &  1  &  \cite{Aaboud:2016xve}  \\
     \midrule
   {\tt ATLAS\_ttZ\_13TeV\_2016} {\bf (*)} & {\bf 13 TeV, 36~fb$^{-1}$}  & total xsec & $\sigma_{\rm tot}(t\bar{t}Z)$  &  1  &  \cite{Aaboud:2019njj}  \\
  \midrule
  \midrule
    {\tt CMS\_ttW\_8\_TeV}  & {\bf 8 TeV, 19.5~fb$^{-1}$}  & total xsec & $\sigma_{\rm tot}(t\bar{t}W)$  &  1  &  \cite{Khachatryan:2015sha}  \\ \midrule
     {\tt CMS\_ttW\_13TeV}  & {\bf 13 TeV, 35.9~fb$^{-1}$}  & total xsec & $\sigma_{\rm tot}(t\bar{t}W)$  &  1  &  \cite{Sirunyan:2017uzs}  \\
   \midrule
   {\tt ATLAS\_ttW\_8TeV}  & {\bf 8 TeV, 20.3~fb$^{-1}$}  & total xsec & $\sigma_{\rm tot}(t\bar{t}W)$  &  1  &  \cite{Aad:2015eua}  \\
    \midrule
   {\tt ATLAS\_ttW\_13TeV}  & {\bf 13 TeV, 3.2~fb$^{-1}$}  & total xsec & $\sigma_{\rm tot}(t\bar{t}W)$  &  1  &  \cite{Aaboud:2016xve}  \\
     \midrule
   {\tt ATLAS\_ttW\_13TeV\_2016} {\bf (*) } & {\bf 13 TeV, 36~fb$^{-1}$} & total xsec & $\sigma_{\rm tot}(t\bar{t}W)$  &  1  &  \cite{Aaboud:2019njj}  \\
\bottomrule
  \end{tabular}
  \caption{\small Same as Table~\ref{eq:input_datasets}, now for the production
    of top quark pairs in association with
    heavy quarks and with weak vector bosons.
     \label{eq:input_datasets2}
  }
\end{table}

%% file: tables/table-input-dataset-top3.tex
\begin{table}[t]
  \centering
  \scriptsize
   \renewcommand{\arraystretch}{1.60}
  \begin{tabular}{c|c|c|c|c|c}
 Dataset   &  $\sqrt{s}, \mathcal{L}$ & Info  &  Observables  & $N_{\rm dat}$ & Ref   \\
\toprule
    {\tt CMS\_t\_tch\_8TeV\_inc}
    & {\bf 8 TeV, 19.7~{\rm \bf fb}$^{-1}$ }
    & $t$-channel
    & $\sigma_{\rm tot}(t),\sigma_{\rm tot}(\bar{t})$  & 2
    & \cite{Khachatryan:2014iya}  \\
\midrule
    {\tt ATLAS\_t\_tch\_8TeV}
    & {\bf 8 TeV, 20.2~{\rm \bf fb}$^{-1}$}
    & $t$-channel
    & $d\sigma(tq)/dy_t$
    & 4
    & \cite{Aaboud:2017pdi}  \\
\midrule
    {\tt CMS\_t\_tch\_8TeV\_dif}
    & {\bf 8 TeV, 19.7~{\rm \bf fb}$^{-1}$}
    & $t$-channel
    & $d\sigma/d|y^{(t+\bar{t})}|$
    & 6
    & \cite{CMS-PAS-TOP-14-004}  \\
\midrule
    {\tt CMS\_t\_sch\_8TeV}
    & {\bf 8 TeV, 19.7~{\rm \bf fb}$^{-1}$}
    & $s$-channel
    & $\sigma_{\rm tot}(t+\bar{t})$
    & 1
    & \cite{Khachatryan:2016ewo}  \\
\midrule
    {\tt ATLAS\_t\_sch\_8TeV}
    & {\bf 8 TeV, 20.3~{\rm \bf fb}$^{-1}$ }
    & $s$-channel
    & $\sigma_{\rm tot}(t+\bar{t})$
    & 1
    & \cite{Aad:2015upn}  \\
\midrule
\midrule
    {\tt ATLAS\_t\_tch\_13TeV}
    & {\bf 13 TeV, 3.2~{\rm \bf fb}$^{-1}$}
    & $t$-channel
    & $\sigma_{\rm tot}(t),\sigma_{\rm tot}(\bar{t})$
    & 2
    & \cite{Aaboud:2016ymp}  \\
    \midrule
    {\tt CMS\_t\_tch\_13TeV\_inc}
    & {\bf 13 TeV, 2.2~{\rm \bf fb}$^{-1}$}
    & $t$-channel
    & $\sigma_{\rm tot}(t),\sigma_{\rm tot}(\bar{t})$
    & 2
    & \cite{Sirunyan:2016cdg}  \\
    \midrule
    {\tt CMS\_t\_tch\_13TeV\_dif}
    & {\bf 13 TeV, 2.3~{\rm \bf fb}$^{-1}$}
    & $t$-channel
    & $d\sigma/d|y^{(t+\bar{t})}|$
    & 4
    & \cite{CMS:2016xnv}  \\
    \midrule
    {\tt CMS\_t\_tch\_13TeV\_2016} {\bf (*)}
    & {\bf 13 TeV, 35.9~{\rm \bf fb}$^{-1}$}
    & $t$-channel
    & $d\sigma/d|y^{(t)}|$
    & 5
    & \cite{Sirunyan:2019hqb}  \\
\bottomrule
  \end{tabular}
  \caption{\small Same as Table~\ref{eq:input_datasets},
    now for inclusive single $t$ production both in the $t$- and the $s$-channels.
     \label{eq:input_datasets3}
  }
\end{table}

%% file: tables/table-input-dataset-top4.tex
\begin{table}[t]
  \centering
  \scriptsize
   \renewcommand{\arraystretch}{1.60}
  \begin{tabular}{c|c|c|c|c|c}
 Dataset   &  $\sqrt{s}, \mathcal{L}$ & Info  &  Observables  & $N_{\rm dat}$ & Ref   \\
\toprule
\multirow{2}{*}{ {\tt ATLAS\_tW\_8TeV\_inc}}      &
\multirow{2}{*}{{\bf 8 TeV, 20.2}~{\rm \bf fb}$^{-1}$}   & \multirow{1}{*}{inclusive}   &
\multirow{2}{*}{$\sigma_{\rm tot}(tW)$}  &  1  &
\multirow{2}{*}{\cite{Aad:2015eto}}  \\
    &
  & \multirow{1}{*}{(dilepton)}   &
  &   &\\
\toprule
\multirow{2}{*}{ {\tt ATLAS\_tW\_inc\_slep\_8TeV} {\bf (*)}}      &
\multirow{2}{*}{{\bf 8 TeV, 20.2}~{\rm \bf fb}$^{-1}$}   & \multirow{1}{*}{inclusive}   &
\multirow{2}{*}{$\sigma_{\rm tot}(tW)$}  &  1  &
\multirow{2}{*}{\cite{Aad:2020zhd}}  \\
    &
  & \multirow{1}{*}{(single lepton)}   &
  &   &\\
\midrule
      \multirow{1}{*}{ {\tt CMS\_tW\_8TeV\_inc}}      &
 \multirow{1}{*}{{\bf 8 TeV, 19.7}~{\rm \bf fb}$^{-1}$}   & \multirow{1}{*}{inclusive}   &
\multirow{1}{*}{$\sigma_{\rm tot}(tW)$}  &  1  &
\multirow{1}{*}{\cite{Chatrchyan:2014tua}}  \\
\midrule
        \multirow{1}{*}{ {\tt ATLAS\_tW\_inc\_13TeV}}      &
 \multirow{1}{*}{{\bf 13 TeV, 3.2}~{\rm \bf fb}$^{-1}$}   & \multirow{1}{*}{inclusive}   &
\multirow{1}{*}{$\sigma_{\rm tot}(tW)$}  &  1  &
\multirow{1}{*}{\cite{Aaboud:2016lpj}}  \\
\midrule
     \multirow{1}{*}{ {\tt CMS\_tW\_13TeV\_inc}}      &
 \multirow{1}{*}{{\bf 13 TeV, 35.9}~{\rm \bf fb}$^{-1}$}   & \multirow{1}{*}{inclusive}   &
\multirow{1}{*}{$\sigma_{\rm tot}(tW)$}  &  1  &
\multirow{1}{*}{\cite{Sirunyan:2018lcp}}  \\
\midrule
\midrule
      \multirow{1}{*}{ {\tt ATLAS\_tZ\_13TeV\_inc}}      &
 \multirow{1}{*}{{\bf 13 TeV, 36.1}~{\rm \bf fb}$^{-1}$}    & \multirow{1}{*}{inclusive}   &
\multirow{1}{*}{$\sigma_{\rm tot}(tZq)$}  &  1  &
\multirow{1}{*}{\cite{Aaboud:2017ylb}}  \\
\midrule
 \multirow{1}{*}{ {\tt ATLAS\_tZ\_13TeV\_run2\_inc} {\bf (*)}}      &
 \multirow{1}{*}{{\bf 13 TeV, 139.1}~{\rm \bf fb}$^{-1}$}    & \multirow{1}{*}{inclusive}   &
\multirow{1}{*}{$\sigma_{\rm fid}(t\ell^+\ell^-q)$}  &  1  &
\multirow{1}{*}{\cite{Aad:2020wog}}  \\
\midrule
       \multirow{1}{*}{ {\tt CMS\_tZ\_13TeV\_inc}}      &
 \multirow{1}{*}{{\bf 13 TeV, 35.9}~{\rm \bf fb}$^{-1}$}   & \multirow{1}{*}{inclusive}   &
\multirow{1}{*}{$\sigma_{\rm fid}(Wb\ell^+\ell^-q)$}  &  1  &
\multirow{1}{*}{\cite{Sirunyan:2017nbr}}  \\
\midrule
       \multirow{1}{*}{ {\tt CMS\_tZ\_13TeV\_2016\_inc}  {\bf (*)}}      &
 \multirow{1}{*}{{\bf 13 TeV, 77.4}~{\rm \bf fb}$^{-1}$ }   & \multirow{1}{*}{inclusive}   &
\multirow{1}{*}{$\sigma_{\rm fid}(t\ell^+\ell^-q)$}  &  1  &
\multirow{1}{*}{\cite{Sirunyan:2018zgs}}  \\
\bottomrule
  \end{tabular}
  \caption{\small Same as Table~\ref{eq:input_datasets},
    now for single top quark production in association with
    electroweak gauge bosons.
     \label{eq:input_datasets4}
  }
\end{table}

%% file: tables/table-input-dataset-HiggsSS.tex
\begin{table}[t]
  \centering
  \scriptsize
   \renewcommand{\arraystretch}{1.65}
  \begin{tabular}{c|c|c|c|c|c}
 Dataset   &  $\sqrt{s},~\mathcal{L}$ & Info  &  Observables  & $n_{\rm dat}$ & Ref.   \\
    \toprule
    \multirow{2}{*}{ {\tt ATLAS\_CMS\_SSinc\_RunI} {\bf (*)}}  &\multirow{2}{*}{ {\bf 7+8 TeV, 20~fb$^{-1}$}}  &
    \multirow{2}{*}{Incl. $\mu_i^f$} &  $gg$F, VBF, $Vh$, $t\bar{t}h$
    &  \multirow{2}{*}{20}    &  \multirow{2}{*}{\cite{Khachatryan:2016vau} } \\
    &   &     & $h\to \gamma\gamma, VV, \tau\tau, b\bar{b}$   &  &    \\ \midrule
    {\tt ATLAS\_SSinc\_RunI} {\bf (*)}  & {\bf 8 TeV, 20~fb$^{-1}$}  &
    Incl. $\mu^f_i$ &  $h\to Z\gamma, \mu\mu$
    & 2    & \cite{Aad:2015gba} \\ \midrule
    \midrule
 \multirow{2}{*}{ {\tt ATLAS\_SSinc\_RunII} {\bf (*)}}  &\multirow{2}{*}{ {\bf 13 TeV, 80~fb$^{-1}$}}  &
 \multirow{2}{*}{Incl. $\mu_i^f$} &  $gg$F, VBF, $Vh$, $t\bar{t}h$
 &  \multirow{2}{*}{16}    &  \multirow{2}{*}{\cite{Aad:2019mbh} } \\
 &   &     & $h\to \gamma\gamma, WW, ZZ, \tau\tau, b\bar{b}$   &  &    \\ \midrule
 \multirow{2}{*}{ {\tt CMS\_SSinc\_RunII} {\bf (*)}}  &\multirow{2}{*}{ {\bf 13 TeV, 36.9~fb$^{-1}$}}  &
 \multirow{2}{*}{Incl. $\mu_i^f$} &  $gg$F, VBF, $Wh$, $Zh$ $t\bar{t}h$
 &  \multirow{2}{*}{24}    &  \multirow{2}{*}{\cite{Sirunyan:2018koj} } \\
 &   &     & $h\to \gamma\gamma, WW, ZZ, \tau\tau, b\bar{b}$   &  &    \\ 
    \bottomrule
    \end{tabular}
  \caption{\small Same as Table~\ref{eq:input_datasets} now for
    the  measurements of the inclusive
    signal strenghts, Eq.~(\ref{eq:signal_strength_def}),
    in Higgs production and decay from the LHC Run I and Run II.
     \label{eq:input_datasets_higgsSS}
  }
\end{table}

%% file: tables/table-input-dataset-Higgs.tex
\begin{table}[t]
  \centering
  \scriptsize
   \renewcommand{\arraystretch}{1.95}
  \begin{tabular}{c|c|c|c|c|c}
 Dataset   &  $\sqrt{s}, \mathcal{L}$ & Info  &  Observables  & $N_{\rm dat}$ & Ref   \\
 \toprule
  \multirow{2}{*}{ {\tt CMS\_H\_13TeV\_2015} {\bf (*)}}  &\multirow{2}{*}{ {\bf 13 TeV, 35.9~fb$^{-1}$}}  &
  $gg$F, VBF, $Vh$, $t\bar{t}h$ & \multirow{2}{*}{$d\sigma/dp_T^h$}    &   \multirow{2}{*}{9}    &  \multirow{2}{*}{ \cite{Sirunyan:2018sgc} } \\
     &  & $h\to ZZ,\gamma\gamma,b\bar{b}$
  &      &     &    \\
  \midrule
   \multirow{2}{*}{ {\tt ATLAS\_ggF\_13TeV\_2015} {\bf (*)}}  &\multirow{2}{*}{ {\bf 13 TeV, 36.1~fb$^{-1}$}}  &
  $gg$F, VBF, $Vh$, $t\bar{t}h$ &  \multirow{2}{*}{$d\sigma/dp_T^h$}    &   \multirow{2}{*}{9}    &  \multirow{2}{*}{ \cite{Aaboud:2018ezd} } \\
     &  & $h\to ZZ(\to 4l)$
  &      &     &    \\
  \midrule
   \midrule
    \multirow{2}{*}{ {\tt ATLAS\_Vh\_hbb\_13TeV} {\bf (*)}}  &\multirow{2}{*}{ {\bf 13 TeV, 79.8~fb$^{-1}$}}  &
    \multirow{2}{*}{$Wh,~Zh$} &  $d\sigma^{\rm (fid)}/dp_T^W$   &  2    &  \multirow{2}{*}{\cite{Aaboud:2019nan} } \\
    &   &     & $d\sigma^{\rm (fid)}/dp_T^Z$   & 3  &    \\ \midrule
     {\tt ATLAS\_ggF\_ZZ\_13TeV} {\bf (*)}  &{\bf 13 TeV, 79.8~fb$^{-1}$}  &
     $gg$F, $h\to ZZ$ &  $\sigma_{\rm ggF}(p_T^h,N_{\rm jets})$    &  6    &  \cite{Aad:2019mbh}  \\
     \midrule
    {\tt CMS\_ggF\_aa\_13TeV} {\bf (*)}  &{\bf 13 TeV, 77.4~fb$^{-1}$}  &
    $gg$F, $h\to \gamma\gamma$ &   $\sigma_{\rm ggF}(p_T^h,N_{\rm jets})$    &  6    &  \cite{CMS:1900lgv}  \\
    \bottomrule
    \end{tabular}
  \caption{\small Same as Table~\ref{eq:input_datasets} for
   differential distributions and STXS for Higgs production and decay.
     \label{eq:input_datasets_higgs}
  }
\end{table}

%% file: tables/table-input-dataset-diboson.tex
\begin{table}[t]
  \centering
  \scriptsize
   \renewcommand{\arraystretch}{1.90}
  \begin{tabular}{c|c|c|c|c|c}
 Dataset   &  $\sqrt{s}, ~\mathcal{L}$ & Info  &  Observables  & $N_{\rm dat}$ & Ref   \\
    \toprule
        {\tt LEP2\_WW\_diff} {\bf (*)}   & {\bf [182,~296] GeV}   & LEP-2 comb   & $d^2\sigma(WW)/dE_{\rm cm}d\cos\theta_{W}$  & 40  &  \cite{Schael:2013ita} \\ \toprule
    {\tt ATLAS\_WZ\_13TeV\_2016} {\bf (*)}  & {\bf 13 TeV, 36.1~fb$^{-1}$}  &
    fully leptonic &  $d\sigma^{\rm (fid)}/dm_T^{WZ}$  &  6    & \cite{Aaboud:2019gxl}  \\
    \midrule
    {\tt ATLAS\_WW\_13TeV\_2016} {\bf (*)}  &{\bf 13 TeV, 36.1~fb$^{-1}$}  &
    fully leptonic &  $d\sigma^{\rm (fid)}/dm_{e\mu}$  &  13    &  \cite{Aaboud:2019nkz} \\
     \midrule
    {\tt CMS\_WZ\_13TeV\_2016} {\bf (*)}  & {\bf 13 TeV, 35.9~fb$^{-1}$}  &
    fully leptonic &  $d\sigma^{\rm (fid)}/dp_T^{Z}$  &  11    &  \cite{Sirunyan:2019bez} \\
    \bottomrule
    \end{tabular}
  \caption{\small Same as Table~\ref{eq:input_datasets} for
    the differential distributions of gauge boson pair
    production from LEP-2 and the LHC.
     \label{eq:input_datasets_diboson}
  }
\end{table}

%% file: tables/table-dataset-overview.tex
\begin{table}[t]
  \centering
  \small
   \renewcommand{\arraystretch}{1.30}
  \begin{tabular}{C{3.9cm}|C{6.6cm}|C{2cm}}
 Category   & Processes    &  $n_{\rm dat}$     \\
    \toprule
    \multirow{6}{*}{Top quark production}   &  $t\bar{t}$ (inclusive)   &  94  \\
    &  $t\bar{t}Z$, $t\bar{t}W$    & 14 \\
    &   single top (inclusive)   & 27 \\
    &  $tZ, tW$   &  9\\
    &  $t\bar{t}t\bar{t}$, $t\bar{t}b\bar{b}$    & 6 \\
    &  {\bf Total}    & {\bf 150 }  \\
    \midrule
    \multirow{3.3}{*}{Higgs production} & Run I signal strengths  &22   \\
    \multirow{3.1}{*}{and decay} & Run II  signal strengths  & 40  \\
    & Run II, differential distributions \& STXS  & 35  \\
    &  {\bf Total}    & {\bf 97}  \\
    \midrule
    \multirow{3}{*}{Diboson production} & LEP-2 &40   \\
     & LHC & 30  \\
    &  {\bf Total}    & {\bf 70}  \\
    \bottomrule
   Baseline dataset     & {\bf Total}      & {\bf 317}  \\
\bottomrule
  \end{tabular}
  \caption{\small The number of data points $n_{\rm dat}$ in our baseline dataset
    for each of the categories of processes considered here.
 \label{eq:table_dataset_overview}
}
\end{table}

%% file: tables/table-processes-theory.tex
\begin{table}[htbp]
  \centering
  \footnotesize
  \renewcommand{\arraystretch}{1.80}
  \begin{tabular}{c|c|c|c|c}
  Category & Process 
  & SM   
  & Code/Ref  
  & SMEFT 
  \\
  \toprule
 \multirow{10}{*}{Top quark}  &   \multirow{2}{*}{$t\bar{t}$ (incl)}  
  & \multirow{2}{*}{NNLO QCD}   
  & {\tt MG5\_aMC} NLO  
  & \multirow{2}{*}{NLO QCD}  
  \\
\multirow{10}{*}{production}   &    
&  & + NNLO $K$-fact  
  &    
\\
\cmidrule(lr{0.7em}){2-5}
  &  \multirow{2}{*}{$t\bar{t}+V$} 
  & \multirow{2}{*}{NLO QCD} 
  & \multirow{2}{*}{{\tt MG5\_aMC} NLO}  
  & LO QCD   
  \\
  &  
  &   
  & & + NLO SM $K$-fact  
  \\
 \cmidrule(lr{0.7em}){2-5}
&    \multirow{2}{*}{single-$t$ (incl)} 
  & \multirow{2}{*}{NNLO QCD} 
  & {\tt MG5\_aMC} NLO  
  & \multirow{2}{*}{NLO QCD}  
  \\
  &   
 & & + NNLO $K$-fact     
  &
  \\
  \cmidrule(lr{0.7em}){2-5}
  &  \multirow{2}{*}{$t+V$} 
  & \multirow{2}{*}{NLO QCD} 
  & \multirow{2}{*}{{\tt MG5\_aMC} NLO}  
  & LO QCD   
  \\
  &  
  &   
 & & + NLO SM $K$-fact  
  \\
  \cmidrule(lr{0.7em}){2-5}
 &   \multirow{2}{*}{$t\bar{t}t\bar{t},~t\bar{b}t\bar{b}$} 
  & \multirow{2}{*}{NLO QCD} 
  & \multirow{2}{*}{{\tt MG5\_aMC} NLO}  
  & LO QCD  
  \\
  &  
  &   
 & & + NLO SM $K$-fact  
  \\
   \midrule
 \multirow{10}{*}{Higgs production}  &   \multirow{2}{*}{$gg\to h$}  
  & \multirow{1}{*}{NNLO QCD +}   
  &\multirow{2}{*}{ HXSWG}
  & \multirow{2}{*}{NLO QCD}  
  \\
\multirow{10}{*}{and decay}   &    
&  \multirow{1}{*}{NLO EW}   & 
  &    
\\
 \cmidrule(lr{0.7em}){2-5}
   &   \multirow{2}{*}{VBF}  
  & \multirow{1}{*}{NNLO QCD +}   
  &\multirow{2}{*}{ HXSWG}
  & \multirow{2}{*}{LO QCD}  
 \\
 &    
&  \multirow{1}{*}{NLO EW}   & 
  &    
 \\
  \cmidrule(lr{0.7em}){2-5}
   &   \multirow{2}{*}{$h+V$}  
  & \multirow{1}{*}{NNLO QCD +}   
  &\multirow{2}{*}{ HXSWG}
  & \multirow{2}{*}{NLO QCD}  
  \\
   &    
&  \multirow{1}{*}{NLO EW}   & 
  &    
\\
 \cmidrule(lr{0.7em}){2-5}
  &   \multirow{2}{*}{$ht\bar{t}$}  
  & \multirow{1}{*}{NNLO QCD +}   
  &\multirow{2}{*}{ HXSWG}
  & \multirow{2}{*}{NLO QCD}  
  \\
   &    
&  \multirow{1}{*}{NLO EW}   & 
  &    
\\
 \cmidrule(lr{0.7em}){2-5}
  &   \multirow{2}{*}{$h\to X$}  
  & \multirow{1}{*}{NNLO QCD +}   
  &\multirow{2}{*}{ HXSWG}
  & \multirow{1}{*}{NLO QCD ($X=b\bar{b}$)}  
 \\
 &    
&  \multirow{1}{*}{NLO EW}   & 
  &    \multirow{1}{*}{LO QCD ($X\ne b \bar{b} $)} 
\\
 \midrule
 \multirow{4}{*}{Diboson}  &   \multirow{2}{*}{$e^+e^- \to W^+W^-$}  
  & \multirow{1}{*}{NNLO QCD +}   
  &\multirow{2}{*}{ LEP EWWG}
  & \multirow{2}{*}{LO QCD}  
  \\
\multirow{4}{*}{production}   &    
&  \multirow{1}{*}{NLO EW}   & 
  &    
\\
\cmidrule(lr{0.7em}){2-5}
\multirow{4}{*}{}  &   \multirow{2}{*}{$pp \to VV'$}  
  & \multirow{2}{*}{NNLO QCD}   
& 
\multirow{2}{*}{ {\tt MATRIX}
}
  & \multirow{2}{*}{NLO QCD}  
  \\
\multirow{4}{*}{}   &    
&     &  
  &    
\\
  \bottomrule
 \end{tabular}
 \caption{\small Summary of the theoretical calculations used for the 
   description various datasets included in the 
   present analysis. We indicate, for both the SM and the SMEFT contributions 
   to the cross-sections, the perturbative accuracy and the codes used to 
   produce the corresponding theoretical predictions.
   In all cases, the EFT cross-sections are evaluated with {\tt MG5\_aMC} interfaced
   to {\tt SMEFT@NLO}.
   See the text for more details and the corresponding references.
 }
  \label{eq:table-processes-theory}
\end{table}

%% file: tables/table-operatorprocess.tex
\begin{table}[p]
 \centering
 \scriptsize
 \renewcommand{\arraystretch}{1.25}
 \begin{tabular}{l|l||C{0.8cm}|C{0.8cm}|C{0.8cm}|C{0.8cm}|C{0.8cm}||C{1.1cm}|C{1.1cm}|C{1.3cm}|C{0.8cm}}
   Class  &
   DoF &
   $\,t\bar{t}\,$
 & $t\bar{t}V$   
 & $\,\, t\,\,$ 
 & $tV$ 
 & $t\bar{t}Q\bar{Q}$ 
 & $h$ ($\mu_i^{f}$, Run-I)
 & $h$ ($\mu_i^{f}$, Run-II)
 & $h$ (STXS, Run-II)
 & $VV$
 \\
 \toprule
\multirow{13.5}{*}{2-heavy-} &  $c_{Qq}^{1,8}$ &  \checkmark  &  \checkmark  &    &    & \checkmark    & \checkmark     &  \checkmark     &    \checkmark  &       \\
\multirow{13.5}{*}{2-light} &  $c_{Qq}^{1,1}$ & (\checkmark)   & (\checkmark)  &    &     &  \checkmark   &  (\checkmark)   & (\checkmark)     & (\checkmark)    &       \\
 &  $c_{Qq}^{3,8}$ &   \checkmark  &  \checkmark  &  (\checkmark)   &  (\checkmark)    &   \checkmark   &  \checkmark    &   \checkmark    &   \checkmark   &       \\
 &  $c_{Qq}^{3,1}$ &   (\checkmark)  &    (\checkmark)   &  \checkmark    &  \checkmark    &  \checkmark   & ( \checkmark)      &   (\checkmark)   &  ( \checkmark)   &  \\
 &  $c_{tq}^{8}$ &  \checkmark   &  \checkmark  &    &      & \checkmark    &  \checkmark    &  \checkmark     &  \checkmark    &       \\
 &   $c_{tq}^{1}$&   (\checkmark)  &  (\checkmark)  &    &     & \checkmark & ( \checkmark)    &   (\checkmark)   & ( \checkmark)     &           \\
 &  $c_{tu}^{8}$ &   \checkmark  &   \checkmark   &     &     &  \checkmark   &    \checkmark   &  \checkmark    &  \checkmark &     \\
 &  $c_{tu}^{1}$ &    (\checkmark) &  (\checkmark)  &    &     &   \checkmark  &  ( \checkmark)   &   (\checkmark)    & ( \checkmark)     &       \\
 &  $c_{Qu}^{8}$ &   \checkmark  &  \checkmark  &    &     &   \checkmark  &  \checkmark    &    \checkmark   &  \checkmark    &       \\
 &  $c_{Qu}^{1}$&   (\checkmark)  &  (\checkmark)  &    &     &   \checkmark  & ( \checkmark)    &  (\checkmark)     &  ( \checkmark)    &       \\
 &  $c_{td}^{8}$ &   \checkmark  &  \checkmark  &    &     &   \checkmark  &   \checkmark   &  \checkmark     &  \checkmark    &       \\
 &  $c_{td}^{1}$ &   (\checkmark)  &   (\checkmark) &    &     &   \checkmark  &  ( \checkmark)   &  (\checkmark)     & ( \checkmark)    &       \\
 &  $c_{Qd}^{8}$ &   \checkmark  &  \checkmark  &    &     &   \checkmark  &   \checkmark   &    \checkmark   &  \checkmark    &       \\
&  $c_{Qd}^{1}$ &   (\checkmark)  &  (\checkmark)  &    &     &   \checkmark  &  ( \checkmark)   &    (\checkmark)   &  ( \checkmark)    &       \\
\midrule
\multirow{5}{*}{4-heavy} & $c_{QQ}^1$ &    &   &    &     &  \checkmark   &     &      &     &       \\
 &  $c_{QQ}^8$ &    &   &    &     &  \checkmark   &     &      &     &       \\
 &  $c_{Qt}^1$ &    &   &    &     &  \checkmark   &     &      &     &       \\
 &  $c_{Qt}^8$ &    &   &    &     &  \checkmark   &     &      &     &       \\
&  $c_{tt}^1$ &    &   &    &     &  \checkmark   &     &      &     &       \\
\midrule
4-lepton &  $c_{ll}$ &    &   & \checkmark   & \checkmark    &    &  \checkmark   & \checkmark     & \checkmark    & \checkmark      \\
\midrule
\multirow{18.5}{*}{2-fermion} &  $c_{t\varphi}$ &    &   &    &     &     &   \checkmark   &   \checkmark    &   \checkmark   &       \\
\multirow{18.5}{*}{+bosonic} &  $c_{tG}$ & \checkmark  & \checkmark   &    &     & \checkmark    &   \checkmark   &   \checkmark    &  \checkmark    &       \\
&  $c_{b\varphi}$ &   &   &    &     &     &  \checkmark   &   \checkmark    & \checkmark (b)    &       \\
&  $c_{c\varphi}$ &    &   &    &     &     &   \checkmark  &  \checkmark    &     &       \\
&  $c_{\tau\varphi}$ &    &   &    &     &     &   \checkmark  &   \checkmark   &     &       \\
&  $c_{tW}$ & \checkmark   &  & \checkmark   &  \checkmark   &     & \checkmark    &  \checkmark    &     &       \\
&  $c_{tZ}$ &    &  \checkmark &    &  \checkmark   &     &   \checkmark  &  \checkmark    &     &       \\[0.1cm]
&  $c_{\varphi Q}^{(3)}$ &    & \checkmark (b)  &  \checkmark  & \checkmark    &     & \checkmark (b)    &  \checkmark (b)    & \checkmark (b)   &       \\[0.1cm]
&  $c_{\varphi Q}^{(-)}$ &    & \checkmark  &    &  \checkmark   &     &  \checkmark   &   \checkmark   &   \checkmark (b)  &       \\
&  $c_{\varphi t}$ &    &  \checkmark   &    & \checkmark    &     & \checkmark    & \checkmark     &     &       \\[0.1cm]
&  $c_{\varphi l_i}^{(1)}$ &    &   &    &     &    &  \checkmark   &  \checkmark    &     & \checkmark      \\[0.1cm]
&  $c_{\varphi l_i}^{(3)}$ &    &   &\checkmark    & \checkmark    &    & \checkmark    &  \checkmark    & \checkmark    &  \checkmark     \\
&  $c_{\varphi e}$ &    &   &    &     &    &   \checkmark  &  \checkmark    &     &    \checkmark   \\
& $c_{\varphi \mu}$  &    &   &    &     &    &  \checkmark    & \checkmark     &    &       \\
& $c_{\varphi \tau}$  &    &   &    &     &    &    \checkmark   &  \checkmark    &   &       \\[0.1cm]
& $c_{\varphi q}^{(3)}$  &    &  \checkmark & \checkmark   &  \checkmark   &    &  \checkmark   &  \checkmark    & \checkmark    &   \checkmark    \\[0.1cm]
&  $c_{\varphi q}^{(-)}$ &    & \checkmark  &    & \checkmark    &    &  \checkmark   &  \checkmark    & \checkmark   &  \checkmark     \\
& $c_{\varphi u}$  &    &  \checkmark &    &  \checkmark   &    &    \checkmark &  \checkmark    &  \checkmark   &   \checkmark    \\
& $c_{\varphi d}$  &    &  \checkmark &    &  \checkmark   &    &  \checkmark   &  \checkmark    &   \checkmark  &    \checkmark   \\
\midrule
\multirow{6.5}{*}{purely} &  $c_{\varphi G}$ &    &   &    &     &     &  \checkmark   & \checkmark     &  \checkmark   &       \\
\multirow{6.5}{*}{bosonic} &  $c_{\varphi B}$ &    &   &    &     &     &   \checkmark  &  \checkmark    &  \checkmark   &       \\
&  $c_{\varphi W}$ &    &   &    &     &     &  \checkmark   &  \checkmark    & \checkmark    &       \\
&  $c_{\varphi d}$ &    &   &    &     &     &   \checkmark  &   \checkmark   &  \checkmark   &       \\
&  $c_{\varphi D}$ &    & \checkmark  & \checkmark   &   \checkmark  &     &  \checkmark   &  \checkmark    &  \checkmark   &    \checkmark   \\
&  $c_{\varphi W B}$ &    & \checkmark  &  \checkmark  &   \checkmark  &     &   \checkmark  & \checkmark     & \checkmark    &   \checkmark    \\
&  $c_{WWW}$  &    &   &    &     &     &     &      &     & \checkmark      \\
 \bottomrule
 \end{tabular}
 \caption{\small Overview indicating which EFT coefficients
   contribute to the theoretical description of each of the processes
   considered in this global analysis.
}
\label{table:operatorprocess}	
\end{table}

%% file: sec-settings.tex
\section{Fitting methodology}
\label{sec:fitsettings}

In this section we describe the fitting methodology that
is used in this work to map the EFT parameter space spanned by 
the Higgs, diboson, and top quark data.
In addition to results obtained with
the Monte Carlo replica fitting (MCfit) method
presented in Ref.~\cite{Hartland:2019bjb},
now we also determine the posterior probability 
distributions in the parameter space
using the {\tt MultiNest} Nested Sampling (NS) 
algorithm~\cite{Feroz:2013hea,Feroz:2007kg},
a robust sampling procedure that is completely orthogonal to
the MCfit method and that is based on Bayesian inference.

We begin with a brief discussion of the 
log-likelihood function and the treatment of uncertainties that is adopted
in the fit.
We then present the individual $\chi^2$ profiles associated to each
EFT coefficient in the quadratic fits
and discuss the eventual presence of (quasi-)degenerate minima.
Subsequently, the main features of the NS and MCFit strategies
used in the global fit are summarized, including 
several improvements that have been
implemented in the latter technique. The results obtained
with the two methods are also benchmarked.
Finally, we carry out a principal component analysis (PCA)
to determine the linear combinations of parameters that have
the highest and lowest variabilities
given our global dataset and
assess the possible presence of flat directions.

\subsection{Log-likelihood}
\label{sec:generalsettings}

The overall fit quality is quantified by the log-likelihood, 
or $\chi^2$ function, defined as
\begin{equation}
  \chi^2\lp {\boldsymbol c} \rp \equiv \frac{1}{n_{\rm dat}}\sum_{i,j=1}^{n_{\rm dat}}\lp 
  \sigma^{(\rm th)}_i\lp {\boldsymbol c} \rp
  -\sigma^{(\rm exp)}_i\rp ({\rm cov}^{-1})_{ij}
\lp 
  \sigma^{(\rm th)}_j\lp {\boldsymbol c}\rp
  -\sigma^{(\rm exp)}_j\rp
 \label{eq:chi2definition2}
    \; ,
\end{equation}
where $\sigma_i^{\rm (exp)}$ and
$\sigma^{(\rm th)}_i\lp {\boldsymbol c}\rp$ are the
central experimental data and corresponding theoretical
predictions for the $i$-th cross-section, respectively.
The total covariance 
matrix, ${\rm cov}_{ij}$, should contain all relevant sources of
experimental and theoretical uncertainties.
Assuming the latter are normally
distributed, and that they are uncorrelated
with the experimental uncertainties,
this total covariance matrix can be expressed as
a sum of the separate experimental and theoretical covariance 
matrices~\cite{AbdulKhalek:2019ihb,AbdulKhalek:2019bux},
\be
\label{eq:covmatsplitting}
{\rm cov}_{ij} = {\rm cov}^{(\rm exp)}_{ij} + {\rm cov}^{(\rm th)}_{ij} \, .
\ee
As usual, the experimental covariance matrix is constructed from all
sources of statistical and systematic uncertainties that are
made available by the experiments (as discussed
in Sect.~\ref{sec:settings_expdata}).
Moreover, the correlated multiplicative uncertainties are treated
via the `$t_0$' prescription~\cite{Ball:2009qv} in the fit, while the standard
experimental definition is used to quote the resulting $\chi^2$ values.

Concerning the theoretical covariance matrix, ${\rm cov}^{(\rm th)}$, its contributions depend
on the specific type of processes considered.
In the case of the top quark and LHC diboson production cross-sections,
we compute the SM predictions using the best possible
theoretical accuracy.
In doing so, we also evaluate the uncertainty associated
to the input PDFs and their correlations, as discussed in Ref.~\cite{Hartland:2019bjb}.
These computations are based on the NNPDF3.1 no-top fit~\cite{Ball:2017nwa}, a global
PDF determination based on a dataset that excludes
all measurements that are used in the present SMEFT analysis.

For the Higgs production
and decay measurements,
we take instead the SM predictions from the experimental publications,
which in turn are collected from the HXSWG reports.
In such a case, the total theory uncertainty is available, which includes both
PDF errors and missing higher order uncertainties (MHOUs).
The total theory uncertainty for Higgs measurements is therefore
included in the fit covariance matrix by means
of the correlation prescription provided in the corresponding ATLAS and CMS
publications.

\subsection{Individual fits from the $\chi^2$ profiles}
\label{sec:quarticfits}

Individual (one-parameter) fits correspond to varying a single EFT coefficient while keeping
the rest fixed to their SM values.
While such fits neglect the correlations between the
different coefficients, they provide a useful 
baseline for the global analysis,
since there the CL intervals will be by construction looser (or at best, similar)
as compared to those of the one-parameters fits.
They are also computationally inexpensive, 
as they can be carried out analytically 
from a scan of the $\chi^2$ profile
without resorting to numerical methods.
Another benefit is that 
they facilitate the comparison between different
EFT analyses, which 
may adopt different fitting bases but whose individual bounds 
should be similar provided they are based on comparable 
data sets and theoretical calculations.

In the scenario where a single EFT coefficient, $c_j$, is allowed to vary while the rest are set to zero,
the theoretical cross-section (for $\Lambda=1$ TeV) given by 
Eq.~(\ref{eq:quadraticTHform}) simplifies to
\be
\label{eq:quadraticTHform_simplified}
\sigma_m^{\rm (th)}(c_j)= \sigma_m^{\rm (sm)} + c_j\sigma^{(\rm eft)}_{m,j} +
c_j^2 \sigma^{(\rm eft)}_{m,jj} \, ,
\ee
which results in a quartic polynomial form for the $\chi^2$ when inserted into 
Eq.~(\ref{eq:chi2definition2}), namely
\be
\label{eq:quartic-chi2}
\chi^2(c_j) = \sum_{k=0}^4 a_k \lp c_j\rp^k \, .
\ee
Restricting the analysis to the linear order in the EFT expansion further
simplifies Eq.~(\ref{eq:quartic-chi2}) to a parabolic form,
\be
\label{eq:quadratic-chi2}
\chi^2(c_j) = \sum_{k=0}^2 a_k \lp c_j\rp^k = \chi^2_0 + b\lp  c_j-c_{j,0}\rp^2 \, ,
\ee
where $c_{j,0}$ is the value of $c_j$ at the minimum of the parabola,
and in this case  linear error propagation (Gaussian statistics) is applicable.

To determine the values of the quartic polynomial coefficients $ a_k$ in Eq.~(\ref{eq:quartic-chi2}),
it is sufficient to fit this functional form to a scan of the $\chi^2$ profile
obtained by varying the EFT coefficient
$c_j$ when all other coefficients are set to their SM value.
The associated
95\% CL interval to the coefficient $c_j$ can then be 
determined by imposing the condition
\be
\label{eq:deltachi2}
\chi^2(c_j)-\chi^2(c_{j,0}) \equiv \Delta \chi^2 \le 
  5.991\, .
\ee
We note that if the size of the quadratic
$\mathcal{O}\lp \Lambda^{-4}\rp$ corrections is sizable,
there will be more than one solution for $c_{j,0}$
and one might end up with  pairwise disjoint CL intervals.

\begin{figure}[htbp]
  \begin{center}
    \includegraphics[width=0.297\linewidth]{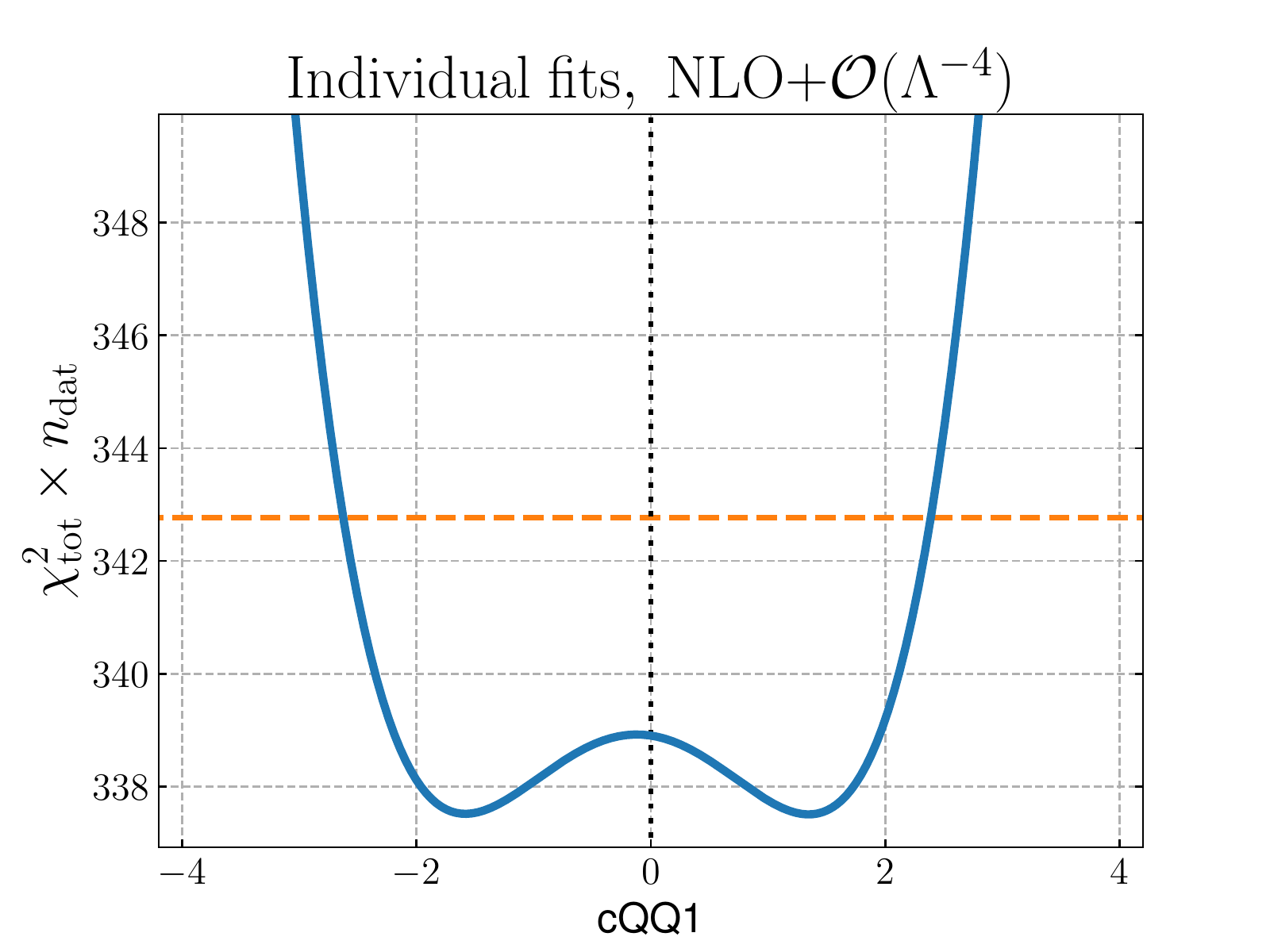}
\includegraphics[width=0.297\linewidth]{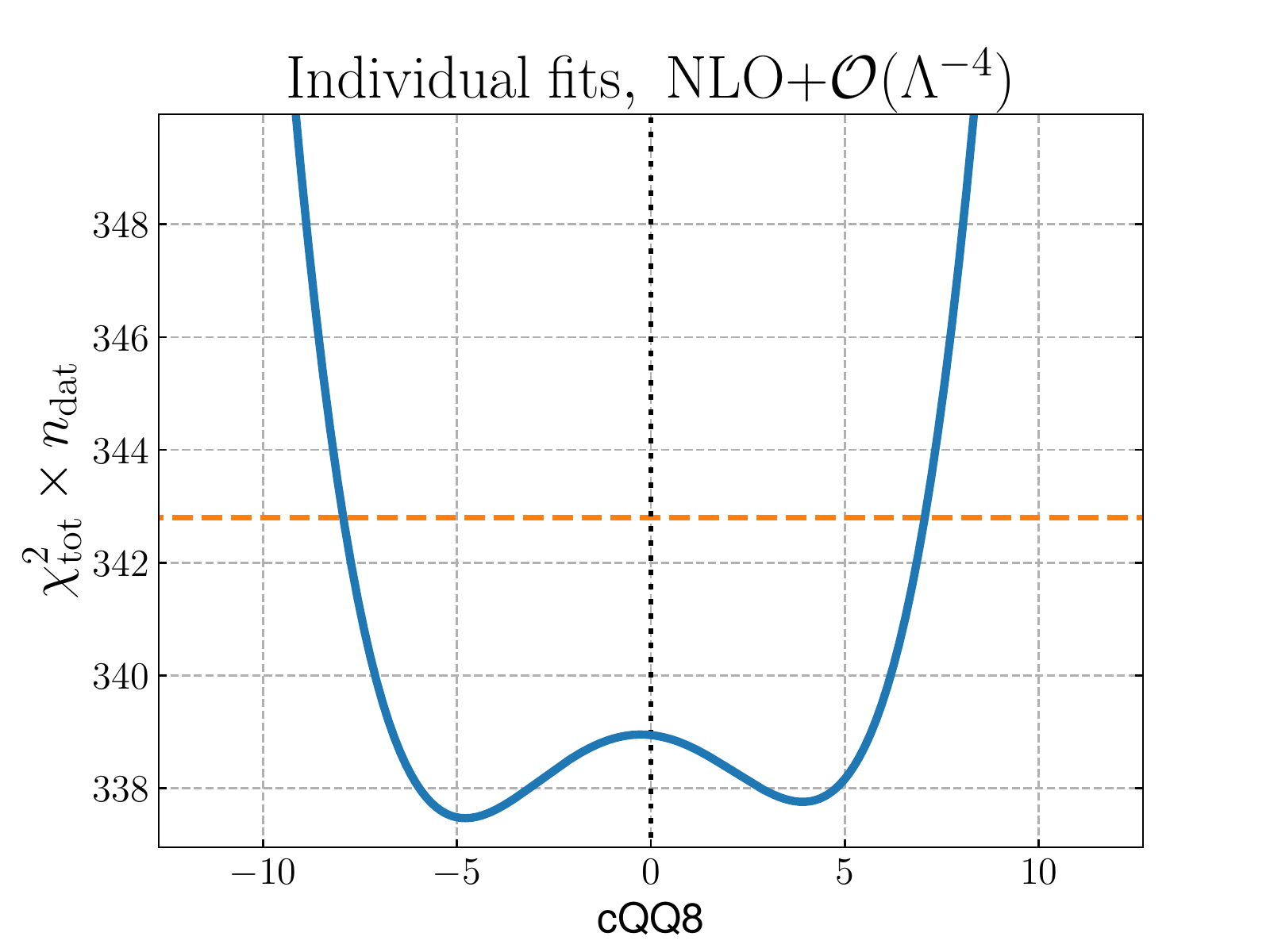}
\includegraphics[width=0.297\linewidth]{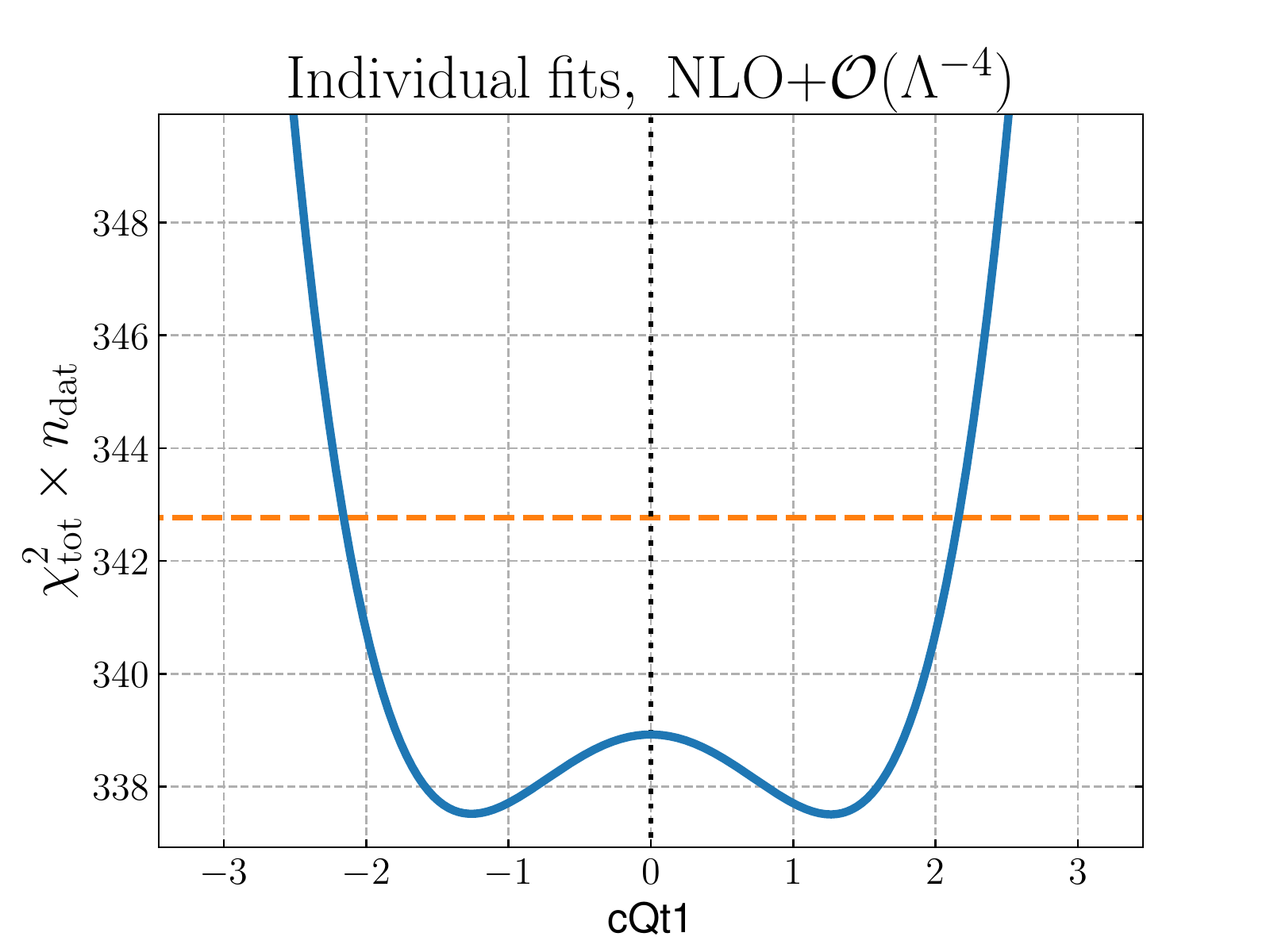}
\includegraphics[width=0.297\linewidth]{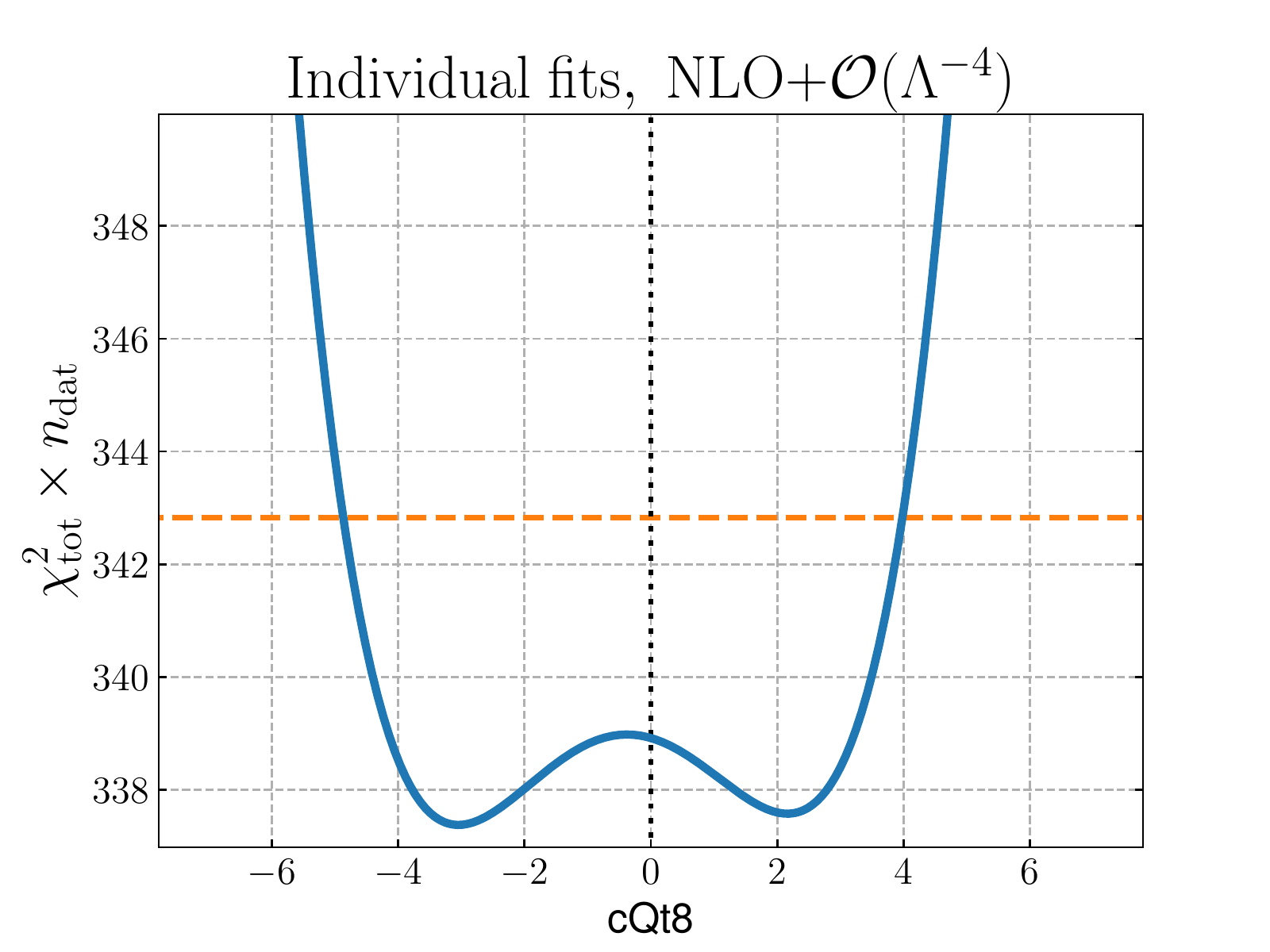}
\includegraphics[width=0.30\linewidth]{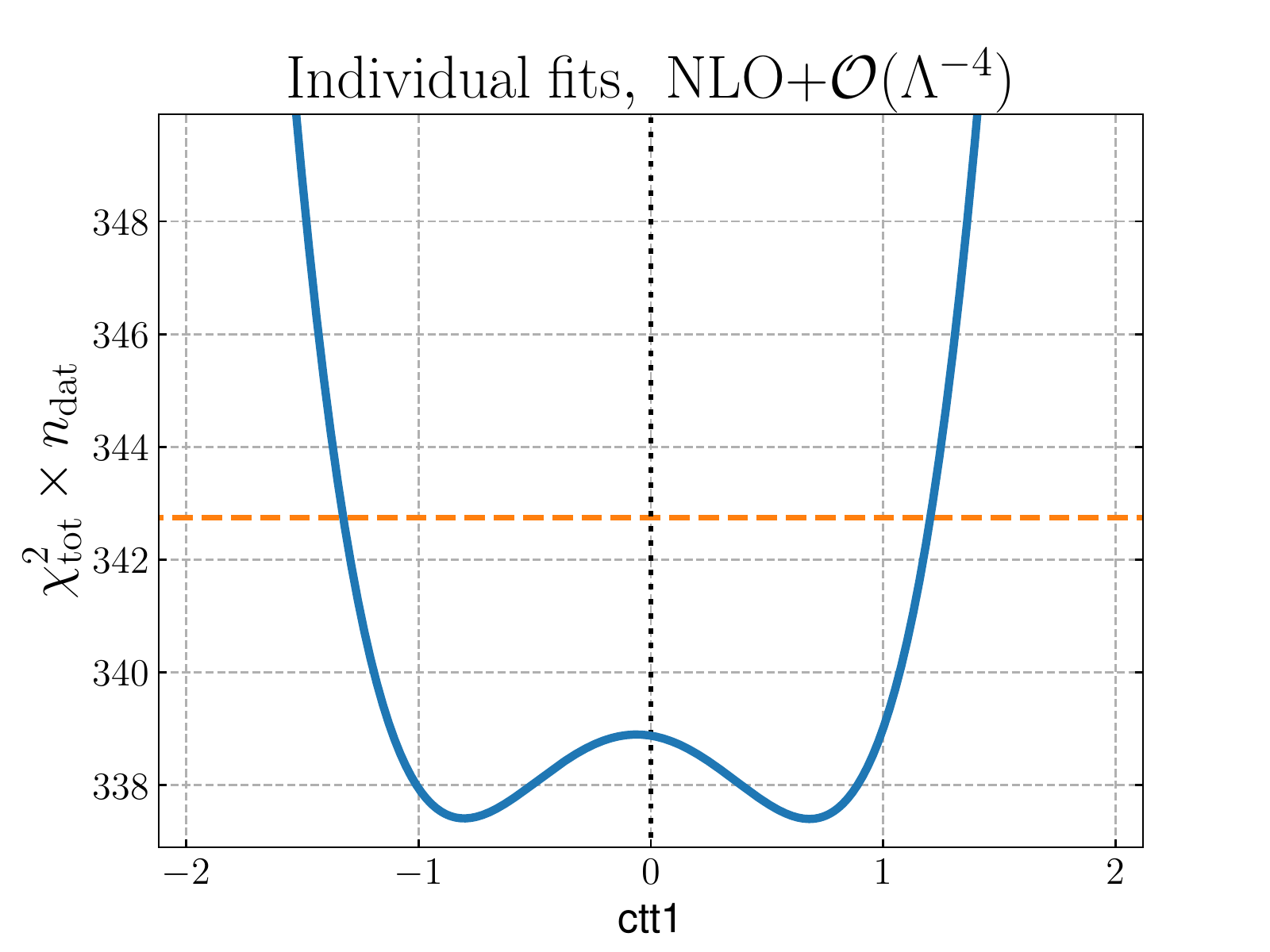}
\includegraphics[width=0.297\linewidth]{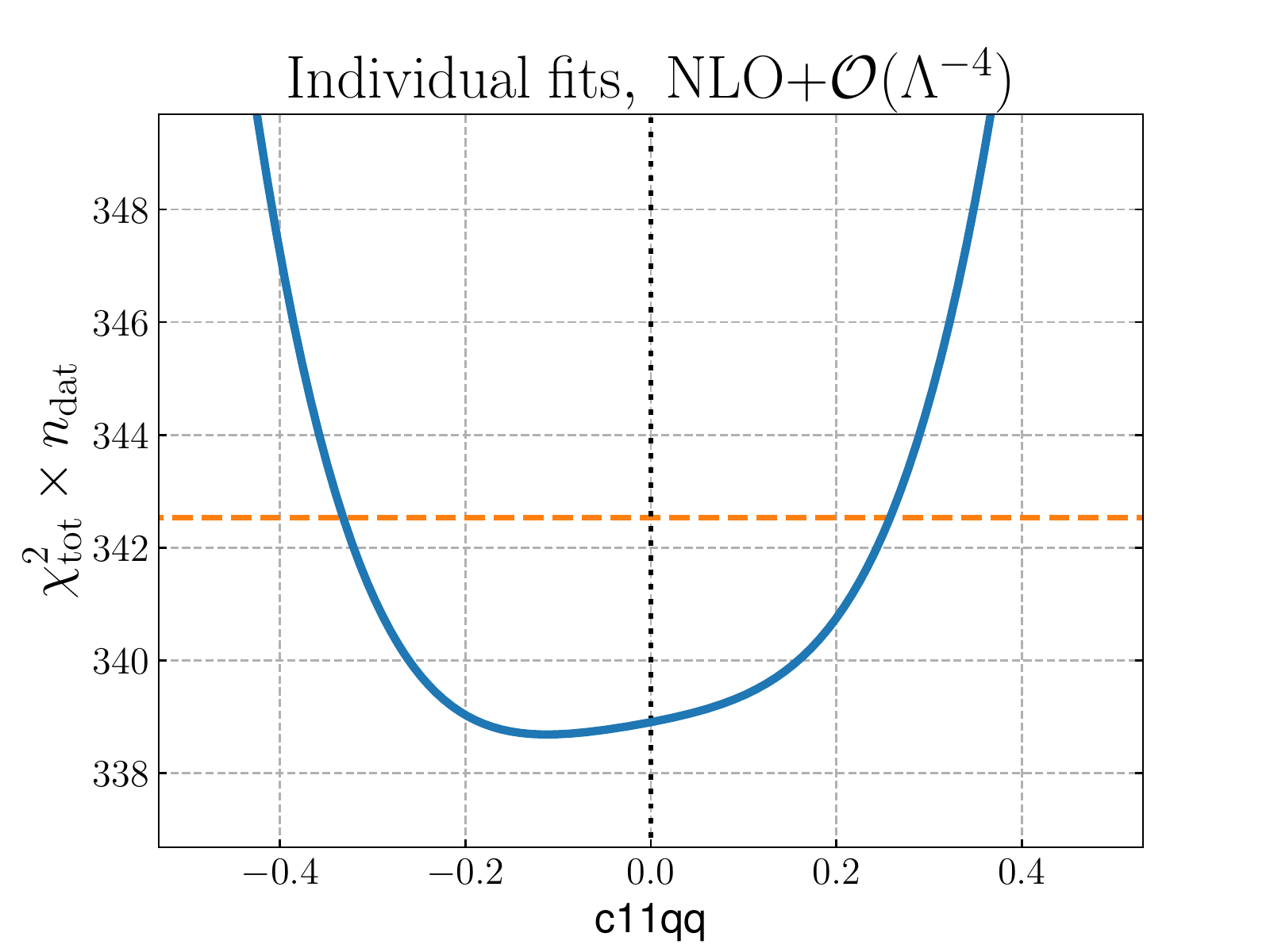}
\includegraphics[width=0.297\linewidth]{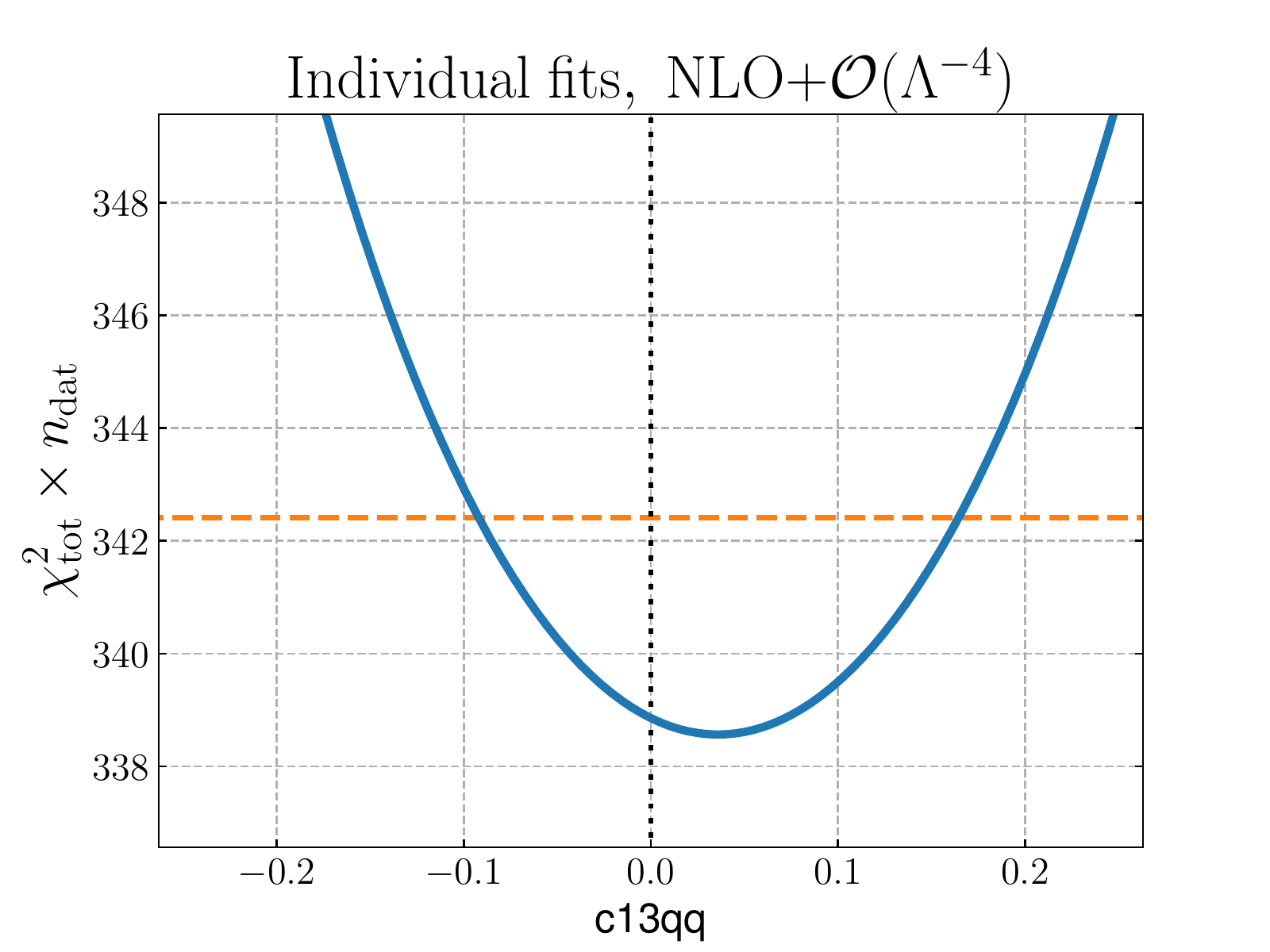}
\includegraphics[width=0.297\linewidth]{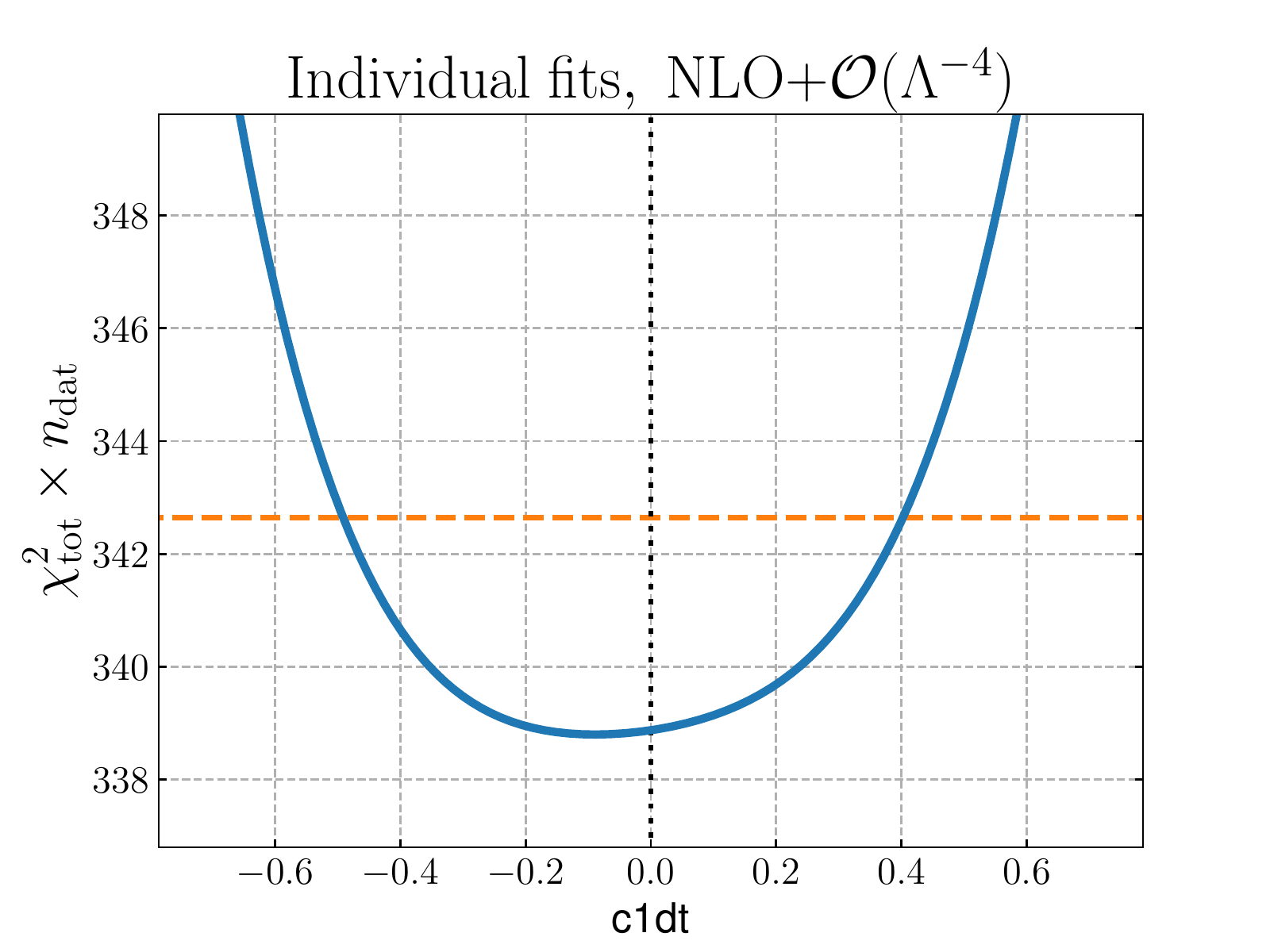}
\includegraphics[width=0.297\linewidth]{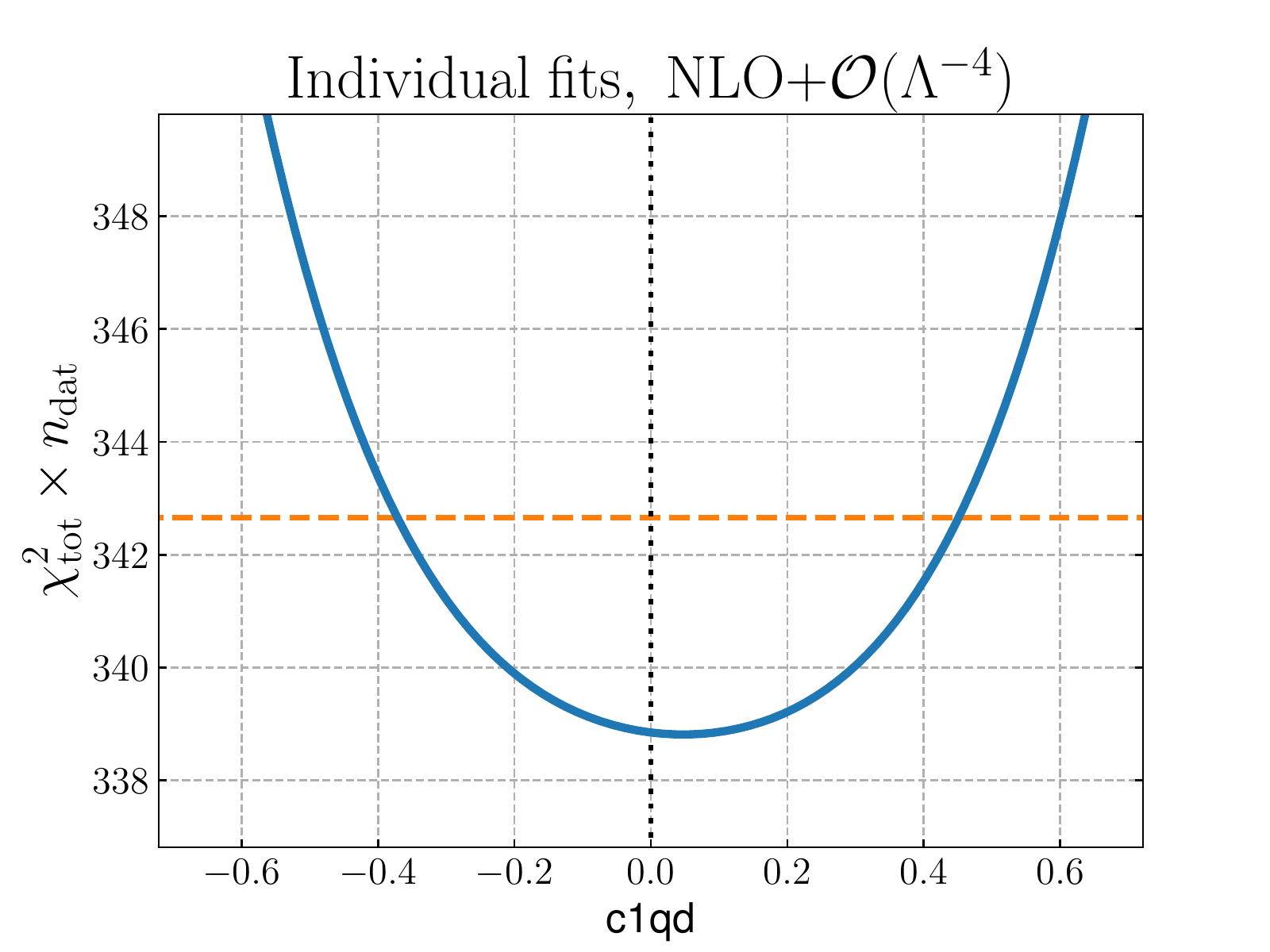}
\includegraphics[width=0.297\linewidth]{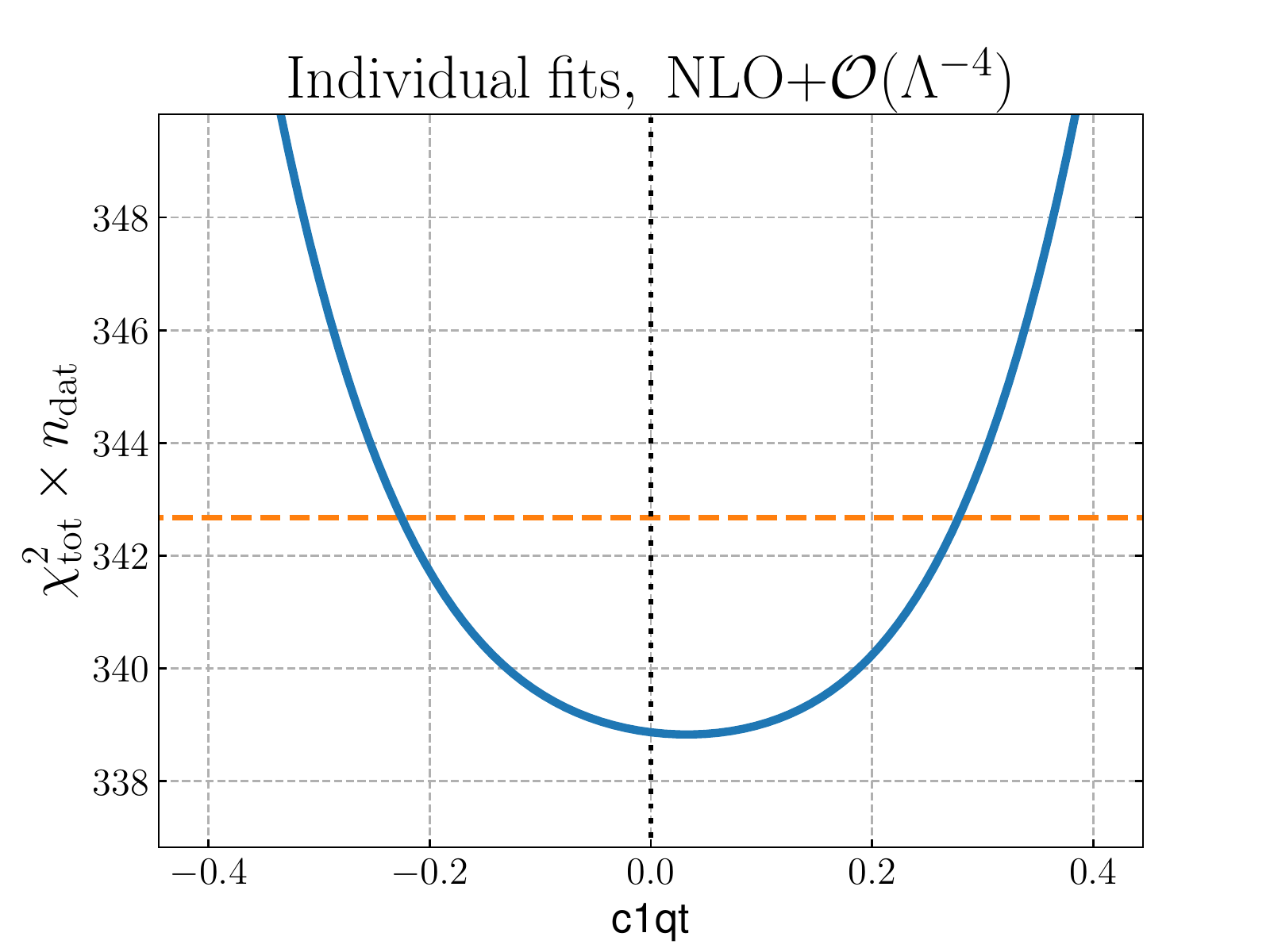}
\includegraphics[width=0.297\linewidth]{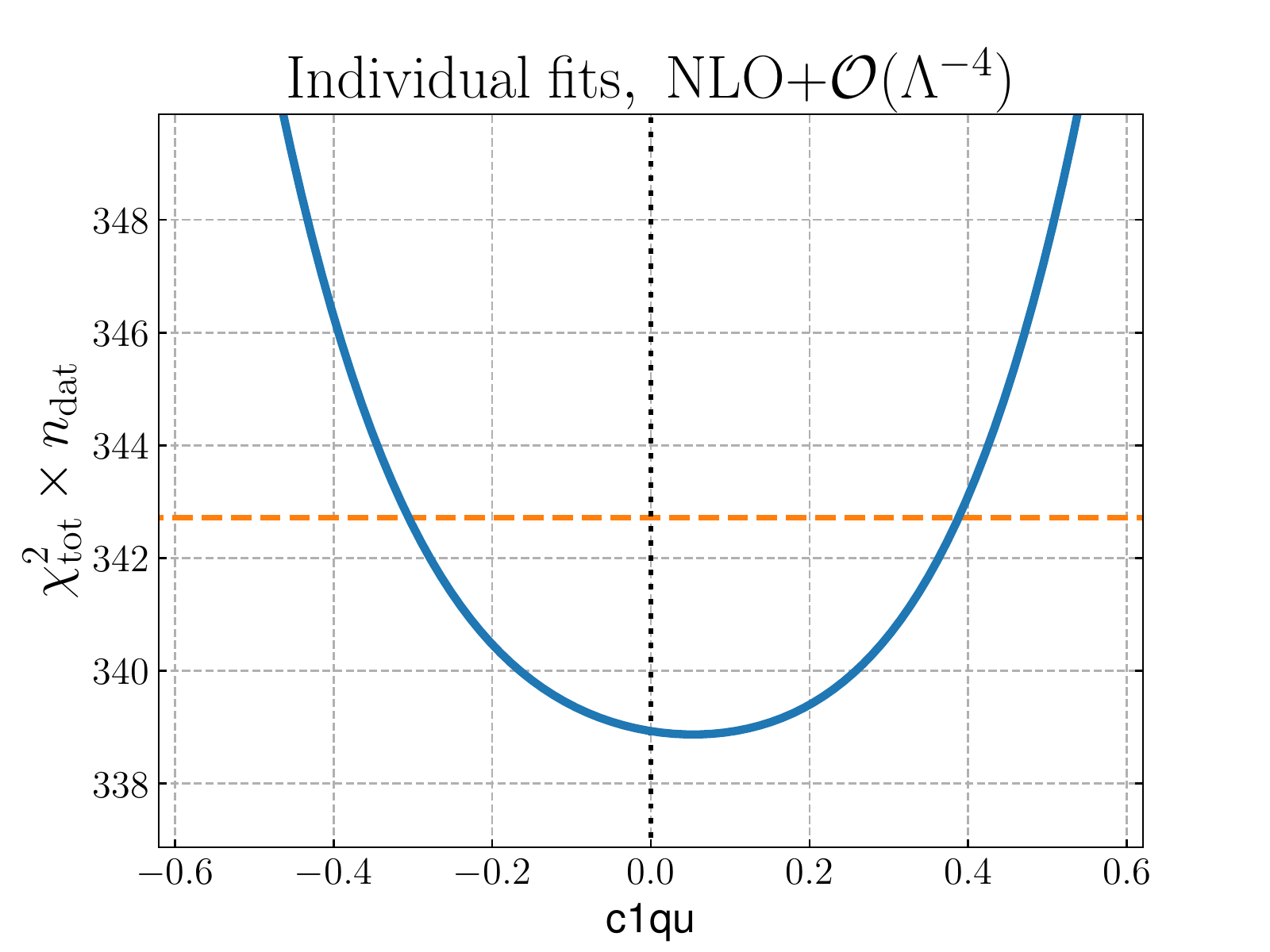}
\includegraphics[width=0.297\linewidth]{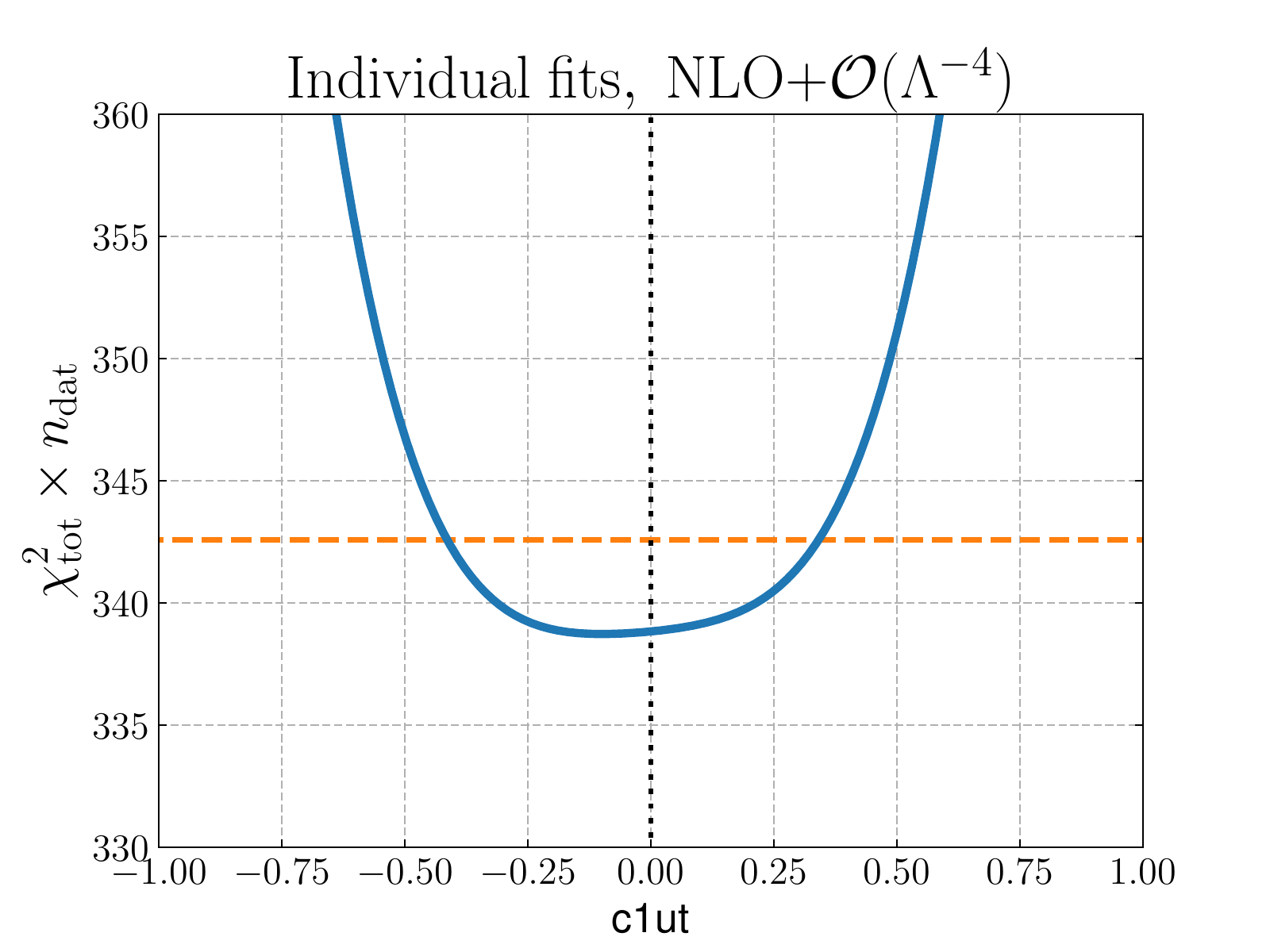}
\includegraphics[width=0.297\linewidth]{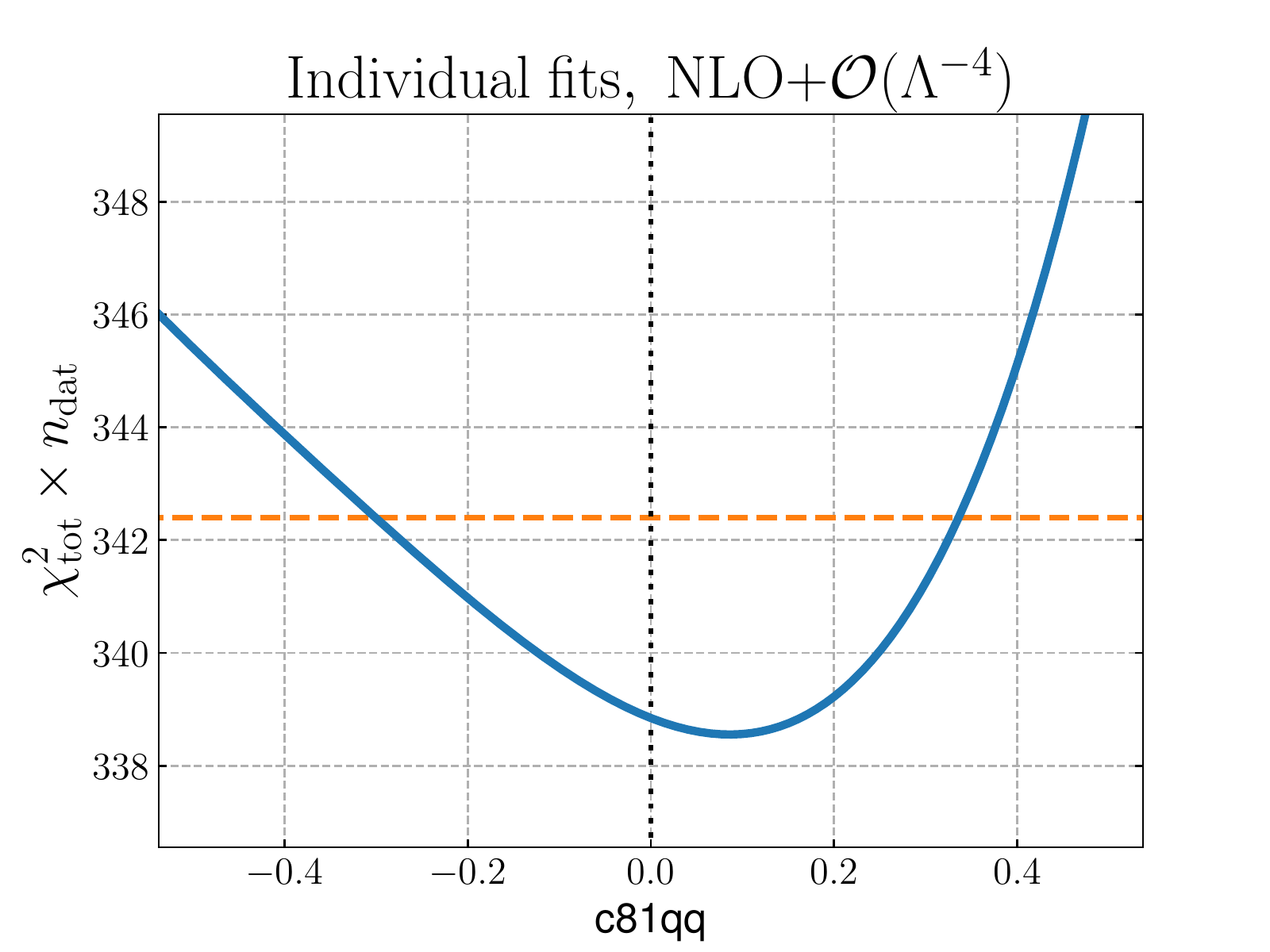}
\includegraphics[width=0.297\linewidth]{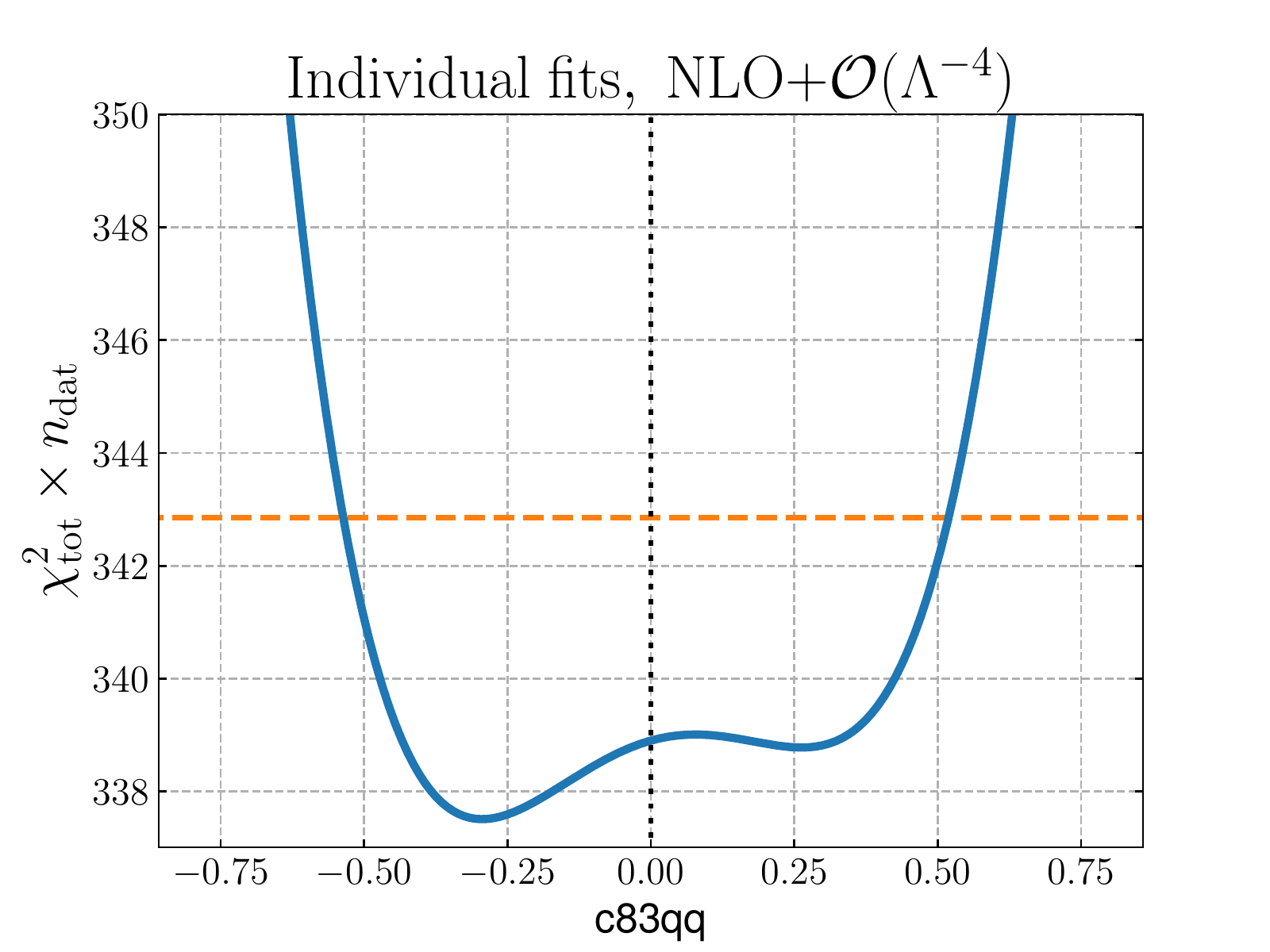}
\includegraphics[width=0.297\linewidth]{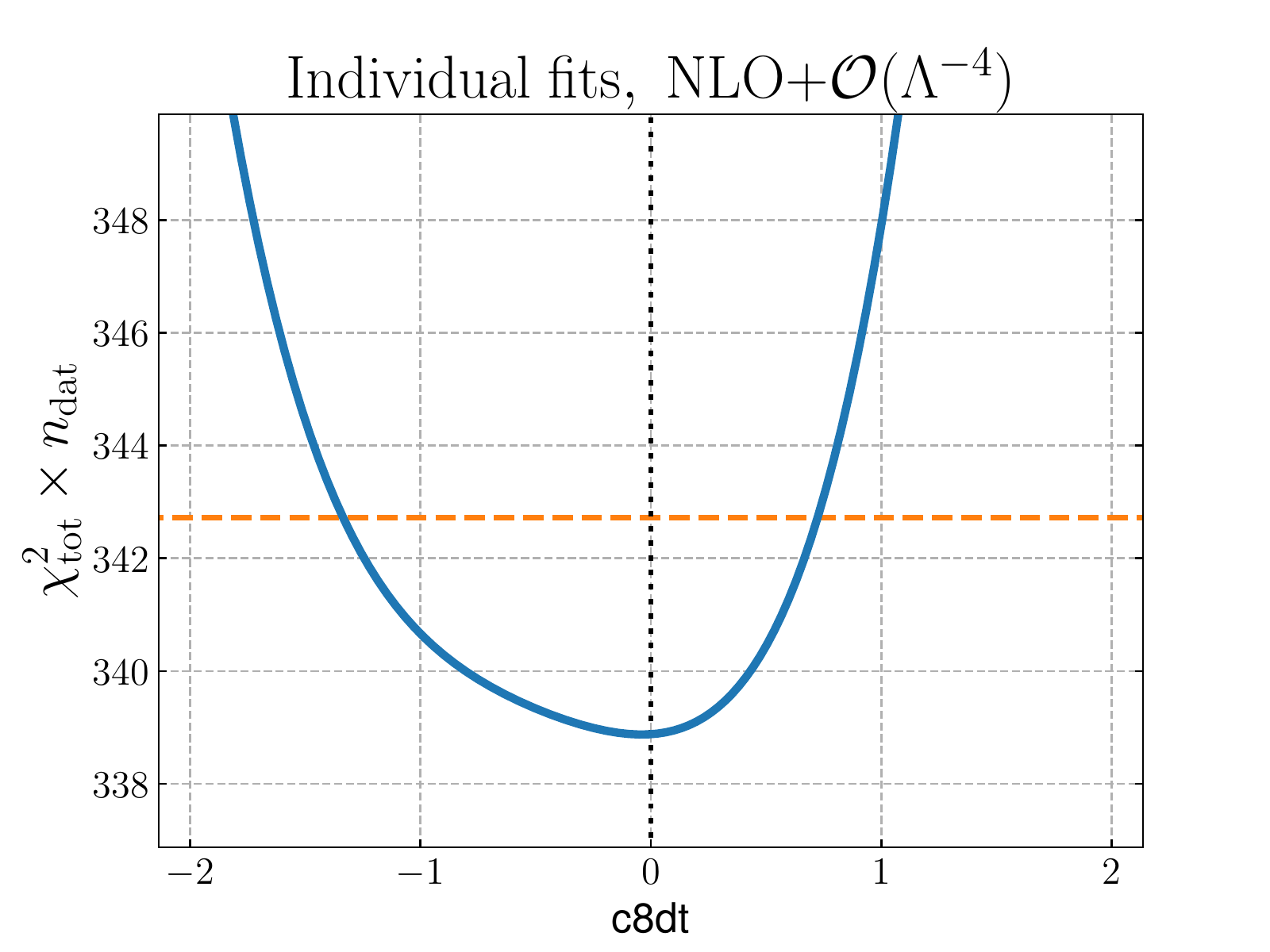}
\includegraphics[width=0.297\linewidth]{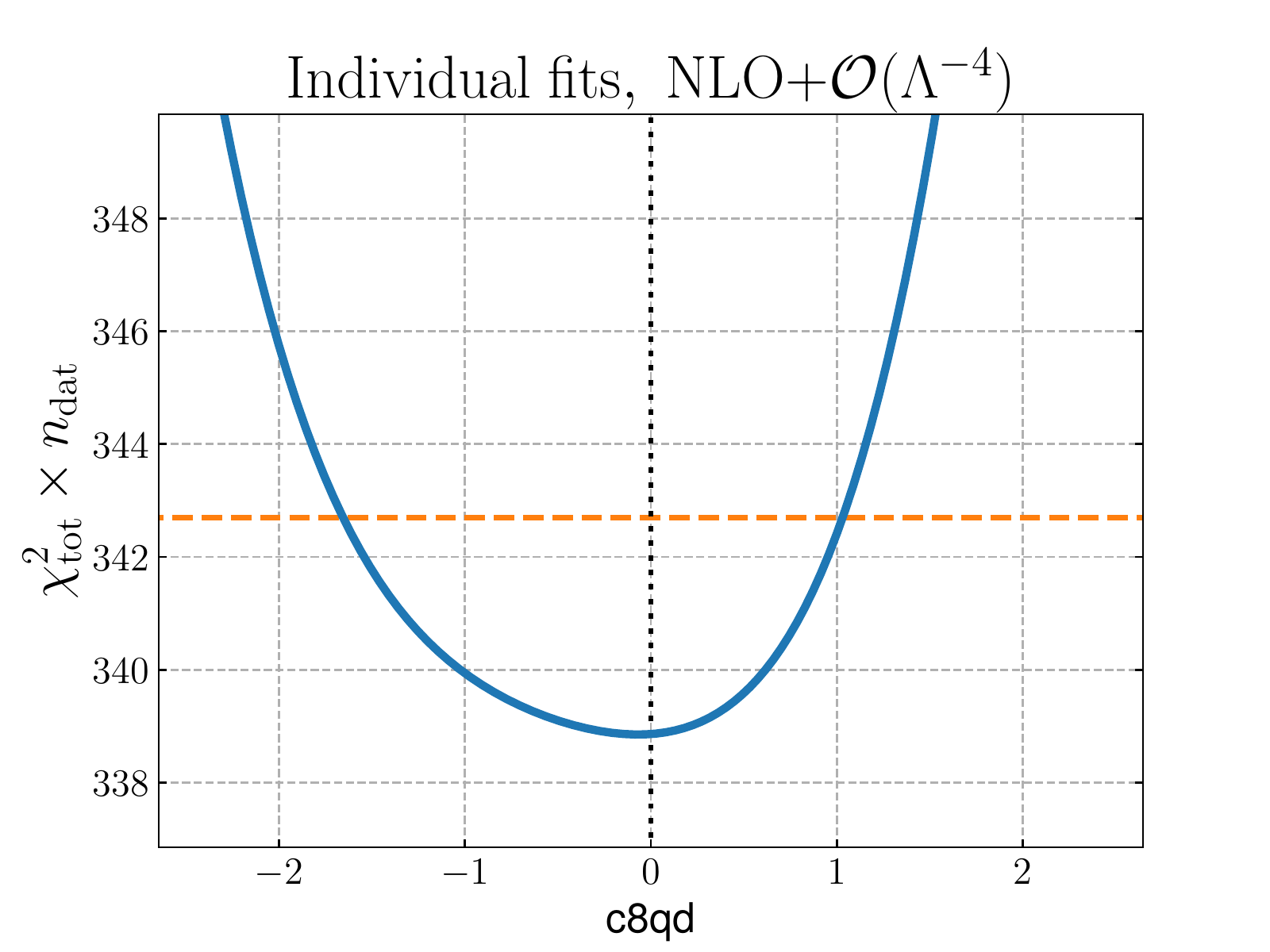}
\includegraphics[width=0.297\linewidth]{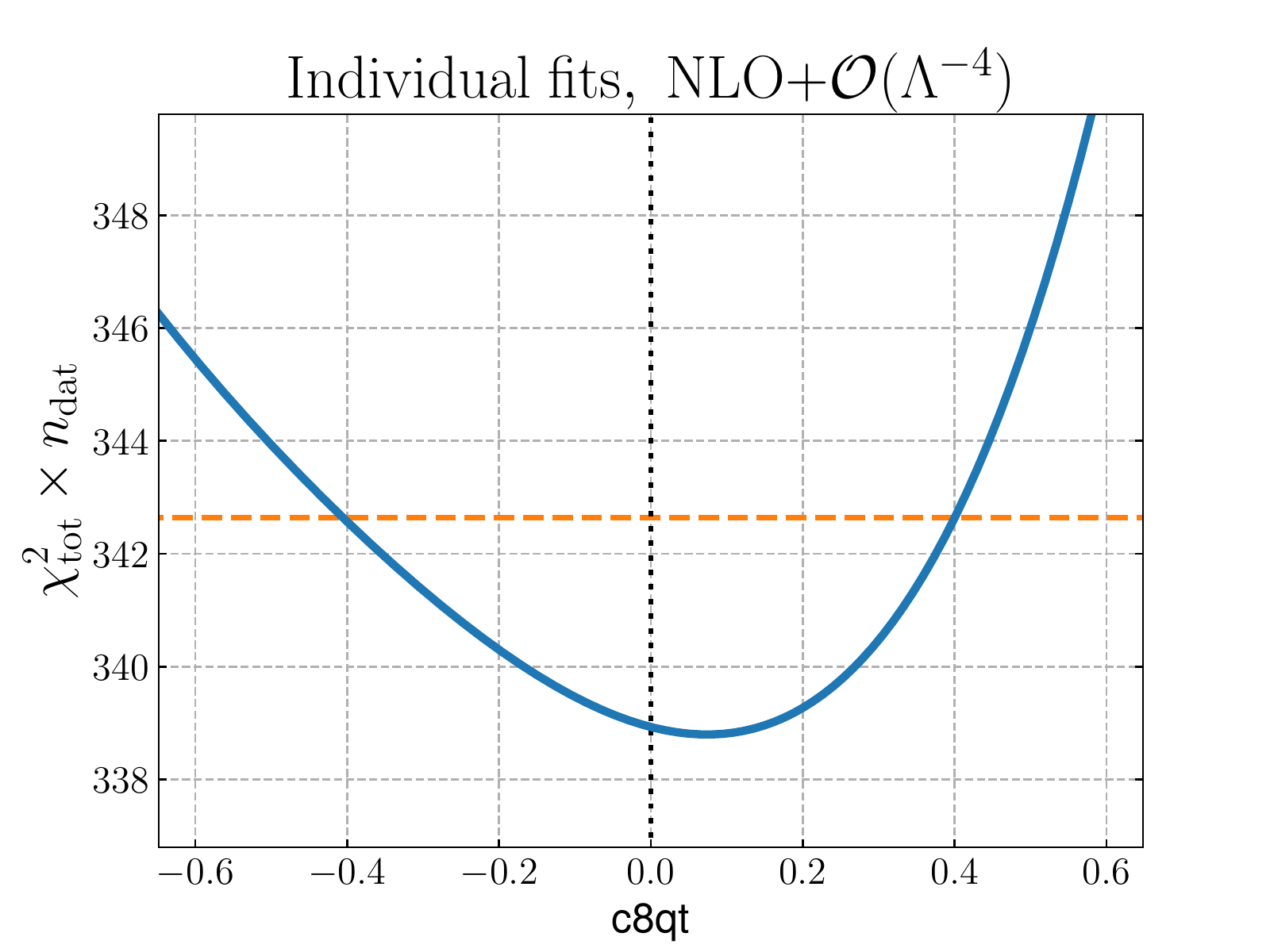}
\includegraphics[width=0.297\linewidth]{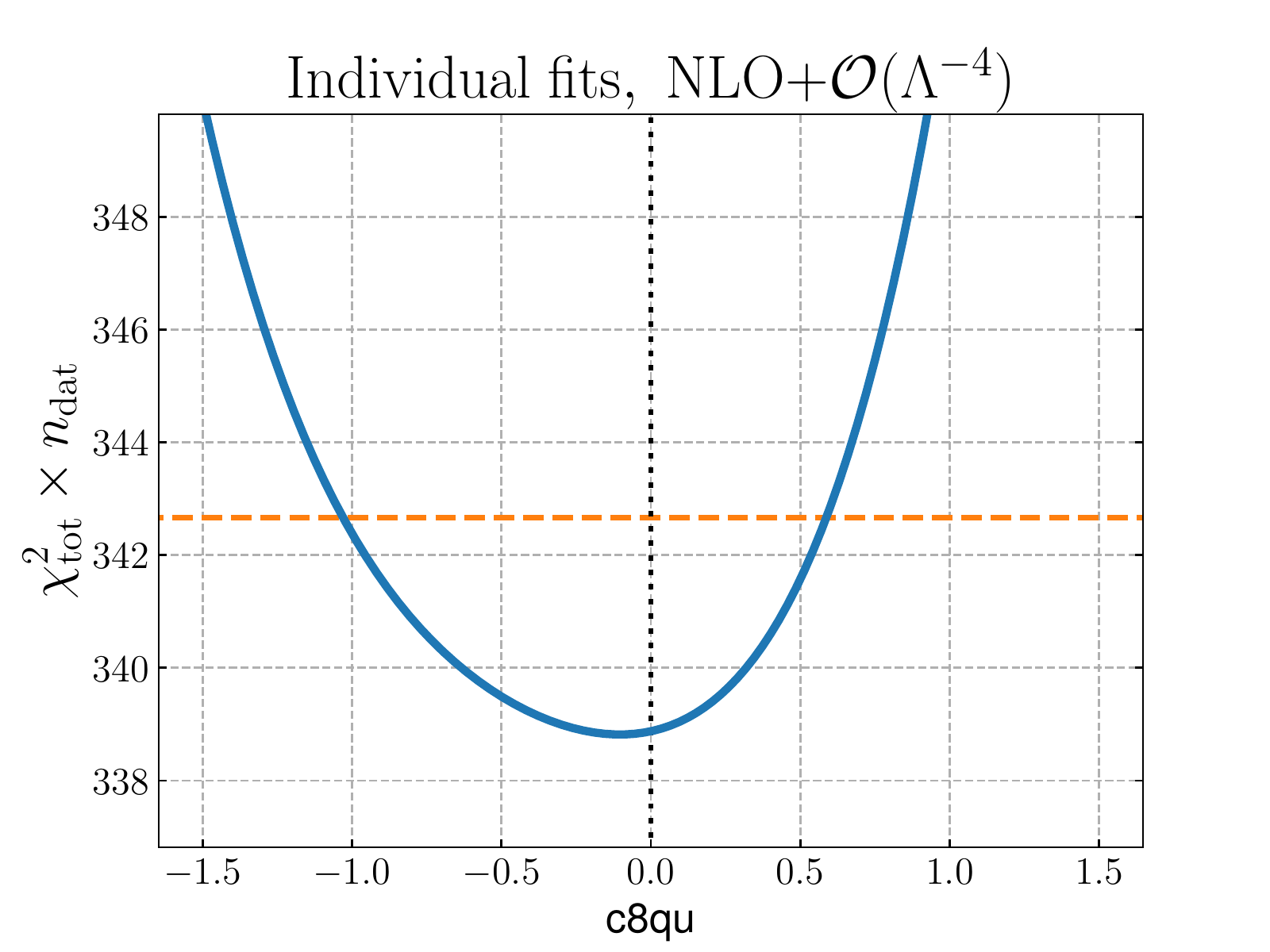}
\caption{\small Results of quartic polynomial fits
  to the $\chi^2$ profiles obtained  in one-parameter scans
  for each EFT coefficient, with all others set to their
  SM values.
  We show the absolute $\chi^2$ for the  $n_{\rm dat}=317$ data points
  of the global dataset calculated with the $t_0$ prescription,
  with the horizontal (vertical) line indicating the
  corresponding 95\% CL ranges (the SM prediction).
     \label{fig:quartic-individual-fits} }
  \end{center}
\end{figure}

\begin{figure}[htbp]
  \begin{center}
    \includegraphics[width=0.297\linewidth]{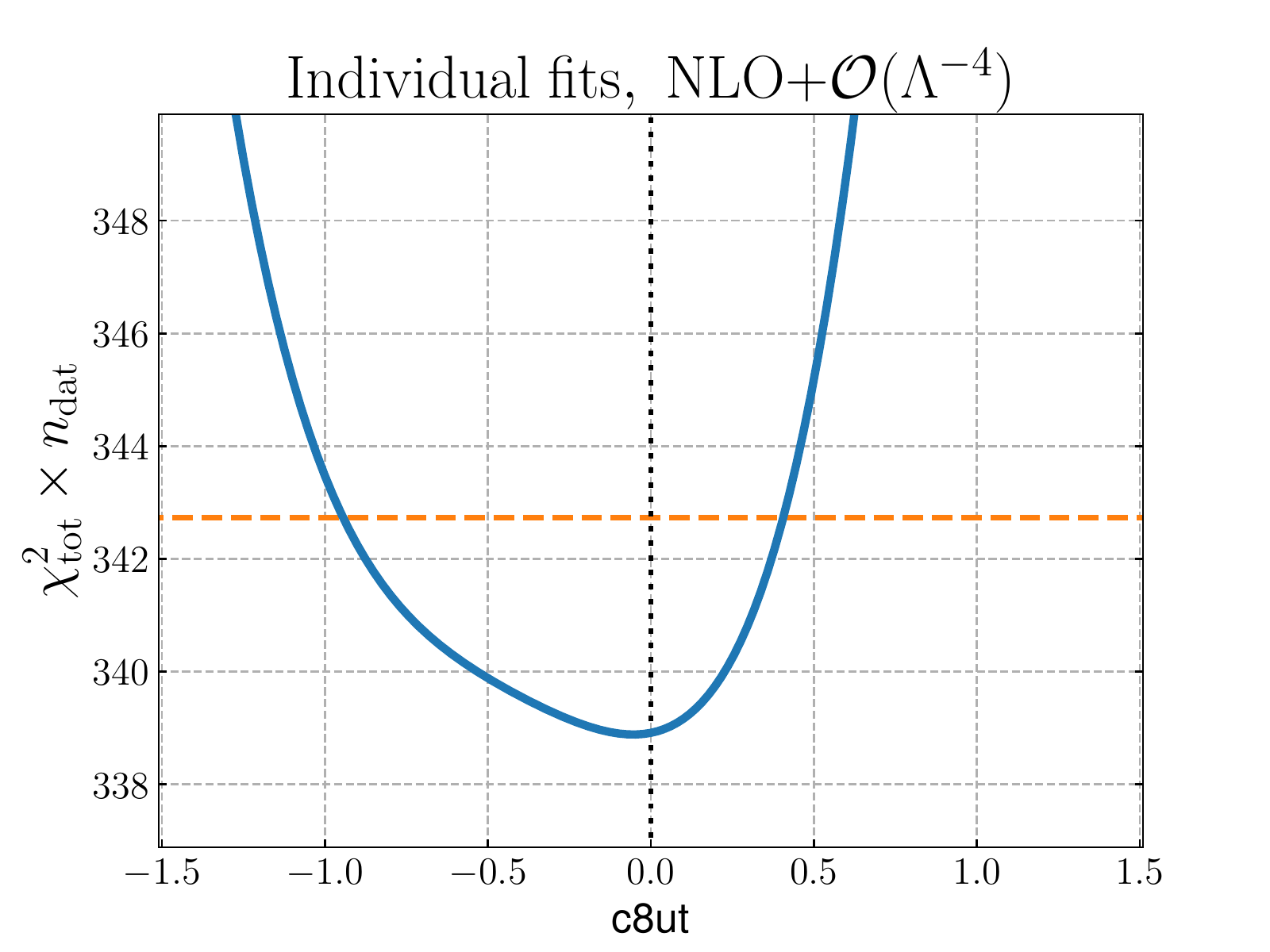}
\includegraphics[width=0.30\linewidth]{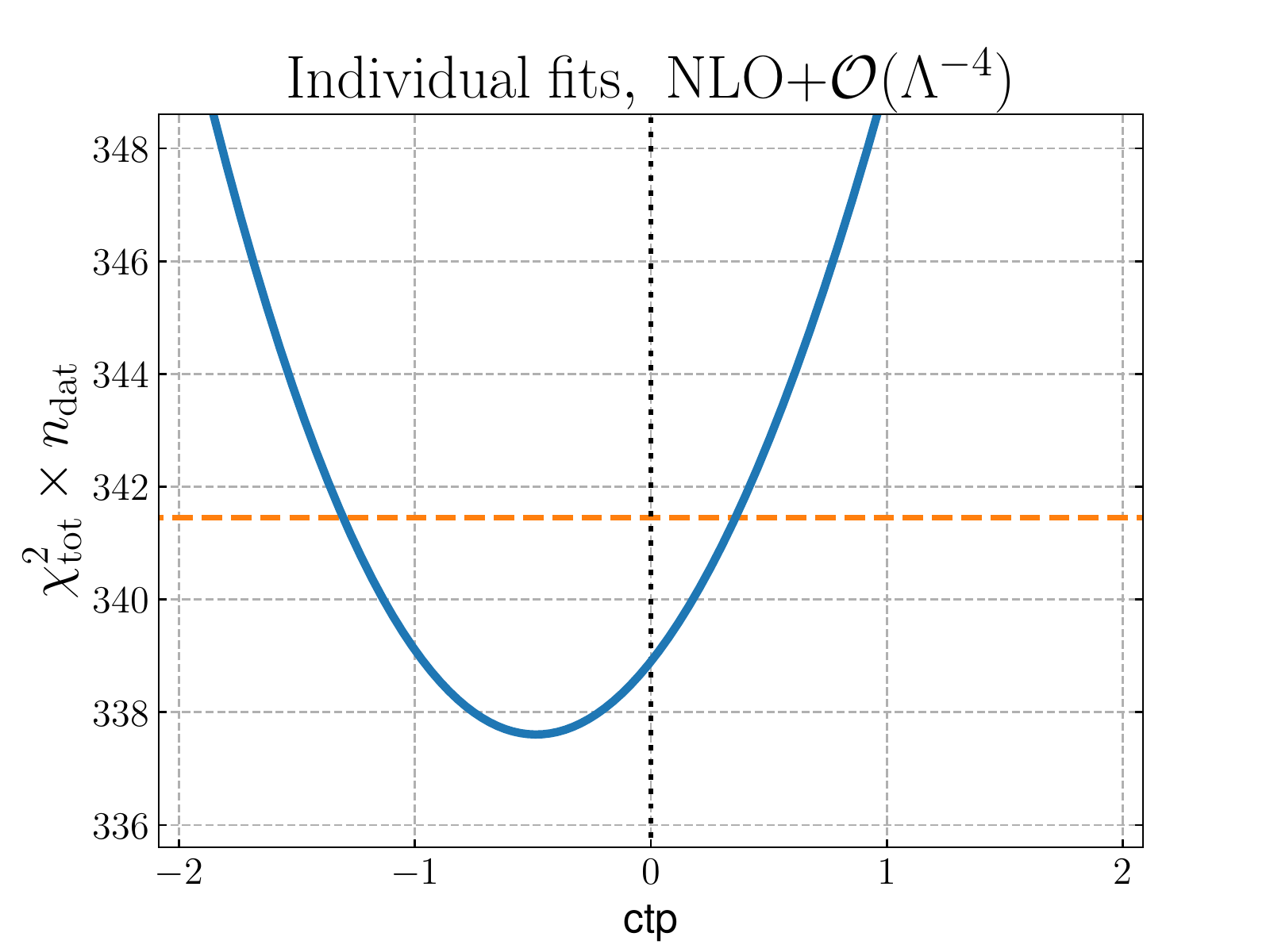}
\includegraphics[width=0.30\linewidth]{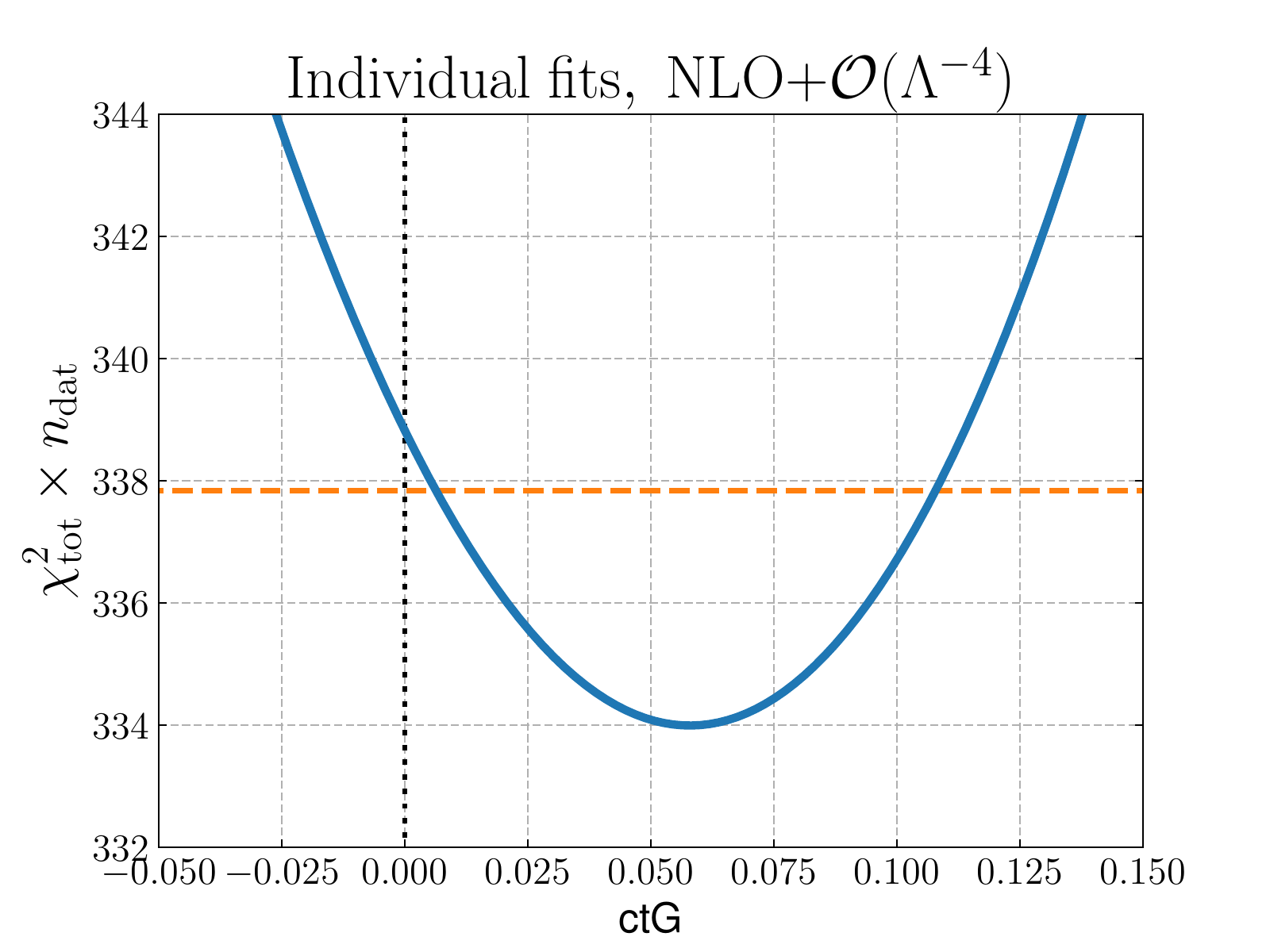}
\includegraphics[width=0.30\linewidth]{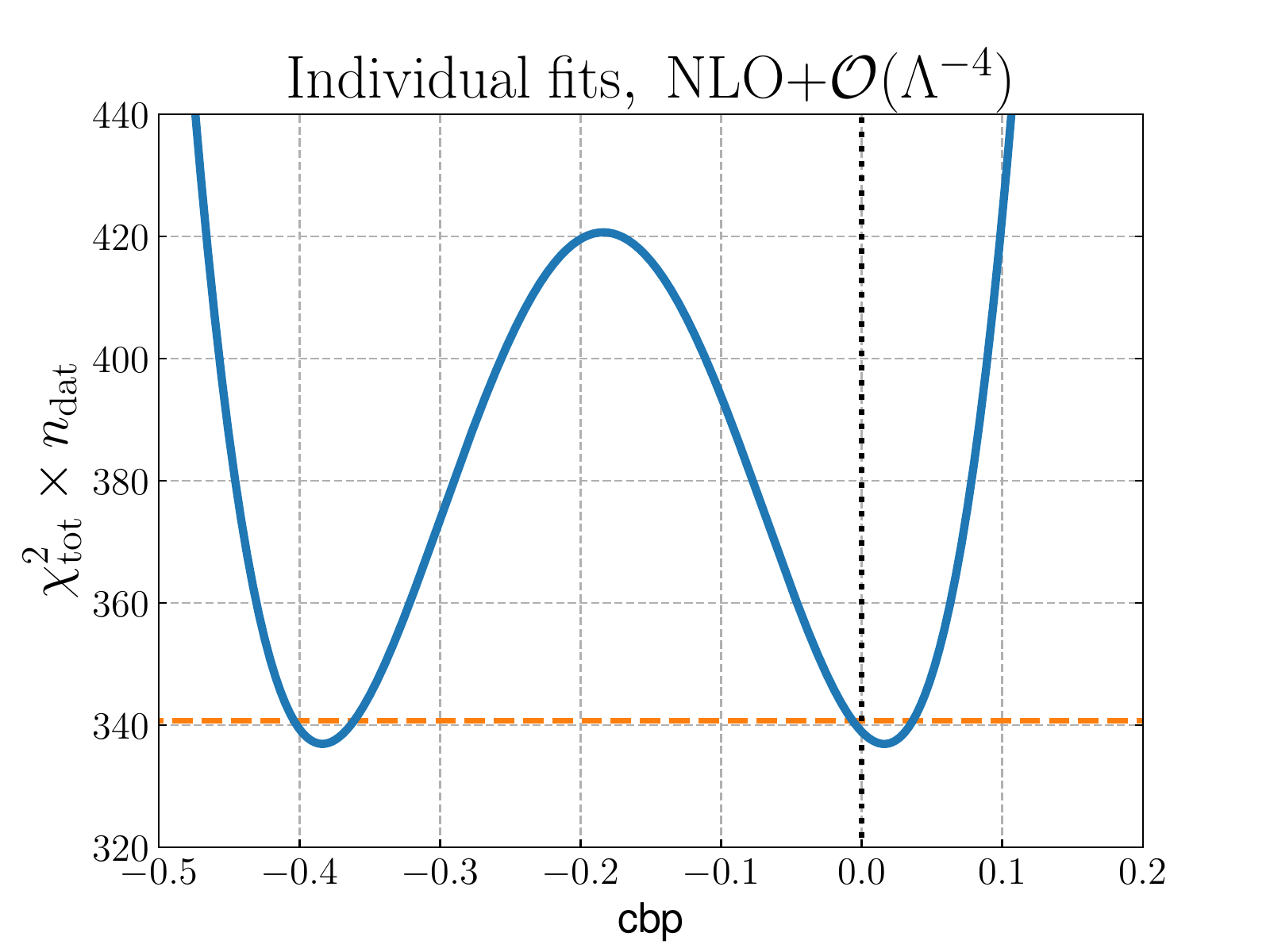}
\includegraphics[width=0.30\linewidth]{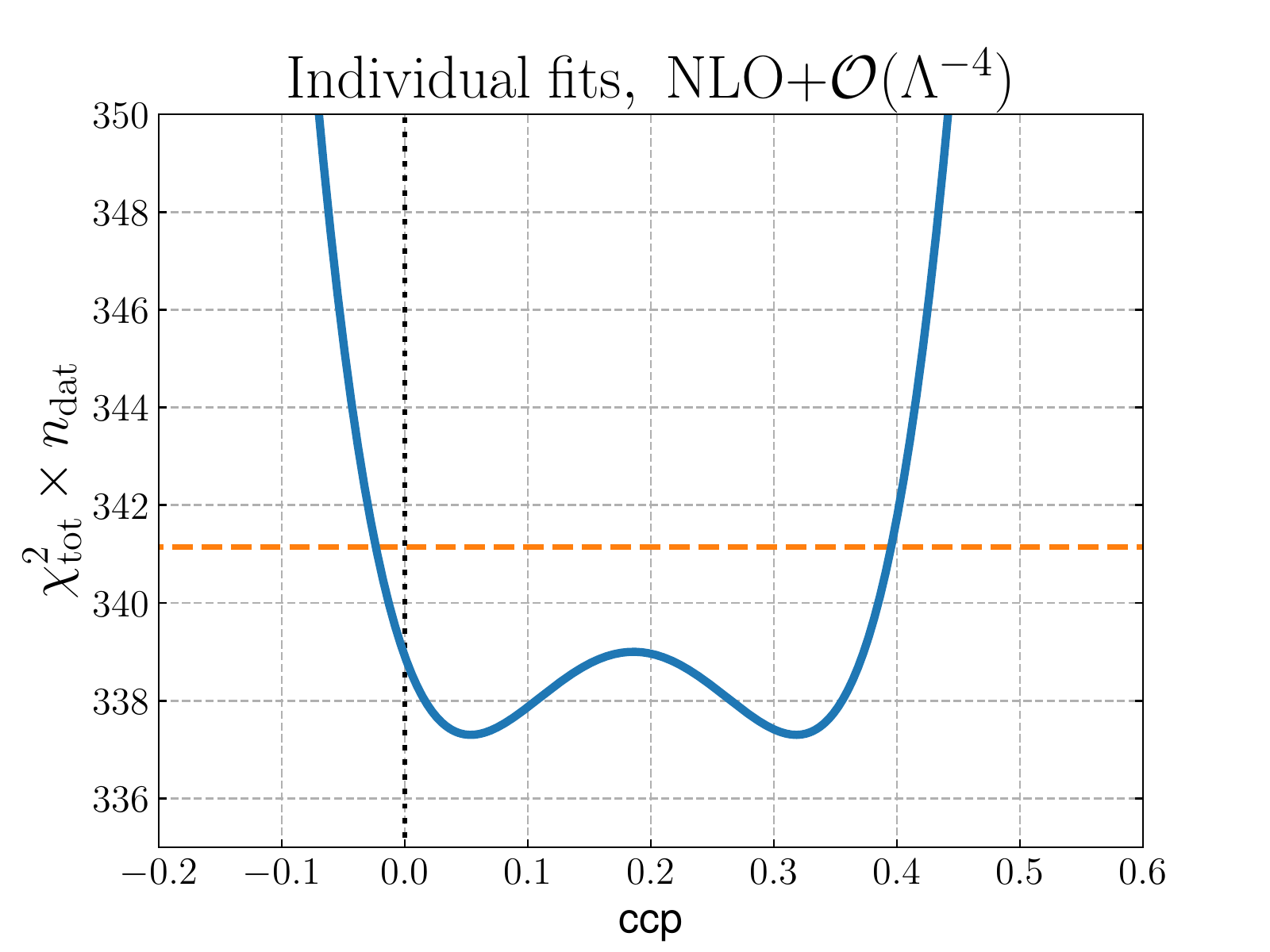}
\includegraphics[width=0.30\linewidth]{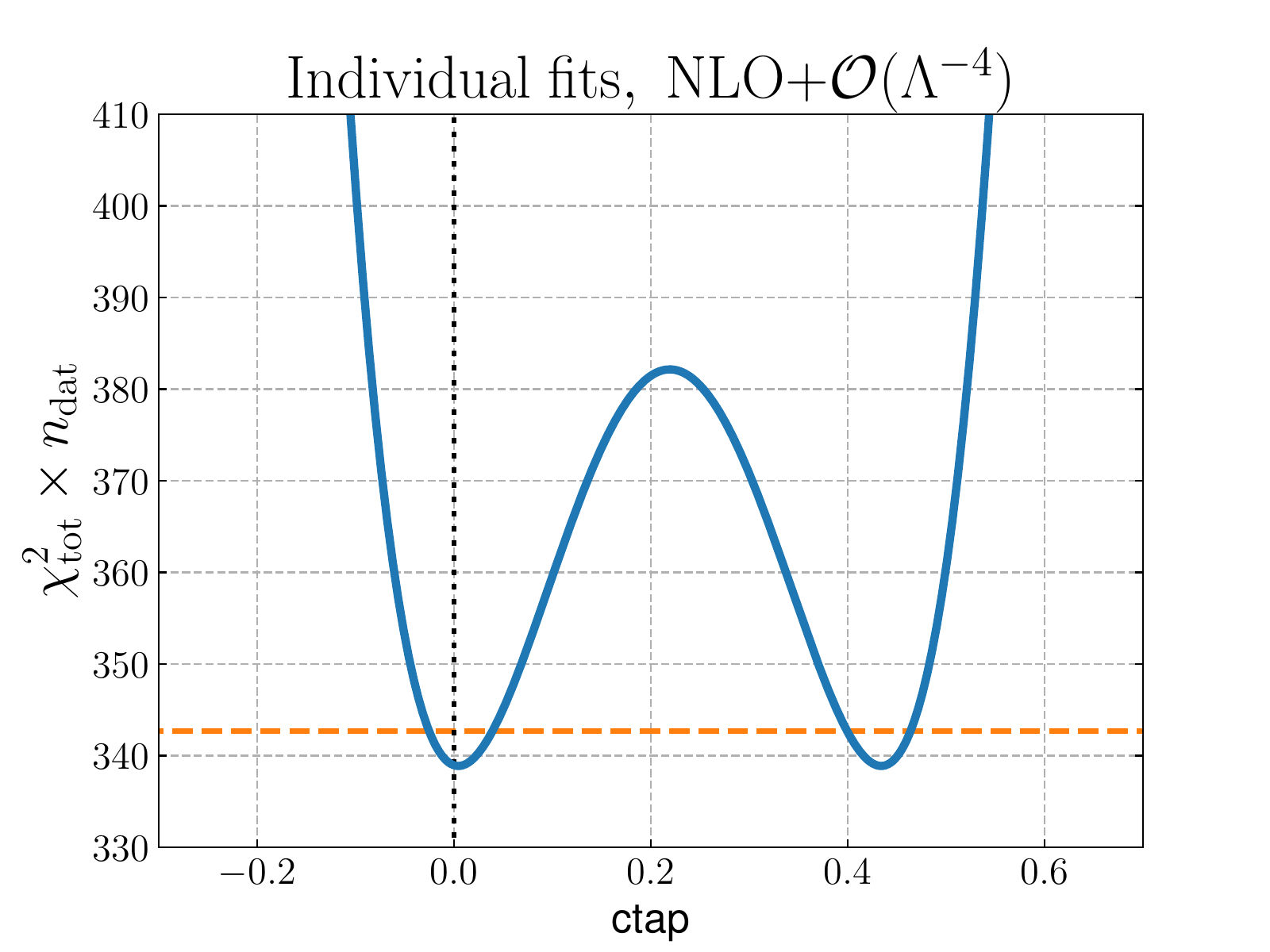}
\includegraphics[width=0.30\linewidth]{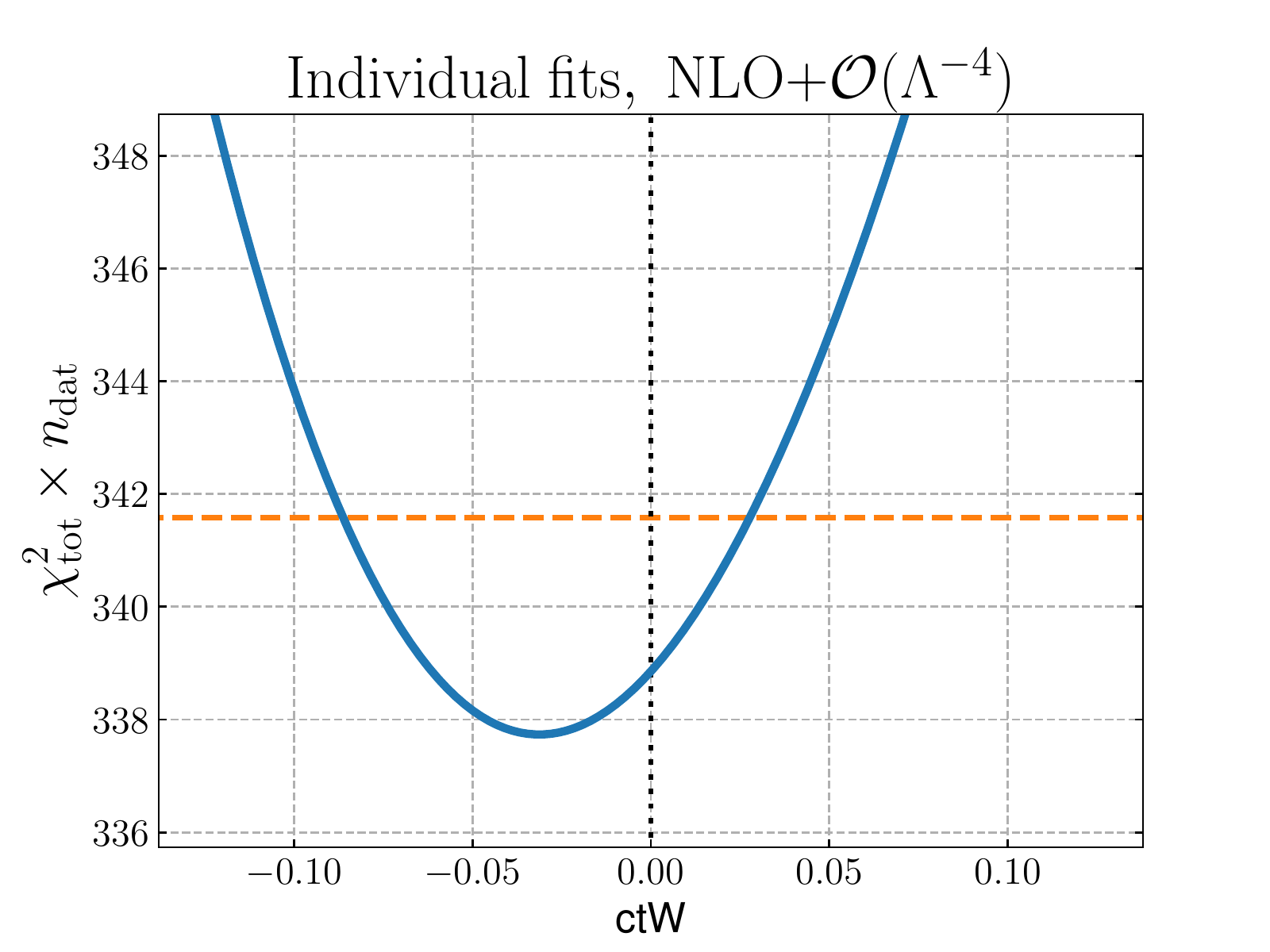}
\includegraphics[width=0.30\linewidth]{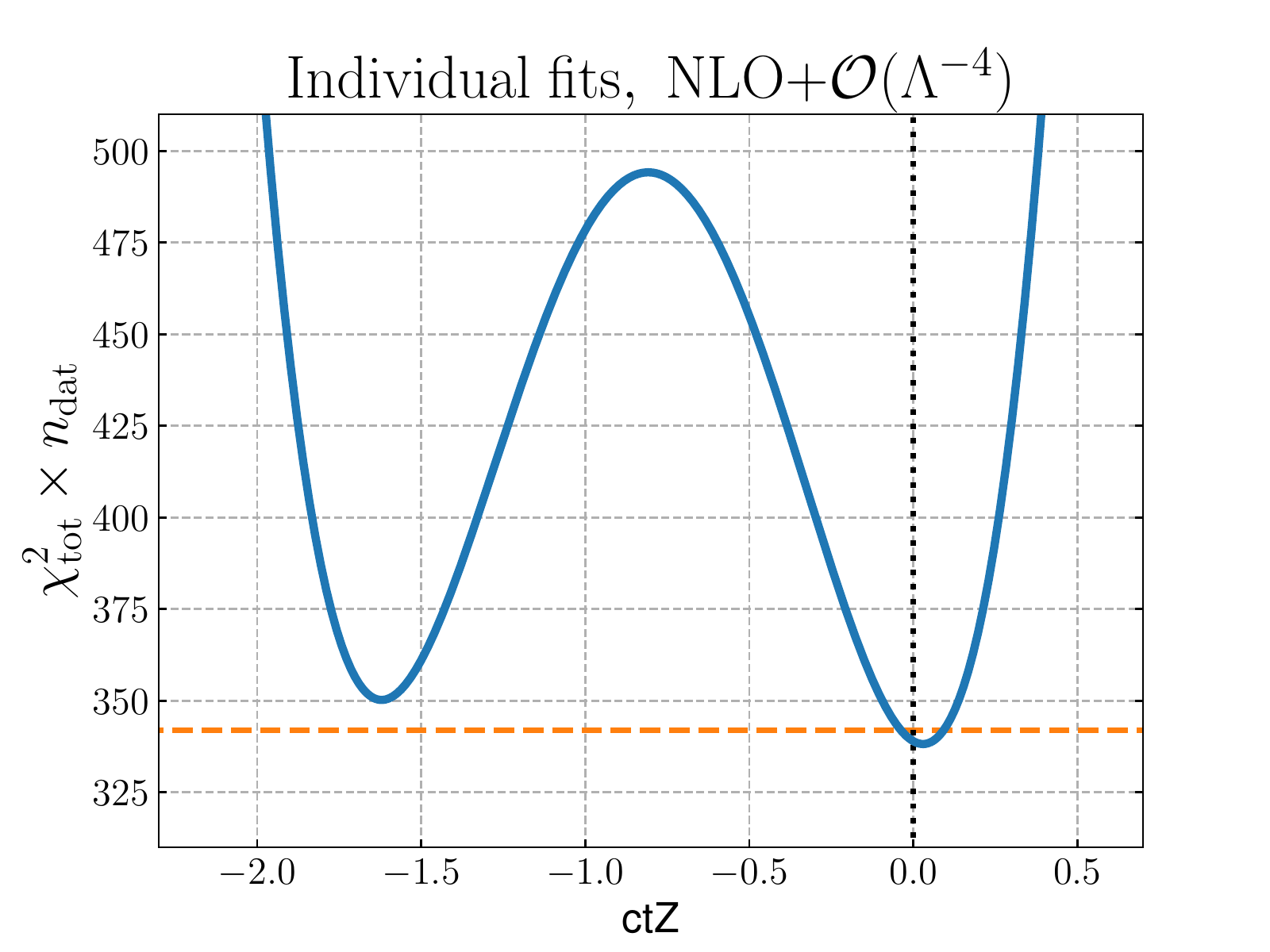}
\includegraphics[width=0.30\linewidth]{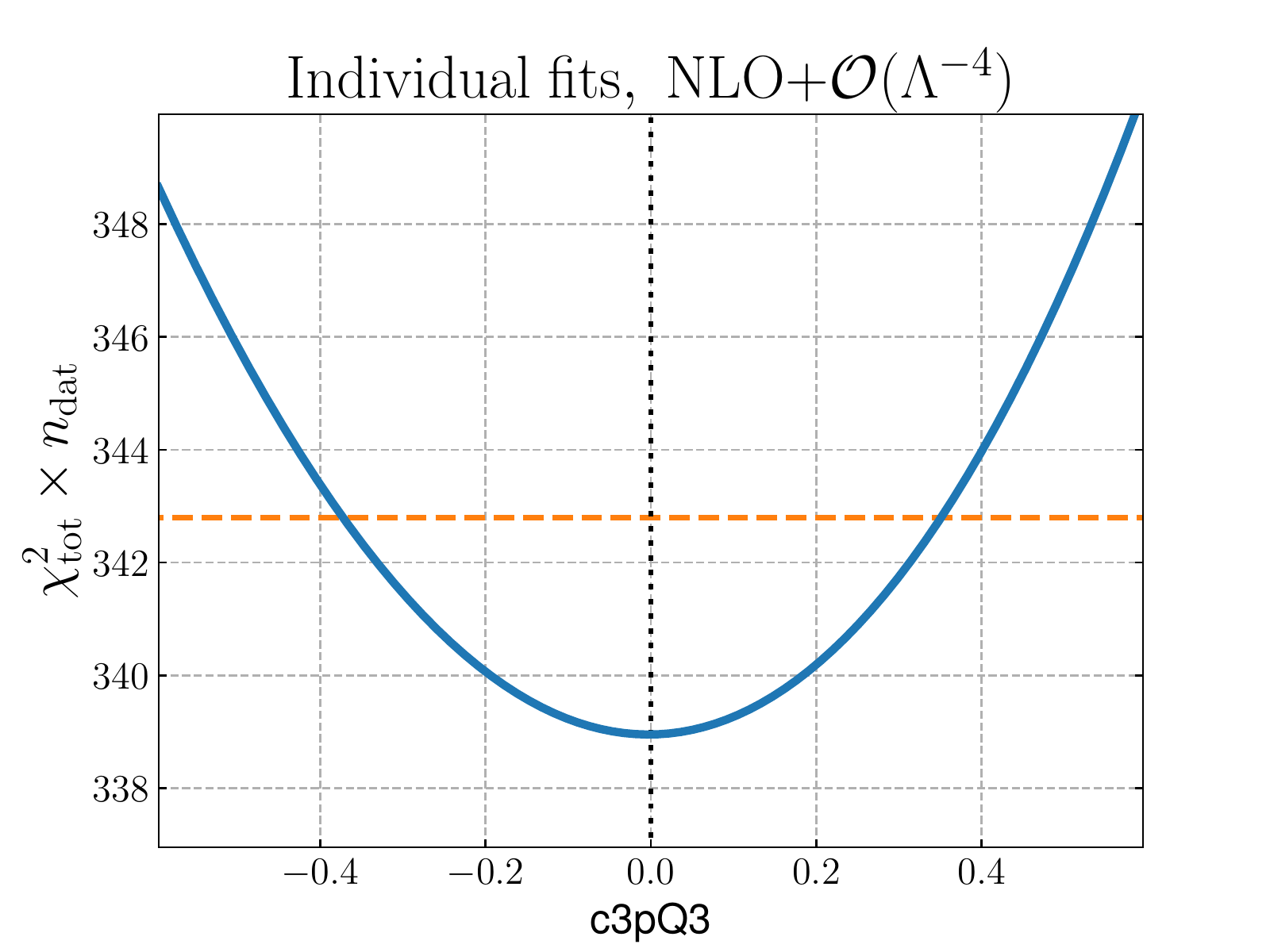}
\includegraphics[width=0.30\linewidth]{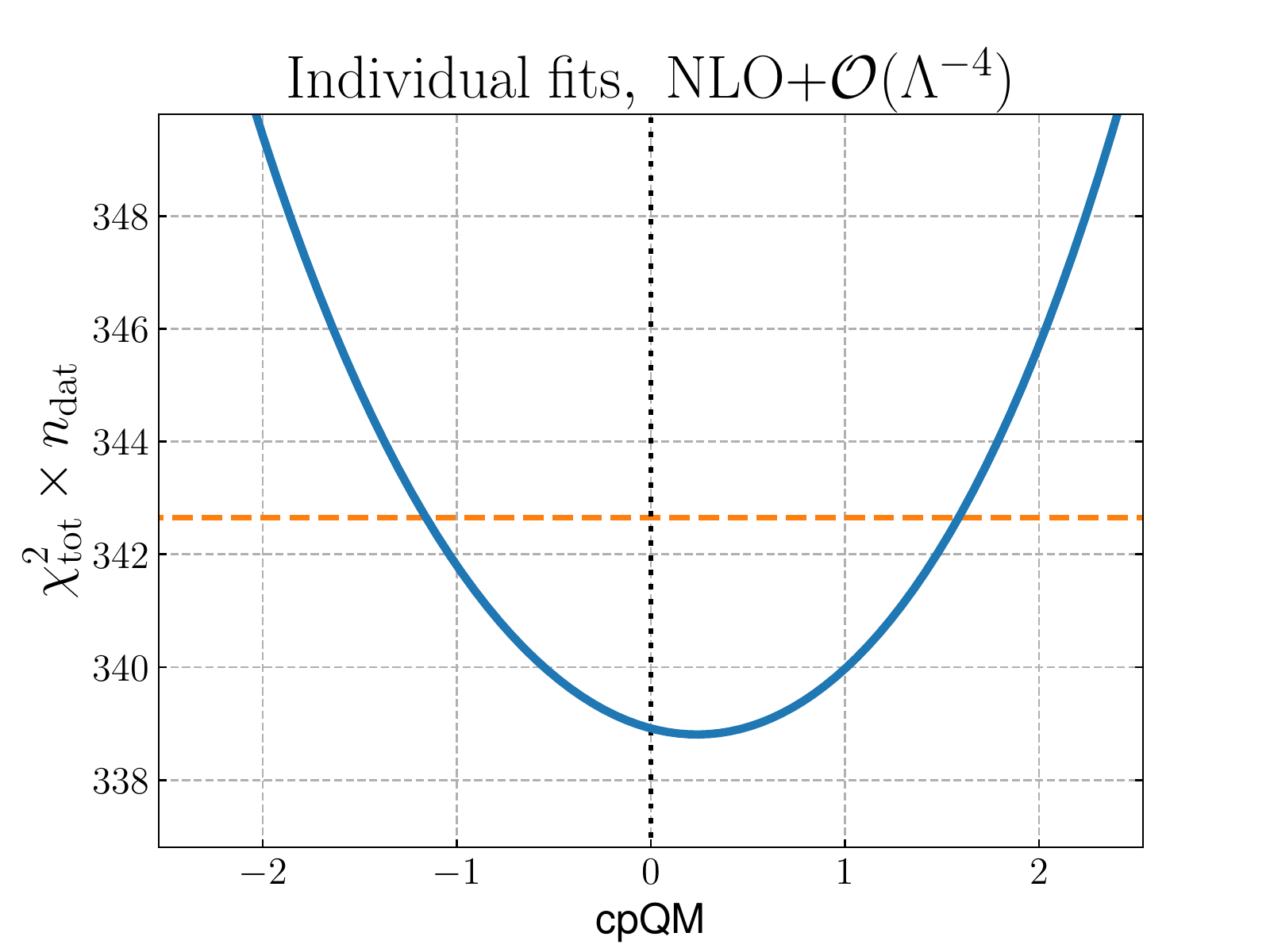}
\includegraphics[width=0.30\linewidth]{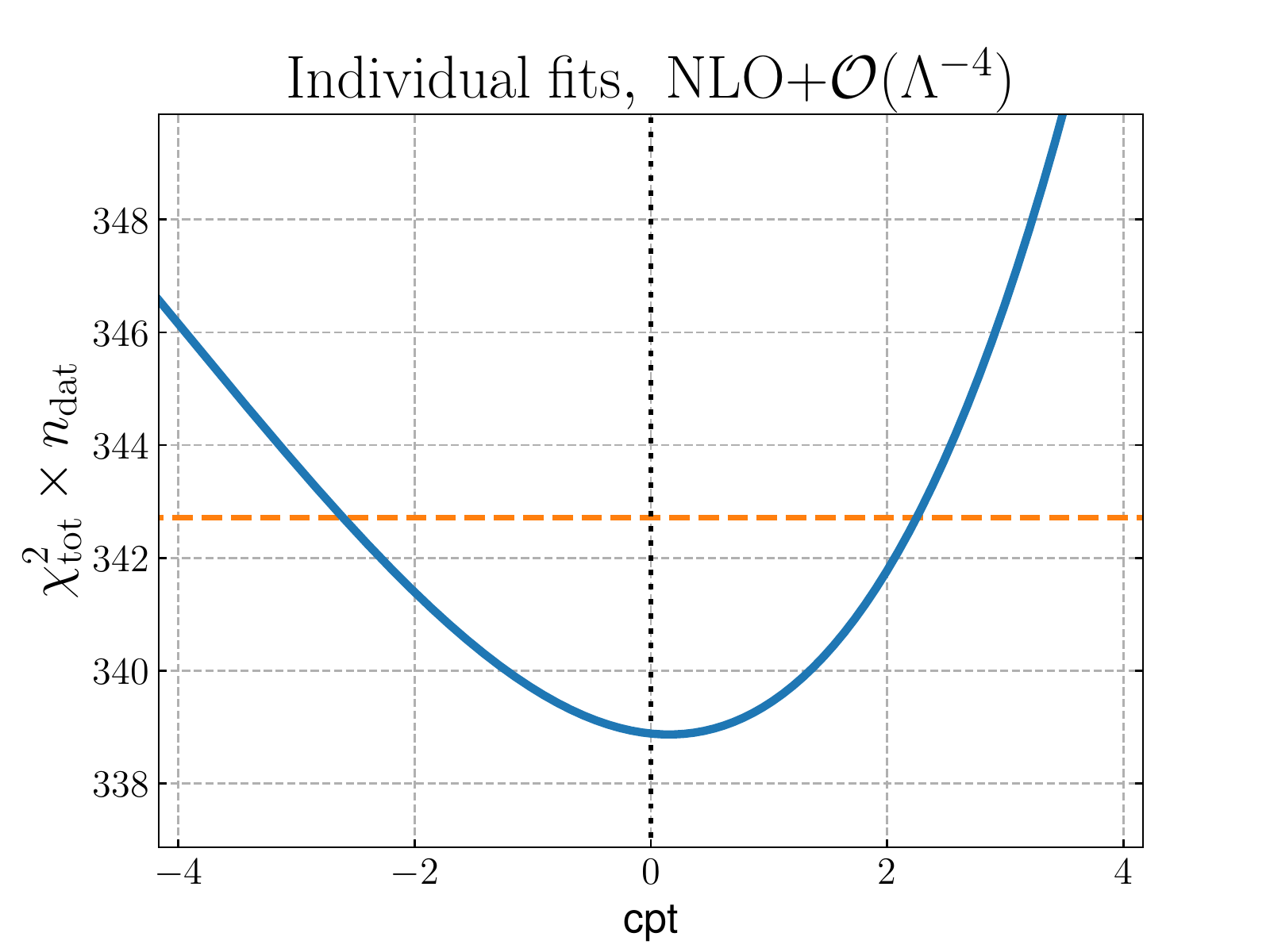}
\includegraphics[width=0.30\linewidth]{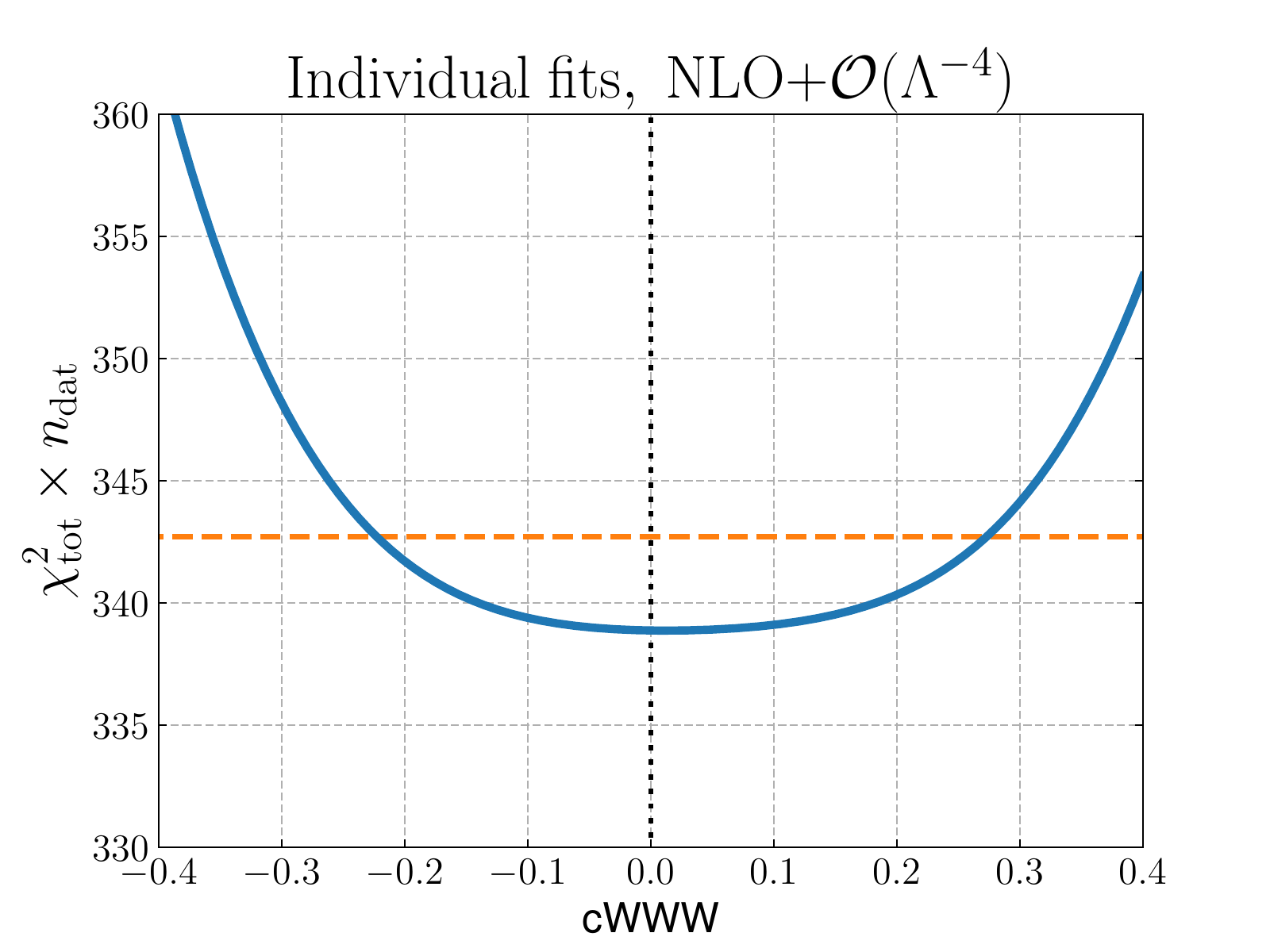}
\includegraphics[width=0.30\linewidth]{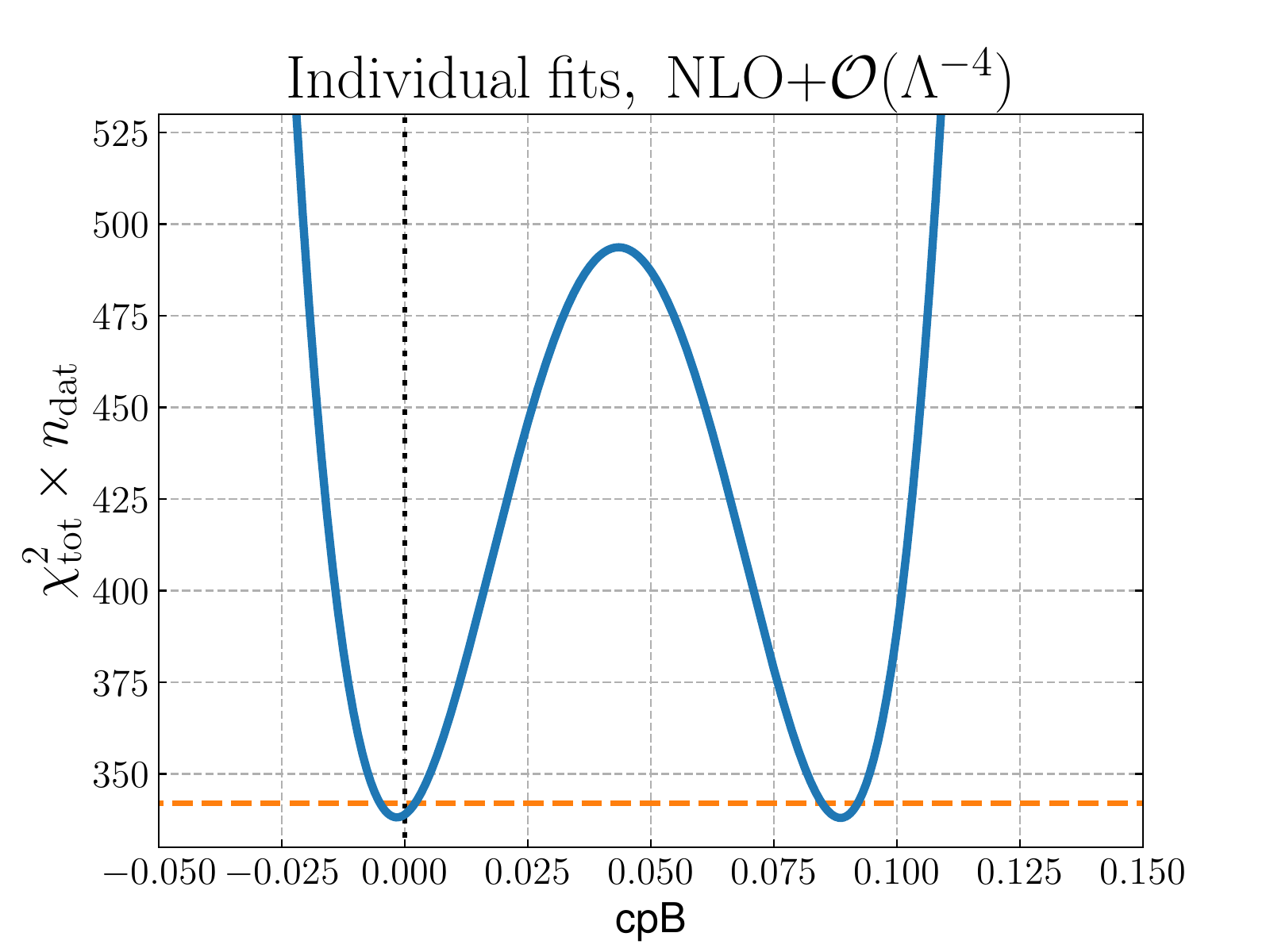}
\includegraphics[width=0.30\linewidth]{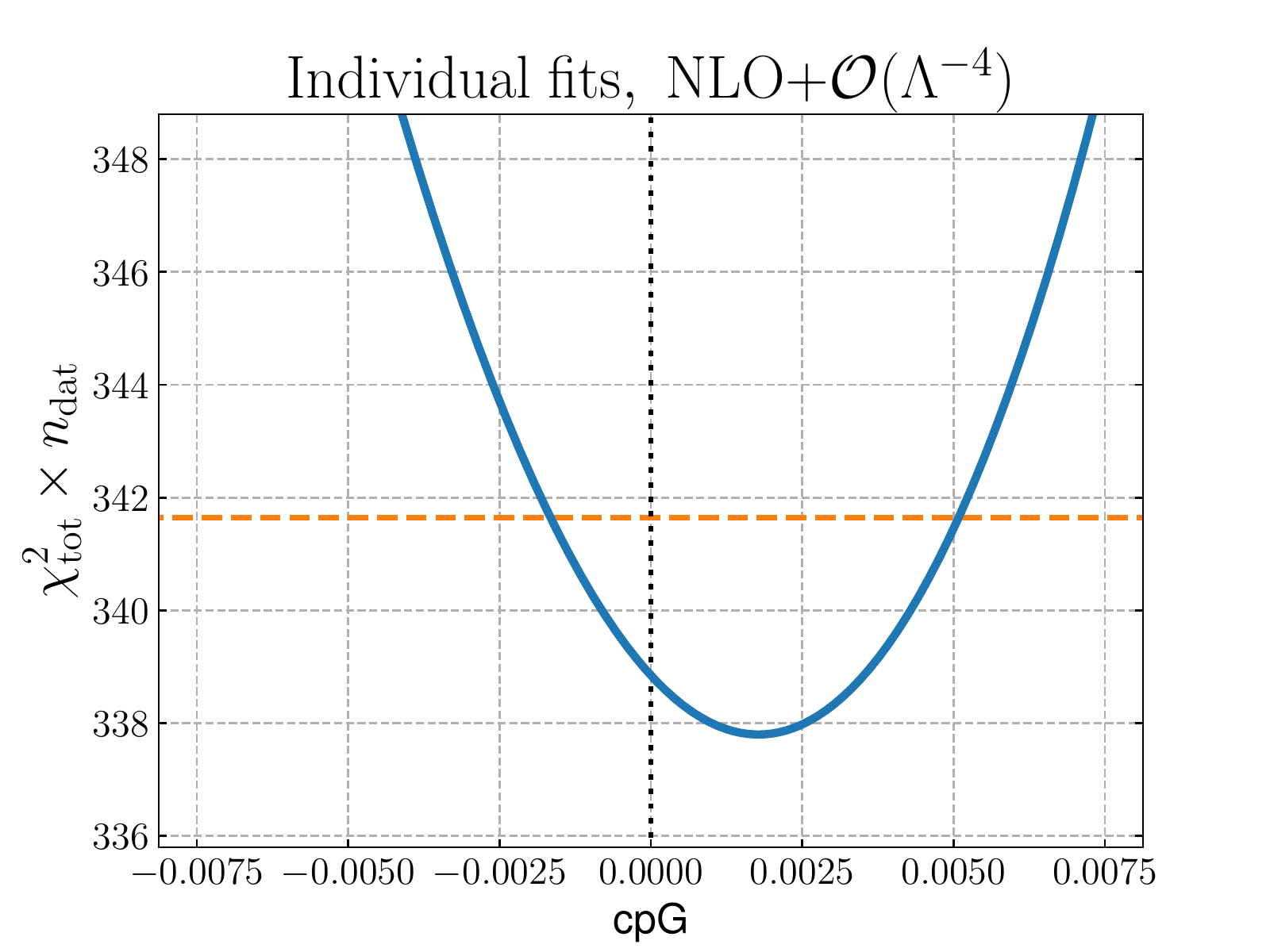}
\includegraphics[width=0.30\linewidth]{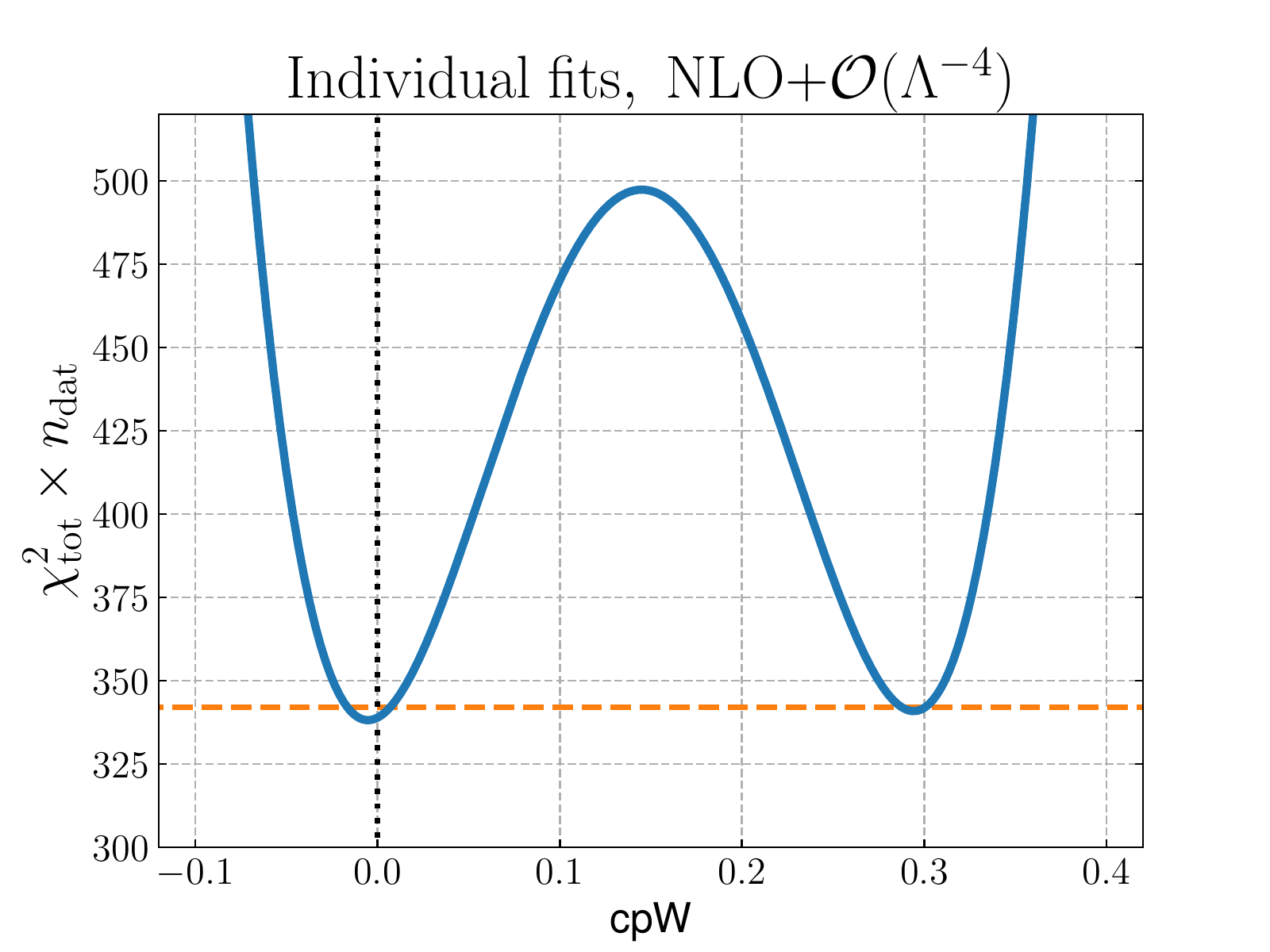}
\includegraphics[width=0.30\linewidth]{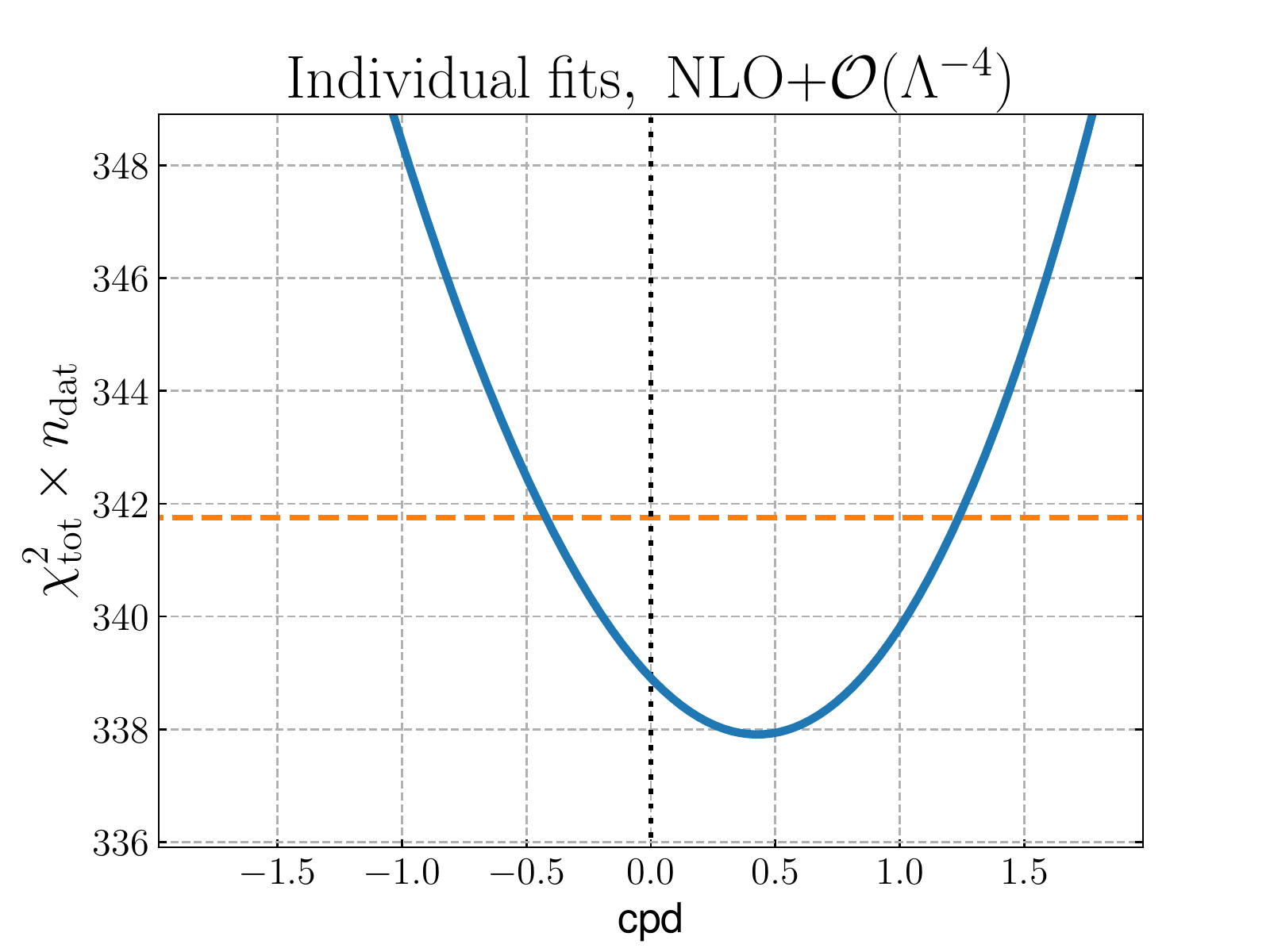}
\caption{\small Fig.~\ref{fig:quartic-individual-fits} continued.
     \label{fig:quartic-individual-fits-2} }
  \end{center}
\end{figure}

Figs.~\ref{fig:quartic-individual-fits} and~\ref{fig:quartic-individual-fits-2} display
the results of quartic polynomial fits
  to the $\chi^2$ profiles obtained  in the one-parameter scans
 of each EFT coefficient,  based on the $n_{\rm dat}=317$ data points
 of the global dataset and the baseline theory settings (where higher-order QCD and EFT
 corrections are accounted for).
 Here the absolute $\chi^2$ is evaluated with the $t_0$ prescription,
 and we also display the 
 corresponding 95\% CL ranges (vertical line) and the SM expectation (horizontal line).
 We show the 34 profiles associated to the independent EFT coefficients
 in Table~\ref{tab:operatorbasis} which are not constrained by the EWPOs.
 These profiles are shown in the following order: four-heavy, two-light-two-heavy, two-fermion,
 and purely bosonic coefficients.

From the $\chi^2$ profiles displayed in Figs.~\ref{fig:quartic-individual-fits} 
and~\ref{fig:quartic-individual-fits-2} one can observe, on the one hand,
how for several of the coefficients  the parabolic approximation 
performs reasonably well, indicating the dominance of the linear EFT corrections.
On the other hand, other coefficients
deviate from the parabolic behaviour in a striking manner, including
several degrees of freedom that exhibit two quasi-degenerate solutions,
one being ``SM-like'' and the other distinctly non-zero.
It is important to identify in particular which coefficients display such degenerate
solutions in the one-parameter fits, since these might lead to a multi-modal posterior distributions
in the case of the global analysis.

From the inspection of these $\chi^2$ profiles,
one can identify three categories of EFT coefficients whose individual profiles
are poorly described by the parabolic approximation.
First of all, one has the case of coefficients such as the four-heavy operators, for which a quartic profile
with two quasi-degenerate solutions distributed symmetrically around the SM value is observed.
Secondly, there are coefficients such as $c_{tZ}$ which display a second solution
far from the SM-like one but which corresponds to higher values of the $\chi^2$, and hence
does not modify the calculation of the CL intervals (at least within these 1D fits).
In both cases,  the resulting CL intervals remain non-disjoint.
Thirdly, one finds coefficients that exhibit quasi-degenerate solutions leading to
disjoint CL intervals, where again one solution is SM-like and the other is far from the SM value.
Examples of this category are the operators that modify the bottom 
 and tau lepton Yukawa interactions, $c_{\varphi b}$ and $c_{\varphi \tau}$, and
 the purely bosonic operators $c_{\varphi B}$ and $c_{\varphi W}$.
 Such degenerate solutions are likely to propagate to the global fit where all operators
 are simultaneously varied,
 and indeed as will be discussed in Sect.~\ref{sec:results} the presence of these quasi-degenerate
 minima on the one-parameter fits has consequences at the level
 of posterior probability distributions in the global case.
 
Another useful application of the parameter bounds
obtained from these individual fits is to help defining in an automated manner
the suitable initial 
sampling ranges for each EFT coefficient  in the global fits based on the MCfit and NS approaches.
With this motivation, the individual bounds
corresponding to a given input dataset and settings of the theoretical calculations
are evaluated by default before each fitting run.
Let us also mention that the bounds on the EFT coefficients obtained
with the method presented here for the one-parameter analyses (quartic fits to the $\chi^2$ profiles) are found
to be in agreement with the corresponding results obtained the NS and MCfit approaches.


\subsection{Nested Sampling}
\label{sec:nestedsampling}

The main approach that is adopted in this work to constrain the EFT parameter
space is Nested Sampling (NS), specifically
the version implemented in the {\tt MultiNest} algorithm~\cite{Feroz:2013hea}.
In comparison to MCfit, which is an optimisation problem aimed
to determine the best-fit values for each of the replicas, NS is based on 
sampling the figure of merit $\chi^2$  to determine its dependence
on the Wilson coefficients and locate the region of maximum likelihood.
Since NS is completely independent from the MCfit procedure,
its availability makes possible validating the robustness of the resulting bounds in EFT parameter
space via  two orthogonal methods. 

The starting point of NS is Bayes' theorem, which allows one to evaluate the
probability distribution of a set of parameters $\boldsymbol{c}$ 
associated to a model $\mathcal{M}(\boldsymbol{c})$
given a set of experimental measurements $\mathcal{D}$,
\be
\label{eq:bayestheorem}
P\lp\boldsymbol{c}| \mathcal{D},\mathcal{M} \rp = \frac{P\lp\mathcal{D}|\mathcal{M},\boldsymbol{c}
  \rp P\lp \boldsymbol{c}|\mathcal{M}  \rp
}{P(\mathcal{D}|\mathcal{M})} \, .
\ee
Here $P\lp\boldsymbol{c}| \mathcal{D},\mathcal{M} \rp $ represents the posterior
probability of the model parameters given the assumed model and the observed
experimental data,
$P\lp\mathcal{D}|\mathcal{M},\boldsymbol{c}
\rp = \mathcal{L}\lp\boldsymbol{c} \rp$ is the likelihood (conditional
probability) of the experimental measurements
given the model and a specific choice of parameters,
and $P\lp \boldsymbol{c}|\mathcal{M}  \rp = \pi \lp  \boldsymbol{c} \rp $
is the prior distribution for the model parameters.
The denominator in Eq.~(\ref{eq:bayestheorem}), $P(\mathcal{D}|\mathcal{M}) = \mathcal{Z}$,
is known as the Bayesian evidence and ensures the normalisation of the posterior
distribution,
\be
\mathcal{Z} = \int \mathcal{L}\lp  \boldsymbol{c} \rp
\pi \lp  \boldsymbol{c} \rp d \boldsymbol{c} \, ,
\ee
where the integration is carried out over the domain of the model parameters $\boldsymbol{c}$.

The key ingredient of Nested Sampling is to utilise the ideas underlying 
Bayesian inference to map the $n$-dimensional integral over the prior density 
in model parameter space $\pi(\boldsymbol{c} )d\boldsymbol{c}$,
where $n$ represents the dimensionality of $\boldsymbol{c}$, into a one-dimensional function
of the form
\be
\label{eq:NS1}
X(\lambda) = \int_{\{ \boldsymbol{c} : \mathcal{L}\lp\boldsymbol{c} \rp > \lambda \}}
\pi(\boldsymbol{c} ) d\boldsymbol{c} \,. 
\ee
In this expression, the prior mass $X(\lambda)$ corresponds to the
(normalised) volume
of the prior density $\pi(\boldsymbol{c} )d\boldsymbol{c}$ associated with values
of the model parameters that lead to a likelihood $\mathcal{L}\lp\boldsymbol{c}\rp $ greater
than the parameter $\lambda$.
Note that by construction, the prior mass $X$ decreases monotonically
from the limiting value $X=1$ to $X=0$ as $\lambda$ is increased.
The integration of $X(\lambda)$ extends over the regions in the model parameter space contained
within the fixed-likelihood contour defined by the condition $\mathcal{L}\lp \boldsymbol{c} \rp=\lambda$.
This property allows the evidence to be expressed as,
\be
\label{eq:bayesianevidence}
\mathcal{Z} = \int_0^1  \mathcal{L}\lp X\rp dX \, ,
\ee
where $  \mathcal{L}\lp X\rp$ is defined as the inverse function of $X(\lambda)$, which
always exists provided the likelihood is a continuous and smooth function
of the model parameters.
Therefore, the transformation from $\boldsymbol{c}$
to $X$ in Eq.~(\ref{eq:NS1}) achieves a mapping of the prior distribution into infinitesimal
elements, sorted by their associated likelihood $\mathcal{L}(\boldsymbol{c})$.

The next step of the NS algorithm is to define a decreasing sequence of values in the prior
volume, that is now parameterised by the prior mass $X$.
In other words, one slices the prior volume into a large number of
small regions
\be
1 = X_0 > X_1 > \ldots X_{\infty} = 0 \, ,
\ee
and then evaluates the likelihood at each of these values, $\mathcal{L}=\mathcal{L}(X_i)$.
This way, all of the $\mathcal{L}_i$ values
can be summed in order to  evaluate the integral 
for the Bayesian evidence, Eq.~(\ref{eq:bayesianevidence}).

Since in general the likelihood $\mathcal{L}({\boldsymbol c})$ exhibits a complex dependence
on the model parameters $\boldsymbol{c}$, the summation
in Eq.~(\ref{eq:bayesianevidence}) must be evaluated
numerically using {\it e.g.} Monte Carlo integration methods.
In practice, one draws $N_{\rm live}$ points from the parameter prior
volume $\pi\lp\boldsymbol{c} \rp$, known as {\it live points}, and orders
the likelihood values from smallest to largest, including the 
starting value of the prior mass at $X_0=1$.
As samples are drawn from the prior volume, 
the live point with the lowest likelihood $\mathcal{L}_i$
is removed from the set and replaced by another live point drawn from the same prior
distribution but now under the constraint that its likelihood is larger than
$\mathcal{L}_i$.
This sampling process is repeated until the entire hyper-volume
$\pi \lp \boldsymbol{c} \rp$ of the prior parameter space has been covered, with
ellipsoids of constrained likelihood being assigned
to the live-points as the prior volume is scanned.

While the end result of the NS procedure is the estimation of the 
Bayesian evidence $\mathcal{Z}$, as a byproduct one also obtains 
a sampling of the posterior distribution
associated to the EFT coefficients expressed as
\be
\{ \boldsymbol{c}^{(k)} \}\, ,\qquad  k=1,\dots,N_{\rm spl}\, ,
\ee
with $N_{\rm spl}$ indicating the number
of samples drawn by the final NS iteration.
One can then compute expectation values, variances, and correlations of the model
parameters by evaluating the MC sum over these posterior samples together 
with their associated weights, in the same
manner as averages are carried out
over the $N_{\rm rep}$ replicas in the MCfit method. 

\paragraph{Prior volume.}
An important input for NS is the choice of prior volume $\pi \lp \boldsymbol{c} \rp$
in the model parameter space.
In this analysis, we adopt flat priors 
defined by ranges in parameter space for the coefficients $\boldsymbol{c}$.
A suitable choice of prior volume where the sampling takes place is important
to speed up the NS algorithm: a range too wide will make the optimisation less
efficient, while a range too narrow might bias the results by cutting 
specific regions of the parameter space that are relevant.
Furthermore, using a common range for all parameters should be avoided,
since the range of intrinsic variation will be rather different for each
of the EFT coefficients, as illustrated also by the one-parameter fits
reported in Figs.~\ref{fig:quartic-individual-fits} 
and~\ref{fig:quartic-individual-fits-2}.

Taking these considerations into account, we adopt here the following strategy.
First, a single model parameter $c_i$ is allowed to vary
while all others are set to their SM value, $c_j=0$ for $j\ne i$.
The $\chi^2 \lp c_i \rp$ is then scanned in this
direction to determine the values $c_i^{\rm (min)}$ and  $c_i^{\rm (max)}$
satisfying the condition $\chi^2/n_{\rm dat}=4$.
We then repeat this procedure 
for all parameters and end up with a hyper-volume 
defined by pairs of values
\be
\pi \lp \boldsymbol{c} \rp = \lc \lp c_i^{\rm (min)},c_i^{\rm (max)} \rp \, ,\quad
i=1,\ldots, n_{\rm op} \rc \, ,
\ee
which then defines the initial prior volume.
At this point, one performs an initial exploratory NS global analysis using this
volume to study the posterior probability distribution
for each EFT coefficient.
Our final analysis is then obtained by manually adjusting the 
initial sampling ranges until the full posterior distributions are captured for the chosen
prior volume. 
For parameters that are essentially unconstrained in the global fit,
such as the four-heavy operators in the case of linear EFT calculations,
a hard boundary of $\lp -50 , 50 \rp$ is imposed (for $\Lambda=1$ TeV) . 

\paragraph{Performance.}
In order to increase the efficiency of the posterior probability estimation by NS, we enable the
``constant efficiency mode'' in MultiNest, which adjusts the
total volume of ellipsoids spanning the live points so that the sampling
efficiency is close to its associated hyperparameter set by the user. 
With 24 cpu cores, 
we are able to achieve an accurate posterior for the linear EFT fits
in around 30 minutes using 500 live points, a target efficiency of 0.05, and
an evidence tolerance of 0.5, which results in $N_{\rm spl}\simeq 5000$ posterior samples.
To ensure the stability of our final results, we chose 
1000 live points and a target efficiency of 
0.005, which yields $\simeq$$1.5\times 10^4$ samples for the
linear analysis and $\simeq$$10^4$ samples for an analysis that includes also the quadratic EFT
corrections.
With these settings, our final global analyses containing the simultaneous
determination of $n_{\rm op}\simeq 36$ coefficients
take $\sim 3.5$ hours running in 24 cpu cores, with a similar performance for
linear and quadratic EFT fits.

The NS method is especially suitable to tackle parameter spaces of moderate dimensionality.
Being based purely on sampling, it is not affected by limitations in minimisation
methods such as ending up in local minima.
It is also more robust upon the presence of fluctuations, and does not
require specifying certain hyperparameters such as the learning rates which are used in MCfit.
The main limitation of NS is that, as in all sampling methods, the execution times
grows exponentially with $n_{\rm op}$, the dimensionality of the model parameter space.
For parameter spaces of dimensionality greater than around 50, 
the current NS implementation that we use
becomes unpractically slow and MCfit becomes the most suitable strategy available.

\subsection{The Monte Carlo replica method revisited}
\label{sec:mcfit}

The {\tt SMEFiT} analysis of Ref.~\cite{Hartland:2019bjb} was based
on the Monte Carlo replica approach (MCfit), which in turn was inspired by the
NNPDF analysis of the quark and gluon substructure of protons.
The MCfit method aims to construct a sampling of the probability
distribution in the space of the experimental data, which then translates 
into a sampling of the probability distribution in the space of the EFT 
coefficients through an optimisation procedure where the best-fit values
of the coefficients for each replica, $\boldsymbol{c}^{(k)}$,
are determined.

Given an experimental measurement of a hard-scattering
cross-section, denoted by $\sigma_i^{\rm (exp)}$, with
total uncorrelated uncertainty $\delta_{i}^{\rm (stat)}$ and $n_{\rm sys}$ 
correlated systematic uncertainties $\delta^{\rm (sys)}_{i,\alpha}$,
the $N_{\rm rep}$ artificial MC replicas of the 
experimental data are generated as
\be
\label{eq:replicas}
\sigma_{i}^{(\art)(k)}
= 
\sigma_{i}^{\rm (\mrexp)}\lp 1
+
r_{i}^{(k)}\delta_{i}^{\rm (stat)}
+
\sum_{\alpha=1}^{n_{\rm sys}}r_{i,\alpha}^{(k)}\delta^{\rm (sys)}_{i,\alpha}\rp
\ , \quad k=1,\ldots,N_{\rep} \ , 
\ee
where the index $i$ runs from 1 to $n_{\rm dat}$ and
$r_{i}^{(k)}$, $r_{i,\alpha}^{(k)}$
are univariate Gaussian random numbers.
Correlations between data points induced by systematic uncertainties 
are accounted for by ensuring that $r^{(k)}_{i,\alpha}=r^{(k)}_{i',\alpha}$.
It can be show that central values, variances, and covariances evaluated
by averaging over the MC replicas reproduce the corresponding
experimental values.

A fit to the $n_{\rm op}$ degrees of freedom $\boldsymbol{c}/\Lambda$
is then performed for each of the MC replicas generated by Eq.~\eqref{eq:replicas}.
These best-fit values are determined from
the minimisation of the cost function
\begin{equation}
  E^{(k)}({\boldsymbol c})\equiv \frac{1}{n_{\rm dat}}\sum_{i,j=1}^{n_{\rm dat}}\lp 
  \sigma^{(\rm th)}_i\lp {\boldsymbol c}^{(k)}\rp-\sigma^{{(\rm art)}(k)}_i\rp ({\rm cov}^{-1})_{ij}
  \lp \sigma^{(\rm th)}_j\lp {\boldsymbol c}^{(k)} \rp-\sigma^{{(\rm art)}(k)}_j\rp
  \label{eq:chi2definition}
    \; ,
\end{equation}
where $\sigma^{(\rm th)}_i( {\boldsymbol c}^{(k)} )$ indicates the theoretical
prediction for the $i$-th cross-section evaluated with the $k$-th set of
EFT coefficients.
This process results in a collection of ${\boldsymbol c}^{(k)}$ best-fit 
coefficient values from which estimators such as expectation values, variances,
and correlations are evaluated.
The overall fit quality is then evaluated using Eq.~(\ref{eq:chi2definition2}),
where the central experimental values are compared to the mean theoretical
prediction computed by the resulting fit replicas.

As mentioned in Sect.~\ref{sec:generalsettings}, 
various theoretical uncertainties
are also included in the $\chi^2$ definition for some datasets.
A consistent treatment
of theoretical uncertainties in the fitting procedure means
that these are not only included in the fit via 
the covariance matrix in Eqs.~(\ref{eq:chi2definition}), 
but also in the corresponding replica generation.
In other words, the replicas are sampled according to a multi-Gaussian distribution
defined by the total covariance matrix Eq.~(\ref{eq:covmatsplitting})
which receives contributions both of experimental and of theoretical origin.
We therefore account for such errors
in the generation of Monte Carlo replicas~\cite{AbdulKhalek:2019ihb} 
using Eq.~(\ref{eq:replicas}).

There are numerous advantages of using the MCfit method for 
global EFT analyses. 
First, it does not require specific 
assumptions about the underlying probability distribution
of the fit parameters, and in particular does not rely
on the Gaussian approximation.
Secondly, the computational cost scales in a much milder way 
with the number of operators 
$n_{\rm op}$ included in the fit as compared to NS. 
Thirdly, it can be used to assess the impact of new datasets in the fit
{\it a posteriori}
with the Bayesian reweighting formalism.\\[-0.3cm]

In comparison with~\cite{Hartland:2019bjb}, several improvements
have been implemented to increase the efficiency and accuracy of the MCfit procedure 
used in this analysis:

\paragraph{Optimisation.}
In the top quark sector analysis of~\cite{Hartland:2019bjb}, the minimisation of 
Eq.~\eqref{eq:chi2definition} was achieved by a gradient descent method which relies 
on local variations of the error function.
This choice is advantageous since $E^{(k)}$ is at most a quartic form
of the fit parameters, see Eq.~(\ref{eq:quartic-chi2}) and its generalisation
to multiple operators, and therefore evaluating its gradient is computationally efficient.

Since in the present analysis our parameter space is more complex, the optimiser that we use now to 
determine the best-fit values of the degrees of freedom ${\boldsymbol c}^{(k)}$ within MCfit is a 
trust-region algorithm {\tt trust-constr} available in the {\tt SciPy} package. 
An advantage of using {\tt trust-constr} in this context is that it allows 
one to provide the optimiser with any combination of constraints on the 
coefficients, including existing bounds.
This is a rather useful feature, since in many cases of interest one would like to restrict
the EFT parameter space based on theoretical considerations, such as when
accounting for the LEP EWPOs or in the top-philic scenario discussed in Sect.~\ref{sec:smefttheory}.

\paragraph{Initial sampling range and bounds.}
For each MC replica fit, the initial values of the fit coefficients ${\boldsymbol c}^{(k)}$
are initialised at random within a pre-defined range.
This sampling range, as well as the boundaries imposed on the minimisation procedure for the poorly constrained
parameters,
are taken to be the same as those used in the NC procedure.
That is, the sampling ranges for the global fits are derived from a one-parameter
$\chi^2$ scanning procedure subsequently
inflated to cover a sufficiently large parameter hyper-volume. 

\paragraph{Cross-validation.}
Given the large dimensionality of the considered EFT parameter space, it is conceivable
that the optimiser algorithm ends up fitting the statistical fluctuations 
of the experimental data rather than the underlying physical law.
One way to prevent the minimiser from over-fitting the data is to use
look-back cross-validation stopping.
In this method, each replica dataset is randomly split with equal probability into two 
disjoint sets, known as the training and validation sets.
Only the data points in the training set are then used to compute the figure of 
merit being minimised, Eq.~(\ref{eq:chi2definition}), while the data points in 
the validation set are monitored alongside the fit.
The random assignment of the data points to the training or validation sets
is different for each MC replica, and the splitting only occurs for experiments
that contain more than 5 bins in the distribution. 
The fit is run for a fixed large number of iterations, and then
the optimal stopping point of 
the fit is then determined as the iteration for which the figure of merit evaluated on the validation set, 
$E^{(k)}_{\rm val}$, exhibits a global minimum.
All in all, it is found that the risk of over-fitting is small and that
MCfit results with and without cross-validation applied are reasonably similar.

\paragraph{Quality selection criteria.}
One disadvantage of optimisation strategies such as MCfit is that as the parameter space
space is increased, the minimiser might sometimes converge on a local,
rather than on the global, minimum.
This is specially problematic in the quadratic EFT fits which often display
quasi-degenerate minima, as illustrated by the $\chi^2$ profiles of
Figs.~\ref{fig:quartic-individual-fits} 
and~\ref{fig:quartic-individual-fits-2}.
For this reason, it is important to implement post-fit quality selection criteria
that indicate when a fitted replica should be kept and when it should be discarded.
Here, a MC
replica is kept if the total error function of the replica dataset, $E_{\rm tot}^{(k)}$, satisfies
$E_{\rm tot}^{(k)}\le 3$.

\begin{figure}[t]
  \begin{center}
    \includegraphics[width=0.86\linewidth]{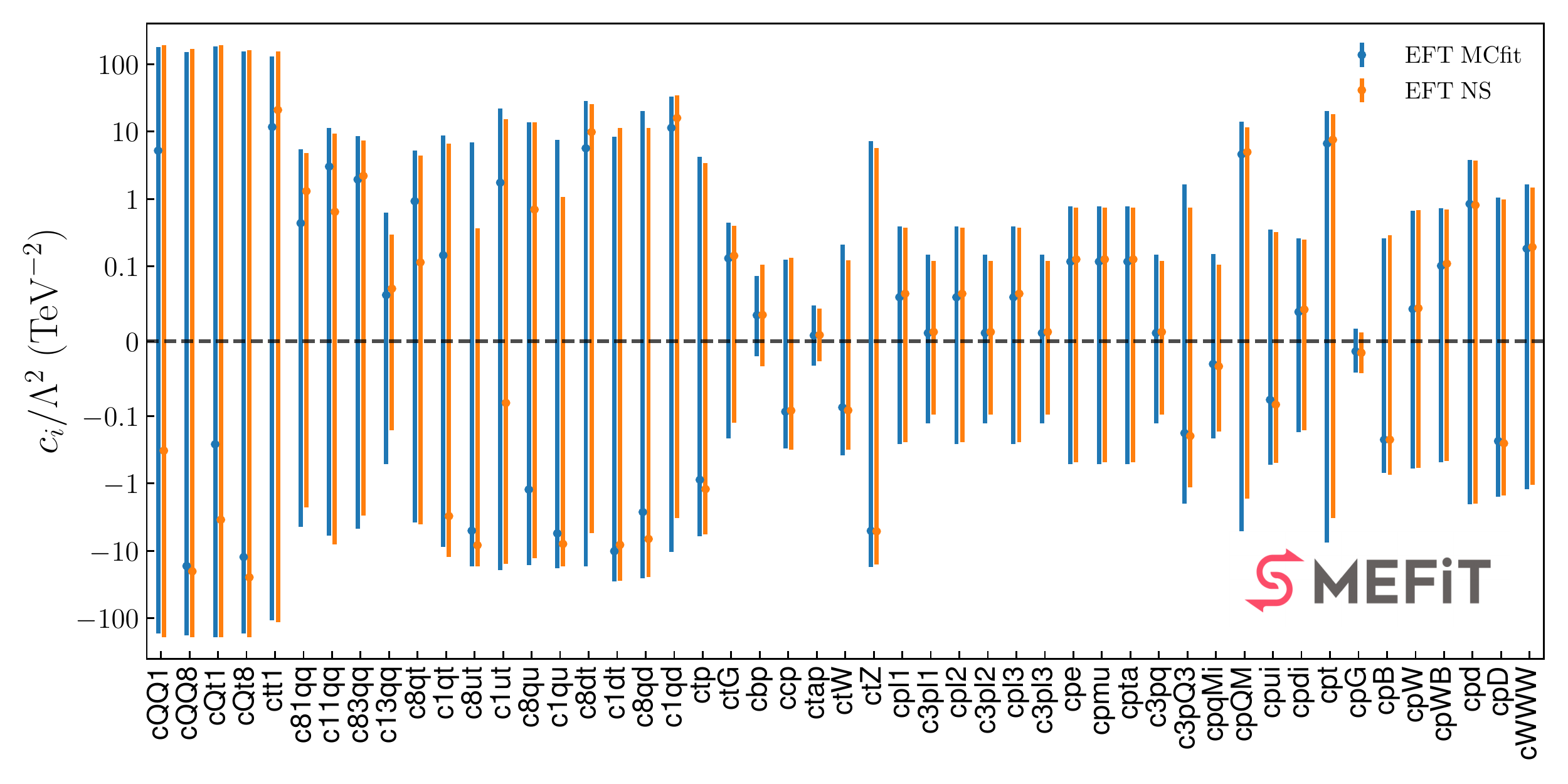}
    \includegraphics[width=0.86\linewidth]{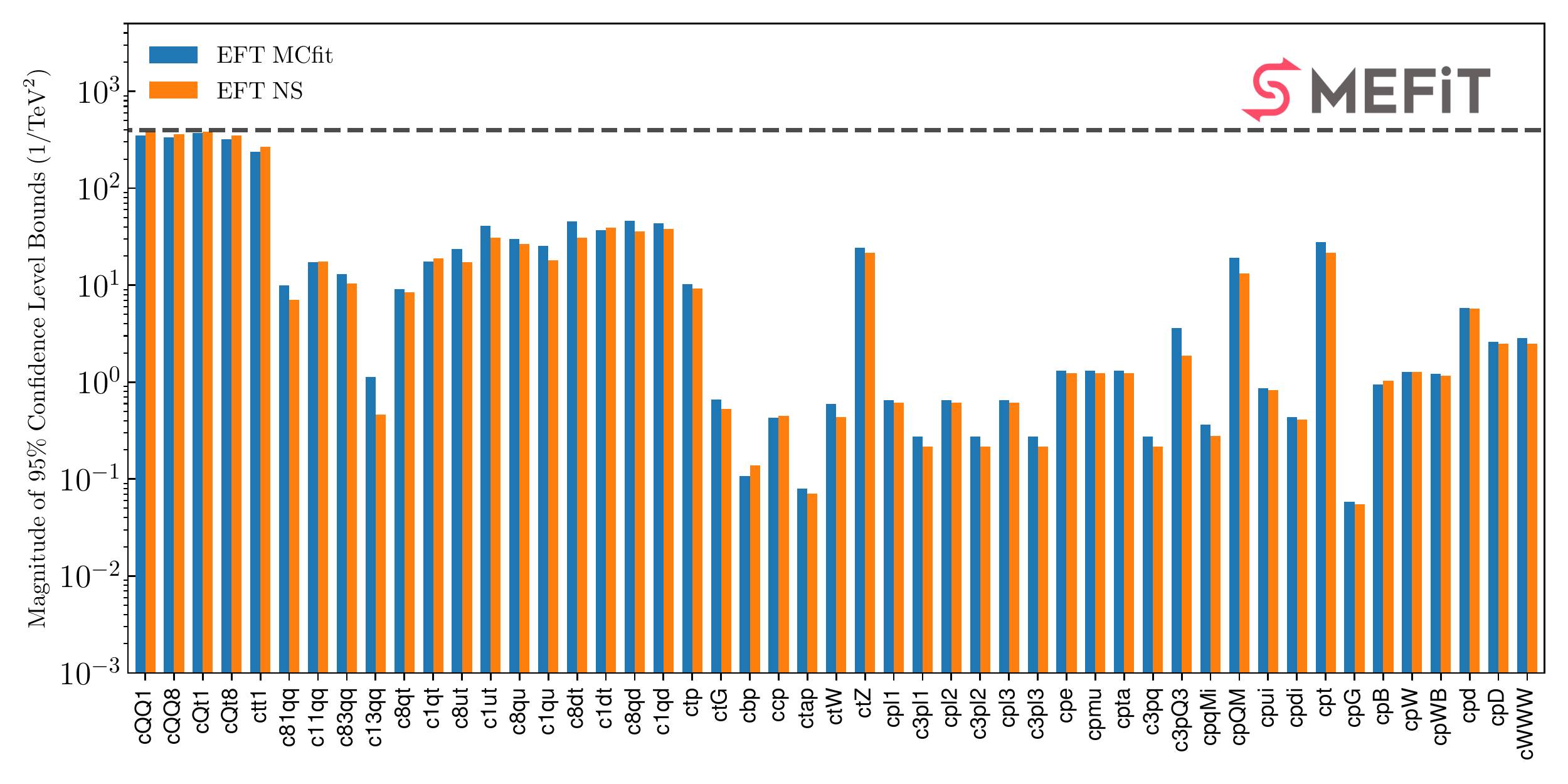}
    \vspace{-0.27cm}
    \caption{\small The best-fit values and 95\% CL intervals for a global fit based on linear EFT
      calculations, comparing the outcome of the NS and MCfit methods.
      We display the results corresponding to the 50 coefficients listed in Table~\ref{tab:operatorbasis}
      (except for $c_{\ell\ell}=0$), of which 36 are independent fit parameters.
      {  The bottom panel displays
        the magnitude of the  95\% CL intervals.}
     \label{fig:smefit-mcfit-vs-ns-global} }
  \end{center}
\end{figure}

\paragraph{Benchmarking.}
Fig.~\ref{fig:smefit-mcfit-vs-ns-global} compares
the outcome of global fits obtained with either the NS or MCfit method,
all other settings identical.
Specifically, here we show the best-fit values and 95\% CL intervals for global fits
based on linear EFT
calculations.
We provide the results corresponding to the 50 coefficients listed in Table~\ref{tab:operatorbasis}
(except for $c_{\ell\ell}$,  which is set to zero by the EWPOs)
of which 36 are independent fit parameters.
We will further discuss the physical interpretation of these
results in Sect.~\ref{sec:results},
here we only aim to establish that the two methods indeed lead to equivalent results.

\begin{figure}[t]
  \begin{center}
    \includegraphics[width=0.80\linewidth]{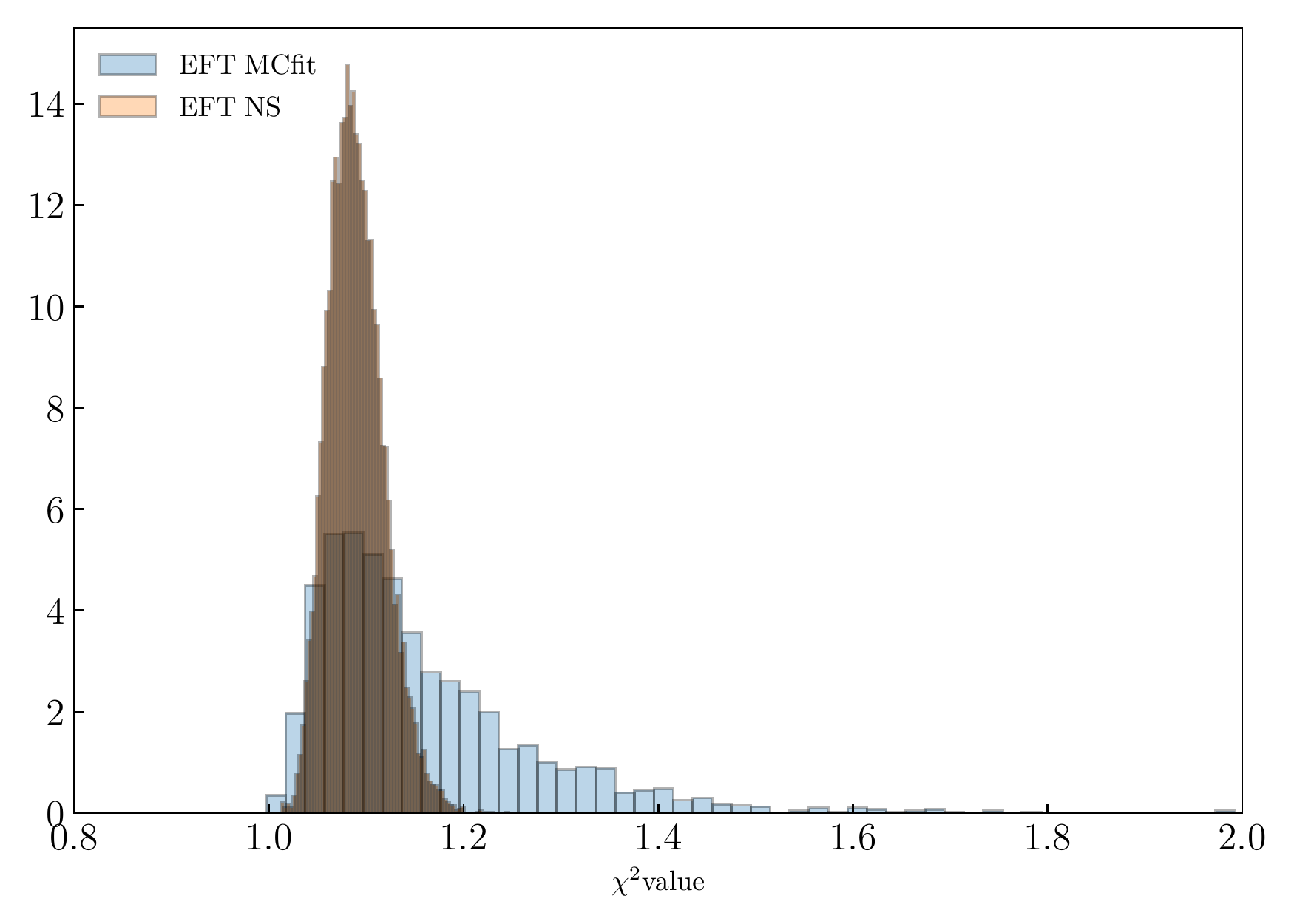}
    \vspace{-0.3cm}
    \caption{\small Comparison of the $\chi^2$ distributions evaluated over replicas
      and posterior samples in global linear EFT fits based on the MCfit
      and NC methods, respectively.
      The corresponding 95\% CL intervals on the EFT coefficients are
      displayed in Fig.~\ref{fig:smefit-mcfit-vs-ns-global}.
     \label{fig:chi2dist_mcfit_vs_ns} }
  \end{center}
\end{figure}

The comparison of Fig.~\ref{fig:smefit-mcfit-vs-ns-global} demonstrates that in general
the two methods are in excellent agreement, both in terms of best-fit values
and of the corresponding uncertainties.
This said, for specific coefficients one observes small differences, with MCfit in general
tending to provide somewhat looser bounds.
The reason for this behaviour is that optimisation-based
methods such as MCfit can be distorted by fitting inefficiencies,
such as when the optimiser finds a local, rather
than global, minimum.
This phenomenon is further illustrated in Fig.~\ref{fig:chi2dist_mcfit_vs_ns}, which compares
the $\chi^2$ distributions evaluated over replicas
and posterior samples in the MCfit
and NC methods respectively.
We observe that the MCfit distribution exhibits broader tails, implying that the
 bounds obtained this way might
in some cases be slightly over-conservative.

Fig.~\ref{fig:smefit-mcfit-vs-ns-global}, as well as the corresponding
benchmark comparison for fits based on quadratic EFT calculations, demonstrates
that results obtained with either NS or MCfit are statistically equivalent.
In the rest of this work, we will adopt NS as the baseline method,
since its not affected by potential inefficiencies in the minimisation procedure
and, as discussed above, can produce global fits within
a reasonable  execution time.


\subsection{Principal Component Analysis}
\label{sec:pca}

Principal Component Analysis (PCA) represents a valuable tool to identify the combinations of
degrees of freedom that exhibit the largest and smallest variabilities in a linear algebra problem.
This identification has many applications,
for instance, a large gap in variability suggests that the effective dimensionality
of the problem is smaller than the nominal one, and thus dimensional reduction methods
are advantageous to simplify the solution.
Furthermore, directions in the parameter space with very small variability are difficult
to constrain from data and are identified with flat directions.
Such flat directions might compromise the reliability of the
obtained results i.e. in Hessian EFT fits.\footnote{The PCA method can also be exploited to
  efficiently carry out linear SMEFT fits~\cite{Bodwin:2019ivc}.}

Here we apply the PCA technique combined with
Singular Value Decomposition (SVD) to global fits based on linear
EFT calculations.
The goal is to ascertain the presence of possible flat directions,
identify  large gaps in variability
between the principal components, and determine the relation between the physical
fitting basis and these principal components.
The starting point is the expression for the cross-section
as a function of the EFT coefficients, Eq.~(\ref{eq:quadraticTHform}), truncated
at the linear order,
\be
\label{eq:linearTHform}
\sigma_m^{\rm (th)}({\boldsymbol c})= \sigma_m^{\rm (sm)} + \sum_{i=1}^{n_{\rm op}}c_i\sigma^{(\rm eft)}_{m,i}\, ,
\qquad m =1\,\ldots, n_{\rm dat} \, ,
\ee
where recall that we have set $\Lambda=1$ TeV.
We then define a matrix $K$ of dimensions $n_{\rm dat}\times~n_{\rm op}$ and
(dimensionless) components $K_{mi}=\sigma^{(\rm eft)}_{m,i}/\delta_{{\rm exp},m}$,
where $\delta_{\rm exp,m}$ is the same total experimental error that appears
in the evaluation of the Fisher information matrix Eq.~(\ref{eq:fisherinformation}).
By means of SVD, we can decompose this matrix $K$ as
\be
\label{eq:SVD}
K = U W V^\dagger \, ,
\ee
where $U~(V)$ is a $n_{\rm dat} \times n_{\rm dat}$~($n_{\rm op} \times n_{\rm op}$) unitary matrix and $W$ is an $n_{\rm dat}\times n_{\rm op}$ diagonal
matrix with semi-positive real entries, called the singular values, which are ordered by decreasing
magnitude.
The larger a singular value, the higher the variability of its principal component
and the higher the likelihood that this component will be well constrained from the fit.

The elements of the symmetric matrix $V$ in Eq.~(\ref{eq:SVD}) contain the principal components associated
to each of the $n_{\rm op}$ singular values.
These correspond to a linear superposition of the original coefficients, that is, we have that
\be
\label{eq:PCdef}
{\rm PC}_k = \sum_{i=1}^{n_{\rm op}} a_{ki}c_i \, , \quad k=1,\ldots,n_{\rm op} \, ,\qquad \lp~ \sum_{i=1}^{n_{\rm op}} a_{ki}^2=1\,~\forall k \rp
\ee
where the larger the value of the squared coefficient $a^2_{kl}$, the larger the relative weight
of the associated EFT coefficient in this specific (normalised) principal component.
By means of the matrix $V$ (and its inverse), one can rotate between
the original fitting basis and the one defined by the principal components.

\begin{figure}[t]
  \begin{center}
    \includegraphics[width=0.95\linewidth]{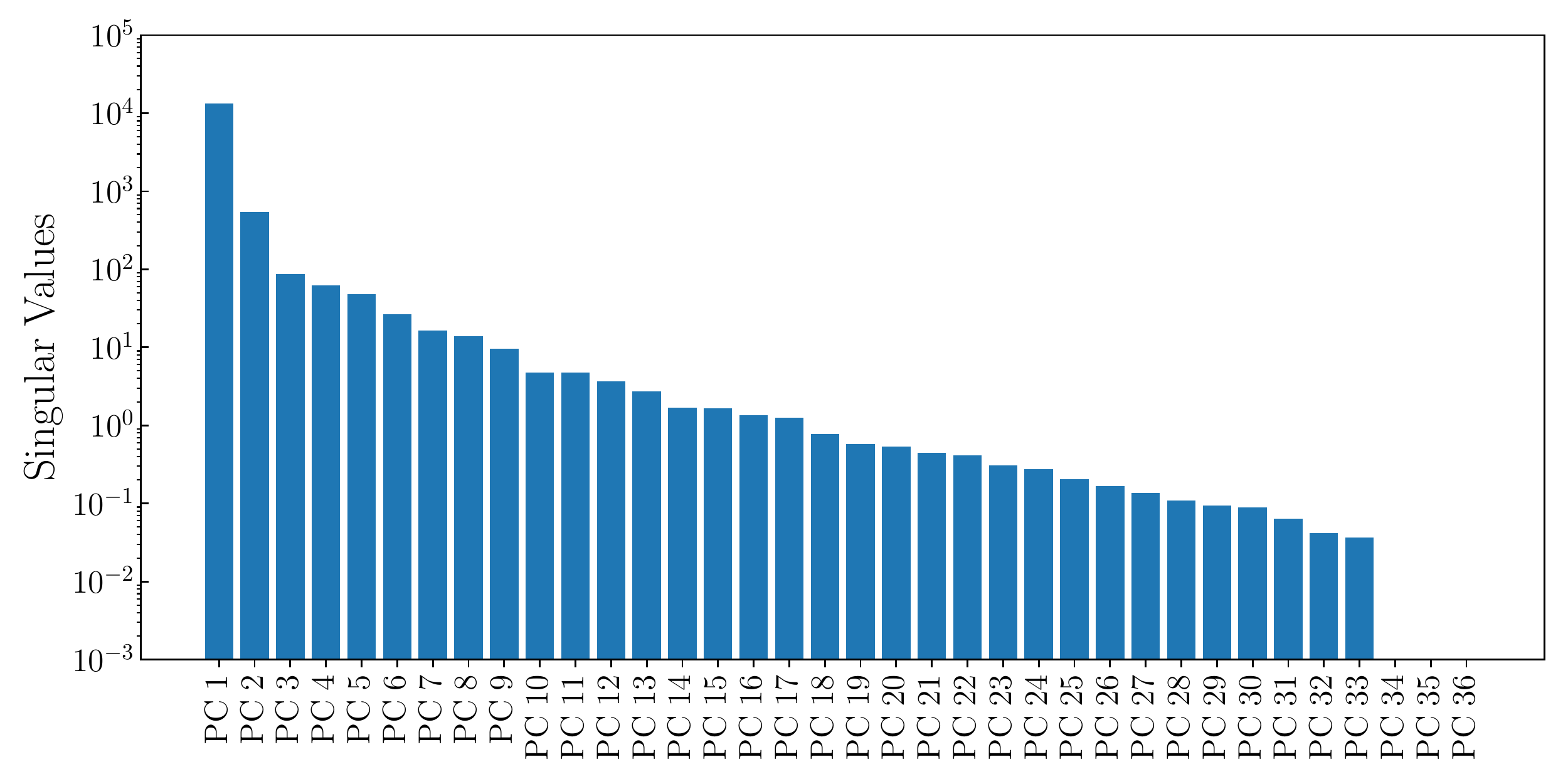}
\caption{\small The distribution of singular values $\lambda_i$ (the elements of the diagonal of the
  matrix $W$), for the principal
  components evaluated for the global fit settings
 summarised in Tables~\ref{tab:operatorbasis} and~\ref{eq:table_dataset_overview}.
  In the linear EFT approximation where the PCA analysis is carried out, there exist three flat
  directions (with vanishing singular values) associated to the four-heavy operators.
  \label{fig:PCA_SVs}
  }
  \end{center}
\end{figure}

Fig.~\ref{fig:PCA_SVs} displays the singular values $\lambda_i$, that is,
the elements of the diagonal
matrix $W$ in the decomposition of Eq.~(\ref{eq:SVD}), for the $n_{\rm op}=36$ principal
components associated to the global fit settings
 summarised in Tables~\ref{tab:operatorbasis} and~\ref{eq:table_dataset_overview}.
From the definition of the matrix $K$, a singular value $\lambda_i\simeq 1$
corresponds to a direction in the parameter space where the magnitude of the (linear) EFT corrections
is of the same size as the associated experimental uncertainties.
We observe that there are three flat directions (principal components with vanishing singular value),
which as shown below can be associated to linear combinations of the four-heavy operators.
Except for these three flat directions, there are no large hierarchies in the distribution
of singular values, indicating that the physical dimensionality of our problem coincides
with that of the chosen fitting basis.
The principal component with the highest singular value is dominated by the bosonic
operator $c_{\varphi D}$, which modifies
the Higgs-gauge interactions and  is well constrained by the EWPOs.

\begin{figure}[t]
  \begin{center}
    \includegraphics[width=0.83\linewidth]{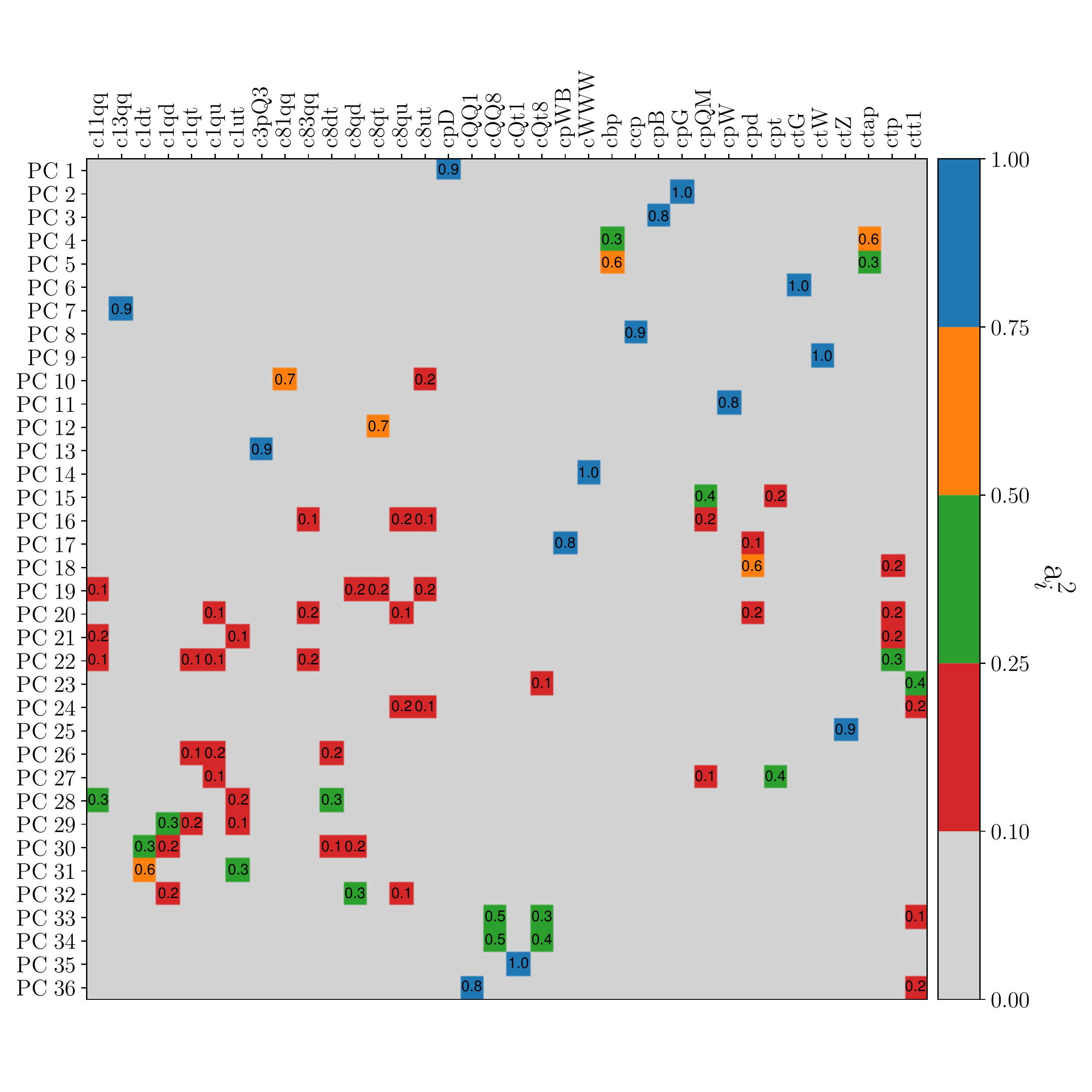}
 \caption{\small Heat map displaying
  the values of the squared coefficients $a_{ki}^2$ that relate the original
  fitting basis to the principal components, Eq.~(\ref{eq:PCdef}), whose
  associated singular values were reported in  Fig.~\ref{fig:PCA_SVs}.
  For the entries with $a_{ki}^2 \ge 0.1$ we also indicate the corresponding numerical value.
  \label{fig:PCA_PCs}
  }
  \end{center}
\end{figure}

Then  Fig.~\ref{fig:PCA_PCs} displays a heat map with
the values of the (squared) coefficients $a_{ki}^2$ that relate the original
fitting basis to the principal components, Eq.~(\ref{eq:PCdef}), and whose
associated eigenvalues are displayed in the upper panel.
For those entries with
$a_{ki}^2 \ge 0.1$ we also indicate the corresponding numerical value.
Since the principal components are normalised, the sum
of the entries associated to a given row in the heat map adds up to unity.
Note also that in this table we have chosen the purely bosonic coefficients
$c_{\varphi W B}$ and $c_{\varphi D}$ to represent the two directions that are left
unconstrained by the EWPOs, see the discussion in Sect.~\ref{sec:operatorbasis}.

From Fig.~\ref{fig:PCA_PCs} one can observe
that some principal components (PC$_k$) are dominated by a single
EFT coefficient from the fitting basis.
Examples of this are $c_{\varphi D}$ (for $k=1$), $c_{\varphi G}$ ($k=2$), $c_{\varphi B}$ ($k=3$),
$c_{t G }$ ($k=4$), and $c_{Qq}^{3,1}$ ($k=7$).
These PCs have associated reasonably large singular values, $\lambda_k\gsim 10$,
and therefore one expects that the corresponding coefficients will be well constrained
from the fit.
Other principal components are instead composed by a superposition
of two or at most three coefficients, for instance $c_{\varphi W B}$ and $c_{\varphi Q}^{(3)}$ are combined
into PC$_{14}$ and PC$_{15}$ with similar weight each.
On the other hand, several PCs arise instead from the combination of a large number of EFT coefficients
without any of them dominating.
This is the case e.g. for PCs associated to combinations involving
of the two-light-two-heavy operators, such $k=22, 23$ and 25, where
no single squared coefficient $a_{ki}^2$ is larger than 0.4.

The three flat directions (vanishing singular values) observed in
Fig.~\ref{fig:PCA_SVs} can be traced back to linear combinations
for four-heavy operators, specifically to the following combinations:
\bea
    {\rm PC}_{34} &=& 0.91 c_{QQ}^{1} - 0.42 c_{tt}^1 \, , \nonumber \\
    {\rm PC}_{35} &=& 0.62 c_{Qt}^{1} - 0.56 c_{Qt}^{8} + 0.49 c_{QQ}^8 \, , \\
    {\rm PC}_{36} &=& 0.78 c_{Qt}^1 -0.50 c_{QQ}^{8}+0.38 c_{tQ}^8 \, ,\nonumber
    \eea
    where we don't indicate the contributions with $a_{ki}^2<0.1$.
    This implies that, in a linear EFT fit, one can only constrain two directions out of the five
four-heavy operators considered.
These flat directions disappear only once we consider the
 quadratic corrections to the $t\bar{t}t\bar{t}$
 and  $t\bar{t}b\bar{b}$ cross-sections.

The results of this PCA indicate that
our choice of fitting basis,
summarised in Table~\ref{tab:operatorbasis}, represents a sensible option
for which well-defined constraints will be obtained from the fit,
up to the previous caveat concerning the four-heavy operators.
Therefore, in our case there is no advantage in carrying out
the fit in the rotated basis spanned by the principal components
 Eq.~(\ref{eq:PCdef}) rather than in the original one.
Furthermore, the lack of large hierarchies in the distribution of singular values
reflects the fact that the true dimensionality of the problem coincides with that
of the original basis.

While for the global dataset genuine flat directions are either absent or removed by quadratic
corrections, this might not be in general the case if we consider
fits to reduced datasets.
In such scenario, one could consider deploying the PCA method to reduce the dimensionality
of the EFT parameter space by removing the directions with singular values below
some threshold before inverting back to the physical basis.
We note that a similar strategy has been successfully applied to
construct compressed Hessian PDF sets in~\cite{Carrazza:2015aoa,Carrazza:2016htc}.
However, in this work we use PCA as a diagnosis tool to guide
the selection of the fitting basis,
and postpone to future work its application to carry out EFT fits
in the PC rotated basis.


%% file: sec-results.tex
\section{Results}
\label{sec:results}

We now present the main results of this work:
the determination of the best-fit values,
confidence level intervals, and posterior probability distributions
associated to the $n_{\rm op}=50$ EFT coefficients (of which 36 are independent) listed in
Table~\ref{tab:operatorbasis} from the global
interpretation of Higgs, top quark, and diboson cross-section measurements.
As motivated in Sect.~\ref{sec:fitsettings}, the results
shown here have been obtained with the NS approach, and we have verified that
equivalent results are obtained with the MCfit method.

First of all, we discuss the quality of the fit,
both for the total and for individual
datasets.
Second, we present the bounds and posterior probability distributions
for the various EFT coefficients, assess their consistency with the
SM hypothesis, and determine their pattern of cross-correlations.
Third, we study the dependence of our results on the choice of input dataset,
in particular with fits based only on top or Higgs data, as well as on that of the theory settings,
where the impact of the NLO QCD corrections to the EFT cross-sections is quantified.
Finally, we present EFT fits  in the top-philic scenario, where the parameter space
is restricted by constraints motivated by specific UV-complete models.
The comparison between SM and SMEFT theory predictions
with the experimental dataset used as input to
the fit is then collected in App.~\ref{sect:app_comparison_data}.

\input{subsec_results_fitquality.tex}

\input{subsec_results_coefficients.tex}

\input{subsec_results_dataset.tex}

\input{subsec_results_loqcd.tex}

\input{subsec_results_topphilic.tex}

%% file: subsec_results_fitquality.tex
\subsection{Fit quality}

To begin with, we investigate the quality of the fit in terms
of the $\chi^2$ values for the individual
datasets as well as for the global one.
The values that will be provided here correspond to
a modified version of Eq.~(\ref{eq:chi2definition2}),
specifically
\begin{equation}
  \chi^2 \equiv \frac{1}{n_{\rm dat}}\sum_{i,j=1}^{n_{\rm dat}}\lp 
  \la \sigma^{(\rm th)}_i\lp {\boldsymbol c}^{(k)} \rp \ra
  -\sigma^{(\rm exp)}_i\rp ({\rm cov}^{-1})_{ij}
\lp 
 \la \sigma^{(\rm th)}_j\lp {\boldsymbol c}^{(k)} \rp \ra
  -\sigma^{(\rm exp)}_j\rp
 \label{eq:chi2definition3}
    \; ,
\end{equation}
where the average over the theory predictions is evaluated over the the $N_{\rm spl}$ samples
provided by NS,
and the covariance matrix is evaluated with the  experimental definition~\cite{Ball:2012wy}.
Note that in general the average over theory predictions does not correspond to the theory
prediction evaluated using the average value of the Wilson coefficients,
\be
\la \sigma^{(\rm th)}_i\lp {\boldsymbol c}^{(k)} \rp \ra \ne
\sigma^{(\rm th)}_i\lp \la  {\boldsymbol c}^{(k)}  \ra \rp \, , 
\ee
due to the presence of the quadratic corrections to the EFT cross-sections.

\begin{figure}[t]
  \begin{center}
    \includegraphics[width=0.99\linewidth]{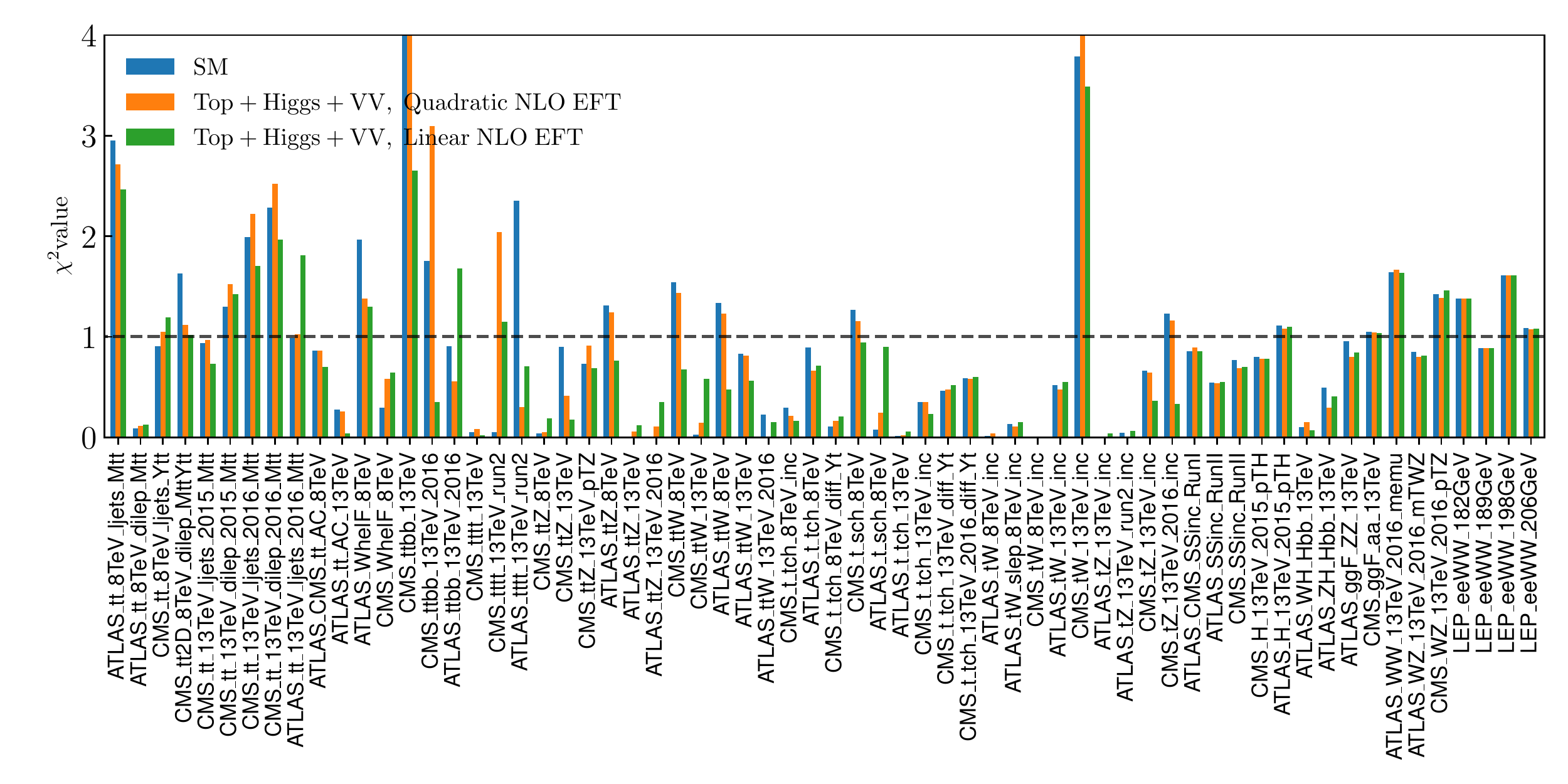}
    \caption{\small Graphical representation of the results of Tables~\ref{eq:chi2-baseline}
      and~\ref{eq:chi2-baseline2},
      displaying the values of the $\chi^2$ per data point, Eq.~(\ref{eq:chi2definition3})
      for the all datasets used as input in the fit.
     The $\chi^2$ values are shown  for the SM and for the two global SMEFT
      baseline fits, based on theory calculations at either the linear $\mathcal{O}\lp \Lambda^{-2}\rp$ or
       quadratic $\mathcal{O}\lp \Lambda^{-4}\rp$ order in the EFT expansion.
     \label{fig:chi2_barplot} }
  \end{center}
\end{figure}

With the figure of merit defined in Eq.~(\ref{eq:chi2definition3}),
we collect in Tables~\ref{eq:chi2-baseline} and~\ref{eq:chi2-baseline2}
the values of the $\chi^2$ per data point corresponding to
the baseline settings of our analysis.
We display both the values based on the SM theory predictions
as well as the best-fit SMEFT results obtained with $\mathcal{O}\lp \Lambda^{-2}\rp$ and
$\mathcal{O}\lp \Lambda^{-4}\rp$ calculations.
Note that, for ease of reference, in these tables
each dataset has associated a hyperlink pointing to the original publication.
For those datasets for which more than one differential distribution is available, we indicate the specific one used in the fit.
Then Table~\ref{eq:chi2-baseline-grouped} presents the
summary of these $\chi^2$ values now indicating the total values for each group of processes
as well as for the global dataset.
Furthermore, the results of Tables~\ref{eq:chi2-baseline}
and~\ref{eq:chi2-baseline2} are graphically represented in Fig.~\ref{fig:chi2_barplot}.

\input{tables/table-chi2-baseline.tex}


\input{tables/table-chi2-grouped.tex}


Let us discuss first the $\chi^2$ results evaluated in terms of the groups
of processes, listed in Table~\ref{eq:chi2-baseline-grouped}.
One can observe that the global $\chi^2$ per data point decreases from 1.05 when using
SM theory to 0.98 (linear) and 1.04 (quadratic) once SMEFT corrections
are accounted for.
Considering the fit quality to the various groups of processes, we find
that the description of the inclusive top-quark pair
cross-sections (without the $A_C$ data),
composed by $n_{\rm dat}=83$ data points, is improved once EFT corrections
are accounted for with $\chi^2=1.46$ in the SM decreasing
to 1.32 and 1.42 in the linear and quadratic EFT fits
respectively.
For the rest of the datasets, and in particular for the Higgs and diboson
measurements, the overall EFT fit quality is similar to that
obtained using SM calculations.

Inspection of the $\chi^2$ values associated to individual datasets
reported in Tables~\ref{eq:chi2-baseline}
and~\ref{eq:chi2-baseline2}, as well
as their graphical representation from Fig.~\ref{fig:chi2_barplot},
reveals that in some cases the agreement
between the prior SM theoretical calculations and the data is poor.
This is the case, in particular, for some of the inclusive $t\bar{t}$ datasets such as
{\tt ATLAS\_tt\_8TeV\_ljets\_mtt} and
{\tt CMS\_tt\_13TeV\_dilep\_2016\_mtt}, binned in terms of the top-quark
pair invariant mass distribution $m_{t\bar{t}}$, with $\chi^2=2.95$ and 2.28
for $n_{\rm dat}=7$ points each.
Such relatively high values of the $\chi^2$ do not necessarily imply the need for
some New Physics effects,
but could also be explained by issues with the modelling of the experimental
systematic correlations in differential distributions, as discussed for the ATLAS 8 TeV
lepton+jets data in~\cite{Czakon:2016olj,Amoroso:2020lgh,ATL-PHYS-PUB-2018-017}.
Nevertheless, when all the inclusive $t\bar{t}$ datasets are considered collectively,
a value of $\chi^2_{\rm SM} = 1.46$ for  the $n_{\rm dat}=83$ data points in the fit is obtained.
In Sect.~\ref{sec:dataset_dependence} we will assess  the stability of the global
fit results by presenting fit variants with the individual datasets
that lead to a poor $\chi^2$ removed.
We will also study variants where distributions sensitive
to the high energy behaviour of the EFT, such as $m_{t\bar{t}}$ in top quark pair
production, have their bins with $m_{t\bar{t}}\gsim 1$ TeV removed from the fit.

Beyond the inclusive $t\bar{t}$ datasets, there are some other instances of a sub-optimal agreement
between SM theory and data.
In all cases, there exist comparable measurements of the same process, either
from the same experiment at a different center-of-mass energy
$\sqrt{s}$ or from a different experiment
at the same value of $\sqrt{s}$, for which the $\chi^2$ reveals good consistency with the SM.
These include {\tt CMS\_ttbb\_13TeV}, when comparing with the same measurement based
on the full 2016 luminosity, and
{\tt CMS\_tW\_13TeV\_inc}, where again the ATLAS measurement at the same $\sqrt{s}$
exhibit as good $\chi^2$.
All in all, one finds a reasonable description of the global input dataset
when using SM cross-sections which is further improved in the EFT fit.

For the rest of this section, when presenting the results of fits corresponding
to variations of the baseline settings, such as fits based on
reduced datasets, we will only indicate the $\chi^2$ values
associated to groups of processes using the same format
as Table~\ref{eq:chi2-baseline-grouped}, rather than to individual
datasets, and comment when required on the results for the latter.

%% file: tables/table-chi2-baseline.tex
\begin{table}[htbp]
  \centering
  \footnotesize
   \renewcommand{\arraystretch}{1.55}
  \begin{tabular}{l|C{1.1cm}|C{2.1cm}|C{2.5cm}|C{2.5cm}}
   \multirow{2}{*}{ Dataset}   & \multirow{2}{*}{$ n_{\rm dat}$} & \multirow{2}{*}{ $\chi^2_{\rm SM}$} &  $\chi^2_{\rm EFT}$   & $\chi^2_{\rm EFT}$     \\
      &   &   & $\mathcal{O}\lp \Lambda^{-2}\rp$ &  $\mathcal{O}\lp \Lambda^{-4}\rp$  \\
 \toprule
 \href{https://arxiv.org/abs/1511.04716}{\tt ATLAS\_tt\_8TeV\_ljets\_mtt} {{\bf (*)}}
 & 7 &  2.95   &  2.46  &  2.71     \\
 \href{https://arxiv.org/abs/1607.07281}{\tt ATLAS\_tt\_8TeV\_dilep\_mtt} & 6 &  0.09   &  0.12  &  0.12   \\
\href{https://arxiv.org/abs/1505.04480}{\tt CMS\_tt\_8TeV\_ljets\_ytt} & 10 &  0.91   &  1.19   &  1.05     \\
\href{https://arxiv.org/abs/1703.01630}{\tt CMS\_tt2D\_8TeV\_dilep\_mttytt} & 16 &  1.63  &  1.01  & 1.12      \\
\href{https://arxiv.org/abs/1610.04191}{\tt CMS\_tt\_13TeV\_ljets\_2015\_mtt} & 8 &  0.94  &  0.72  &  0.97     \\
\href{https://arxiv.org/abs/1708.07638}{\tt CMS\_tt\_13TeV\_dilep\_2015\_mtt} & 6 &  1.30   &  1.42   & 1.52      \\
\href{https://arxiv.org/abs/1803.08856}{\tt CMS\_tt\_13TeV\_ljets\_2016\_mtt}  {{\bf (*)}} & 10 &  1.99  &  1.70  & 2.22      \\
\href{https://arxiv.org/abs/1811.06625}{\tt CMS\_tt\_13TeV\_dilep\_2016\_mtt}  {{\bf (*)}} & 7 & 2.28  & 1.96   & 2.52      \\
\href{https://arxiv.org/abs/1908.07305}{\tt ATLAS\_tt\_13TeV\_ljets\_2016\_mtt} & 7 & 0.99   & 1.81   & 1.02       \\
\href{https://arxiv.org/abs/1709.05327}{\tt ATLAS\_CMS\_tt\_AC\_8TeV} & 6 &  0.86  &  0.70  & 0.86       \\
\href{https://inspirehep.net/literature/1743677}{\tt ATLAS\_tt\_AC\_13TeV} & 5 &  0.03  & 0.32   &  0.26     \\
\href{https://arxiv.org/abs/1612.02577}{\tt ATLAS\_WhelF\_8TeV}  {{\bf (*)}} & 3 &  1.97  & 1.30   & 1.38       \\
\href{https://arxiv.org/abs/1605.09047}{\tt CMS\_WhelF\_8TeV} & 3 &  0.30  & 0.64   &  0.58    \\
\midrule
\href{https://arxiv.org/abs/1509.05276}{\tt ATLAS\_ttZ\_8TeV} & 1 & 1.31   &  0.76     &  1.24      \\
\href{https://arxiv.org/abs/1609.01599}{\tt ATLAS\_ttZ\_13TeV} & 1 &  0.01    & 0.12   &  0.05     \\
\href{https://arxiv.org/abs/1901.03584}{\tt ATLAS\_ttZ\_13TeV\_2016} & 1 & 0.001   &  0.35   &  0.10     \\
\href{https://arxiv.org/abs/1510.01131}{\tt CMS\_ttZ\_8TeV} & 1 &  0.04    &  0.19   &  0.05      \\
\href{https://arxiv.org/abs/1711.02547}{\tt CMS\_ttZ\_13TeV} & 1 & 0.90    & 0.17   &  0.41      \\
\href{https://arxiv.org/abs/1907.11270}{\tt CMS\_ttZ\_13TeV\_pTZ} & 4 &  0.73  &  0.69   & 0.91      \\
\href{https://arxiv.org/abs/1509.05276}{\tt ATLAS\_ttW\_8TeV} & 1 & 1.33   & 0.47    & 1.22      \\
\href{https://arxiv.org/abs/1609.01599}{\tt ATLAS\_ttW\_13TeV} & 1 &  0.83   &  0.56   & 0.81      \\
\href{https://arxiv.org/abs/1901.03584}{\tt ATLAS\_ttW\_13TeV\_2016} & 1 &   0.23   & 0.14    &  0.00      \\
\href{https://arxiv.org/abs/1510.01131}{\tt CMS\_ttW\_8TeV} & 1 &  1.54    &  0.68     &  1.43     \\
\href{https://arxiv.org/abs/1711.02547}{\tt CMS\_ttW\_13TeV} & 1 &  0.03   &   0.57    &  0.14     \\
\midrule
\href{https://arxiv.org/abs/1705.10141}{\tt CMS\_ttbb\_13TeV}  {{\bf (*)}} & 1 &  4.96    &  2.65     &  6.66     \\
\href{https://arxiv.org/abs/1909.05306}{\tt CMS\_ttbb\_13TeV\_2016} & 1 &  1.75    &  0.35     &  3.09     \\
\href{https://arxiv.org/abs/1811.12113}{\tt ATLAS\_ttbb\_13TeV\_2016} & 1 & 0.91     &  1.68     &  0.55     \\
\href{https://arxiv.org/abs/1710.10614}{\tt CMS\_tttt\_13TeV} & 1 &  0.05   & 0.02      &  0.08     \\
\href{https://arxiv.org/abs/1908.06463}{\tt CMS\_tttt\_13TeV\_run2} & 1 &  0.05    & 1.15      &  2.04     \\
\href{https://arxiv.org/abs/2007.14858}{\tt ATLAS\_tttt\_13TeV\_run2}  {{\bf (*)}} & 1 &  2.35   &  0.70     &  0.30     \\
\bottomrule
\end{tabular}
\caption{\small The values of the $\chi^2$ per data point corresponding to
  the baseline settings of our analysis.
  We indicate the results for the $t\bar{t}$ datasets, both in inclusive
  production and in association with vector bosons or heavy quarks.
  We display the SM values and then the best-fit SMEFT results obtained
  in analyses based on theory predictions at either $\mathcal{O}\lp \Lambda^{-2}\rp$ or 
$\mathcal{O}\lp \Lambda^{-4}\rp$ accuracy.
Each dataset has a hyperlink pointing to the original publication.
For those datasets for which more than one differential distribution is available,
we indicate the specific
ones used in the fit.
    {Datasets indicated with {\bf (*)} are excluded from the ``conservative'' EFT
    fit to be discussed in Sect.~\ref{sec:dataset_dependence}.}
\label{eq:chi2-baseline}
}
\end{table}

\begin{table}[htbp]
  \centering
  \footnotesize
  \renewcommand{\arraystretch}{1.50}
  \begin{tabular}{l|C{1.1cm}|C{2.1cm}|C{2.5cm}|C{2.5cm}}
       \multirow{2}{*}{ Dataset}   & \multirow{2}{*}{$ n_{\rm dat}$} & \multirow{2}{*}{ $\chi^2_{\rm SM}/n_{\rm dat}$} &  $\chi^2_{\rm EFT}/n_{\rm dat}$   & $\chi^2_{\rm EFT}/n_{\rm dat}$     \\
      &   &   & $\mathcal{O}\lp \Lambda^{-2}\rp$ &  $\mathcal{O}\lp \Lambda^{-4}\rp$  \\
       \toprule
 \href{https://arxiv.org/abs/1403.7366}{\tt CMS\_t\_tch\_8TeV\_inc} & 2 &  0.29    &  0.17    & 0.21     \\
 \href{https://arxiv.org/abs/1702.02859}{\tt ATLAS\_t\_tch\_8TeV} & 4 &  0.89     & 0.71     & 0.66     \\
 \href{https://cds.cern.ch/record/1956681}{\tt CMS\_t\_tch\_8TeV\_diff\_Yt} & 6 &  0.20    &  0.11    &  0.16    \\
 \href{https://arxiv.org/abs/1603.02555}{\tt CMS\_t\_sch\_8TeV} & 1 &  1.26     &  0.94     &  1.16    \\
 \href{https://arxiv.org/abs/1511.05980}{\tt ATLAS\_t\_sch\_8TeV} & 1 &  0.08     &  0.90    &  0.25    \\
 \href{https://arxiv.org/abs/1609.03920}{\tt ATLAS\_t\_tch\_13TeV} & 2 &  0.01     &  0.06    &  0.02    \\
 \href{https://arxiv.org/abs/1610.00678}{\tt CMS\_t\_tch\_13TeV\_inc} & 2 &  0.35    & 0.24     & 0.35     \\
 \href{https://cds.cern.ch/record/2151074}{\tt CMS\_t\_tch\_13TeV\_diff\_Yt} & 4 &   0.52   & 0.47     &  0.47    \\
 \href{https://arxiv.org/abs/1907.08330}{\tt CMS\_t\_tch\_13TeV\_2016\_diff\_Yt} & 5 &  0.60  & 0.59  & 0.59      \\
\midrule
 \href{https://arxiv.org/abs/1712.02825}{\tt ATLAS\_tZ\_13TeV\_inc} & 1 &  0.00     & 0.04     &  0.00    \\
 \href{https://arxiv.org/abs/2002.07546}{\tt ATLAS\_tZ\_13TeV\_run2\_inc} & 1 &  0.05    &  0.07    &  0.01    \\
 \href{https://arxiv.org/abs/1712.02825}{\tt CMS\_tZ\_13TeV\_inc} & 1 &  0.66    & 0.36     & 0.64     \\
 \href{https://arxiv.org/abs/1812.05900}{\tt CMS\_tZ\_13TeV\_2016\_inc} & 1 &   1.23    &  0.33    &  1.16    \\
 \href{https://arxiv.org/abs/1510.03752}{\tt ATLAS\_tW\_8TeV\_inc} & 1 &  0.02     &  0.01    &  0.05    \\
 \href{https://arxiv.org/abs/2007.01554}{\tt ATLAS\_tW\_slep\_8TeV\_inc} & 1 &   0.13    &  0.15    & 0.11    \\
 \href{https://arxiv.org/abs/1401.2942}{\tt CMS\_tW\_8TeV\_inc} & 1 &  0.00     & 0.00     & 0.00     \\
 \href{https://arxiv.org/abs/1612.07231}{\tt ATLAS\_tW\_13TeV\_inc} & 1 &  0.52     &  0.55    &  0.47    \\
 \href{https://arxiv.org/abs/1805.07399}{\tt CMS\_tW\_13TeV\_inc}  {{\bf (*)}} & 1 &  3.79     & 3.49     &  4.33    \\
 \midrule
 \href{https://arxiv.org/abs/1606.02266}{\tt ATLAS\_CMS\_SSinc\_RunI} & 22 &   0.86    & 0.86     & 0.89     \\
 \href{https://arxiv.org/abs/1909.02845}{\tt ATLAS\_SSinc\_RunII} & 16 &   0.54   &  0.55    &  0.54    \\
 \href{https://arxiv.org/abs/1809.10733}{\tt CMS\_SSinc\_RunII} & 24 &  0.77     &  0.70    &  0.68    \\
 \href{https://arxiv.org/abs/1909.02845}{\tt ATLAS\_ggF\_ZZ\_13TeV} & 6 &  0.96     &   0.84   &  0.81    \\
 \href{https://inspirehep.net/literature/1725274}{\tt CMS\_ggF\_aa\_13TeV} & 6 & 1.05   & 1.04   & 1.05     \\
 \href{http://cdsweb.cern.ch/record/2682844/files/ATLAS-CONF-2019-032.pdf}{\tt ATLAS\_H\_13TeV\_2015\_pTH} & 9 & 1.11      & 1.10     & 1.08     \\ 
 \href{https://arxiv.org/abs/1812.06504}{\tt CMS\_H\_13TeV\_2015\_pTH} & 9 &  0.80     &  0.78    & 0.78     \\
 \href{https://arxiv.org/abs/1903.04618}{\tt ATLAS\_WH\_Hbb\_13TeV} & 2 &   0.10    &  0.07    &  0.15    \\
 \href{https://arxiv.org/abs/1903.04618}{\tt ATLAS\_ZH\_Hbb\_13TeV} & 3 &   0.50    &  0.41    &  0.30    \\
 \midrule
 \href{https://arxiv.org/abs/1905.04242}{\tt ATLAS\_WW\_13TeV\_2016\_memu} & 13 & 1.64   &  1.64    &  1.67    \\
 \href{https://arxiv.org/abs/1902.05759}{\tt ATLAS\_WZ\_13TeV\_2016\_mTWZ} & 6 &  0.81   &  0.81  & 0.80     \\
 \href{https://arxiv.org/abs/1901.03428}{\tt CMS\_WZ\_13TeV\_2016\_pTZ} & 11 &  1.46   &  1.44    &  1.39    \\
 \href{https://arxiv.org/abs/1302.3415}{\tt LEP\_eeWW\_182GeV} & 10 &  1.38  &  1.38   & 1.38     \\
 \href{https://arxiv.org/abs/1302.3415}{\tt LEP\_eeWW\_189GeV} & 10 & 0.88    & 0.88   &  0.89    \\
 \href{https://arxiv.org/abs/1302.3415}{\tt LEP\_eeWW\_198GeV} & 10 &  1.61   &  1.61    &  1.61    \\
 \href{https://arxiv.org/abs/1302.3415}{\tt LEP\_eeWW\_206GeV} & 10 &  1.09   &  1.08   &  1.08    \\
\bottomrule
\end{tabular}
  \caption{\small Same as Table~\ref{eq:chi2-baseline} now for the single top
    datasets (inclusive and in association with gauge bosons), the Higgs production and decay measurements (signal streghts and differential
    distributions), and the LEP and LHC
    diboson cross-sections.
\label{eq:chi2-baseline2}
}
\end{table}

%% file: tables/table-chi2-grouped.tex
\begin{table}[htbp]
  \centering
  \small
   \renewcommand{\arraystretch}{1.70}
   \begin{tabular}{l|C{1.0cm}|C{1.8cm}|C{2.7cm}|C{2.7cm}}
        \multirow{2}{*}{ Dataset}   & \multirow{2}{*}{$ n_{\rm dat}$} & \multirow{2}{*}{ $\chi^2_{\rm SM}$} &  $\chi^2_{\rm EFT}$   & $\chi^2_{\rm EFT}$     \\
      &   &   & $\mathcal{O}\lp \Lambda^{-2}\rp$ &  $\mathcal{O}\lp \Lambda^{-4}\rp$  \\
        \toprule
 $t\bar{t}$ inclusive  & 83 &  1.46    &  1.32    &  1.42       \\
 $t\bar{t}$ charge asymmetry  & 11 &  0.60    &  0.39    &  0.59       \\
 $t\bar{t}+V$  & 14  &  0.65     &  0.48     &  0.65       \\
 single-top inclusive  &  27  &   0.43    &  0.44     &  0.41       \\
 single-top $+V$ & 9  &  0.71     &  0.55     &  0.75       \\
 $t\bar{t}b\bar{b}$ \& $t\bar{t}t\bar{t}$   & 6   &  1.68     &  1.09     &  2.12       \\
 Higgs signal strenghts (Run I)  &  22  &   0.86     &  0.85     &  0.90       \\
 Higgs signal strenghts (Run II)  &  40 &   0.67    &  0.64     &  0.63       \\
 Higgs differential \& STXS  &  35  & 0.88     &  0.85     &   0.83      \\
 Diboson (LEP+LHC)  & 70  &  1.31     &  1.31     &  1.30       \\
 \midrule
 {\bf Total}  & {\bf 317}  &  {\bf 1.05 }   & {\bf 0.98 }  & {\bf 1.04 } \\
\bottomrule
\end{tabular}
\caption{\small Summary of the $\chi^2$ results listed in Tables~\ref{eq:chi2-baseline}
  and~\ref{eq:chi2-baseline2}.
  We indicate the total values for each group of processes
  as well as for the global dataset.
\label{eq:chi2-baseline-grouped}
}
\end{table}

%% file: subsec_results_coefficients.tex
\subsection{Constraints on the EFT parameter space}

Following this assessment of the fit quality,
we move to present the constraints on the SMEFT parameter space
that can be derived from the present global fit.
We will present results for the $n_{\rm op}=50$ Wilson
coefficients listed in Table~\ref{tab:operatorbasis},
with the understanding that only 36 of them are linearly
independent.\footnote{We note that the EWPO constraints of Eq.~(\ref{eq:2independents}) set
  the four-lepton operator to zero, $c_{\ell\ell}=0$, and hence we exclude this coefficient
from the plots and tables of this section.}
Specifically, we provide the 95\% confidence level intervals for each
EFT coefficients,
study their posterior probability distributions, evaluate the pattern
of their correlations, and compare the marginalised bounds with those
obtained in individual fits where only one coefficient is varied at a time.
We will also assess the overall consistency of the fit results with respect
to the Standard Model hypothesis.
The results discussed here correspond to the global dataset with the baseline theory
settings for both $\mathcal{O}\lp \Lambda^{-2}\rp$ and $\mathcal{O}\lp \Lambda^{-4}\rp$
theory calculations.
Fits based on either reduced datasets or alternative theory settings
are then discussed in Sects.~\ref{sec:dataset_dependence}
and~\ref{subsec:loqcd}  respectively.

\begin{figure}[t]
  \begin{center}
     \includegraphics[width=0.99\linewidth]{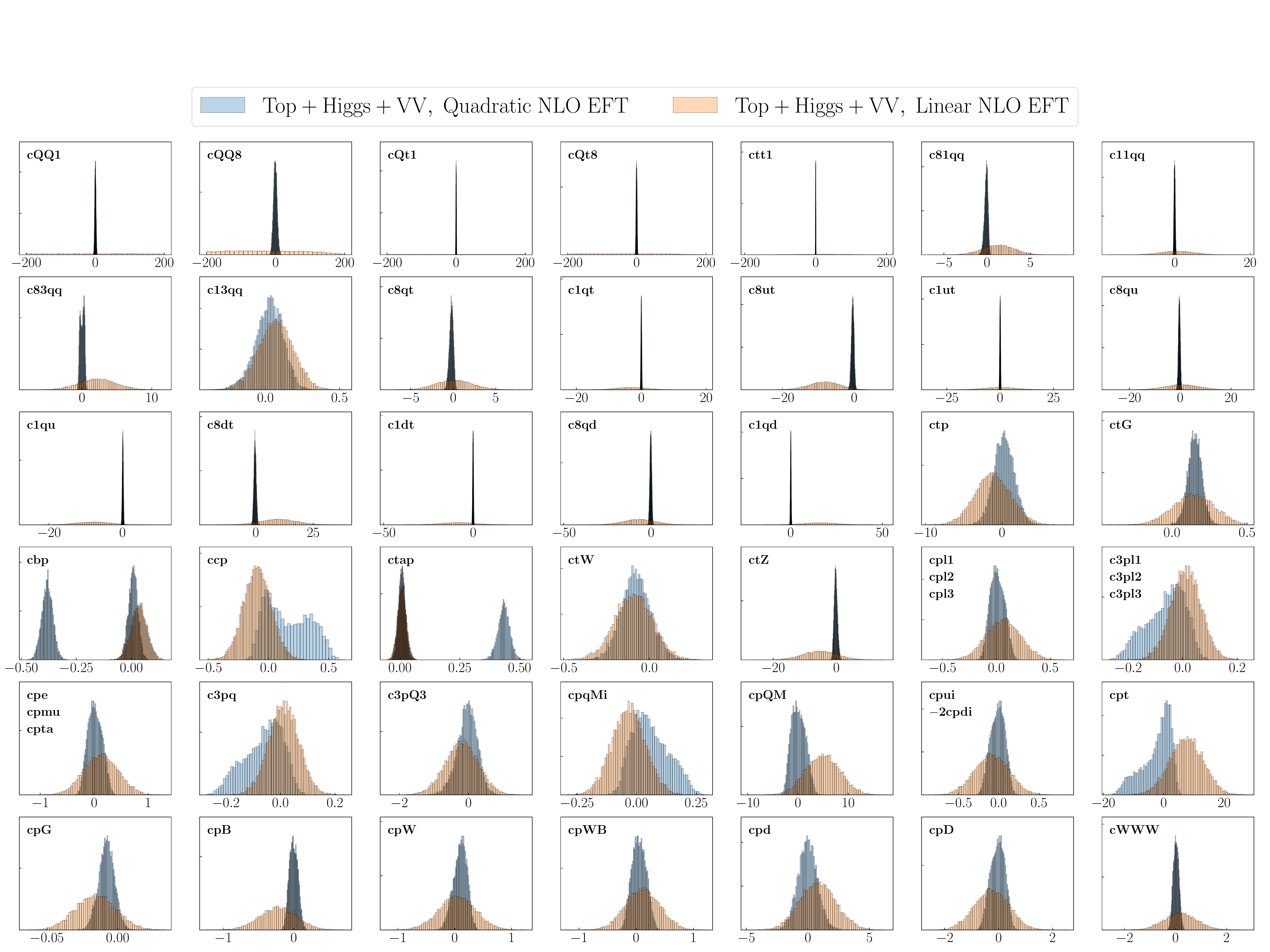}
    \caption{\small The normalised posterior probability distributions associated
      to each of the $n_{\rm op}=50$ fit coefficients considered in the present
      analysis, for both the linear and quadratic EFT fits.
      Note that the $x$-axis range is different in each case.
      From top to bottom and from left to right, we display
      the four-heavy, two-light-two-heavy,
      two-fermion, and purely bosonic coefficients.
      Only 36 of these coefficients are independent
      as indicated in Table~\ref{tab:operatorbasis}.
     \label{fig:posterior_coeffs} }
  \end{center}
\end{figure}

\paragraph{Posterior distributions.}
Fig.~\ref{fig:posterior_coeffs} displays
the normalised posterior probability distributions associated
to each of the $n_{\rm op}=50$ fit coefficients considered in the present
analysis, for the linear (blue) and quadratic (orange) EFT fits.
As discussed in Sect.~\ref{sec:nestedsampling},
the NS prior sampling volumes have been optimised to ensure
that the posterior distribution associated to each coefficient
is fully contained within them.
One can observe how in general
the $\mathcal{O}\lp \Lambda^{-4}\rp$ corrections modify significantly
 the distributions that is obtained from the linear fits, for instance
 by shifting its median or by decreasing its variance.
 For several coefficients, the posterior distributions would be poorly
 described in the Gaussian approximation,
 and in some cases one finds multi-modal distributions
 such as for the Yukawa operators $c_{\varphi c}$, $c_{\varphi b}$, and
 $c_{\varphi \tau}$.
 Such double-humped distributions can be traced back to the (quasi)-degenerate
 minima in the individual
 $\chi^2$ profiles reported in
 Figs.~\ref{fig:quartic-individual-fits} and~\ref{fig:quartic-individual-fits-2}.
 We can also observe how the four-heavy coefficients can only be meaningfully
 constrained in the quadratic fit.
 All in all, inclusion of the quadratic EFT corrections modifies
 in a significant manner the posterior distributions associated
 to most of the fit coefficients as compared to the linear approximation.

\begin{figure}[t]
  \begin{center}
    \includegraphics[width=0.99\linewidth]{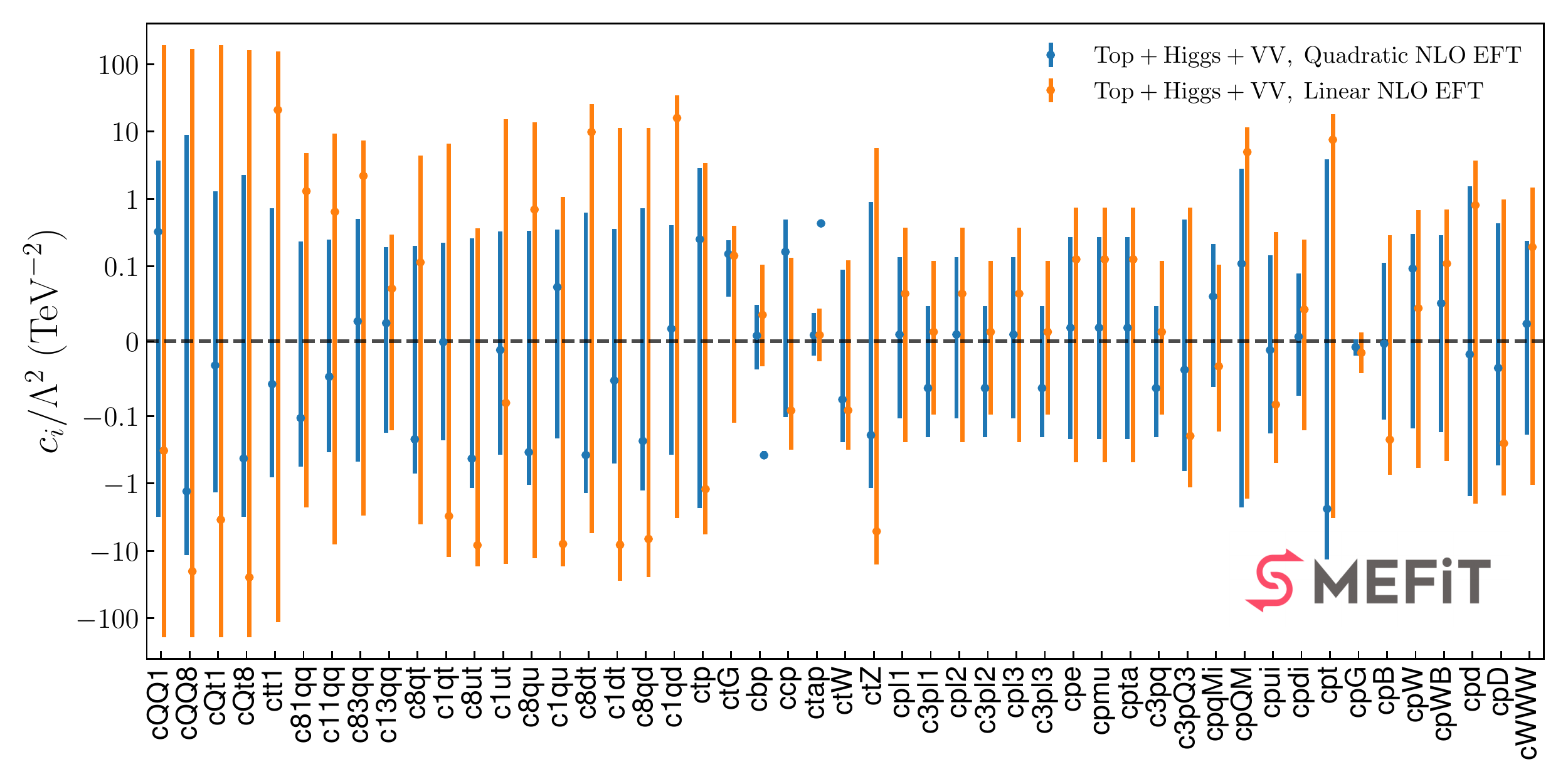}
    \vspace{-0.1cm}
    \caption{\small The best-fit (median) value of the EFT coefficients $c_i/\Lambda^2$
      and their associated 95\% CL intervals for the  global fits
      based on either linear or quadratic EFT calculations,
      whose posterior distributions are represented in
      Fig.~\ref{fig:posterior_coeffs}.
      The dashed horizontal line indicates the SM expectation.
     \label{fig:globalfit-baseline-coeffsabs-lin-vs-quad} }
  \end{center}
\end{figure}

\paragraph{Confidence level intervals.}
From the posterior probability distributions displayed in Fig.~\ref{fig:posterior_coeffs},
one can derive the marginalised 95\% CL intervals on the 
EFT coefficients both for the linear and quadratic fits.
These results are collected in Table~\ref{tab:coeff-bounds-baseline}
(for $\Lambda=1$ TeV)
and represented graphically in
Fig.~\ref{fig:globalfit-baseline-coeffsabs-lin-vs-quad}.
In addition,  Table~\ref{tab:coeff-bounds-baseline} also includes
the corresponding obtained in  individual NS fits, where only one
operator is varied at a time and the rest are set
to their SM values (recall the $\chi^2$ profiles from
Figs.~\ref{fig:quartic-individual-fits} and~\ref{fig:quartic-individual-fits-2}).
We will further discuss the outcome of these individual fits below.

\input{tables/coeff-bounds-baseline.tex}

From the marginalised bounds displayed in Fig.~\ref{fig:globalfit-baseline-coeffsabs-lin-vs-quad},
one can observe how the uncertainties associated to the fit coefficients are in all cases
reduced in the quadratic fit in comparison to the linear one.
The 95\% CL interval is disjoint for the Yukawa coefficients $c_{b\varphi}$
and $c_{\tau\varphi}$ in the quadratic fit,
with both a SM-like solution and a second one far from the SM.
For the linear fit, we find that all EFT coefficients agree with the SM expectation
at the 95\% CL level.
For the quadratic fit instead, this is not the case only for
the chromo-magnetic operator $c_{tG}$.
We will trace back below the origin of this discrepancy,
here we only point out that at the level of individual fits
$c_{tG}$ exhibits the same trend but there
agrees with the SM at the  95\% CL
as indicated in Fig.~\ref{fig:quartic-individual-fits-2}.
{  We note that for unconstrained operators, such as the four-heavy operators
  in the linear fit, the best-fit value (median) should be ignored since
the underlying posterior is essentially flat.}

\begin{figure}[t]
  \begin{center}
 \includegraphics[width=0.85\linewidth]{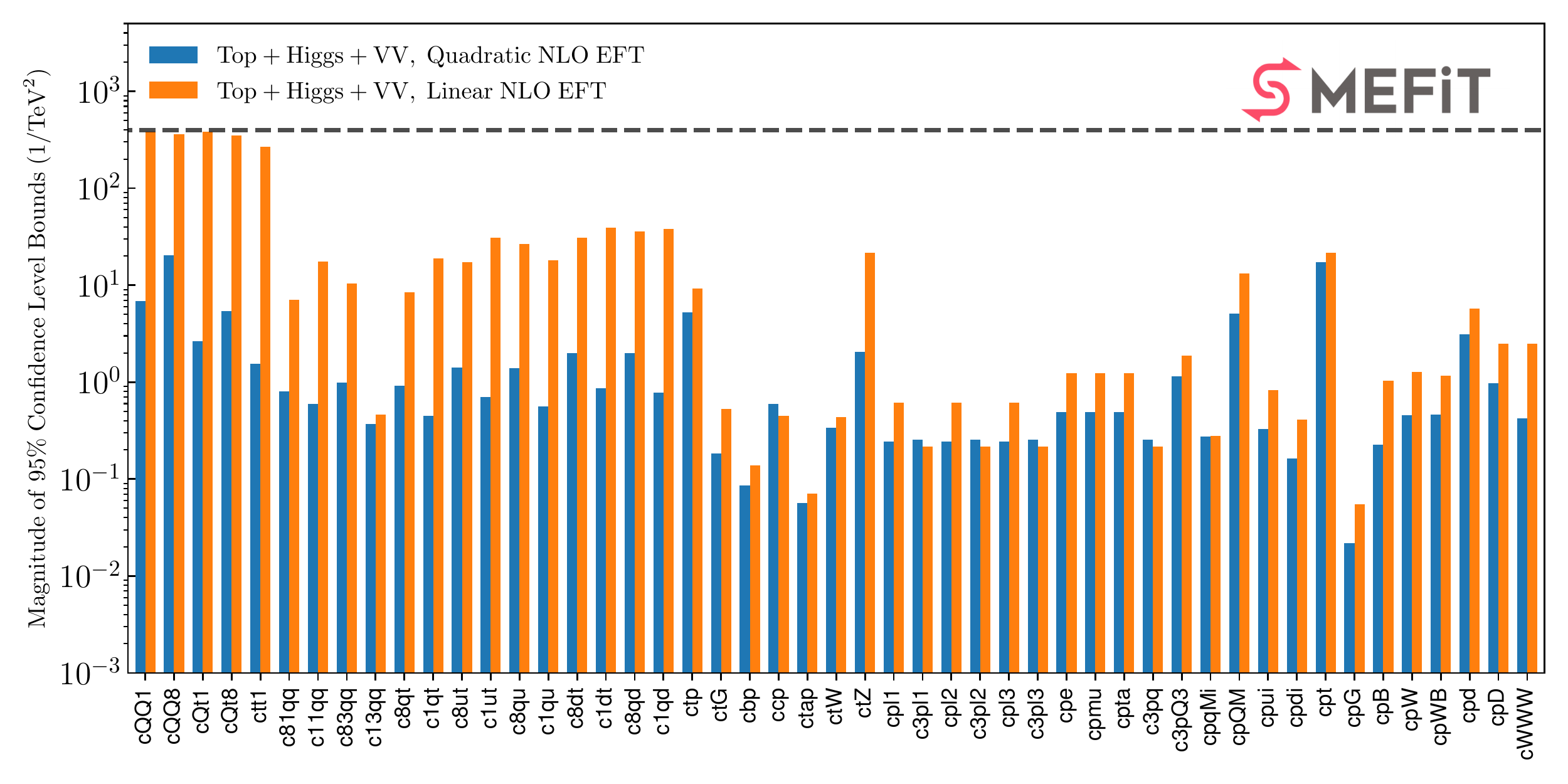}
 \includegraphics[width=0.85\linewidth]{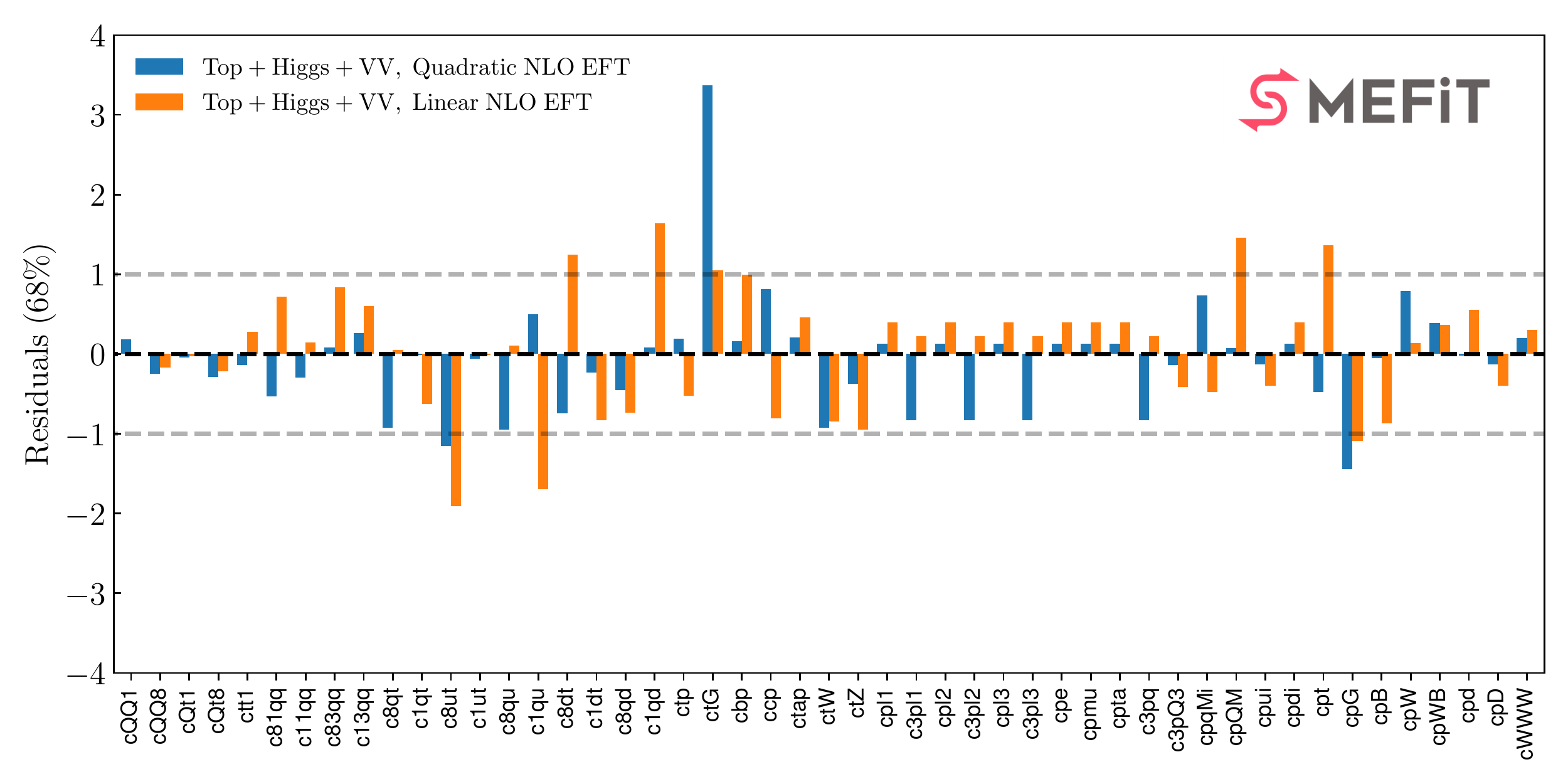}
 \vspace{-0.3cm}
 \caption{\small The magnitude of the 95\% CL intervals (top)
   and the value of the 68\% CL residuals compared to the SM hypothesis (bottom panel)
   corresponding to the global fit results
   displayed in Fig.~\ref{fig:globalfit-baseline-coeffsabs-lin-vs-quad}.
   In the upper plot, the dashed horizontal line indicates the maximum
   prior volume used for the sampling of unconstrained coefficients.
         \label{fig:globalfit-baseline-bounds-lin-vs-quad} }
  \end{center}
\end{figure}

The global fit results of Fig.~\ref{fig:globalfit-baseline-coeffsabs-lin-vs-quad}
are further scrutinized in Fig.~\ref{fig:globalfit-baseline-bounds-lin-vs-quad},
which displays both the magnitude of the 95\% CL intervals 
   and the 68\% CL residuals compared to the SM hypothesis
   associated to the linear and quadratic EFT fits.
   In the upper panel,
      the horizontal line indicates the boundaries of the sampling volume
      used for the poorly-constrained coefficients
as explained in Sect.~\ref{sec:nestedsampling}.
From these comparisons, one can observe how the inclusion of quadratic corrections
leads to markedly more stringent bounds for most of the fit coefficients,
a trend which is specially significant for the four-heavy (unconstrained
in the linear fit) and two-light-two-heavy operators which modify the properties
of the top quark.
The only exception is the charm Yukawa coefficient $c_{\varphi c}$, since there the quadratic
corrections introduce a second degenerate solution thus enlarging the magnitude
of the CL interval.

\begin{figure}[t]
  \begin{center}
\includegraphics[width=0.70\linewidth]{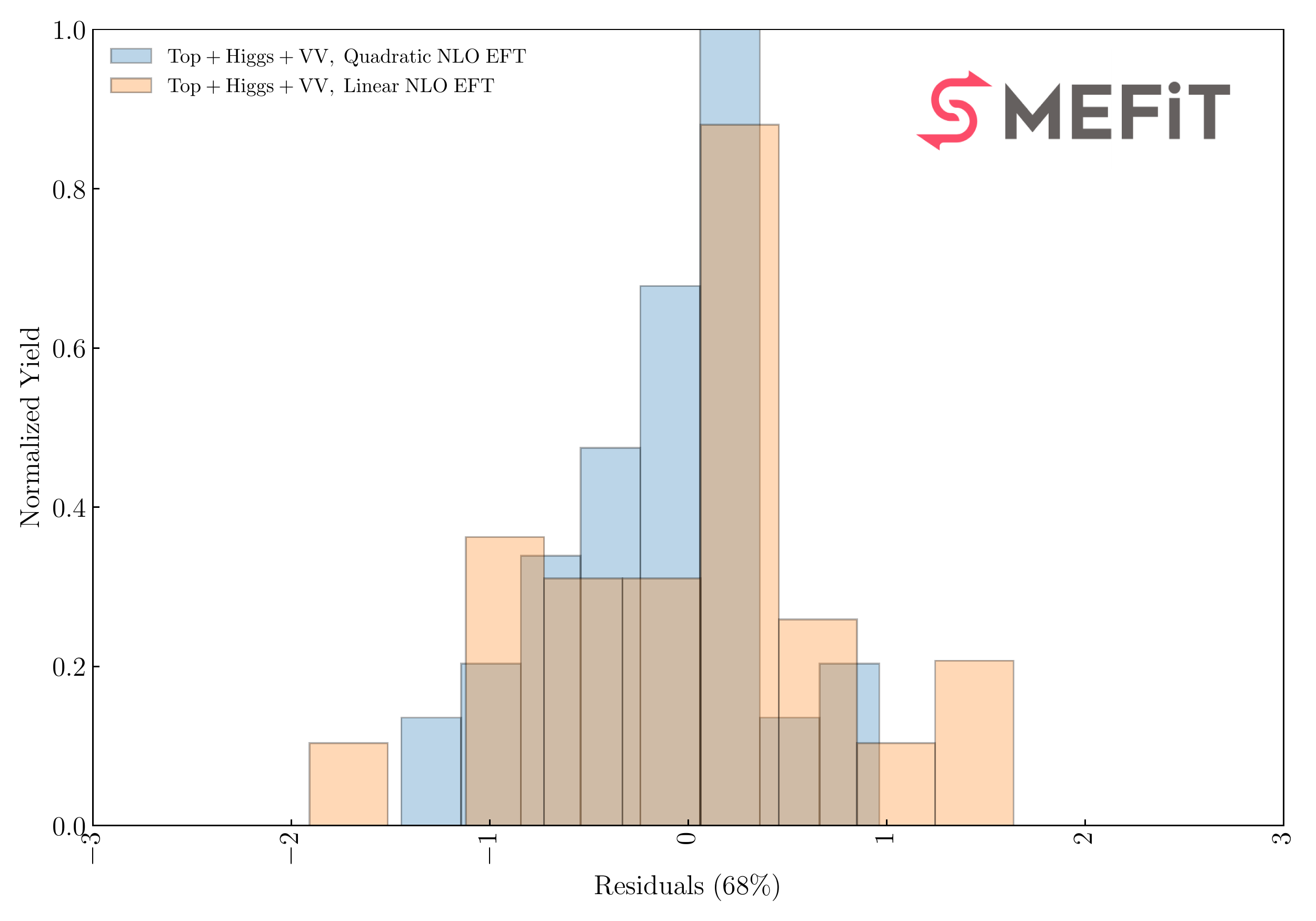}
    \vspace{-0.2cm}
    \caption{\small The (normalised) distribution of
      the fit residuals shown in the bottom panel of
      Fig.~\ref{fig:globalfit-baseline-bounds-lin-vs-quad}.
     \label{fig:residuals-histo} }
  \end{center}
\end{figure}

The 68\% CL residuals displayed in the bottom panel of
Fig.~\ref{fig:globalfit-baseline-bounds-lin-vs-quad} are defined by
\be
\label{eq:fit_residual}
R(c_i)\equiv \frac{\lp c_i|_{\rm EFT} -c_i|_{\rm SM} \rp }{\delta c_i} \, ,\qquad
i=1,\ldots,n_{\rm op} \, ,
\ee
with $c_i|_{\rm EFT}$ being the median of the posterior distribution
from the EFT fit, $c_i|_{\rm SM}=0$, and $\delta c_i$ is the total fit uncertainty
for this parameter.
We can observe that $|R(c_i)|\lsim 1$ for most of the fit coefficients,
both for the linear and quadratic cases.
The only exception is $c_{tG}$, where a residual of
$R(c_{tG})\simeq 3.5$ is found in the quadratic fit.
Nevertheless, for a large enough number of EFT coefficients
one would expect a fraction of these residuals to be larger than
unity, even if the SM is the
underlying theory.
Fig.~\ref{fig:residuals-histo} then displays
the normalised distribution of
these fit residuals.
While these coefficients are correlated among them (see the
following discussion) and thus
cannot be treated as independent variables,
the shapes of these distributions are reasonably close
to a Gaussian, specially for the linear fit, highlighting
again the overall consistency of the fit results
with the SM expectations.

\begin{figure}[htbp]
  \begin{center}
  \includegraphics[width=0.49\linewidth]{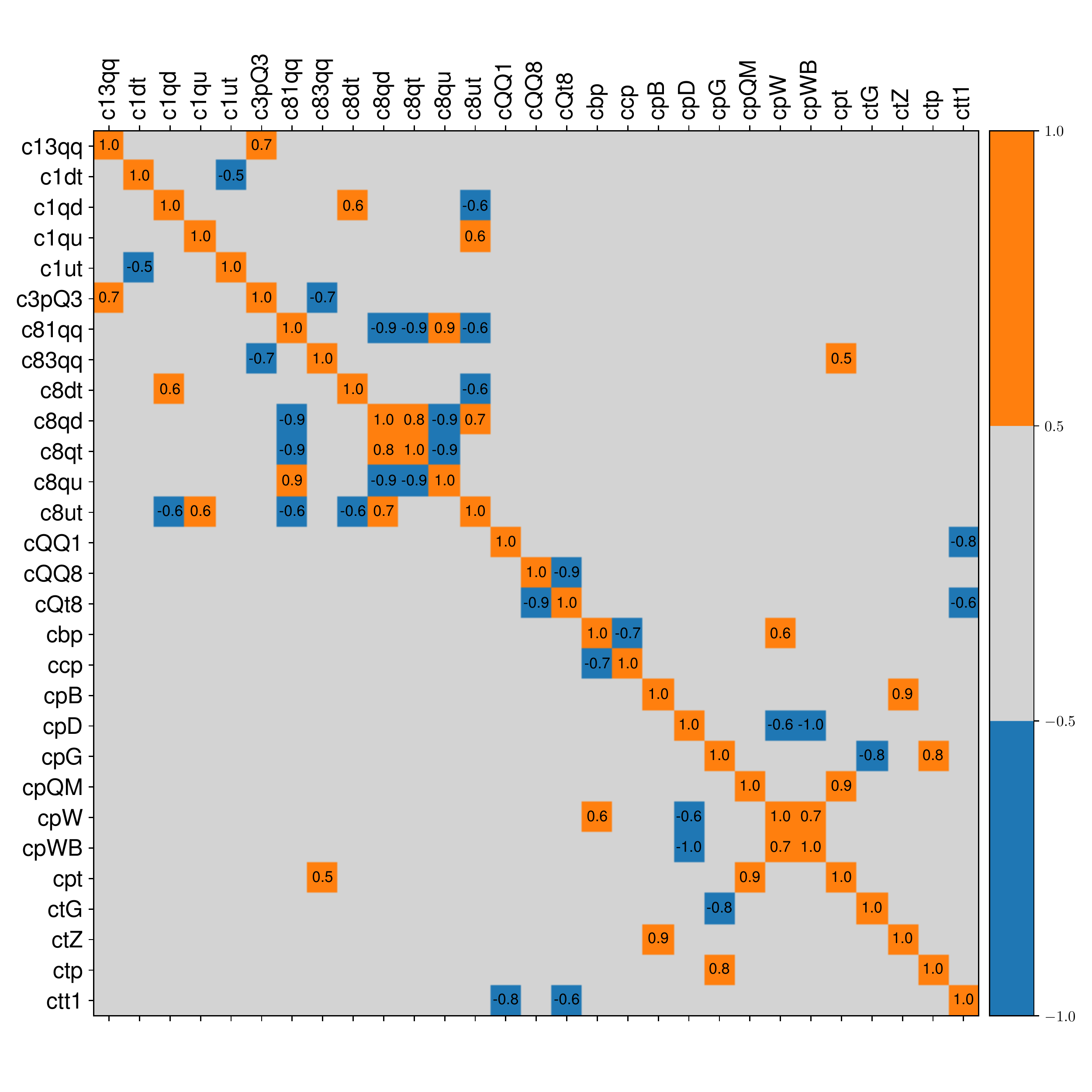}
  \includegraphics[width=0.49\linewidth]{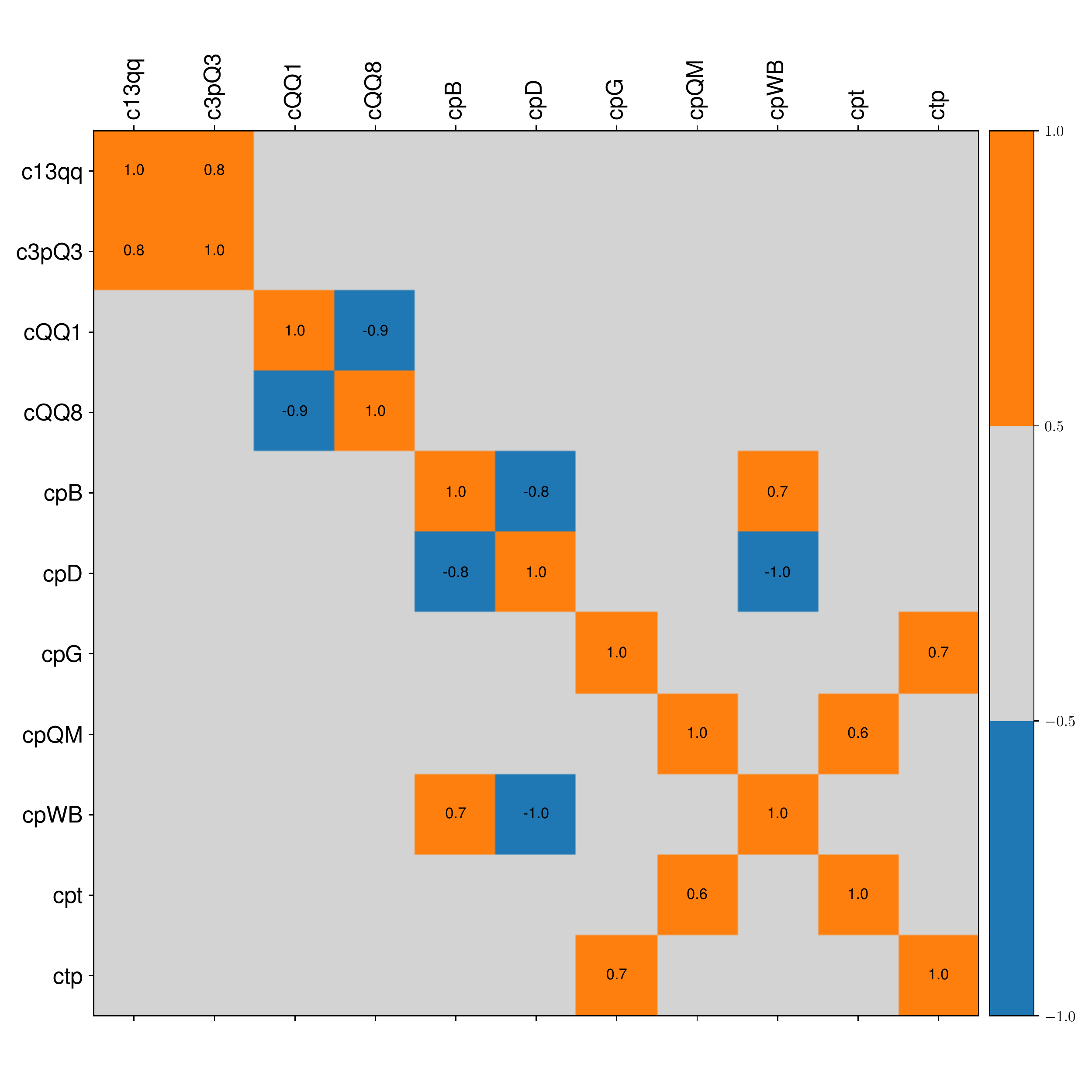}
  \vspace{-0.3cm}
  \caption{\small The correlation coefficients $\rho\lp c_i,c_j\rp$
    between the EFT coefficients
      in the linear (left) and quadratic (right panel) fits.
      We only display the  entries with significant (anti)-correlation,
      $|\rho|\ge 0.5$.
      Pairs of coefficients $(c_i,c_j)$ that do not displayed here have a correlation
coefficient below this threshold.
     \label{fig:globalfit-correlations} }
  \end{center}
\end{figure}

\paragraph{Correlations.}
The correlation coefficient between any two fit coefficients
$c_i$ and $c_j$ can be evaluated as follows,
\be
\label{eq:correlationL2CT}
\rho\lp c_i,c_j\rp=\frac{\lp \frac{1}{N_{\rm spl}}\sum_{k=1}^{N_{\rm spl}}
c_i^{(k)} c_j^{(k)}\rp -\la c_i\ra \la c_j\ra
}{\delta c_i \delta c_j} \, ,\qquad i,j=1,\ldots,n_{\rm op} \, ,
\ee
where $N_{\rm spl}$ denotes the number of samples produced by NS, $\la c_i\ra$ indicates
the mean value of this coefficient, and, as in Fig.~\ref{eq:fit_residual},
$\delta c_i$ is the corresponding uncertainty (standard deviation).
The values of Eq.~(\ref{eq:correlationL2CT})
are displayed in Fig.~\ref{fig:globalfit-correlations} 
separately for the linear  and quadratic fits.
We display only the numerical values for the pair-wise
coefficient combinations for which the correlation coefficient
is numerically significant,
$|\rho(c_i,c_j)|\ge 0.5$.
The pairs $(c_i,c_j)$ that do not appear in Fig.~\ref{fig:globalfit-correlations}  have a correlation
coefficient below this threshold.

We observe how the  majority of
the fit coefficients are  loosely correlated among them, that is,
their correlations being $|\rho| \le 0.5$.
One  also finds that while several of the two-light-two-heavy coefficients turn out to be
strongly correlated
at the linear EFT level, this pattern disappears once   the quadratic  corrections
are accounted for.
Concerning the two-fermion operators,
the correlation patterns present at the linear level
are also often reduced  in the quadratic fits.
For instance, $c_{tZ}$ displays a strong correlation with $c_{\varphi B}$ at the linear level
which is then washed out  by the quadratic effects.
The purely bosonic operators exhibit
in general more stable correlations, for example $c_{\varphi W B}$ 
is strongly anti-correlated  with
$c_{\varphi D}$ in a manner which is similar in the linear and the quadratic fits.
Furthermore,
we do not find any pair of fit coefficients where the quadratic corrections
flip the sign of their correlation.

In general, from Fig.~\ref{fig:globalfit-correlations} one can
conclude that only a moderate subset of Wilson coefficients end
up being strongly (anti-)correlated
among them after the fit, specially so once quadratic EFT corrections are taken into account.
This finding is partially explained by our wide input dataset,
which makes possible constraining independently most if not all
the EFT degrees of freedom.

{  For completeness, App.~\ref{sec:fullcovmat}
  provides the
  correlation matrices for the complete set of operators
considered in this analysis.}

\begin{figure}[t]
  \begin{center}
    \includegraphics[width=0.91\linewidth]{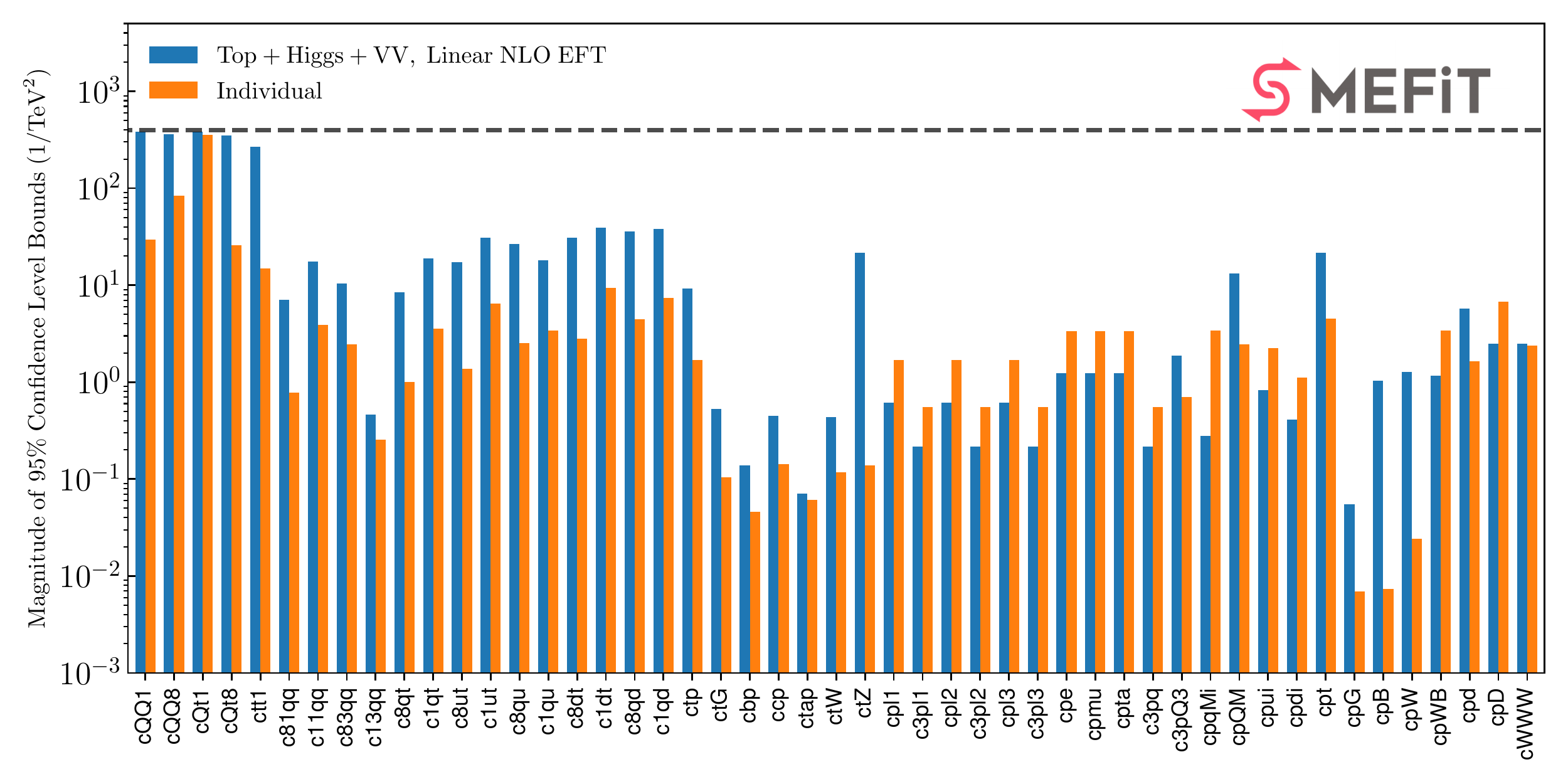}
    \includegraphics[width=0.91\linewidth]{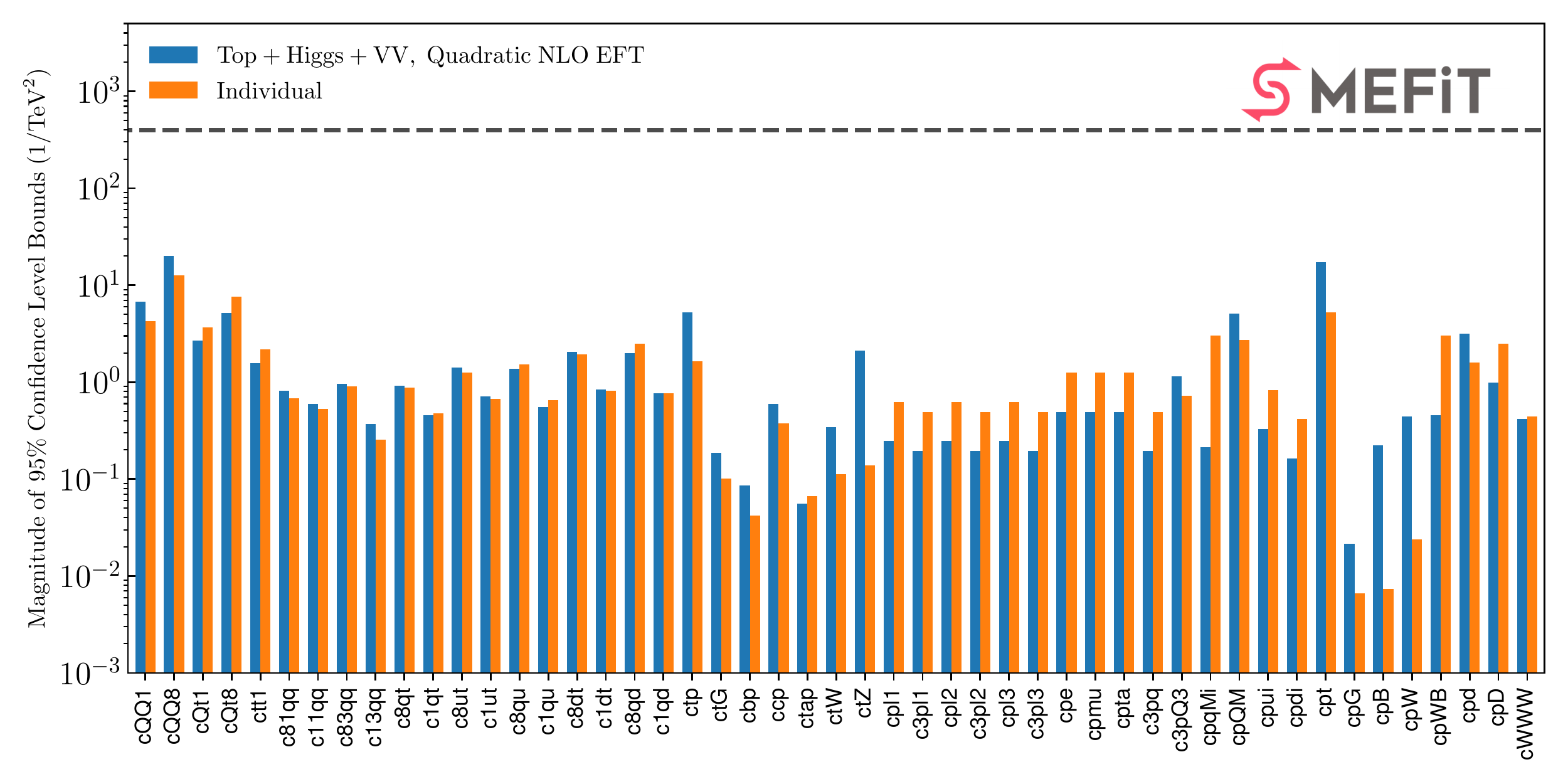}
    \vspace{-0.4cm}
    \caption{\small Comparison of the magnitude of 95\% CL intervals in the global
      (marginalised) and individual fits at the linear (top) and quadratic
      (bottom) level, see also Table~\ref{tab:coeff-bounds-baseline}.
     \label{fig:globalfit-baseline-bounds-lin-NS-ind-vs-marg} }
  \end{center}
\end{figure}

\paragraph{Individual fits.}
As motivated in Sect.~\ref{sec:quarticfits}, individual
(one-parameter) fits have several useful applications.
These include representing a benchmark reference for the global fit
results, where the obtained bounds can only loosen as compared
to one-parameter fits.
The 95\% CL bounds associated to the one-parameter linear and quadratic EFT
fits were reported in Table~\ref{tab:coeff-bounds-baseline},
and the corresponding graphical comparison with the marginalised
global fit results is displayed in
Fig.~\ref{fig:globalfit-baseline-bounds-lin-NS-ind-vs-marg}.
{  While the expectation is that individual bounds
are comparable or more stringent than the marginalised ones, }
this property does not necessarily hold for the coefficients
constrained by the EWPOs, for which an individual fit is not meaningful.
{  Indeed, one-parameter fits are ill-defined in
  the case of the coefficients constrained by EWPOs since
  these coefficients cannot be determined independently from each other.
  Hence the comparison between marginalised and individual bounds
  is only meaningul for the 34 independent coefficients listed
  in Table~2.5.
}

Considering first the results of the linear analysis,
one can observe how for the fitted degrees of freedom the individual bounds are tighter
(or at most comparable) than the marginalised
ones by a large amount, around a factor ten or more in most cases.
These differences are particularly striking for some of the two-fermion operators,
in particular for $c_{tZ}$, as well as for bosonic operators such as
$c_{\varphi B}$ and $c_{\varphi W}$,
for which the differences between the individual and marginalised results can be as large
as two orders of magnitude.
Specifically, in the cases of $c_{tZ}$ and $c_{\varphi B}$, the 95\% CL intervals
found in the linear EFT analysis are
increased as follows when going from the individual to the marginalised fits:
\bea
\nonumber
c_{tZ}:\qquad [-0.04,0.10]~~{\rm (individual)}  \quad&{\rm vs}&         \quad [-17,5.6]~~{\rm (marginalised)} \, ,\\
c_{\varphi B}:~~ \quad [-0.005,0.002] ~~{\rm (individual)}  \quad&{\rm vs}&   \quad  [-0.7,0.3]~~{\rm (marginalised)}\, .
\nonumber
\eea
This effect clearly emphasizes the importance of adopting a fitting basis as wide as possible,
in order to avoid obtaining artificially stringent bounds simply because one is being
blind to other relevant directions of the parameter space.
One important exception of this rule would be those cases where one is guided by 
specific UV-complete models, which motivate the reduction in the parameter space
to a subset of operators.
We also note that the triple gauge operator $c_W$ is one of the few coefficients whose individual
and marginalised bounds are identical: this can be traced back
to the fact that this operator is very weakly correlated with
other coefficients (see also  Fig.~\ref{fig:globalfit-correlations}), being
constrained exclusively by the diboson data.

Inspection of the corresponding results
from the quadratic fits, bottom panel of Fig.~\ref{fig:globalfit-baseline-bounds-lin-NS-ind-vs-marg}, 
 reveals that the differences between individual and marginalised bounds are in general
smaller as compared to the linear case.
This effect is particularly visible
for the two-light-two-heavy and the four-heavy operators, for which one finds
that the individual fits underestimate the magnitude of the 95\% CL interval by around
a factor two on average, rather than by a factor 10 as in the linear case.
The situation is instead similar to the linear fits for the two-fermion and the purely
bosonic operators, and for example now also for  $c_{tZ}$, $c_{\varphi B}$ and $c_{\varphi W}$ one finds
large differences between marginalised and individual fits.
One should point out, however,
that even on those cases where the magnitude of the bound
does not vary much, the central best-fit values can still shift in a non-negligible manner.

\begin{figure}[t]
  \begin{center}
    \includegraphics[width=0.32\linewidth]{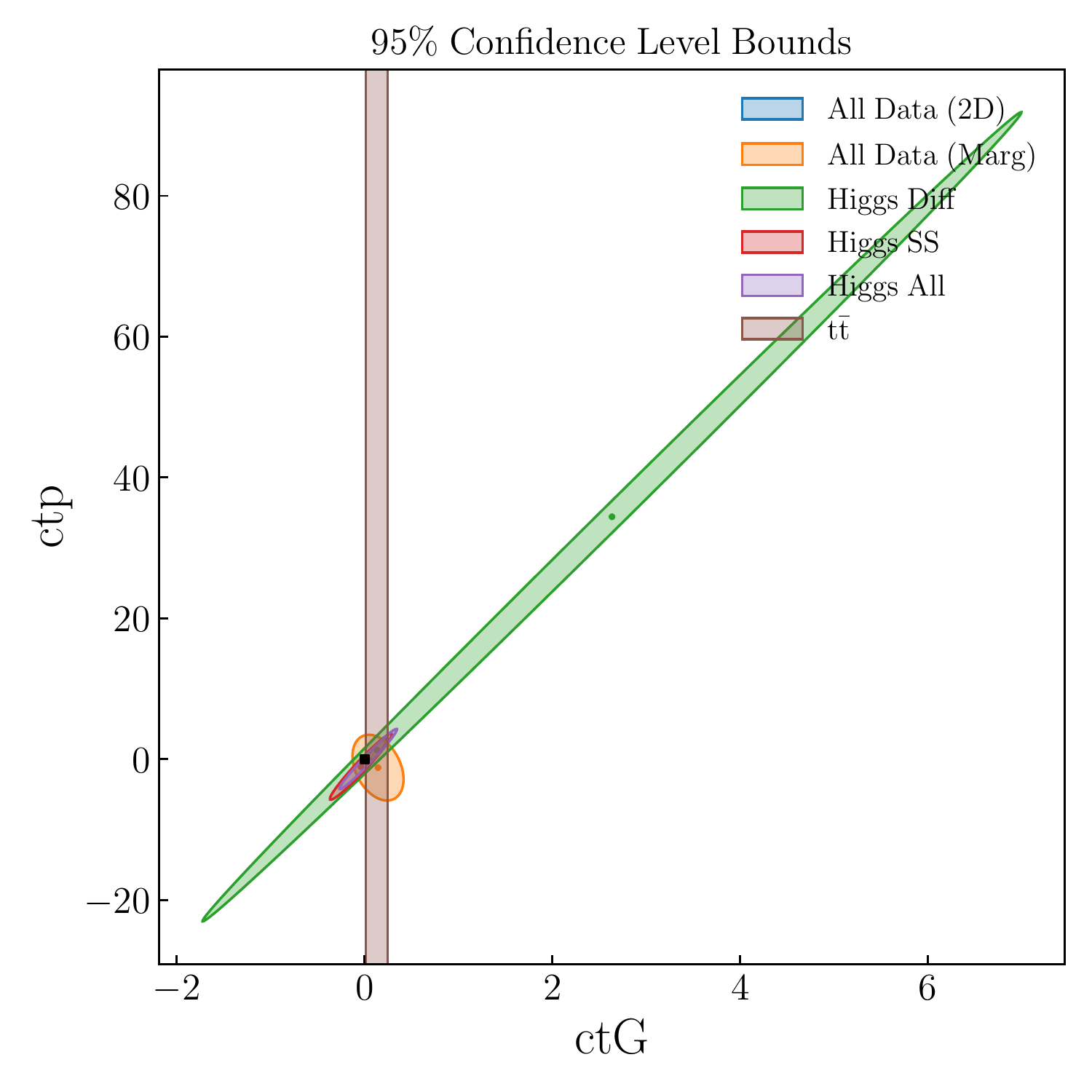}
    \includegraphics[width=0.32\linewidth]{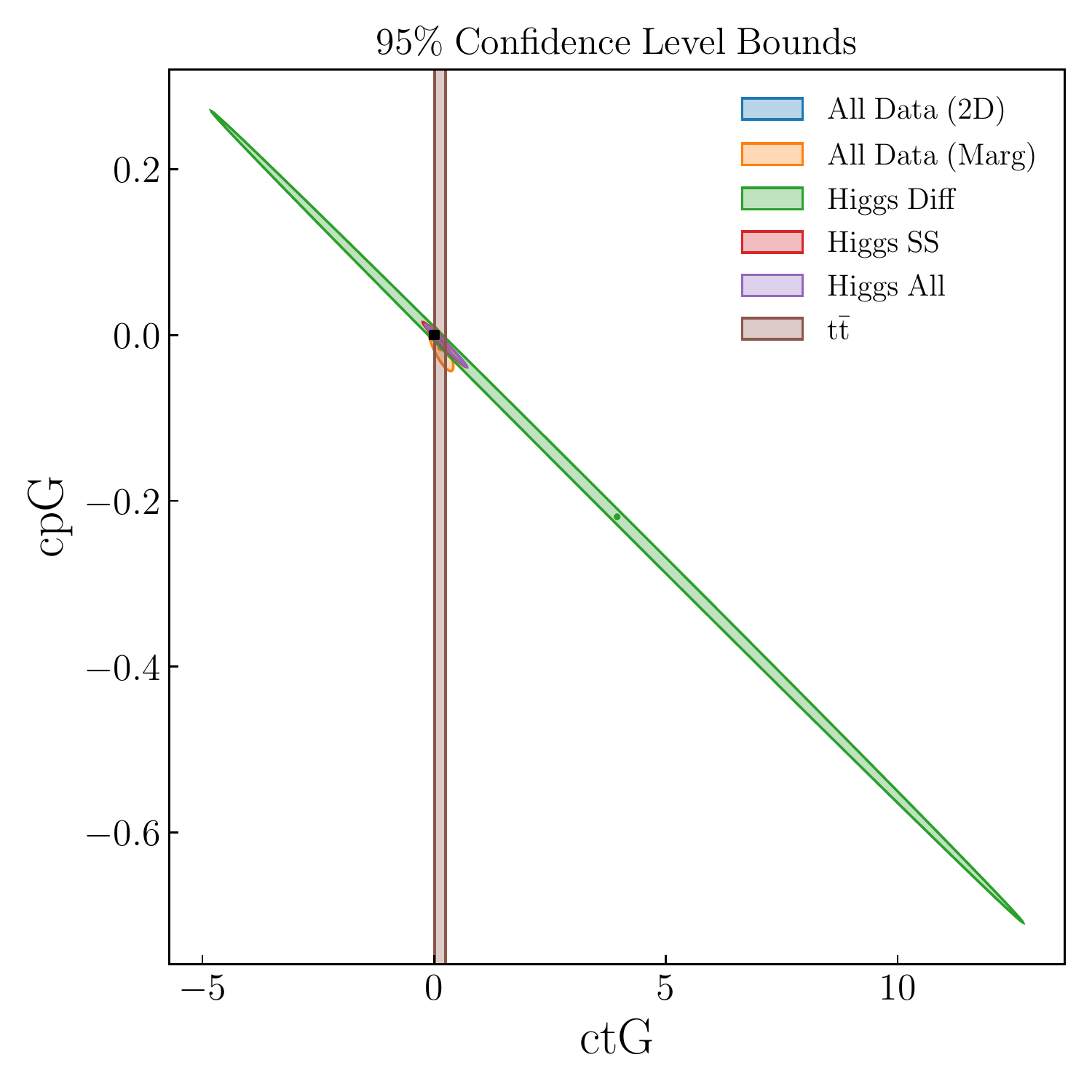}
    \includegraphics[width=0.32\linewidth]{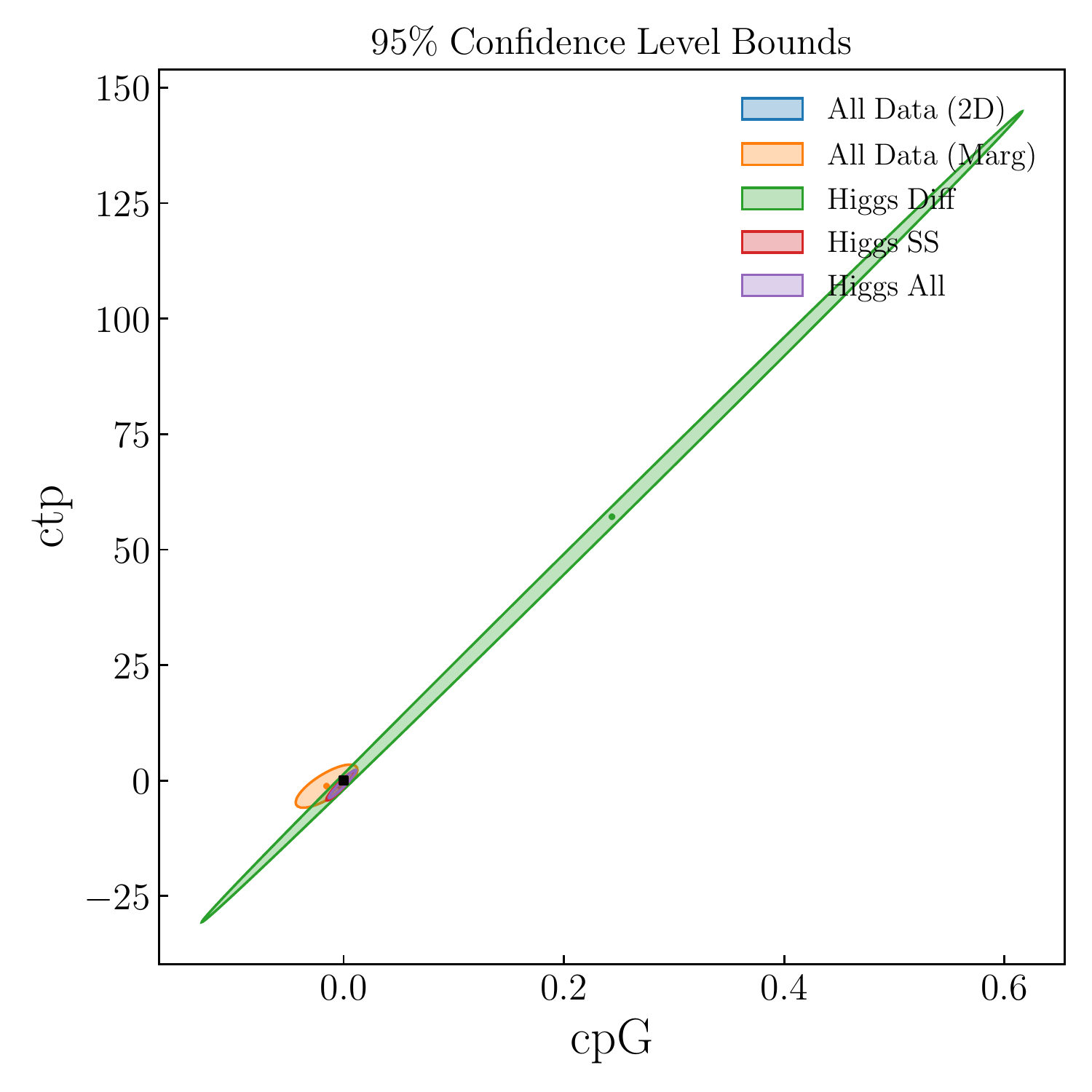}
    \includegraphics[width=0.32\linewidth]{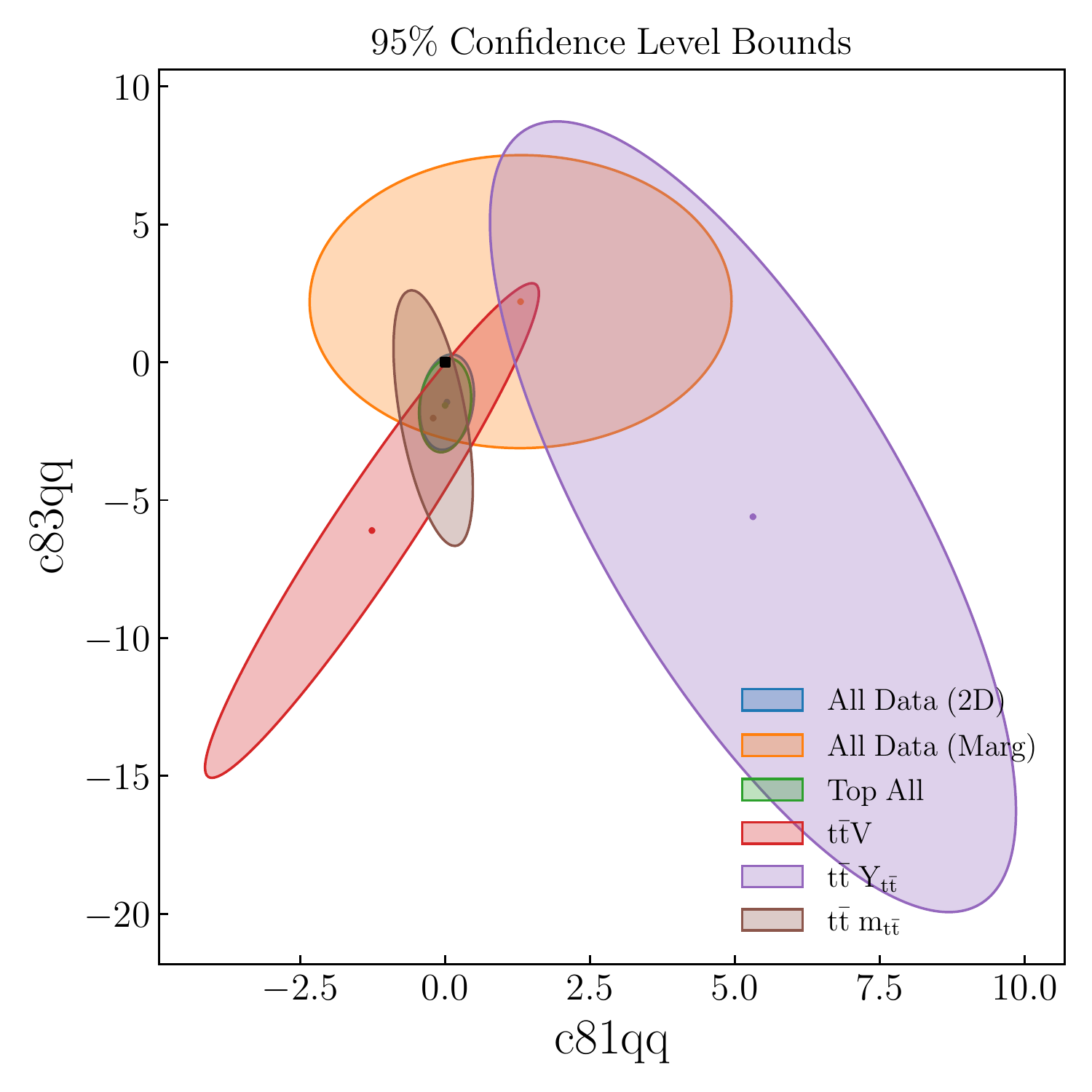}
    \includegraphics[width=0.32\linewidth]{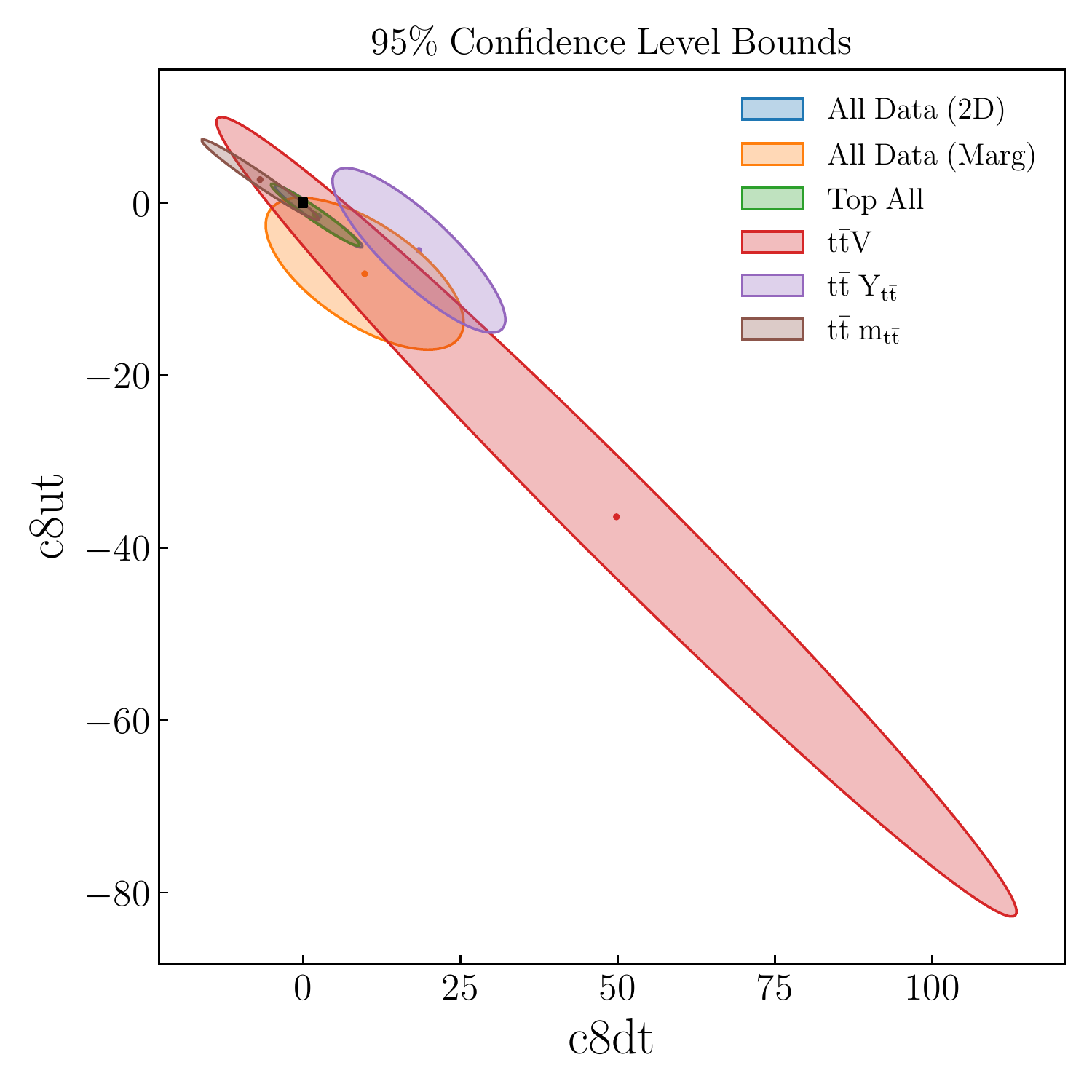}
    \includegraphics[width=0.32\linewidth]{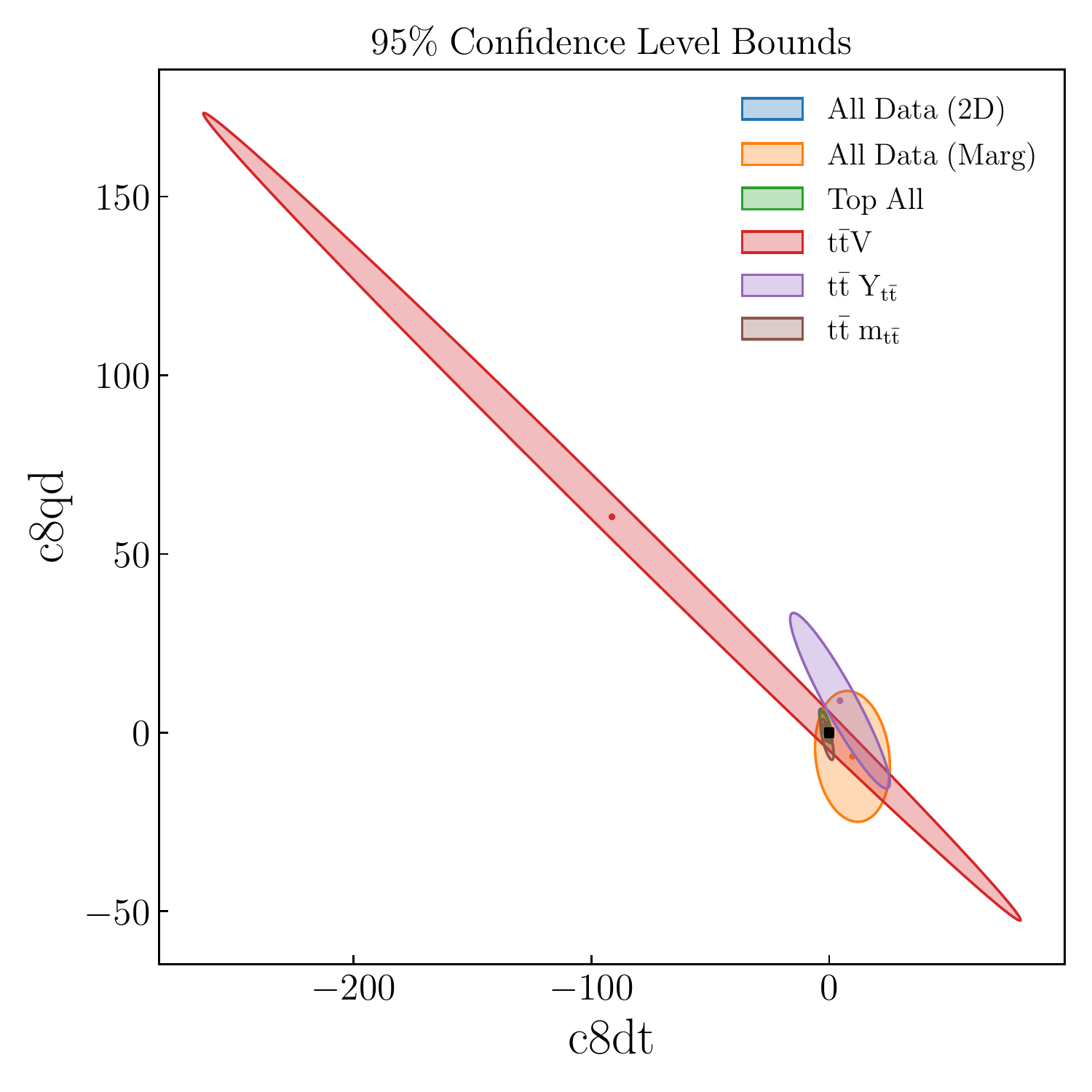}
    \vspace{-0.2cm}
    \caption{\small Representative results for two-parameter fits carried out
      at linear order in the EFT.
      We display the 95\% CL ellipses obtained for different data subsets
      and for the complete dataset, labelled as ``All Data (2D)''.
      For reference, we also show the marginalised bounds obtained
      in the global fit.
      The black square in the center of the plot indicates the SM value.
     \label{fig:2Dfits} }
  \end{center}
\end{figure}

\paragraph{Two-parameter fits.}
To complement the insights provided by individual fits, it can also be instructive
to carry out two-parameter fits, specially to investigate the relative
interplay between specific pairs of EFT coefficients.
In such fits, two coefficients are allowed to vary simultaneously
while the rest are set to zero.
To illustrate the information that can be provided
by such two-parameter fits, Fig.~\ref{fig:2Dfits} displays
representative results for fits performed at the linear order.
We display the 95\% CL ellipses obtained when different subsets of data are used as input,
as well as for the complete dataset, labelled as ``All Data (2D)''.
For reference, we also show here the marginalised bounds obtained
from the global fit.

To begin with, the upper panels of Fig.~\ref{fig:2Dfits} display two-parameter fits
for the three possible pair-wise combinations of the $c_{t \varphi}$, $c_{tG}$, and $c_{\varphi G}$
coefficients,
which connect Higgs production in gluon fusion with top quark pair production, see also
the Fisher information table of  Fig.~\ref{fig:FisherMatrix}.
These comparisons illustrate the relative impact of the various dataset in constraining
each coefficient.
For example, from the $\lp c_{t\varphi},c_{tG}\rp$ fit we see that the sensitivity of
$c_{tG}$ is driven by $t\bar{t}$ data, while the Higgs differential
measurements have a flat direction
resulting in a elongated ellipse.
The overlap between $t\bar{t}$ data and  Higgs differential
measurements results in similar constraints as compared to those
provided by the Higgs signal strengths alone.
Note that, as in the case of the individual fits reported in
Fig.~\ref{fig:globalfit-baseline-bounds-lin-NS-ind-vs-marg}, also for two-parameter fits
the obtained bounds are more stringent as compared to the global marginalised results.
Similar considerations apply to the $\lp c_{\varphi G},c_{tG}\rp$ fit, while
from the $\lp c_{\varphi G},c_{t\varphi}\rp$ one learns that the sensitivity
is still dominated by the Higgs signal strengths rather than by the differential cross-section
measurements.

Then the bottom panels of Fig.~\ref{fig:2Dfits} display two-parameter fits involving the
two-light-two-heavy coefficients $c_{Qq}^{1,8}$, $c_{Qq}^{3,8}$, $c_{tu}^8$, $c_{td}^8$, and $c_{tq}^8$,
all of which are constrained mostly from top quark pair differential distributions as indicated
by the Fisher information matrix.
Here the scope is to illustrate the relative sensitivity provided by some of the $t\bar{t}$
datasets that enter the fit: single-inclusive $m_{t\bar{t}}$ distributions, the double-differential
$(m_{t\bar{t}},y_{t\bar{t}})$ distributions, and $t\bar{t}V$ measurements.
The results confirm both that the $m_{t\bar{t}}$ distributions completely dominate
the fit of these coefficients, and that the marginalised CL ellipses are rather broader
than for the two-dimensional fits.
The latter is again in agreement with the results of the individual
linear fits, reported in the upper panel of
Fig.~\ref{fig:globalfit-baseline-bounds-lin-NS-ind-vs-marg}.


%% file: tables/coeff-bounds-baseline.tex
\begin{table}[htbp]
  \centering
  \scriptsize
   \renewcommand{\arraystretch}{1.24}
   \begin{tabular}{l|C{0.8cm}|C{2.3cm}|C{2.3cm}|C{4.0cm}|C{4.0cm}}
     \multirow{2}{*}{Class}   &  \multirow{2}{*}{DoF}
     &  \multicolumn{2}{c|}{ 95\% CL bounds, $\mathcal{O}\lp \Lambda^{-2}\rp$} &
     \multicolumn{2}{c}{95\% CL bounds, $\mathcal{O}\lp \Lambda^{-4}\rp$,} \\ 
 &  & Individual & Marginalised &  Individual & Marginalised  \\ \toprule
 \multirow{5}{*}{4H}
 &{\tt cQQ1}& [-6.132,23.281] & [-190,189] & [-2.229,2.019] & [-2.995,3.706] \\ \cline{2-6}
 & {\tt cQQ8}  & [-26.471,57.778] & [-190,170] & [-6.812,5.834] & [-11.177,8.170] \\ \cline{2-6}
 & {\tt cQt1}& [-195,159] & [-190,189] & [-1.830,1.862] & [-1.391,1.251] \\ \cline{2-6}
 & {\tt cQt8}& [-5.722,20.105] & [-190,162] & [-4.213,3.346] & [-3.040,2.202] \\ \cline{2-6}
 & {\tt ctt1}& [-2.782,12.114] & [-115,153] & [-1.151,1.025] & [-0.791,0.714] \\ \hline
\multirow{14}{*}{2L2H}
 & {\tt c81qq}& [-0.273,0.509] & [-2.258,4.822] & [-0.373,0.309] & [-0.555,0.236] \\ \cline{2-6}
 & {\tt c11qq}& [-3.603,0.307] & [-8.047,9.400] & [-0.303,0.225] & [-0.354,0.249] \\ \cline{2-6}
 & {\tt c83qq}& [-1.813,0.625] & [-3.014,7.365] & [-0.470,0.439] & [-0.462,0.497] \\ \cline{2-6}
 & {\tt c13qq}& [-0.099,0.155] & [-0.163,0.296] & [-0.088,0.166] & [-0.167,0.197] \\ \cline{2-6}
 & {\tt c8qt}& [-0.396,0.612] & [-4.035,4.394] & [-0.483,0.393] & [-0.687,0.186] \\ \cline{2-6}
 & {\tt c1qt}& [-0.784,2.771] & [-12.382,6.626] & [-0.205,0.271] & [-0.222,0.226] \\ \cline{2-6}
 & {\tt c8ut}& [-0.774,0.607] & [-16.952,0.368] & [-0.911,0.347] & [-1.118,0.260] \\ \cline{2-6}
 & {\tt c1ut}& [-6.046,0.424] & [-15.565,15.379] & [-0.380,0.293] & [-0.383,0.331] \\ \cline{2-6}
 & {\tt c8qu}& [-1.508,1.022] & [-12.745,13.758] & [-1.007,0.521] & [-1.002,0.312] \\ \cline{2-6}
 & {\tt c1qu}& [-0.938,2.462] & [-16.996,1.072] & [-0.281,0.371] & [-0.207,0.339] \\ \cline{2-6}
 & {\tt c8dt}& [-1.458,1.365] & [-5.494,25.358] & [-1.308,0.638] & [-1.329,0.643] \\ \cline{2-6}
 & {\tt c1dt}& [-9.504,-0.086] & [-27.673,11.356] & [-0.449,0.371] & [-0.474,0.347] \\ \cline{2-6}
 & {\tt c8qd}& [-2.393,2.042] & [-24.479,11.233] & [-1.615,0.888] & [-1.256,0.715] \\ \cline{2-6}
 & {\tt c1qd}& [-0.889,6.459] & [-3.239,34.632] & [-0.332,0.436] & [-0.370,0.384] \\ \hline
\multirow{23}{*}{2FB}
 & {\tt ctp}& [-1.331,0.355] & [-5.739,3.435] & [-1.286,0.348] & [-2.319,2.797] \\ \cline{2-6}
 & {\tt ctG}& [0.007,0.111] & [-0.127,0.403] & [0.006,0.107] & [0.062,0.243] \\ \cline{2-6}
 & {\tt cbp}& [-0.006,0.040] & [-0.033,0.105]& [-0.007,0.035]$\cup$ [-0.403,-0.360] & [-0.035,0.047]$\cup$ [-0.430,-0.338] \\ \cline{2-6}
 & {\tt ccp}& [-0.025,0.117] & [-0.316,0.134] & [-0.004,0.370] & [-0.096,0.484] \\ \cline{2-6}
 & {\tt ctap}& [-0.026,0.035] & [-0.027,0.044] & [-0.027,0.040]$\cup$ [0.395,0.462] & [-0.019,0.037]$\cup$ [0.389,0.480] \\ \cline{2-6}
 & {\tt ctW}& [-0.093,0.026] & [-0.313,0.123] & [-0.084,0.029] & [-0.241,0.086] \\ \cline{2-6}
 & {\tt ctZ}& [-0.039,0.099] & [-15.869,5.636] & [-0.044,0.094] & [-1.129,0.856] \\ \cline{2-6}
 & {\tt cpl1}& [-0.664,1.016] & [-0.244,0.375] & [-0.281,0.343] & [-0.106,0.129] \\ \cline{2-6}
 & {\tt c3pl1}& [-0.472,0.080] & [-0.098,0.120] & [-0.432,0.062] & [-0.209,0.046] \\ \cline{2-6}
 & {\tt cpl2}& [-0.664,1.016] & [-0.244,0.375] & [-0.281,0.343] & [-0.106,0.129] \\ \cline{2-6}
 & {\tt c3pl2}& [-0.472,0.080] & [-0.098,0.120] & [-0.432,0.062] & [-0.209,0.046] \\ \cline{2-6}
 & {\tt cpl3}& [-0.664,1.016] & [-0.244,0.375] & [-0.281,0.343] & [-0.106,0.129] \\ \cline{2-6}
 & {\tt c3pl3}& [-0.472,0.080] & [-0.098,0.120] & [-0.432,0.062] & [-0.209,0.046] \\ \cline{2-6}
 & {\tt cpe}& [-1.329,2.033] & [-0.487,0.749] & [-0.562,0.687] & [-0.213,0.258] \\ \cline{2-6}
 & {\tt cpmu}& [-1.329,2.033] & [-0.487,0.749] & [-0.562,0.687] & [-0.213,0.258] \\ \cline{2-6}
 & {\tt cpta}& [-1.329,2.033] & [-0.487,0.749] & [-0.562,0.687] & [-0.213,0.258] \\ \cline{2-6}
 & {\tt c3pq}& [-0.472,0.080] & [-0.098,0.120] & [-0.432,0.062] & [-0.209,0.046] \\ \cline{2-6}
 & {\tt c3pQ3}& [-0.350,0.353] & [-1.145,0.740] & [-0.375,0.344] & [-0.615,0.481] \\ \cline{2-6}
 & {\tt cpqMi}& [-2.905,0.490] & [-0.171,0.106] & [-2.659,0.381] & [-0.060,0.216] \\ \cline{2-6}
 & {\tt cpQM}& [-0.998,1.441] & [-1.690,11.569] & [-1.147,1.585] & [-2.250,2.855] \\ \cline{2-6}
 & {\tt cpui}& [-1.355,0.886] & [-0.499,0.325] & [-0.458,0.375] & [-0.172,0.142] \\ \cline{2-6}
 & {\tt cpdi}& [-0.443,0.678] & [-0.162,0.250] & [-0.187,0.229] & [-0.071,0.086] \\ \cline{2-6}
 & {\tt cpt}& [-2.087,2.463] & [-3.270,18.267] & [-3.028,2.195] & [-13.260,3.955] \\ \hline
\multirow{7}{*}{B}
 & {\tt cpG}& [-0.002,0.005] & [-0.043,0.012] & [-0.002,0.005] & [-0.019,0.003] \\ \cline{2-6}
 & {\tt cpB}& [-0.005,0.002] & [-0.739,0.289] & [-0.005,0.002]$\cup$ [0.085,0.092] & [-0.114,0.108] \\ \cline{2-6}
 & {\tt cpW}& [-0.018,0.007] & [-0.592,0.677] & [-0.016,0.007]$\cup$ [0.281,0.305] & [-0.145,0.303] \\ \cline{2-6}
 & {\tt cpWB}& [-2.905,0.490] & [-0.462,0.694] & [-2.659,0.381] & [-0.170,0.273] \\ \cline{2-6}
 & {\tt cpd}& [-0.428,1.214] & [-2.002,3.693] & [-0.404,1.199]$\cup$ [-34.04,-32.61] & [-1.523,1.482] \\ \cline{2-6}
 & {\tt cpD}& [-4.066,2.657] & [-1.498,0.974] & [-1.374,1.124] & [-0.516,0.425] \\ \cline{2-6}
 & {\tt cWWW}& [-1.057,1.318] & [-1.049,1.459] & [-0.208,0.236] & [-0.182,0.222] \\ \bottomrule
\end{tabular}
   \caption{\small The 95\% CL bounds for all the
     EFT coefficients
     considered in this analysis, for  both individual and global (marginalised) fits
     obtained using either linear or quadratic EFT calculations.
\label{tab:coeff-bounds-baseline}
}
\end{table}

%% file: subsec_results_dataset.tex
\subsection{Dataset dependence}
\label{sec:dataset_dependence}

The discussion so far has focused on the output of the global fits
obtained for the baseline dataset summarised in
Tables~\ref{eq:input_datasets} to~\ref{eq:input_datasets4}.
Here we aim to assess the dependence of these results with respect to the choice
of input dataset.
With this purpose, we consider here fits for the following variations:
\begin{itemize}

\item A fit which includes only top quark measurements.
  This fit makes possible quantifying the interplay
  between the top and the Higgs data in the global fit.

\item A fit which includes only Higgs boson production and decay data,
  which provides complementary information as compared to
  the top-only fit.

\item  A fit which includes only top quark measurements, but now restricted to
  the same dataset as in our original study from~\cite{Hartland:2019bjb}.
  This comparison allows one to assess the impact
  in the top-only EFT fit of the new LHC top quark measurements that have
  become available in the last two years.

\item A fit where the diboson data is removed, to determine how much weight 
  the diboson cross-sections carry in the global fit results.

\item A fit where all high-energy bins, defined as those bins
  probing the region $E\gsim 1 $ TeV, are removed.
  The motivation for such a fit is to study how important are the constraints
  provided by the high-energy region in the global fit results,
  which in turn is an important input to  establish the validity
of the EFT approximation.

\item A fit where those datasets displaying poor agreement with the SM cross-sections
  are removed.
  Specifically, here one removes the datasets whose $\chi^2$ differs by more
than $3\sigma$ from their statistical expectation assuming the SM hypothesis.
While such disagreements between data and SM theory
could very well indicate hints of BSM physics, they can also be explained
by for example issues with the  experimental correlation models.
Hence, this fit  allows us to verify to which extent the baseline results
are driven from the datasets that disagree the most with the SM predictions.
{  The datasets indicated with {\bf (*)} in Tables~\ref{eq:chi2-baseline} and~\ref{eq:chi2-baseline2} are those
  excluded from this ``conservative'' EFT fit.}

\end{itemize}
Note that, as explained in Sect.~\ref{sec:settings_expdata}, for the purposes
of categorisation into datasets
the $t\bar{t}h$ cross-sections are considered part of the Higgs measurements.
Furthermore, we note that all these fits are based on  quadratic EFT calculations
and that the constraints provided by the EWPOs on
the EFT parameter space are always accounted for.

\input{tables/table-chi2-datasetvariations.tex}


To begin with,  Table~\ref{eq:chi2-datasetvariations}
collects the values of the $\chi^2$ per data points for EFT fits obtained from
variations of the input dataset.
We list the results of the various fits described above:
including only  top quark measurements (either from the current or the 2018
dataset); with a Higgs-only dataset; without the diboson cross-sections; with the high-energy bins excluded;
and with the datasets with a poor $\chi^2_{\rm sm}$ excluded.
The numbers in parentheses indicate the number of data points, in the case that these are different
from those of the baseline settings listed in the second column.
We observe how the description of the Higgs cross-sections is essentially
unaffected in these fits with reduced datasets.
Concerning the total $\chi^2$ for the top data, we see that it is stable in the fit
where the high-energy bins are removed, but that is markedly improved (from 1.10 to 0.82)
in the fit where the datasets with poor $\chi^2_{\rm sm}$ are excluded
and the number of top-quark points in the fit decreases from $n_{\rm dat}=150$
to 123.

\begin{figure}[t]
  \begin{center}
    \includegraphics[width=0.90\linewidth]{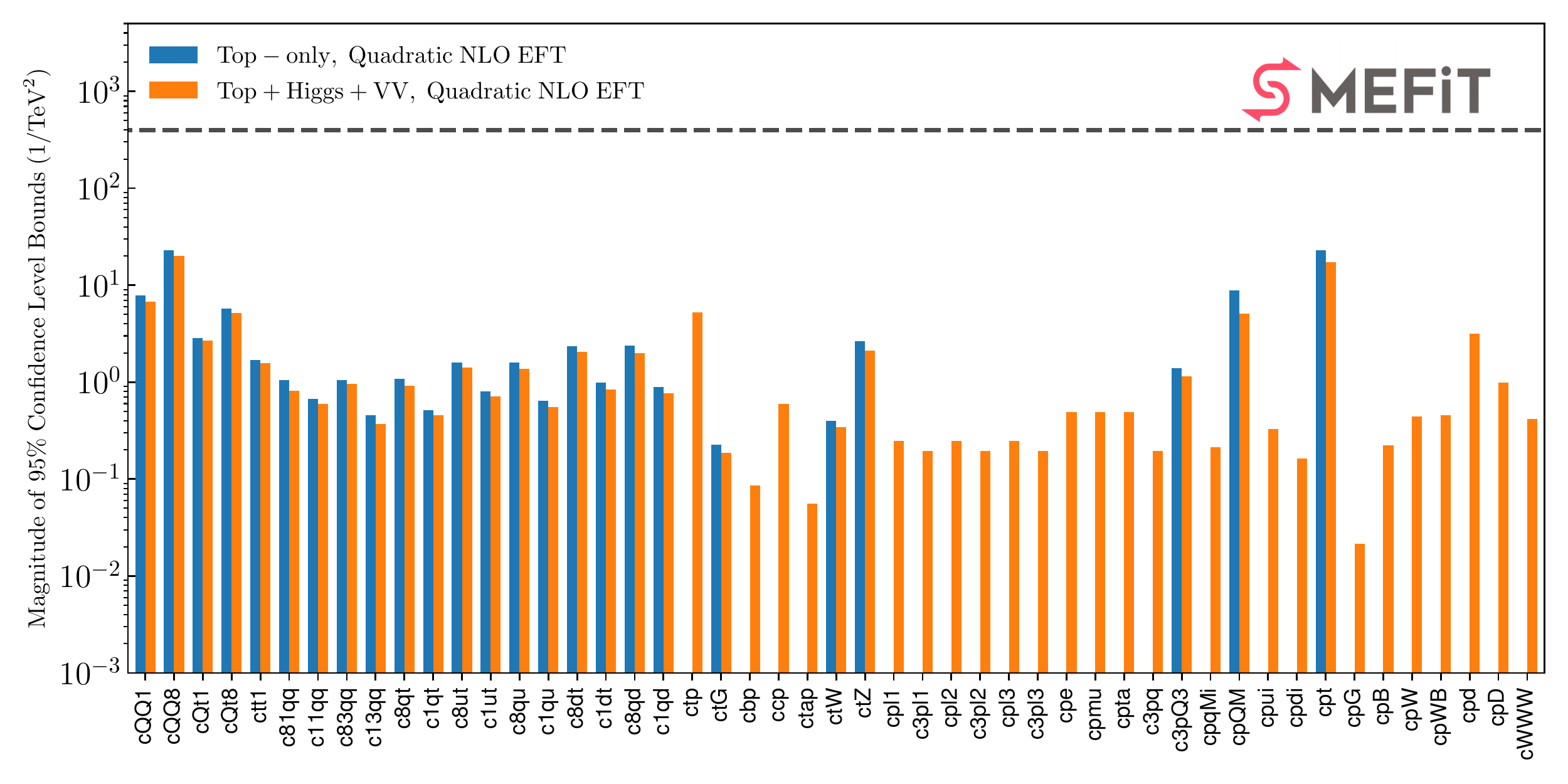}
    \includegraphics[width=0.90\linewidth]{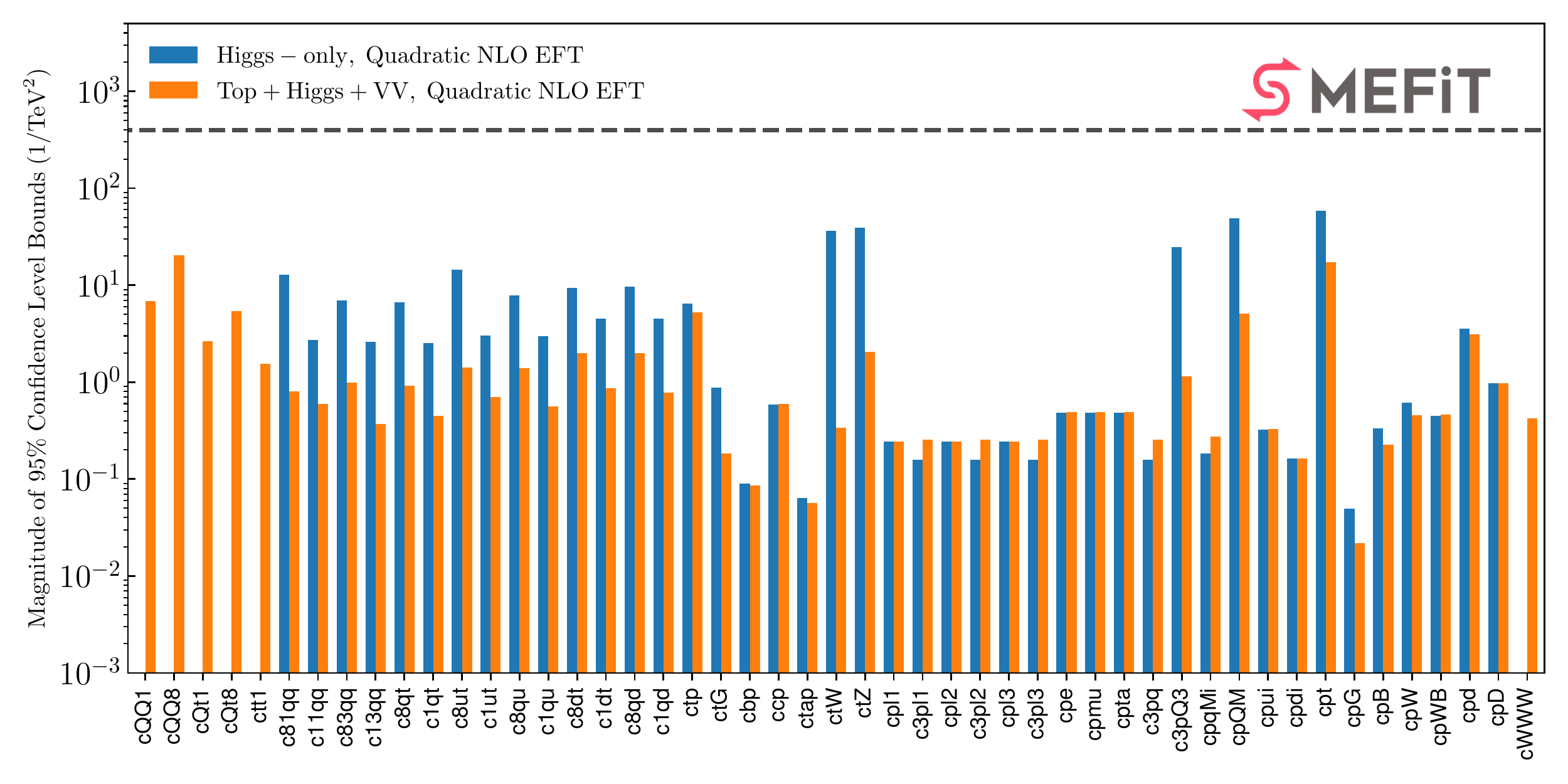}
    \caption{\label{fig:global_vs_toponly} \small
      Same as upper panel of Fig.~\ref{fig:globalfit-baseline-bounds-lin-vs-quad}
      now comparing the global fit results with those obtained in a top-only (upper)
    and Higgs-only (lower panel) fits.}
  \end{center}
\end{figure}

Then in Fig.~\ref{fig:global_vs_toponly} we compare the magnitude of the 95\% CL bounds,
same as in the upper panel of Fig.~\ref{fig:globalfit-baseline-bounds-lin-vs-quad},
between the global fit results with those obtained in the top-only 
and Higgs-only fits.
As mentioned above, these fits
allow us to  assess the interplay
between the top and the Higgs data in the global analysis, in other words,
to identify what are the main benefits of the simultaneous mapping of the EFT parameter space
as compared to carrying out separate fits to each group of processes.
First of all, we note that the global fit bounds are more stringent for
all the EFT coefficients than in either the top-only or Higgs-only fit, highlighting the overall
consistency of the two datasets.
Secondly, the cross-talk of the top and Higgs data is found to be most
relevant for the two-fermion coefficients $c_{\varphi t}$
and $c_{\varphi Q}^{(-)}$, whose bounds are improved
by around a factor 2 in the global fit as compared to the top-only fit.
Another operator that benefits from the global fit is
$c_{\varphi G}$, which is unconstrained in the top-only fit but
whose bound in the global fit is clearly improved as compared to the Higgs-only fit.
These comparisons show how by breaking degeneracies one gains information in the global fit as compared to the partial ones, sometimes in unexpected directions in the parameter space such as for $c_{\varphi G}$  in this case.
The bottom panel of Fig.~\ref{fig:global_vs_toponly} also indicates that in a Higgs-only fit
a large number of EFT coefficients are poorly constrained, in particular
those involving fermion bilinears.

\begin{figure}[t]
  \begin{center}
    \includegraphics[width=0.90\linewidth]{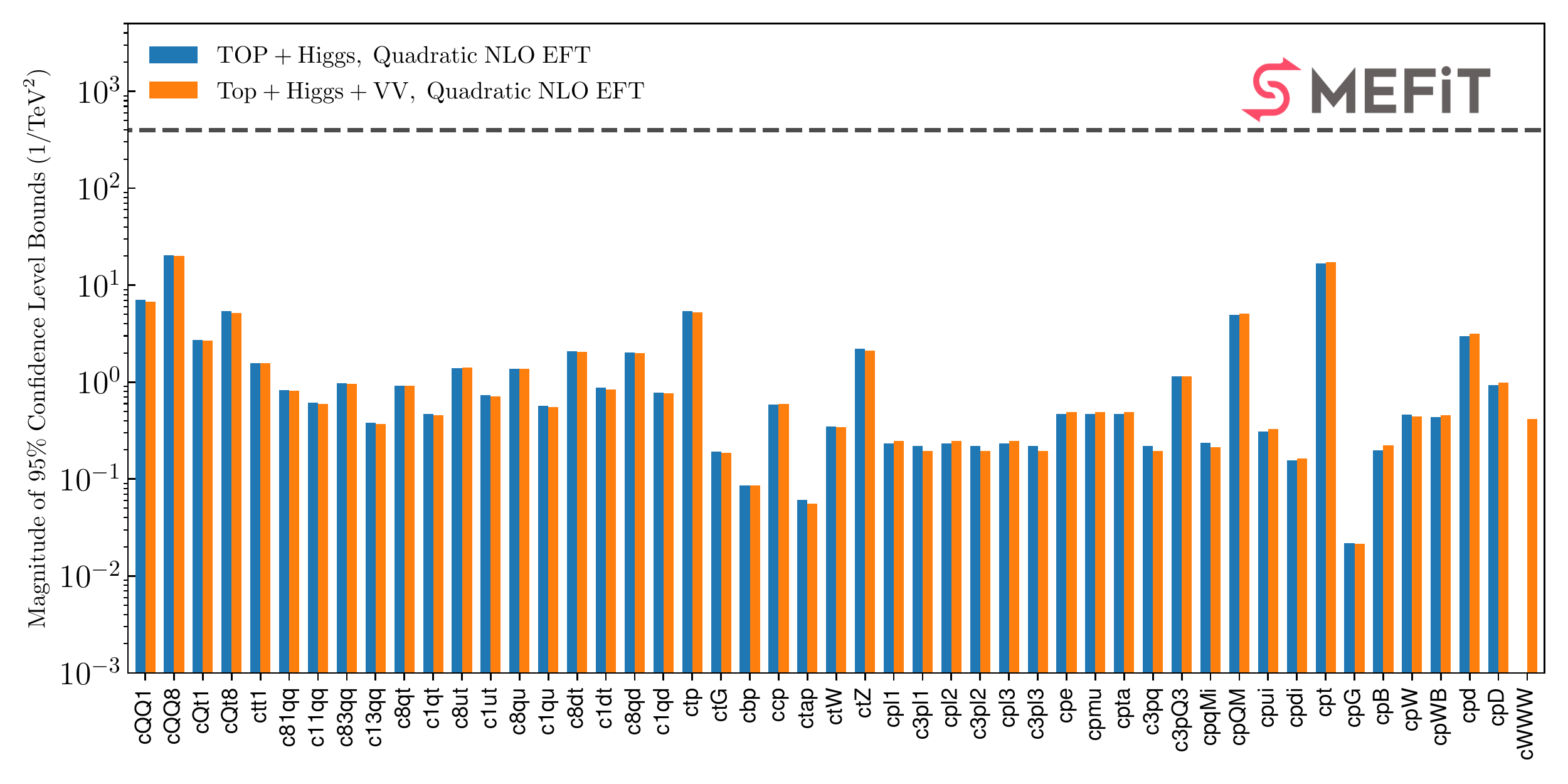}
   \includegraphics[width=0.90\linewidth]{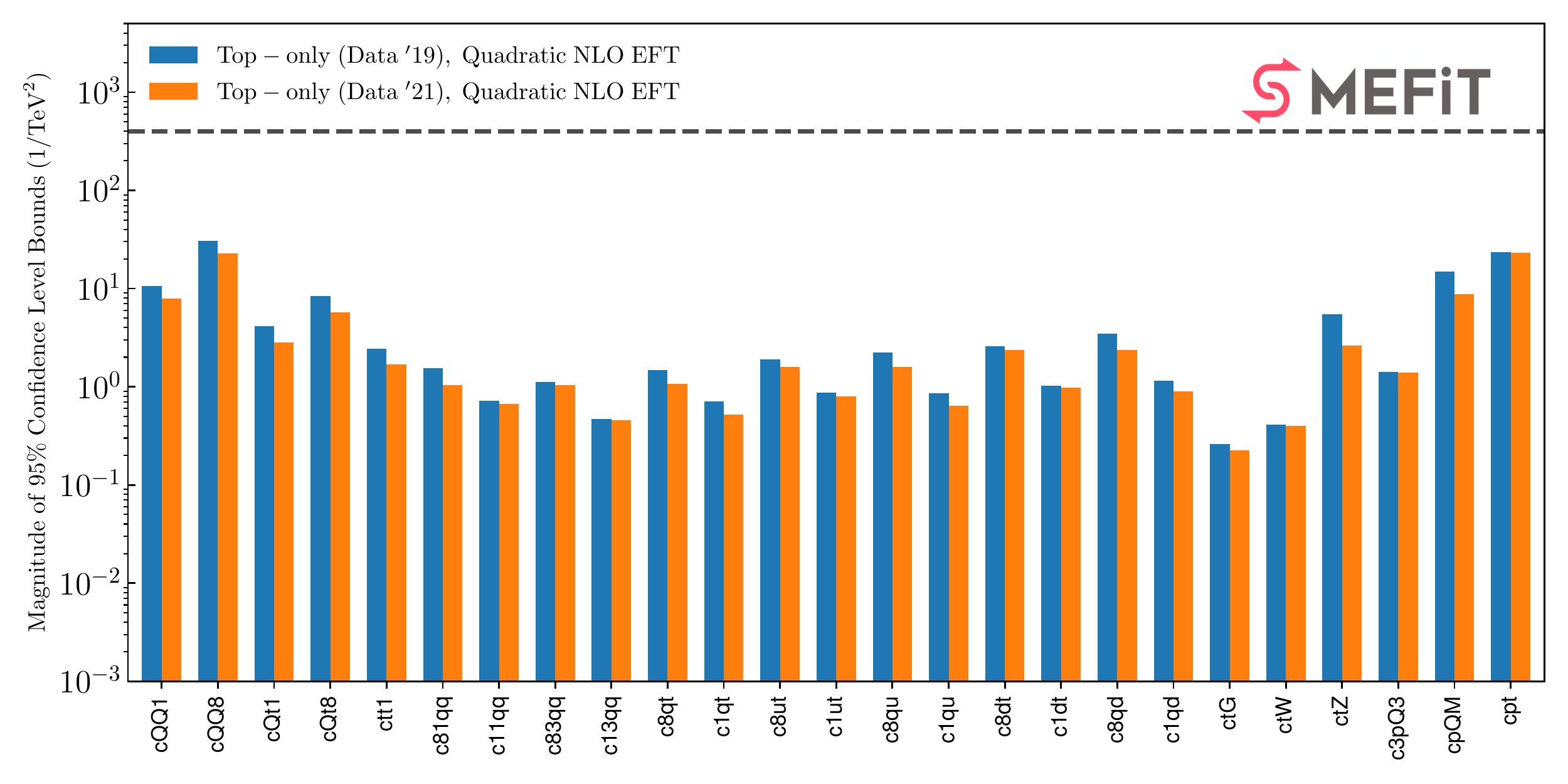}
    \caption{\label{fig:global_vs_top2019} \small
      Same as Fig.~\ref{fig:global_vs_toponly},
      now comparing the global fit with a no-diboson fit (upper)
    and the two top-only fits with different datasets (lower panel).}
  \end{center}
\end{figure}

Next, Fig.~\ref{fig:global_vs_top2019} displays a
similar comparison as in Fig.~\ref{fig:global_vs_toponly}
now comparing first the outcome of the global fit with that of a fit
where the diboson cross-sections have been removed,
and second comparing two top-only fits, namely the fit displayed
in the upper panel of Fig.~\ref{fig:global_vs_toponly} with  a
fit based on the same dataset as our previous study from~\cite{Hartland:2019bjb}.
The fit without diboson data demonstrates that the constraints provided by the diboson
cross-sections are negligible in comparison with those provided by the Higgs data
(and the EWPOs) for all coefficients considered in the fit, except for the triple
gauge operator $c_{W}$.
This result is consistent with the Fisher information
analysis of Fig.~\ref{fig:FisherMatrix}, and indicates that, apart from $c_W$, the diboson
data does not provide competitive information on the EFT parameter space in the context
of a global fit.

The comparison of the two top-only fits in the bottom
panel of Fig.~\ref{fig:global_vs_top2019} illustrates how for all coefficients
the bounds  are improved thanks to the more recent LHC measurements.\footnote{Recall that now we consider
  the top Yukawa coefficient $c_{t\varphi}$ as part of the Higgs dataset.}
The improvement is consistent across the board, quantifies the additional information
brought in by the new top-quark cross-section measurements
(see Table~\ref{eq:chi2-datasetvariations}),
and confirms that the broader and more diverse the input dataset is,
 the more stringent the resulting constraints
on the EFT parameter space that will be obtained.

\begin{figure}[t]
  \begin{center}
    \includegraphics[width=0.80\linewidth]{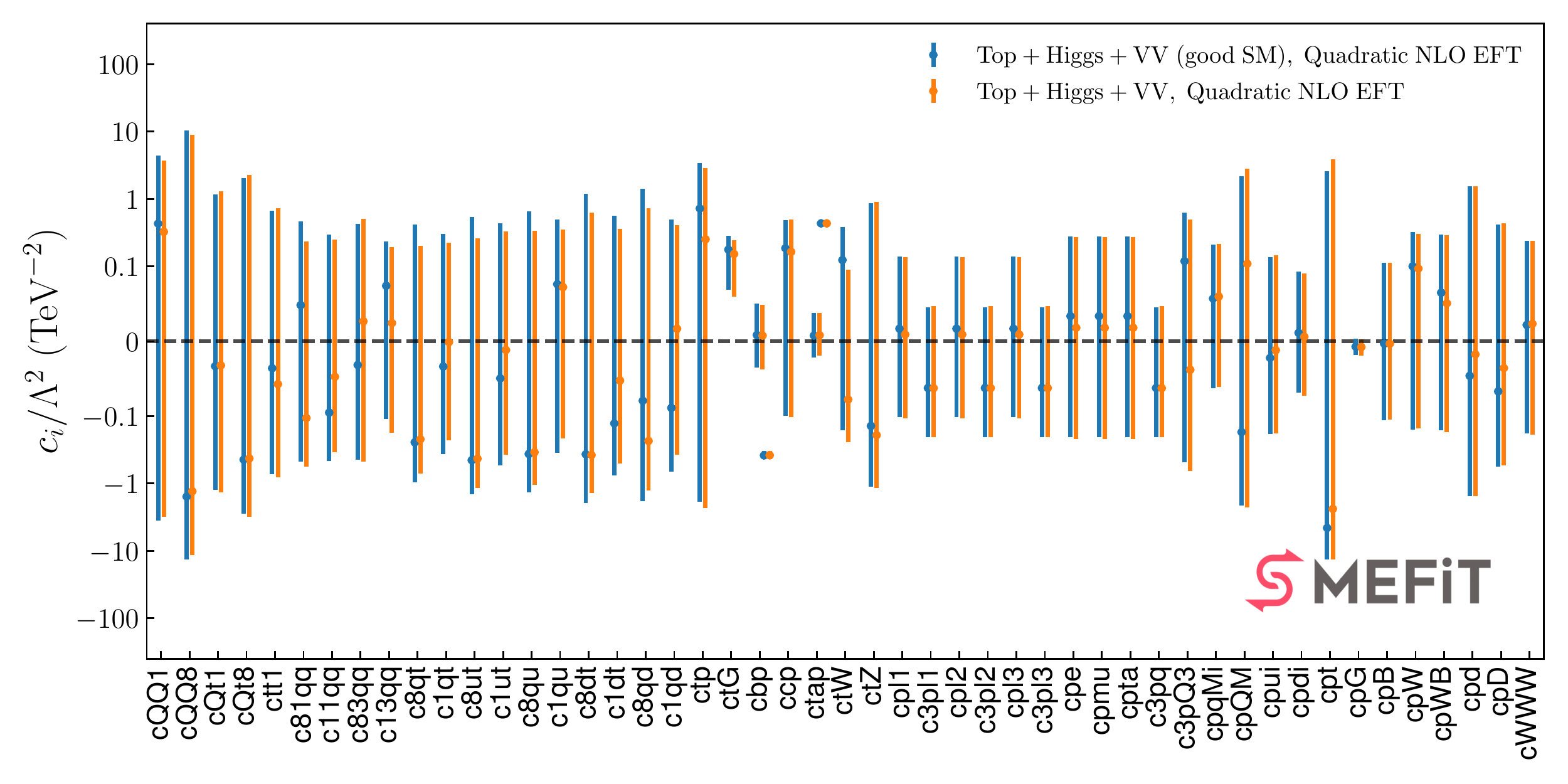}
   \includegraphics[width=0.80\linewidth]{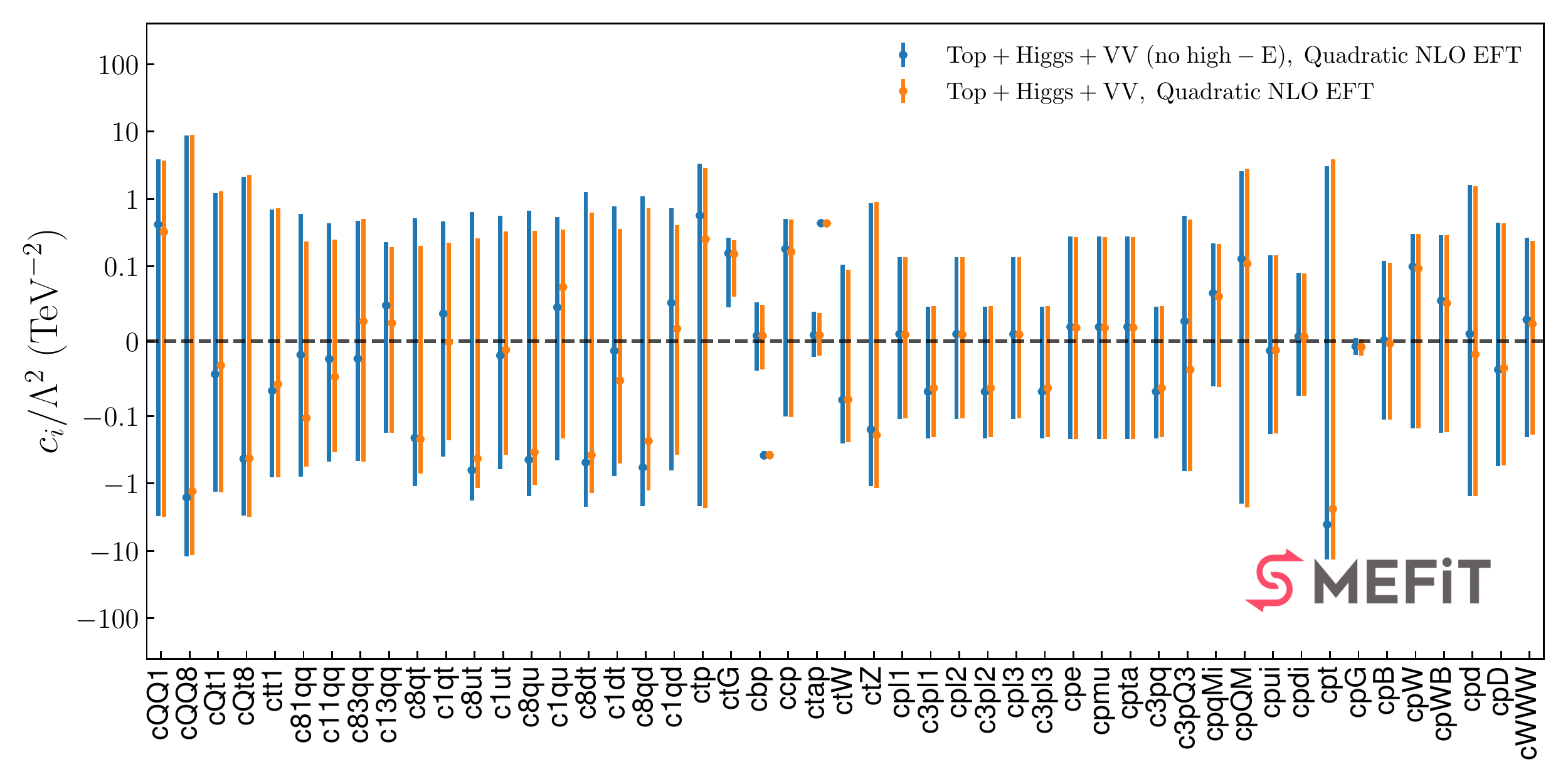}
   \caption{\label{fig:global_vs_gooddata} \small
     Same as  Fig.~\ref{fig:globalfit-baseline-coeffsabs-lin-vs-quad}
     comparing the global fit results with those of the fit excluding datasets with poor $\chi^2_{\rm sm}$
   (upper) and with the fit where the bins with $E\gsim 1$ TeV are removed (bottom panel).}
  \end{center}
\end{figure}

To continue with this discussion of the dataset dependence of our results, we consider now
the outcome of
two more fits: first, one where the datasets exhibiting poor agreement with the SM predictions
are excluded, and second, another where all bins sensitive to the high-energy region,
defined as $E\gsim 1 $ TeV, are removed.
The best-fit values and 95\% CL intervals of these two fits are compared
with the baseline results in Fig.~\ref{fig:global_vs_gooddata}.
As indicated in Table~\ref{eq:chi2-datasetvariations}, in the fit where those
datasets with poor $\chi^2_{\rm sm}$
have been removed, one is essentially cutting away 27 points from top quark production, mostly
from the inclusive $t\bar{t}$ category.
The only coefficients that are affected by this reduction in the dataset are some of the
two-light-two-heavy operators, whose bounds are mildly enlarged consistently
with the loss of experimental information.
This comparison highlights the stability of the global fit results, whose
outcome is unchanged when potentially problematic datasets with high $\chi^2_{\rm sm}$
are excluded from the fit.
Concerning the outcome of the fit without the high-energy bins, as expected the only differences
are observed again for the two-light-two-heavy coefficients, with a similar
outcome as in the previous fit.
From this analysis, one can conclude that the global fit is not dominated by the high-energy regions
where the EFT validity could be questioned, and hence that results are stable
upon removal of these high-energy bins.

\begin{figure}[t]
  \begin{center}
    \includegraphics[width=0.80\linewidth]{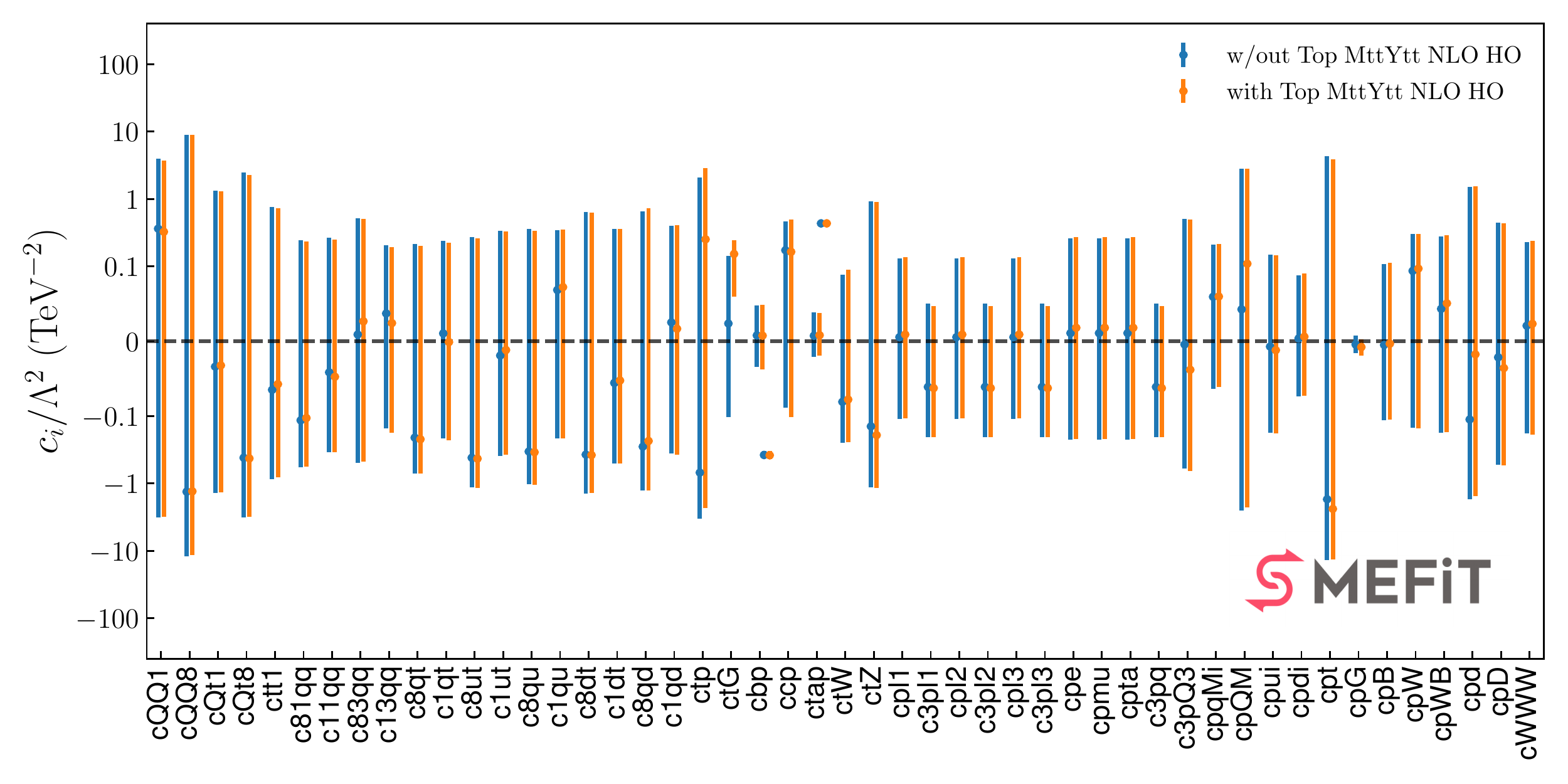}
   \caption{\label{fig:wo_CMS2Dttbar} \small
    The outcome of quadratic EFT fits with and without
   the CMS top-quark pair double-differential $(m_{t\bar{t}},y_{t\bar{t}})$ distributions.}
  \end{center}
\end{figure}

Finally, we show in Fig.~\ref{fig:wo_CMS2Dttbar} a comparison
of the outcome of quadratic EFT fits with and without
the CMS top-quark pair double-differential $(m_{t\bar{t}},y_{t\bar{t}})$ distributions.
We have identified this dataset as the one being responsible for
driving upwards the fit value of the chromo-magnetic
operator $c_{tG}$.
Indeed, 
one can observe how once this dataset is removed
then $c_{tG}$ agrees with the SM at the 95\% CL.
Given that both in the global linear and the individual quadratic fits
$c_{tG}$ also agrees with the SM (even in fits where {\tt CMS\_tt2D\_8TeV\_dilep\_mttytt}
is included), the pull found in the global quadratic case must arise from a non-trivial
interplay between different EFT degrees of freedom.
Further studies are required to elucidate why this specific dataset has such as strong
pull on $c_{tG}$ in the quadratic fits.

%% file: tables/table-chi2-datasetvariations.tex
\begin{table}[t]
  \centering
  \tiny
   \renewcommand{\arraystretch}{1.90}
  \begin{tabular}{l|C{0.8cm}|C{0.7cm}|C{1.0cm}|C{1.0cm}|C{1.0cm}|C{1.5cm}|C{0.9cm}|C{1.2cm}|C{1.2cm}}
    \multirow{3}{*}{Dataset}   & \multirow{3}{*}{$ n_{\rm dat}$} & \multirow{3}{*}{$\chi^2_{\rm SM}$} &  \multicolumn{6}{c}{$\chi^2_{\rm eft}$}   \\\cmidrule(lr{0.7em}){4-10}
    &   &  &  baseline  &  top-only  & top-only & Higgs-only &   diboson  & high-$E$   &  poor $\chi^2_{\rm sm}$  \\
       &   &  &    &  (2021)  & (2018) & &  excluded  & excluded  &  excluded  \\
 \toprule
$t\bar{t}$ incl.        &  83   & 1.46  & 1.42  &  1.44 & 1.52~(63) & --- & 1.42   &  1.40~(67)   &  0.95~(67)   \\
$t\bar{t}$ charge asym.           &  11   & 0.60  & 0.59  &  0.58 & ---       & --- & 0.60   &  0.58   &  0.56   \\
$t\bar{t}V$             &  14   &  0.65 & 0.65  & 0.69  & 0.64~(8)  & --- & 0.65   & 0.72    &   0.68     \\
single-$t$ incl.        &  27   & 0.43  & 0.41  &  0.40 & 0.36~(22) & --- & 0.41   & 0.41    &  0.46  \\
$tV$                    &  9    & 0.71  & 0.75  &  0.65 & 0.76~(6)  & --- & 0.75   & 0.80    &  0.31~(8)   \\
 $t\bar{t}Q\bar{Q}$     &  6    & 1.68  & 2.12  &  2.29 &  4.73~(2) & --- & 2.12   & 2.40    &  1.54~(4)   \\
{\bf Top total}         & 150   & 1.10  & 1.09  &  1.10 & 1.22~(101)& --- & 1.09   & 1.06~(134) &  0.82~(123)   \\
\midrule
Higgs $\mu_i^f$ (RI)    &  22   & 0.86  & 0.90  &  ---  &  ---      &   0.90  &  0.89   & 0.89  &  0.89     \\
Higgs $\mu_i^f$ (RII)   &  40   & 0.67  & 0.63  &  ---  &  ---      &   0.63  &  0.62   & 0.63  &  0.62    \\
Higgs  STXS             &  35   & 0.88  & 0.83  &  ---  &  ---      &   0.82  &  0.83   & 0.83  &  0.83   \\
{\bf Higgs  total}      &  97   & 0.78  & 0.76  &  ---  &  ---      &   0.76   & 0.76   & 0.76  &  0.76   \\
\midrule
{\bf Diboson}           &  70   & 1.31  & 1.30  &  ---  &  ---      &  ---  &   ---  &  1.31   &  1.30    \\
\bottomrule
    {\bf Total $n_{\rm dat}$}    & 317    &  317   &  317   &  150  & 101  & 97   &247  & 301  & 287    \\
    {\bf Total $\chi^2$}    & ---    &  {\bf 1.05}    &  {\bf 1.04}  &  {\bf 1.10}  &  {\bf 1.22}  &
    {\bf 0.75}  &  {\bf 0.96} &  {\bf 1.02}   & {\bf 0.89}    \\ 
\bottomrule
\end{tabular}
  \caption{\small Same as Table~\ref{eq:chi2-baseline-grouped}  for EFT fits obtained from
    variations of the baseline dataset.
    We list the results of the following fits: including only  top quark measurements (either for the 2018 or the current
    dataset); a Higgs-only dataset; without the diboson cross-sections; with the high-energy bins excluded;
    and with the datasets with a poor $\chi^2_{\rm sm}$ excluded.
    In all cases, the  quadratic EFT corrections are accounted for.
    The numbers in parentheses indicate the number of data points, in the case that these are different
    from those of the baseline settings (listed in the second column).
\label{eq:chi2-datasetvariations}
}
\end{table}

%% file: subsec_results_loqcd.tex
\clearpage
\subsection{Impact of NLO QCD corrections in the EFT cross-sections}
\label{subsec:loqcd}

In addition to the choice of input dataset, another important
factor that determines the outcome of a global analysis
such as the present one is the  accuracy
of the EFT theoretical calculations.
Here we assess the role played at the level of the fit results
by the inclusion of NLO QCD corrections to the EFT cross-sections,
both in the linear and in the quadratic fits.
As indicated in Table~\ref{eq:table-processes-theory},
our baseline fit includes these NLO corrections
to the EFT calculations whenever available, so now we switch them off deliberately
to quantify how much they affect the fit outcome.\footnote{This study
  is also motivated by the fact that many EFT fits
rely on LO QCD for the EFT cross-sections.}
In the following, the theoretical predictions
for the SM cross-sections, based on the state-of-the-art calculations,
remain unchanged, and only the EFT ones are modified as compared to the baseline
settings.
First of all, Table~\ref{eq:chi2-theoryvariations}
compares the values of the $\chi^2$ for the various groups of processes
in quadratic fits with and without NLO QCD corrections to  the  EFT cross-sections,
as well as for the associated SM results.
One can observe how the overall fit quality
is similar whether or not NLO QCD effects are not accounted for.
Nevertheless, as will be discussed next, this does
not imply that the fit posterior distributions
are likewise unchanged.

\input{tables/table-chi2-theoryvariations.tex}


\begin{figure}[htbp]
  \begin{center}
    \includegraphics[width=0.99\linewidth]{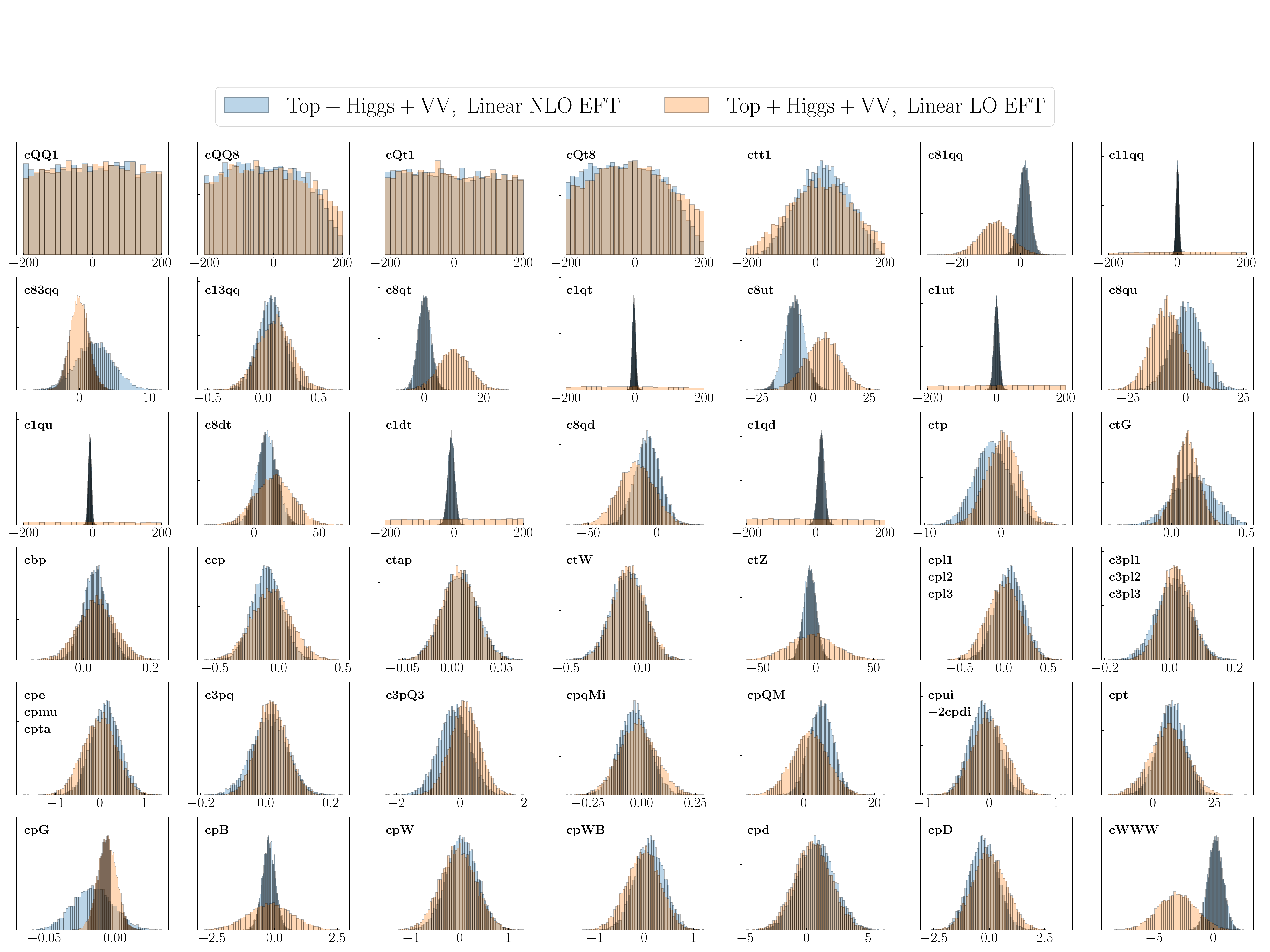}
    \includegraphics[width=0.99\linewidth]{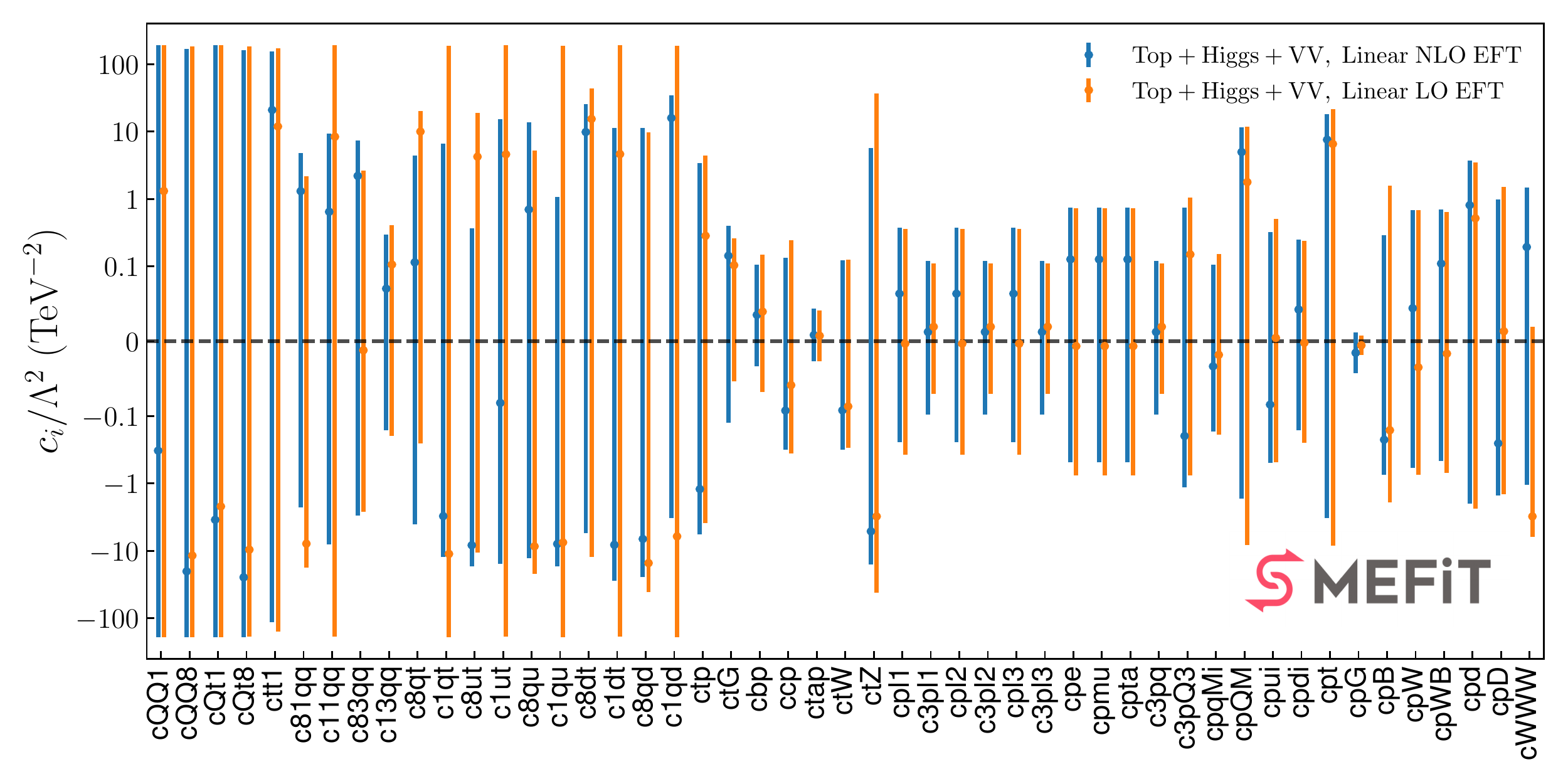}
    \caption{\small Top: comparison of the posterior probability
      distributions of the Wilson coefficients
      between linear fits with and without NLO QCD corrections to the EFT cross-sections.
      Bottom: the corresponding 95\% CL intervals, compared to the SM expectation.
     \label{fig:posterior-distributions-NLO-vs-LO-linear} }
  \end{center}
\end{figure}

\begin{figure}[htbp]
  \begin{center}
    \includegraphics[width=0.99\linewidth]{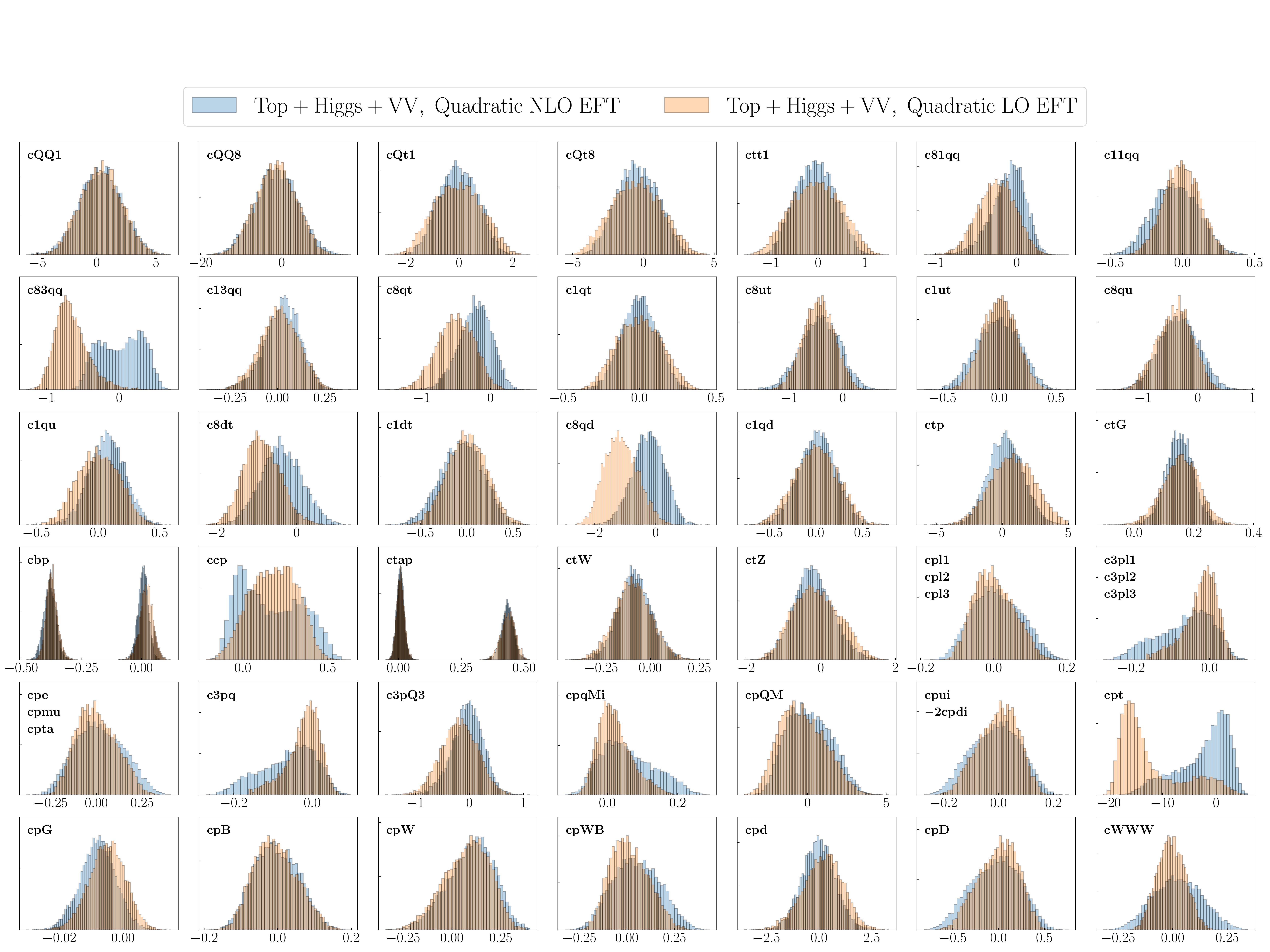}
    \includegraphics[width=0.99\linewidth]{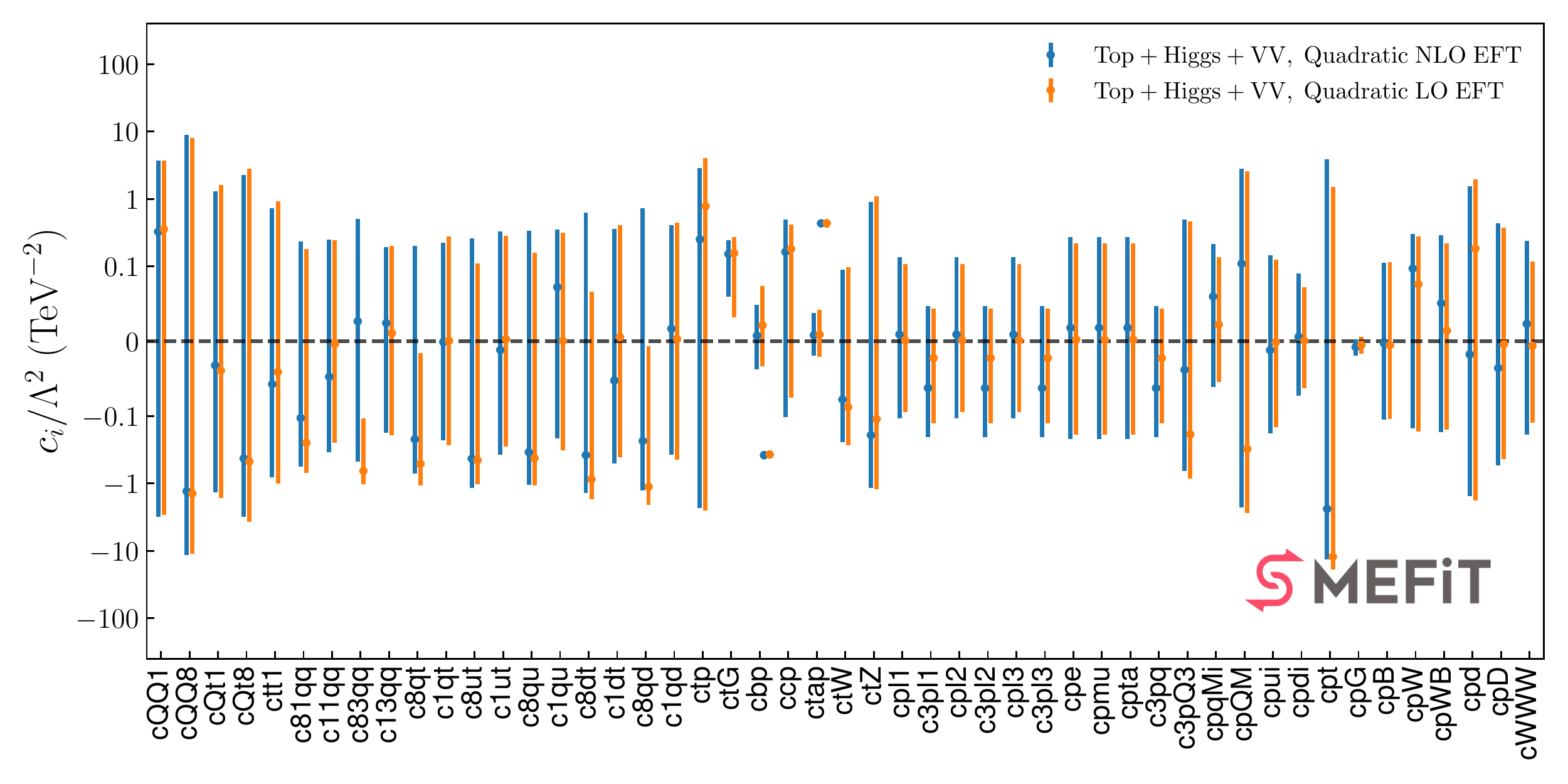}
    \caption{\small
      Same as Fig.~\ref{fig:posterior-distributions-NLO-vs-LO-linear}
      for the quadratic EFT fits.
      \label{fig:posterior-distributions-NLO-vs-LO-quad} }
  \end{center}
\end{figure}

Figs.~\ref{fig:posterior-distributions-NLO-vs-LO-linear}
and~\ref{fig:posterior-distributions-NLO-vs-LO-quad} then display the
posterior probability distributions 
and the corresponding 95\% CL intervals for the Wilson coefficients,
comparing the results of linear and quadratic fits respectively
with and without NLO corrections to the EFT cross-sections.
Scrutinizing first the linear fit results collected in
Fig.~\ref{fig:posterior-distributions-NLO-vs-LO-linear},
one can observe that these posterior distributions can be
severely distorted when LO EFT calculations are used as compared
to the baseline,
for instance in terms of a shift in the best-fit values and/or due
to an increase in the width of the Gaussian distributions.
Also in the LO linear fit, all considered coefficients
agree with the SM expectation at the 95\% CL.
Note that the two-light-two-heavy singlet operators do not
interfere with the SM at LO,
and hence the corresponding coefficients turn out to be unconstrained
in the linear LO fit.
Remarkably, for several fit coefficients
such as $c_{tZ}$, $c_{\varphi B}$, and $c_{W}$, one finds that a marked
improvement in the obtained bounds is achieved upon the inclusion
of the NLO QCD corrections to the EFT cross-sections.
One would conclude that, at least in the global linear EFT fit,
the inclusion of NLO QCD corrections
is of clear importance to obtain both more accurate and more precise results
for the Wilson coefficients.
Alternatively, one could account for the missing
higher-order uncertainties (MHOUs) in the EFT cross-sections, which
are usually neglected, using for instance the approach advocated
in~\cite{AbdulKhalek:2019bux,AbdulKhalek:2019ihb}.
Implementing MHOUs systematically is expected to further improve
the overall compatibility of EFT fits performed with and
without NLO QCD corrections.

Moving to the associated comparisons in the case of the quadratic fits summarised in
Fig.~\ref{fig:posterior-distributions-NLO-vs-LO-quad},
also here we find that the parameter distributions can be modified in a marked
way depending on whether or not NLO QCD calculations are adopted.
As an illustration, the operator that modifies the charm Yukawa interaction,
$c_{c\varphi}$, exhibits
a bimodal distribution once NLO effects are accounted for,
while the dominant solution for the $c_{\varphi t}$  coefficient
is far from the SM in the LO fit but SM-like in the NLO case
(though the 95\% CL interval itself remains stable).
As opposed to the case of the linear fits,
in the quadratic case one finds that the addition of NLO corrections
does not in general reduce the uncertainties
on the fit coefficients, but rather distorts
the posterior distributions and shifts the central values.
As an illustration,
if NLO QCD corrections are removed, the posterior
distribution for the two-light-two-heavy coefficient
$c_{Qq}^{3,8}$ is shifted such that it does not agree anymore
with the SM at the 95\% CL.

\begin{figure}[t]
  \begin{center}
    \includegraphics[width=0.41\linewidth]{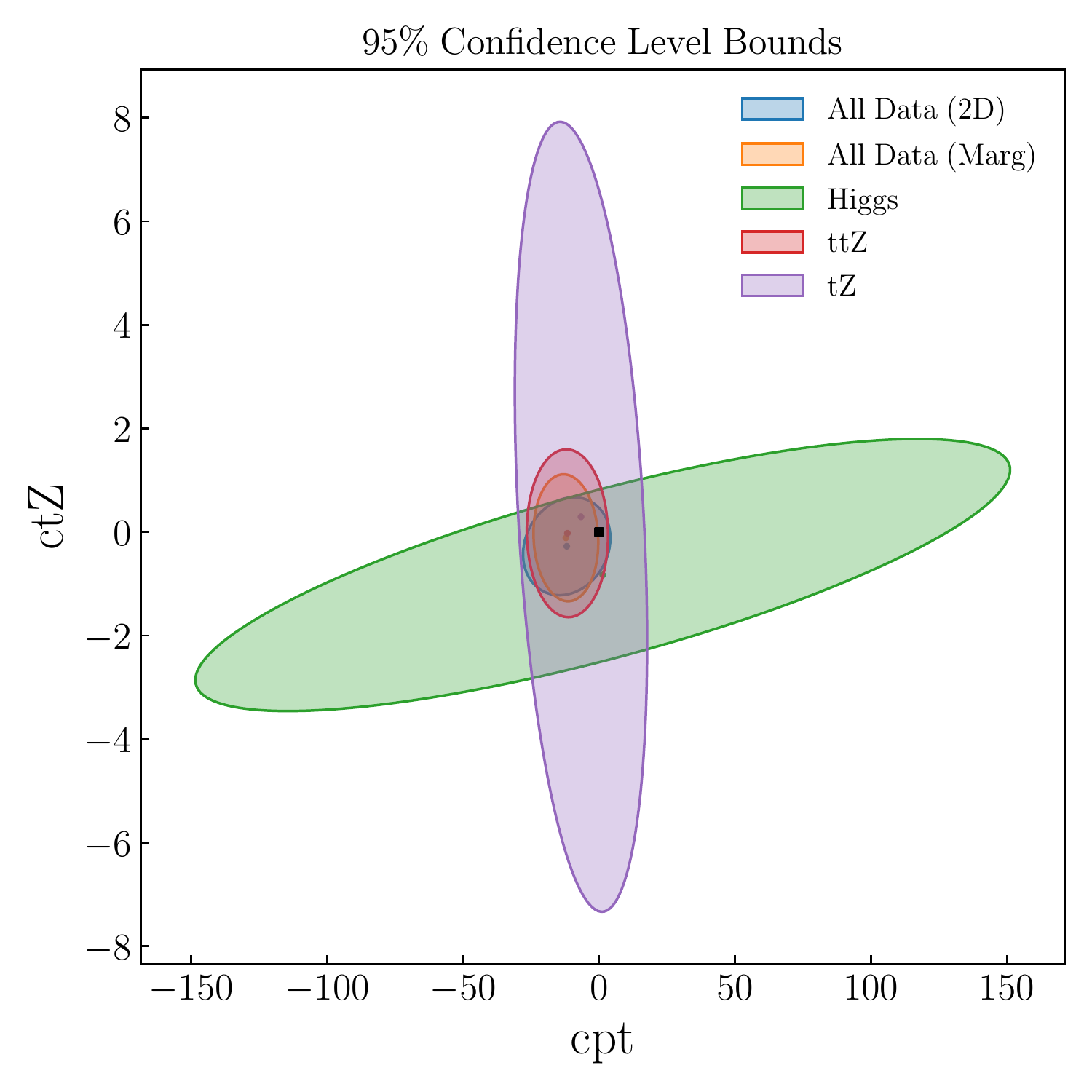}
    \includegraphics[width=0.41\linewidth]{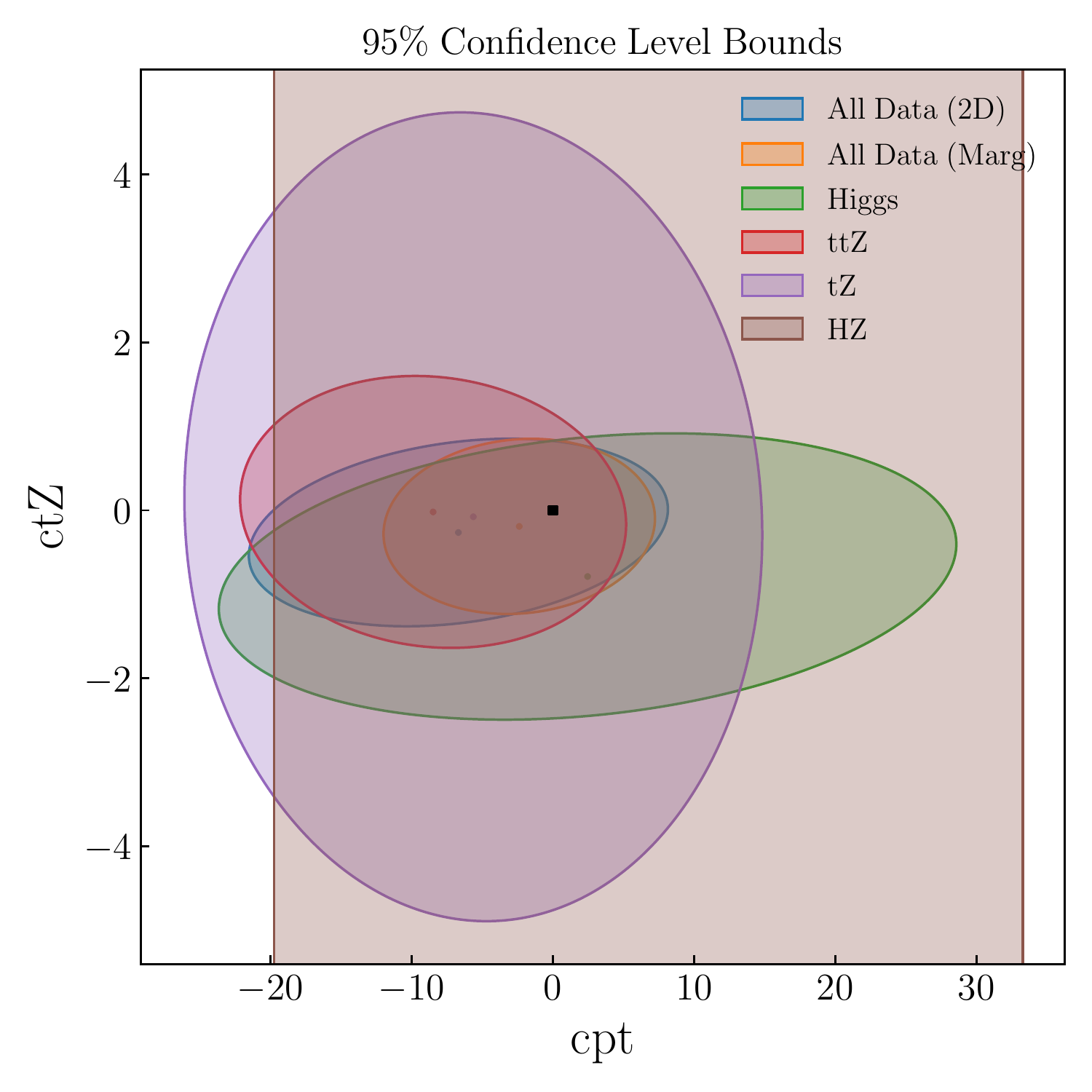}
    \includegraphics[width=0.41\linewidth]{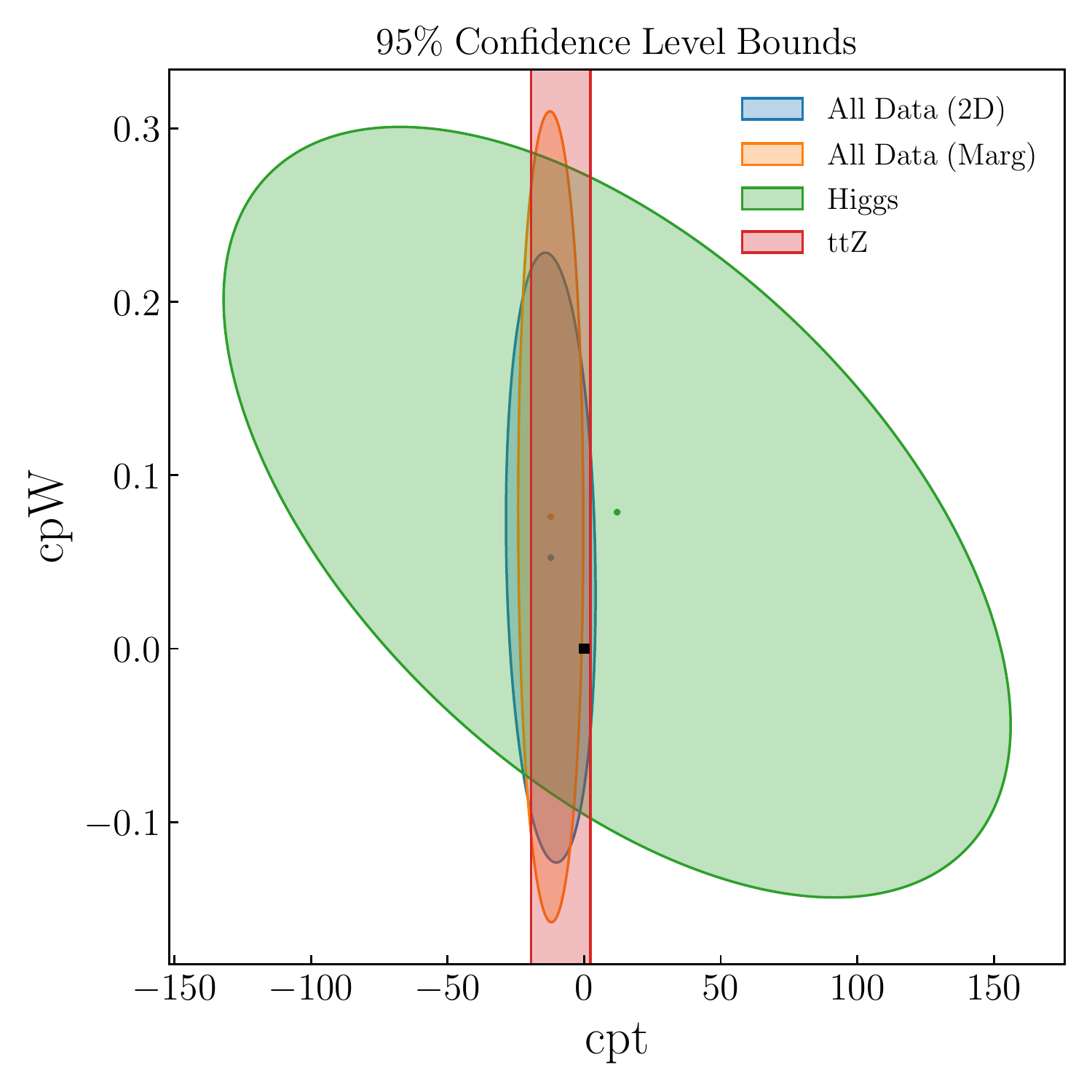}
    \includegraphics[width=0.41\linewidth]{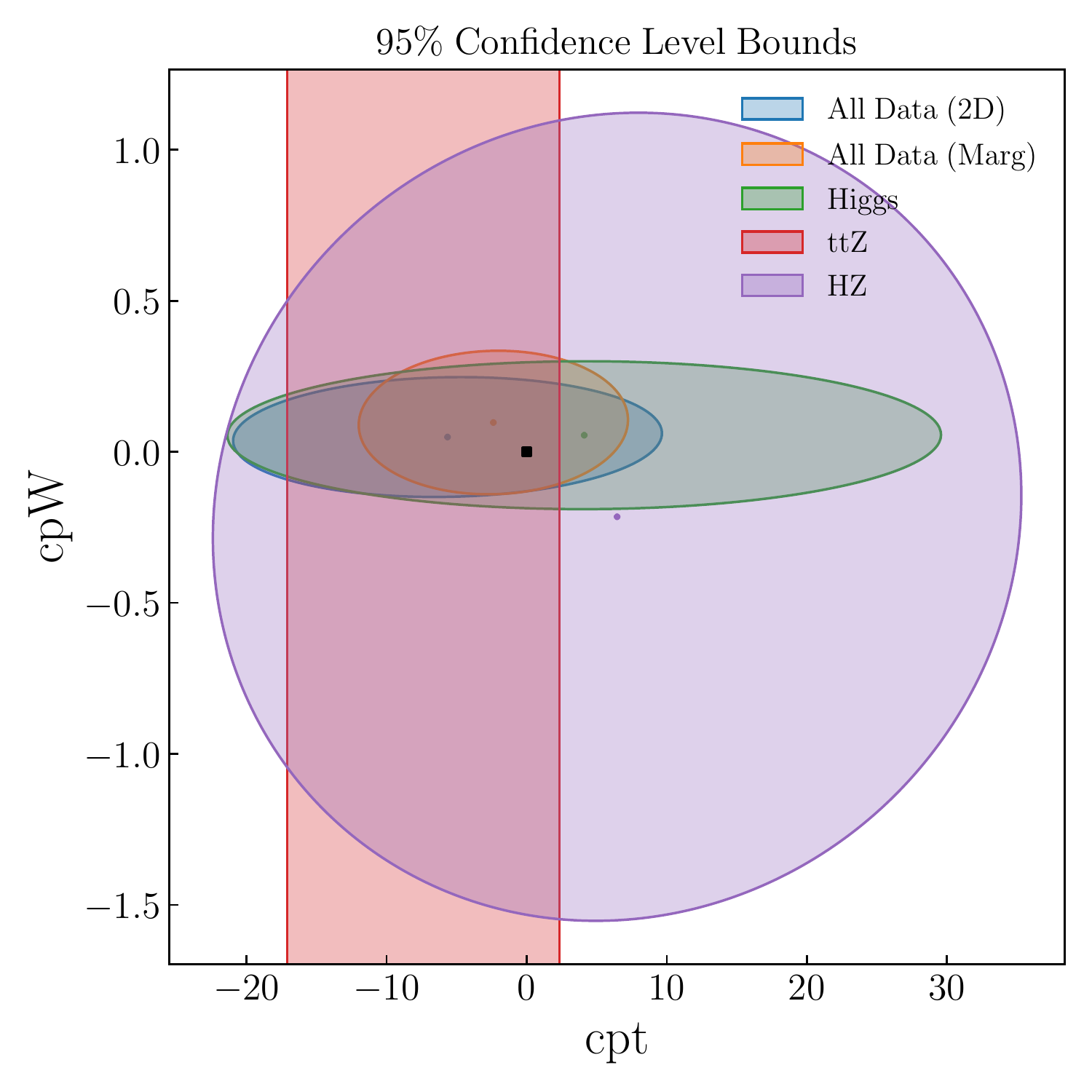}
    \caption{\small
      Same as Fig.~\ref{fig:2Dfits} for the two-parameter quadratic fits
      of $(c_{\varphi t},c_{tZ})$ (upper) and $(c_{\varphi t},c_{\varphi W})$ (lower panels)
      comparing the results of the LO EFT fit (left) with its NLO counterpart (right panels)
      \label{fig:Ellipse_Opt_OtZ_LO_HO} }
  \end{center}
\end{figure}

In the specific case of the $c_{\varphi t}$  coefficient, one can verify that the corresponding individual $\chi^2$ profile
(analog of Fig.~\ref{fig:quartic-individual-fits-2} for LO fits) does not
exhibit this second solution, and hence it must be induced by the cross-talk
with other coefficients in the fit.
To validate this hypothesis, Fig.~\ref{fig:Ellipse_Opt_OtZ_LO_HO} displays the outcome of 
 two-parameter quadratic fits for
     $(c_{\varphi t},c_{tZ})$  and $(c_{\varphi t},c_{\varphi W})$ 
 comparing the results of the LO EFT fit  with its NLO counterpart.
 In both cases, the LO two-parameter fits based on the full dataset
 favour the solution far from the SM, while the NLO ones
 instead favour the SM-like one.
 The explanation for this behaviour can be traced back
 to the fact that the non-SM solution is disfavored
 by the NLO EFT corrections to $hZ$ associated production,
 in particular those related to gluon-induced contributions.

\begin{figure}[t]
  \begin{center}
    \includegraphics[width=0.49\linewidth]{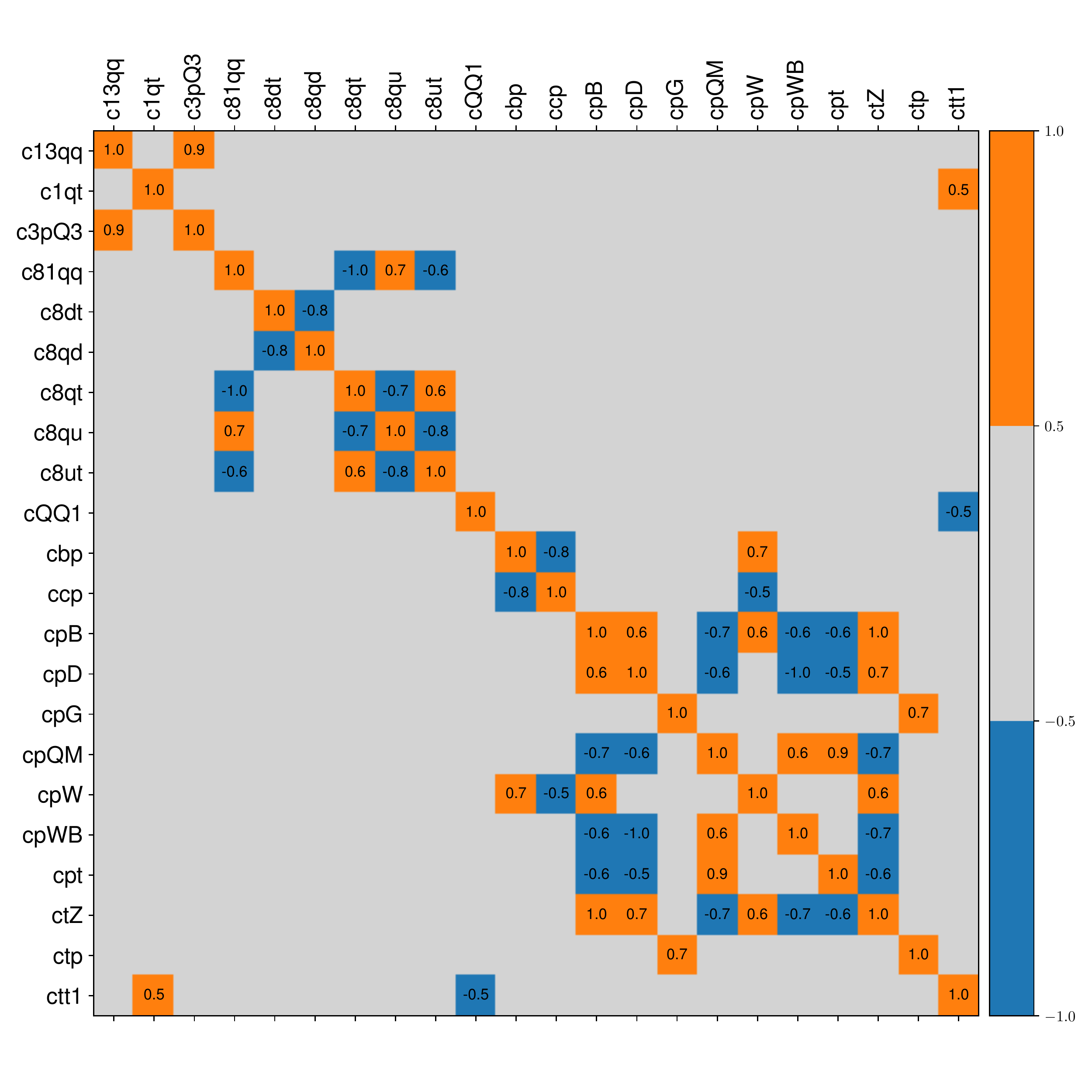}
    \includegraphics[width=0.49\linewidth]{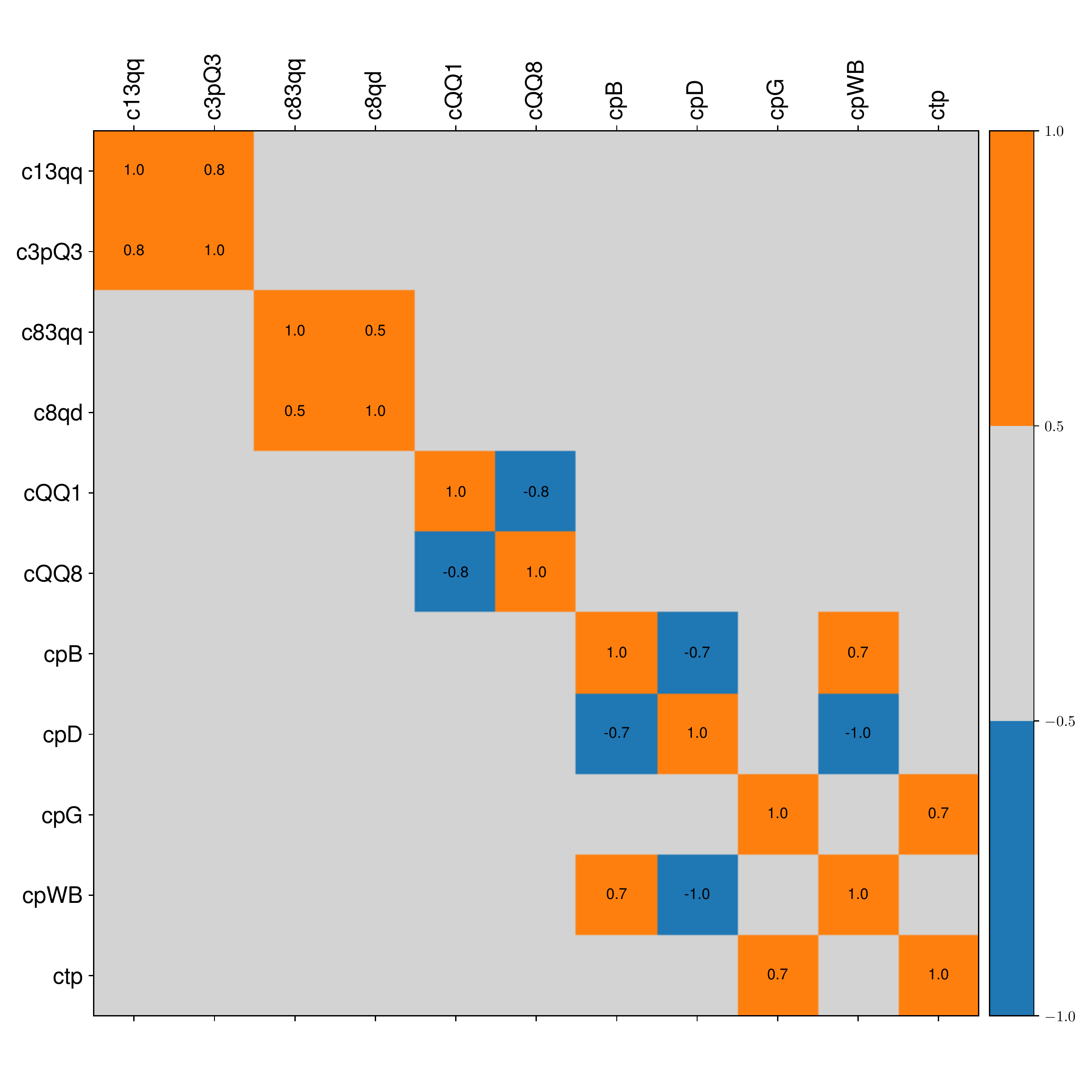}
     \caption{\small
       Same as Fig.~\ref{fig:globalfit-correlations} for
       LO EFT calculations in linear (left) and quadratic (right) fits.
      \label{fig:Coeffs_Corr_260121-NS_GLOBAL_LO} }
  \end{center}
\end{figure}

Another remarkable effect of the  NLO QCD corrections
to the EFT cross-sections can be observed in the modified
correlation patterns.
Fig.~\ref{fig:Coeffs_Corr_260121-NS_GLOBAL_LO} displays
the same correlations maps as in Fig.~\ref{fig:globalfit-correlations} now
for global fits based
on LO EFT calculations at  the linear and quadratic level.
Specially for the linear fits, we observe that correlations
become more sizable in general
for the two-fermion and purely bosonic operators,
while these are
reduced once NLO corrections are accounted for.
This feature demonstrates how NLO QCD effects may reduce parameter correlations
by introducing additional sensitivity to the fit coefficients for the same input dataset.

%% file: tables/table-chi2-theoryvariations.tex
\begin{table}[t]
  \centering
  \footnotesize
   \renewcommand{\arraystretch}{1.40}
  \begin{tabular}{l|C{1.0cm}|C{1.2cm}|C{2.0cm}|C{2.7cm}|C{2.2cm}}
    \multirow{2}{*}{Dataset}   & \multirow{2}{*}{$ n_{\rm dat}$} & $\chi^2_{\rm SM}$ &  $\chi^2_{\rm EFT}$   & $\chi^2_{\rm EFT}$    & $\chi^2_{\rm EFT}$    \\
      &   &  &  (baseline)  &  (LO QCD in EFT)  & (top-philic)   \\
 \toprule
$t\bar{t}$ inclusive        &  83    &  1.46  &  1.42      &   1.39     &   1.41      \\
$t\bar{t}$ AC                & 11    &  0.60  &  0.59      &   0.57     &   0.60      \\
$t\bar{t}V$                 &  14    &  0.65  &   0.65     &   0.54     &  0.68     \\
single top inclusive        &  27    &  0.43  &   0.41     &   0.42     &  0.41      \\
$tV$                        &  9     &  0.71  &   0.75     &   0.68     &  0.78      \\
 $t\bar{t}Q\bar{Q}$         &  6     & 1.68  &  2.12       &   2.24     &  2.16       \\
{\bf Top quark total}       &  150   & 1.10  &  1.09       &   1.06     &  1.09      \\
\midrule
Higgs $\mu_i^f$  (Run I)    &  22    & 0.86  &  0.90       &  0.95     &  0.90       \\
Higgs $\mu_i^f$  (Run II)   &   40   & 0.67  &  0.63       &  0.67     &  0.63       \\
Higgs differential \& STXS  &  35    & 0.88  &  0.83       &  0.78     &  0.83       \\
{\bf Higgs  total}          &   97   & 0.78  &  0.76       &  0.77     &  0.76     \\
\midrule
{\bf Diboson}               &  70    & 1.31  &  1.30       &   1.32    &   1.30      \\
\midrule
{\bf Global dataset}        & {\bf 317}   & {\bf 1.05}  &  {\bf 1.04}   &   {\bf 1.03}     &   {\bf 1.04}      \\
\bottomrule
\end{tabular}
  \caption{\small Same as Table~\ref{eq:chi2-baseline-grouped} now for fits based on
    variations of the theory settings as compared to the baseline ones.
    Specifically, we provide the results of a fit where the EFT cross-sections are evaluated at LO in the QCD
    expansion, as well as those of the top-philic scenario where the  parameter space
    has been restricted as described in Sect.~\ref{sec:topphilic}.
    In both cases, quadratic EFT corrections are being included.
    Note that the SM cross-sections are always evaluated using state-of-the-art theory
    calculations.
\label{eq:chi2-theoryvariations}
}
\end{table}

%% file: subsec_results_topphilic.tex
\subsection{The top-philic scenario}

To conclude this section, we present results for a global EFT fit
carried out in the top-philic scenario defined in Sect.~\ref{sec:topphilic}.
In this scenario, we have the 9 equations of Eq.~(\ref{eq:topphilic}) that 
relate a subset of the 14
two-heavy-two-light coefficients listed in Table~\ref{tab:operatorbasis}
among them,  leaving 5 independent parameters to be constrained
in the fit.
Given the more constraining assumptions associated
to the top-philic scenario, one expects to find an improvement
in the bounds of the two-light-two-heavy EFT operators due to the fact
that the parameter space is being restricted by theoretical
considerations, rather than by data in this case.

\begin{figure}[t]
  \begin{center}
     \includegraphics[width=0.99\linewidth]{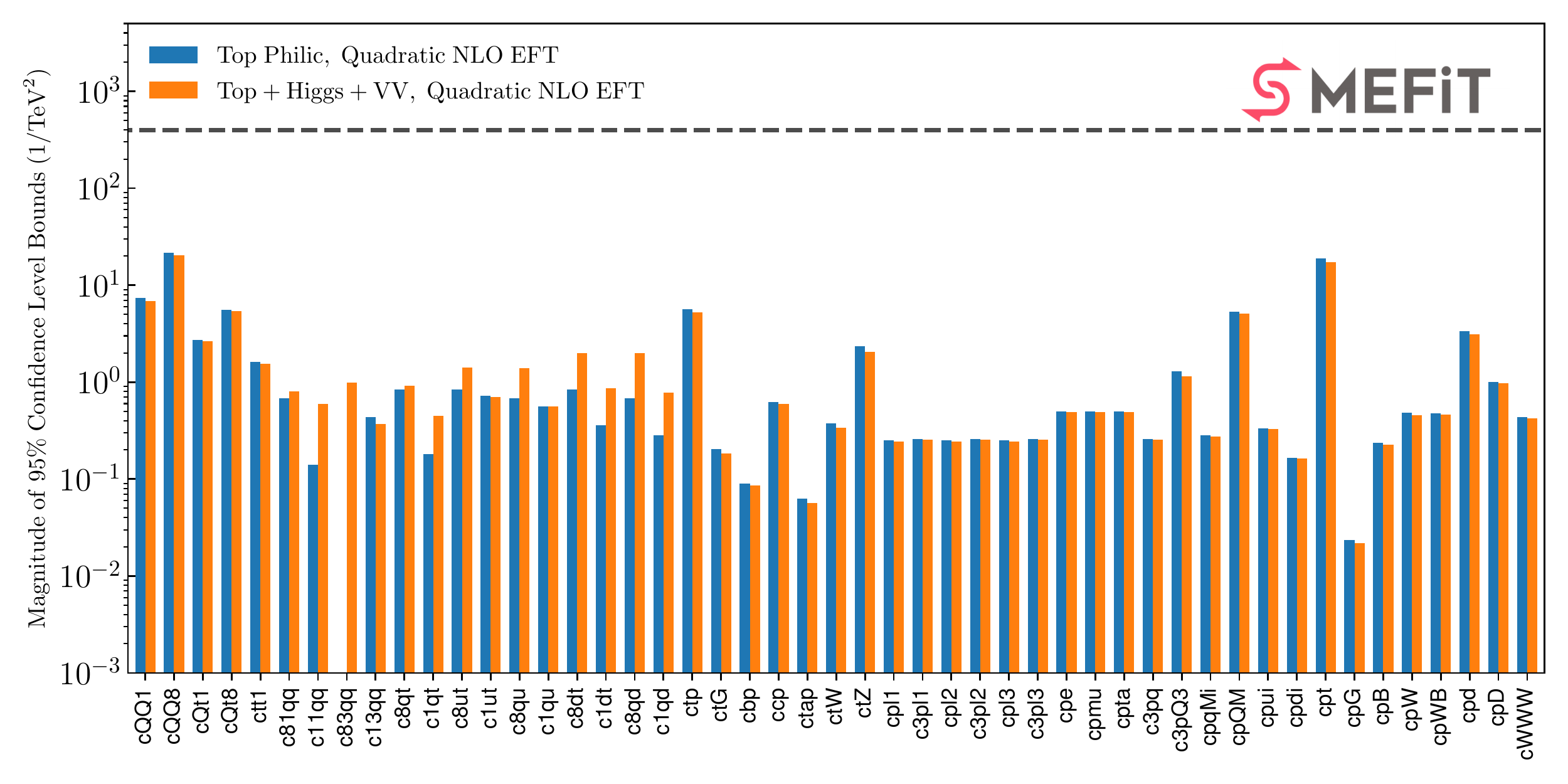}
     \caption{\small Same as Fig.~\ref{fig:posterior_coeffs}
       comparing the global fit results with the same fit in the top-philic
       scenario defined by the relations in Eq.~(\ref{eq:topphilic}).
     \label{fig:Coeffs_Bar_TopPhilic} }
  \end{center}
\end{figure}

The values of the $\chi^2$ for each group of datasets in the top-philic scenario
were reported in Table~\ref{eq:chi2-theoryvariations},
where we see that the fit quality is very similar to the fit with the baseline settings.
Fig.~\ref{fig:Coeffs_Bar_TopPhilic} then displays the 95\% CL
intervals for the EFT coefficients comparing the
global fit results with those of the top-philic scenario.
The only operators that are affected in a significant manner
turn out to be the two-light-two-heavy
operators, with the bounds in several of them such as
$c_{td}^1$, $c_{Qq}^{1,1}$, and $c_{tq}^1$  improving by almost an order of magnitude.
The fact that only the bounds on the two-light-two-heavy operators are modified
is consistent with the top-philic scenario, given that only
this specific group of EFT coefficients is being constrained by its model assumptions.

It is worth emphasizing at this point that,
from the technical point of view, carrying out global EFT fits with specific restrictions
in the parameter space motivated by UV-completions, such as those arising
in the top-philic scenario and leading to Fig.~\ref{fig:Coeffs_Bar_TopPhilic},
is relatively straightforward.
Indeed, the most efficient fitting strategy would be to start from the broadest possible
parameter space, and once the corresponding fit has been performed,
introduce model assumptions relating EFT coefficients  in a systematic manner.
This way one can connect with specific models for  UV-completions of the SM,
which typically result in a rather smaller number of EFT coefficients to be constrained
from data.

%% file: sec-summary.tex
\section{Summary and outlook}
\label{sec:summary}

In this work we have presented an extensive interpretation of Higgs, diboson, and top
quark measurements from the LHC in the framework of the Standard Model Effective Field Theory.
By combining the most updated  experimental data with state-of-the-art theory calculations,
both in the SM and in the EFT,
we have provided bounds on 50 directions in the SMEFT parameter space
of which 36 correspond to independent parameters.
We have quantified in detail
the relative impact that the different types of processes have in the results
of this global EFT analysis, both in terms of fits with dataset variations
and by means of statistical diagnosis tools such as information geometry techniques
and principal component analysis.
Our analysis 
highlights the overall complementarity of the various input processes,
further motivating the need
for a global interpretation of LHC measurements.
We have also demonstrated how, within such
a global EFT analysis, genuinely flat directions are essentially
absent since each process and each kinematic bin of a distribution
constrains separate combinations of the fit parameters.
The robustness  of our fitting methodology has been cross-validated by deploying
two completely independent methods, MCfit and NS, for mapping the EFT parameter space.

We have also extensively quantified the role played in the global analysis
by the inclusion of NLO QCD corrections to the EFT cross-sections,
whose automation has been recently achieved.
We find that the posterior probability distributions of the fit parameters
can be modified in non-trivial ways by these NLO QCD effects, shifting the best-fit
value, modifying the magnitude of the 95\% CL intervals, and even inducing multi-modal
distributions.
These findings demonstrate that available LHC data is already sensitive to
NLO effects in the EFT cross-sections, further highlighting the
importance of accounting for them in a systematic manner to achieve both
accurate and precise results.

One could consider several directions in which the present study could be extended.
To begin with, one would like to include directly the constraints provided
by LEP's electroweak precision observables, rather than in an 
approximate manner as done in the present work.
Furthermore, it would be beneficial to add to the global analysis new  high-$p_T$ observables
providing complementary information on the Higgs and gauge sectors of the SMEFT,
such as for instance vector boson scattering (VBS),  $Z$ production in vector boson fusion (VBF),
or high-mass Drell-Yan production, all of them constraining the electroweak interactions.
Other processes that one might consider in this context
are  single-inclusive jet, dijet,
and multijet production, which are sensitive to several directions
in the parameter space not covered by other processes, specifically to a large
number of four-fermion operators.

From the point of view of theoretical calculations, it would be important to systematize the study
of higher-order terms in the EFT expansion, considering in particular double insertions
of dimension-six operators and representative subsets of dimension-eight operators.
We point out that, within our fitting methodology, accounting for these higher order terms
is technically straightforward.
Along the same lines, one could extend the fitting formalism to account
for all sources of theoretical uncertainties and their correlations in a systematic
manner, both for the SM and for the EFT calculations, something which is done
only partially here.
In addition, it should be interesting to develop statistically optimal observables
for EFT analyses, such as those based on deep
learning~\cite{Brehmer:2018kdj,DAgnolo:2018cun,Chen:2020mev},
making possible complementing
 the constraints obtained at the level of unfolded
 cross-sections with those extracted directly at the detector level.

Another promising research direction is that of combining
the global EFT interpretation of high-$p_T$ observables
at the LHC presented here with that of flavour data from LHCb and from other $B$-factories
such as Belle.
The urgency of a simultaneous EFT analyses of high-$p_T$ and flavour data
has been further highlighted by the recent evidence reported by the LHCb
experiment for the violation
of lepton flavour universality (LFU) in $B$-meson decays~\cite{Aaij:2021vac}.
These findings demand exploiting the flexibility of the EFT framework
in order to  comprehensively map the allowed signatures of eventual LFU violation
in  high-$p_T$ cross-sections at the LHC and elsewhere.

Likewise, it would be important to account for the constraints
provided by low energy processes in the SMEFT parameter space,
from neutrino data and electric dipole moment measurements
to the anomalous magnetic moment of the muon.
In the latter case, a $4.1\sigma$ deviation
with respect to the SM expectation has recently been reported~\cite{Abi:2021gix},
confirming and strengthening one of the most puzzling anomalies in particle
physics.
Indeed, the ultimate goal of our program would be a truly global EFT interpretation
including all processes sensitive to the sought-for UV completion of the SM,
making sure no stone is left unturned in the ongoing quest to unravel the new particles
and interactions that lie beyond the Standard Model.

\paragraph{Acknowledgments.}
F.~M. has received funding from the European Union's Horizon 2020 research and innovation programme as part of the Marie Sk\l{}odowska-Curie Innovative Training Network MCnetITN3 (grant agreement no. 722104) and by the F.R.S.-FNRS under the `Excellence of Science` EOS be.h project n. 30820817.
Computational resources have been provided by the supercomputing facilities of the Universit\'e catholique de Louvain (CISM/UCL) and the Consortium des \'Equipements de Calcul Intensif en F\'ed\'eration Wallonie Bruxelles (C\'ECI). 
J.~R. and E.~S. have been partially supported by the European Research Council Starting
Grant ``PDF4BSM''.
E.~S. has also received funding from the European Research Council (ERC) under the European Union’s Horizon 2020 research and innovation programme (grant agreement No 788223, PanScales).
J.~R. is also partially supported by the Netherlands Organization for Scientific
Research (NWO).
E.~R.~N. is supported by the UK Science and Technology Facility Council through
the UK STFC grant ST/T000600/1 and was supported by the European Commission
through the Marie  Sk\l{}odowska-Curie Action ParDHonS FFs.TMDs
(grant number 752748).
E.~V. is supported by a Royal Society University Research Fellowship through Grant URF/R1/201553.
C.~Z.~is supported by IHEP under contract number Y7515540U1 and by the National Natural
Science Foundation of China under grant number 12035008 and 12075256.

%% file: app-comparisondata.tex
\section{Comparison with experimental data}
\label{sect:app_comparison_data}

In this appendix, we present a systematic comparison between the
experimental data used as input to the fit with  the corresponding theoretical cross-sections
based both on the SM and on the best-fit SMEFT results, either
 at  linear $\mathcal{O}\lp \Lambda^{-2}\rp$ or at quadratic $\mathcal{O}\lp \Lambda^{-4}\rp$ accuracy.
In these comparisons, the experimental measurements will be presented both
in terms of unshifted central values, where the error band represents the total uncertainty, and once
the best-fit systematic shifts have been subtracted, such that the error band contains only the statistical component.
Note that the evaluation of the shifted data is only possible whenever the full breakup of the experimental
systematic uncertainties is made available by in {\tt HepData}.
If this is not the case, for example when only the full experimental correlation matrix is provided
or no information on correlations is released, we will display only the unshifted data.

\begin{figure}[htbp]
  \begin{center}
    \includegraphics[width=0.325\linewidth]{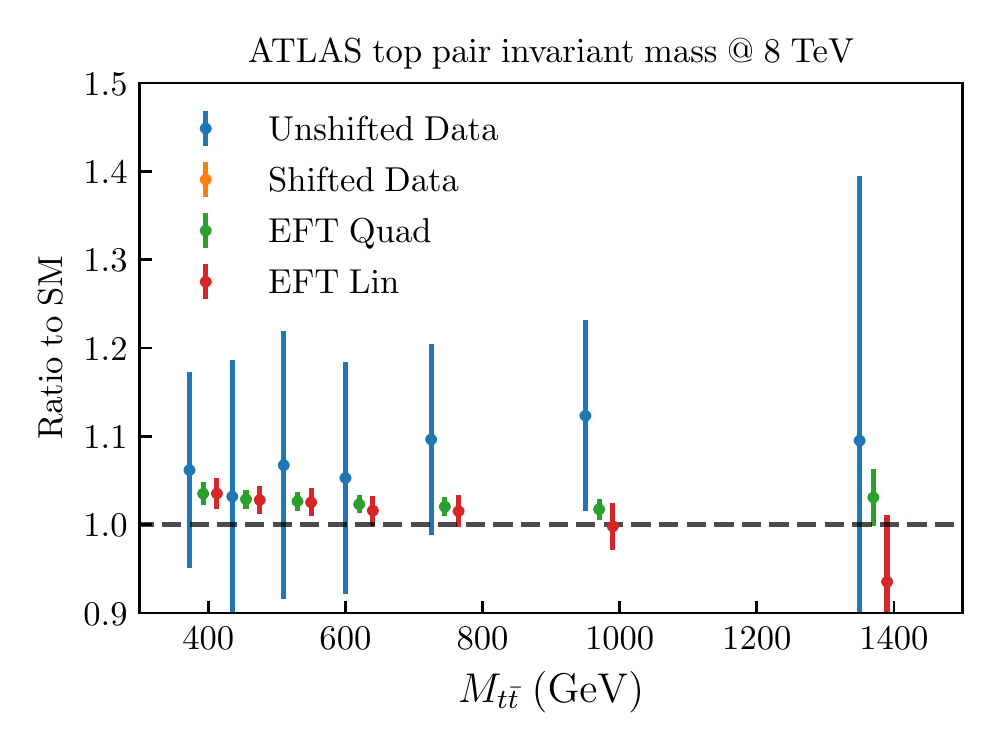}
    \includegraphics[width=0.325\linewidth]{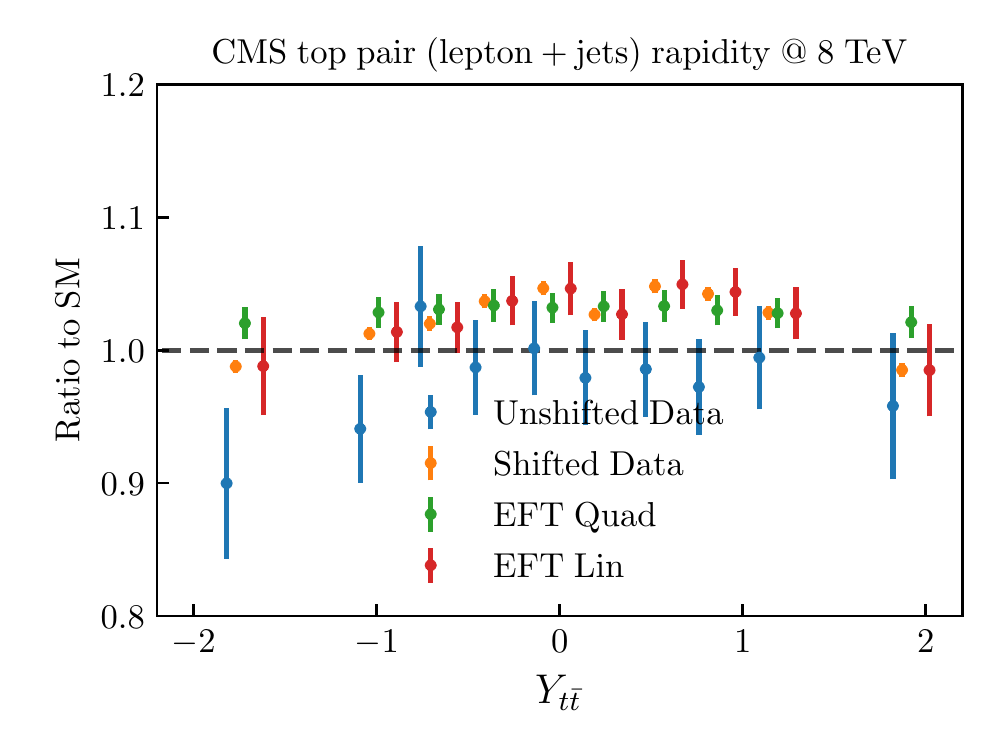}
    \includegraphics[width=0.325\linewidth]{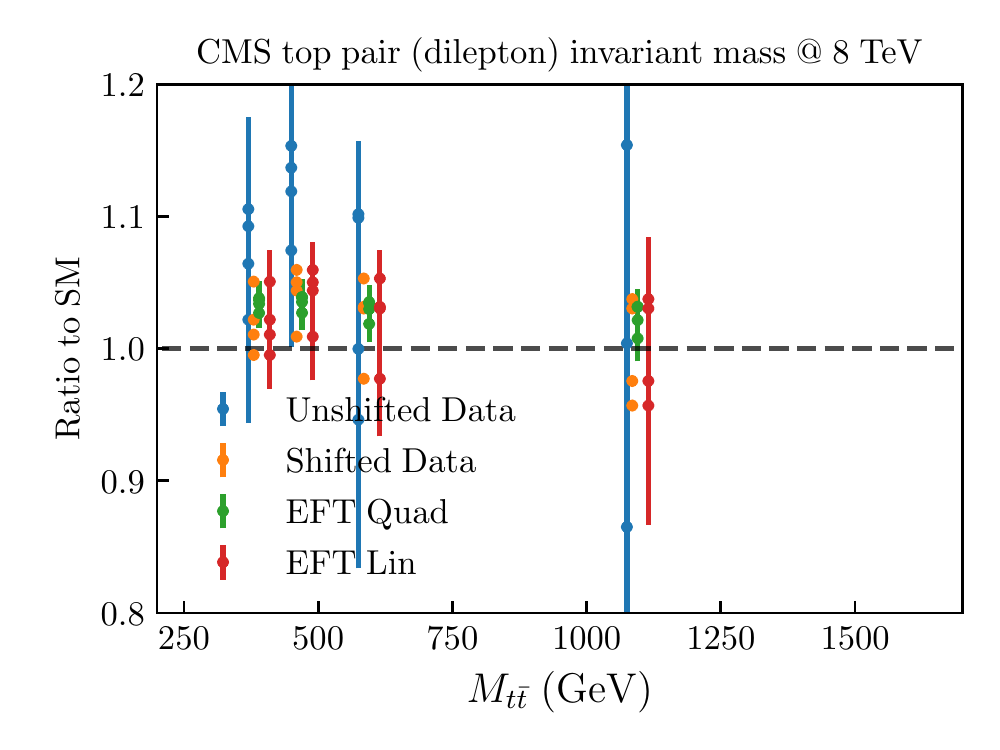}
    \includegraphics[width=0.325\linewidth]{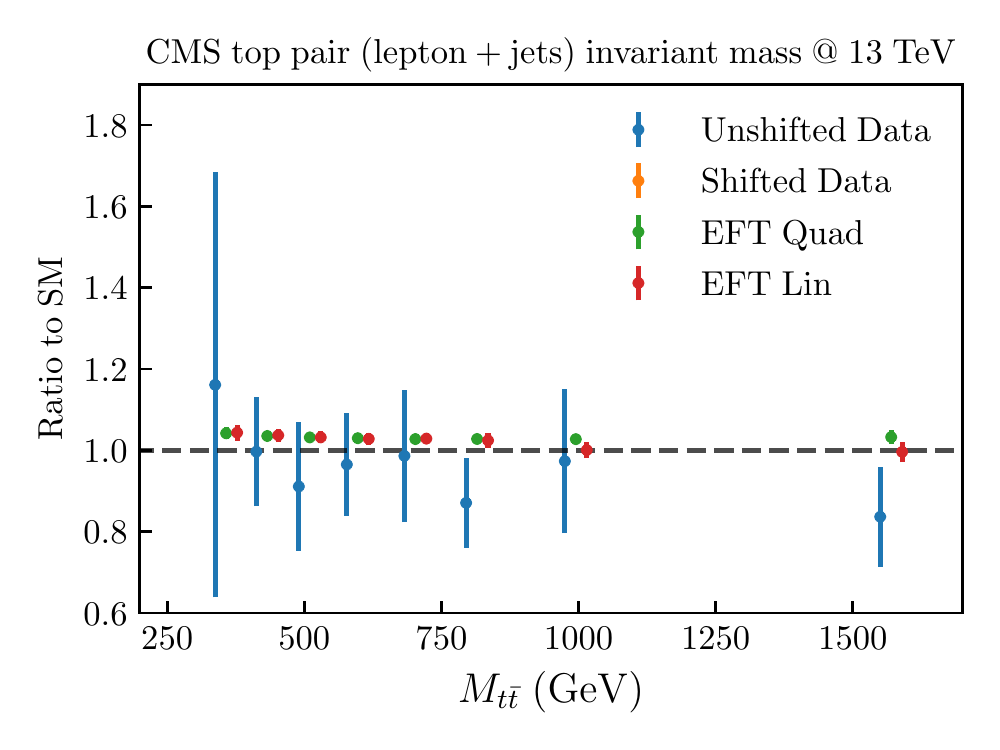}
    \includegraphics[width=0.325\linewidth]{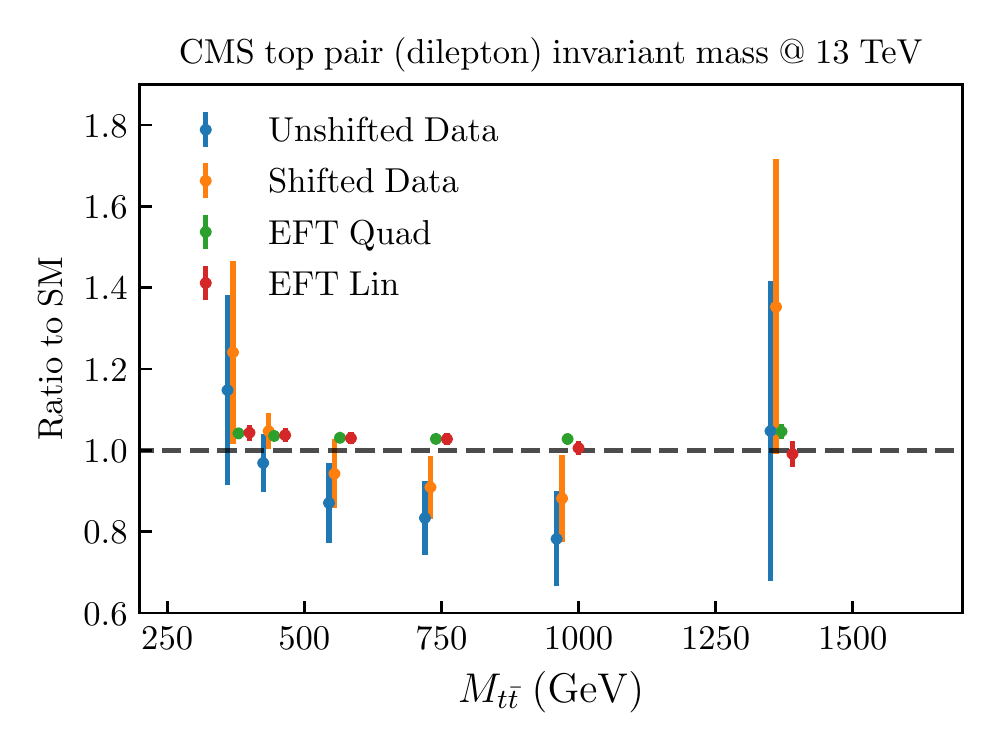}
    \includegraphics[width=0.325\linewidth]{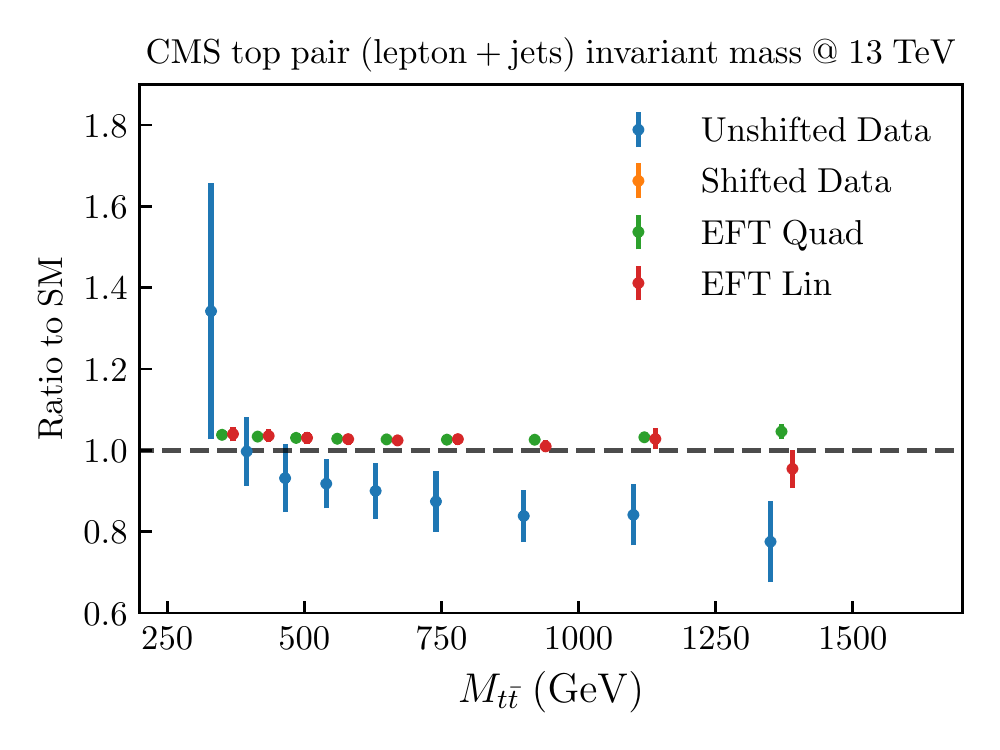}
    \includegraphics[width=0.325\linewidth]{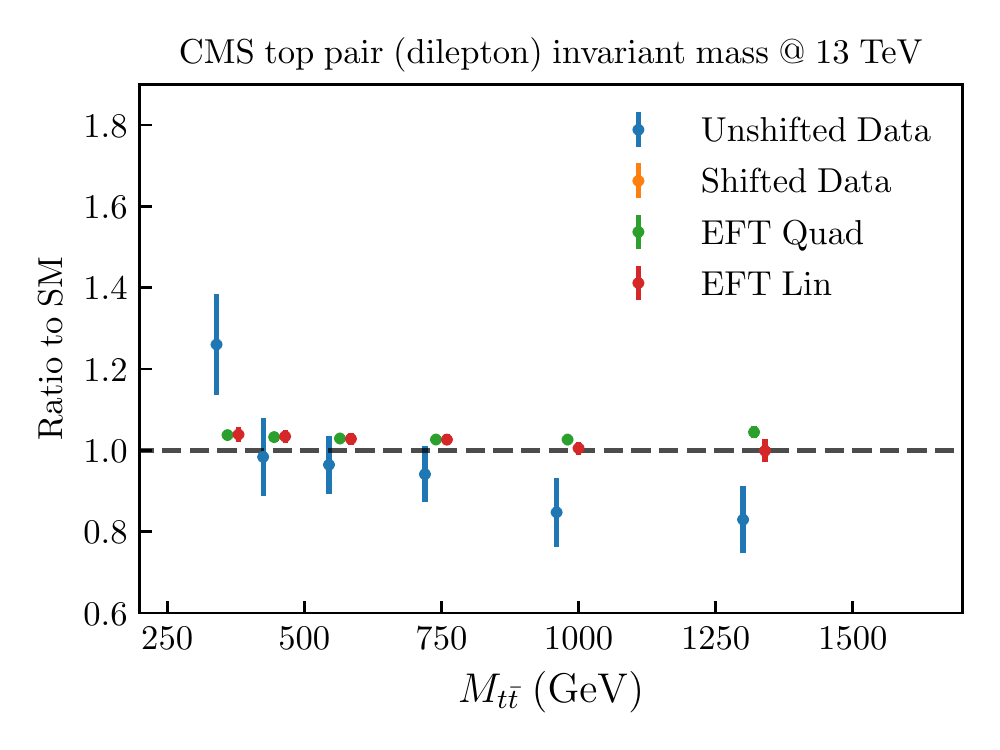}
    \includegraphics[width=0.325\linewidth]{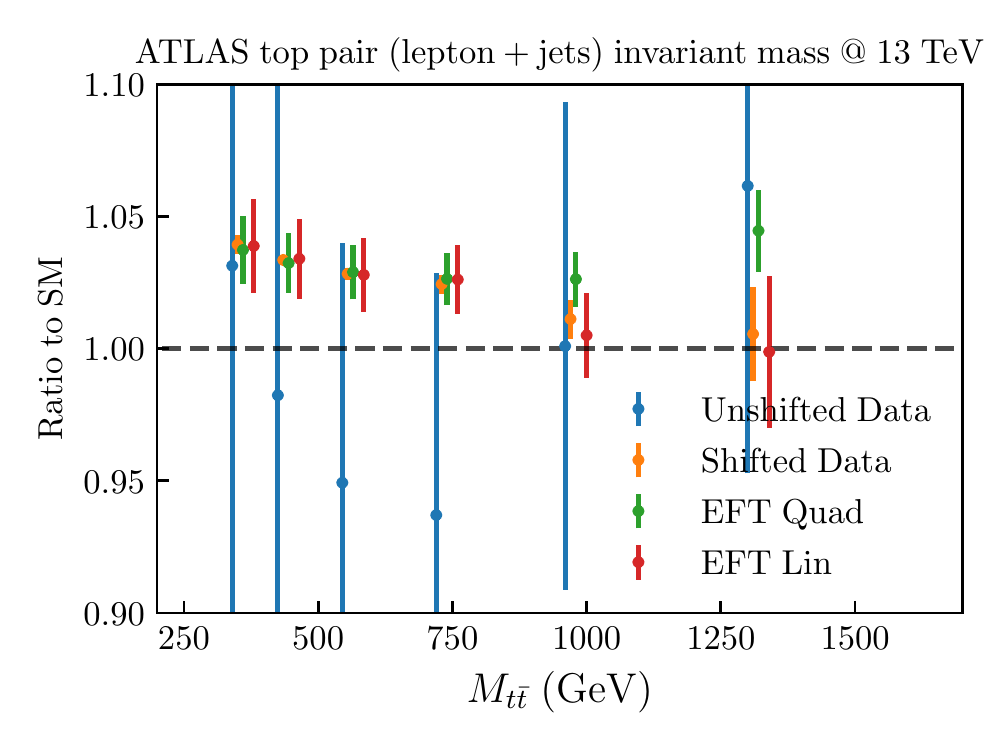}
    \includegraphics[width=0.325\linewidth]{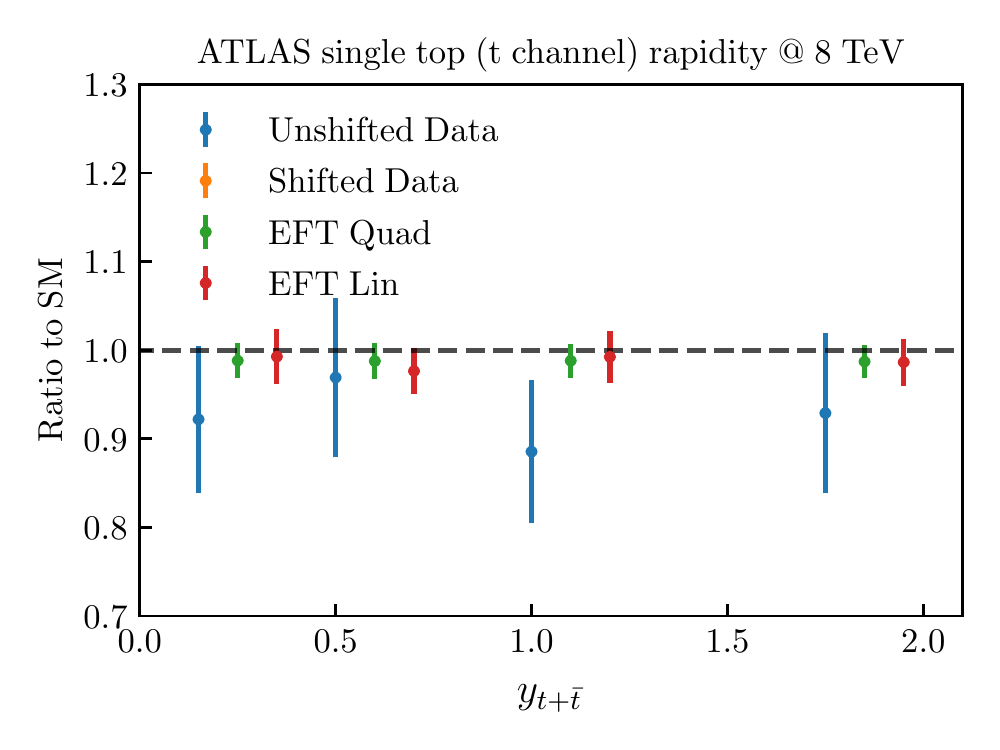}
    \includegraphics[width=0.325\linewidth]{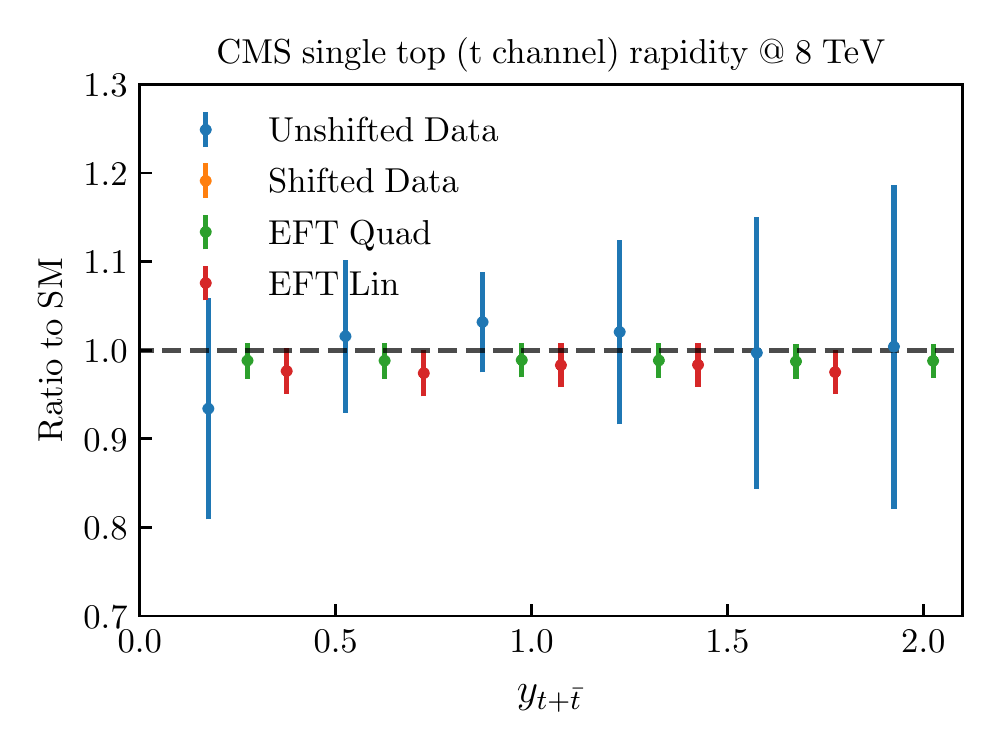}
    \includegraphics[width=0.325\linewidth]{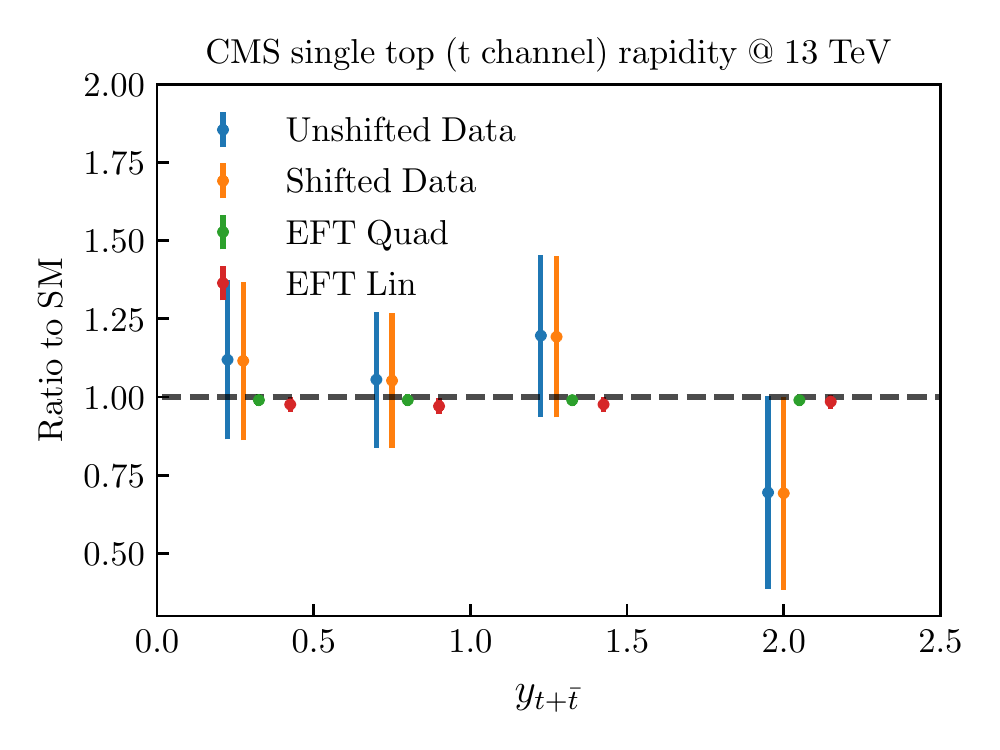}
    \includegraphics[width=0.325\linewidth]{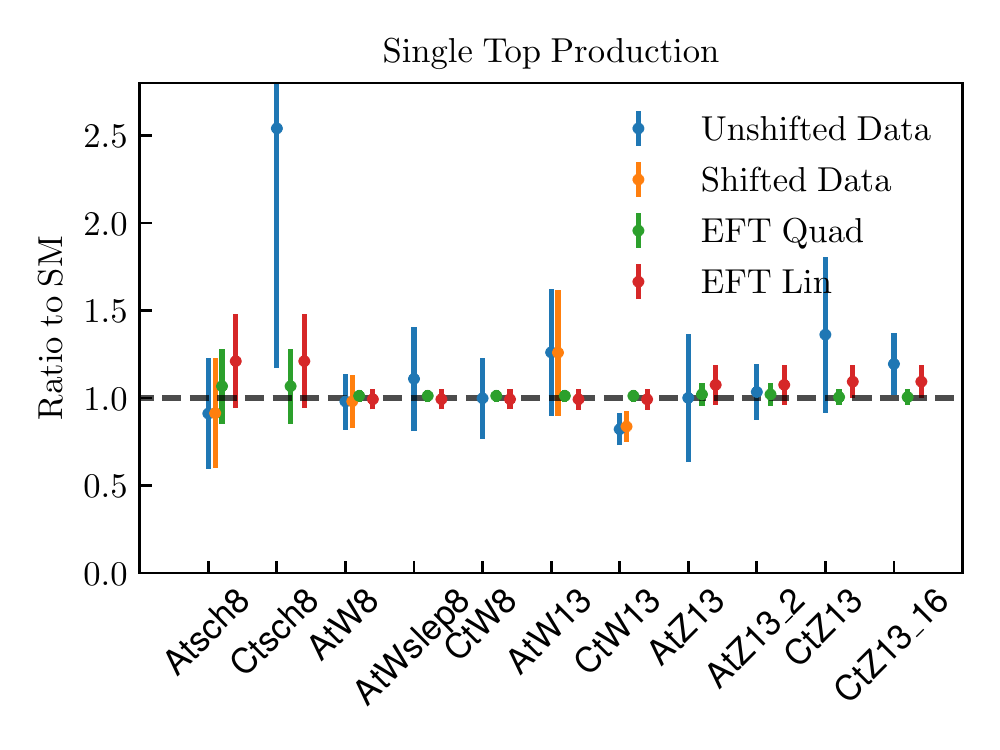}
    \includegraphics[width=0.325\linewidth]{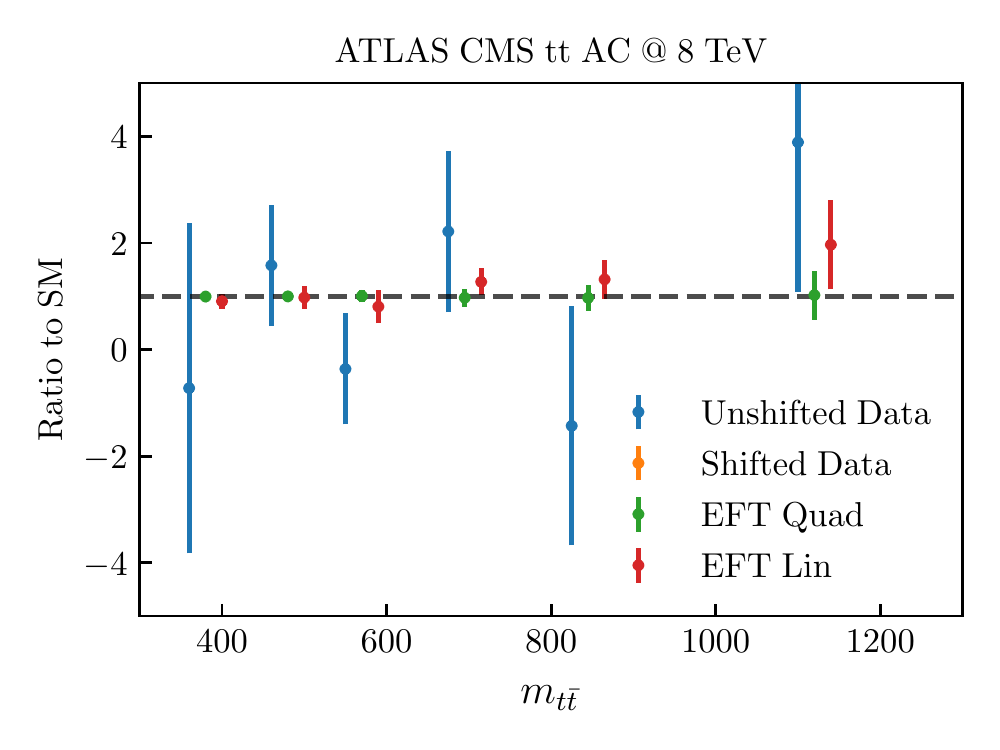}
    \includegraphics[width=0.325\linewidth]{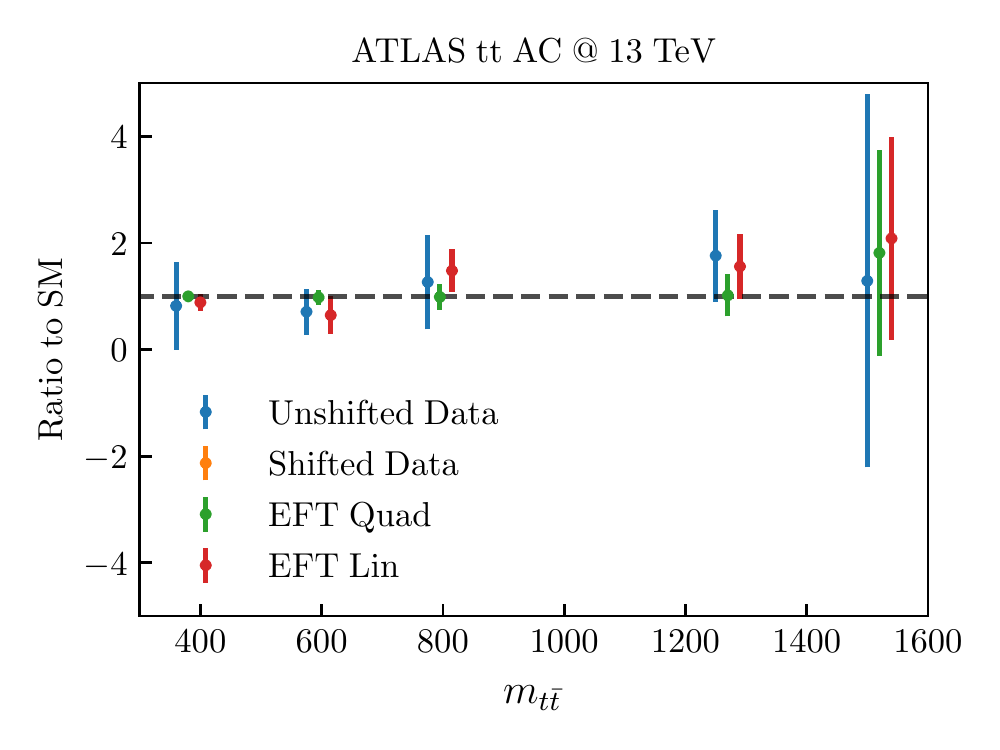}
    \caption{\small Comparison between experimental data and best-fit EFT theory
      predictions (both for the linear and the quadratic fits)
      for representative differential top quark pair
      and single top quark production datasets.
      Both the data and the EFT fit results are normalised to the central value of the SM cross-section.
      The data is presented both with unshifted central values
      (where the band represents the total experimental error) and once
      the best-fit systematic shifts have been subtracted (so that the error band
      contains only the statistical component).
      The error band in the EFT prediction indicates the 95\% CL interval evaluated
      over the $N_{\rm spl}$ samples produced by the NS method.
     \label{fig:data_vs_theory_top_diff} }
  \end{center}
\end{figure}

To begin with, Fig.~\ref{fig:data_vs_theory_top_diff} displays
the comparison between experimental data and the best-fit EFT theory
predictions (for linear and quadratic fits)
in the case of representative differential top quark pair
and single top quark production datasets.
Both the data and the EFT fit results are normalised to the central value of the SM theory prediction.
This implies that the more the fit results deviate from unity, the larger the best-fit EFT effects
are for this specific observable.
Furthermore, the error band in the EFT prediction indicates the associated 95\% CL interval evaluated
over the $N_{\rm spl}$ samples produced by the NS method, {  that is, the 95\% interval
of the corresponding marginalised posterior distributions shown in Fig.~\ref{fig:posterior_coeffs}.}

From these comparisons, one can observe how for some datasets the best-fit EFT results
move in the direction of the experimental data, for instance in the case of the $m_{t\bar{t}}$
distributions at large invariant masses for inclusive $t\bar{t}$ production.
This is an important kinematic region in the fit, since energy-growing effects increase the EFT
sensitivity.
Interestingly, in the highest $m_{t\bar{t}}$ bins for some of the 13 TeV top datasets
the 95\% CL interval associated to the EFT prediction does not include the SM expectation.
In the case of the single-top $t$-channel differential cross-sections, the EFT fit results
are very close to the SM predictions, indicating that EFT effects are well constrained
for this process at the scale of the present experimental uncertainties.
We also note that the uncertainty band associated to the EFT prediction
can turn out to be rather different in the $\mathcal{O}\lp \Lambda^{-2}\rp$ fits
as compared to the $\mathcal{O}\lp \Lambda^{-4}\rp$ ones, with the latter in general
being more precise than the former for the processes considered here.
{  We note that the CMS $t\bar{t}$ double differential distributions at 8 TeV
  (upper right plot in Fig.~\ref{fig:data_vs_theory_top_diff}) are provided in bins
  of both $m_{t\bar{t}}$ and $y_{t\bar{t}}$, and hence there is more than one data point
for each $m_{t\bar{t}}$ bin.}

\begin{figure}[t]
  \begin{center}
    \includegraphics[width=0.325\linewidth]{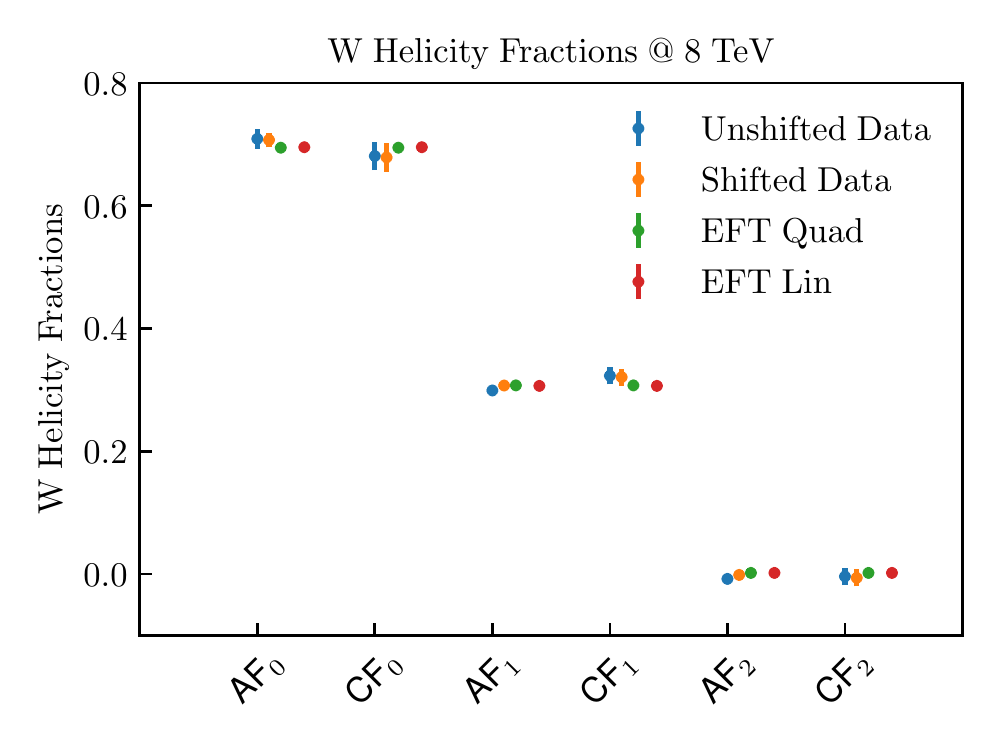}
  \includegraphics[width=0.325\linewidth]{plots_v2/DvT12_SingleTop.pdf}
  \includegraphics[width=0.325\linewidth]{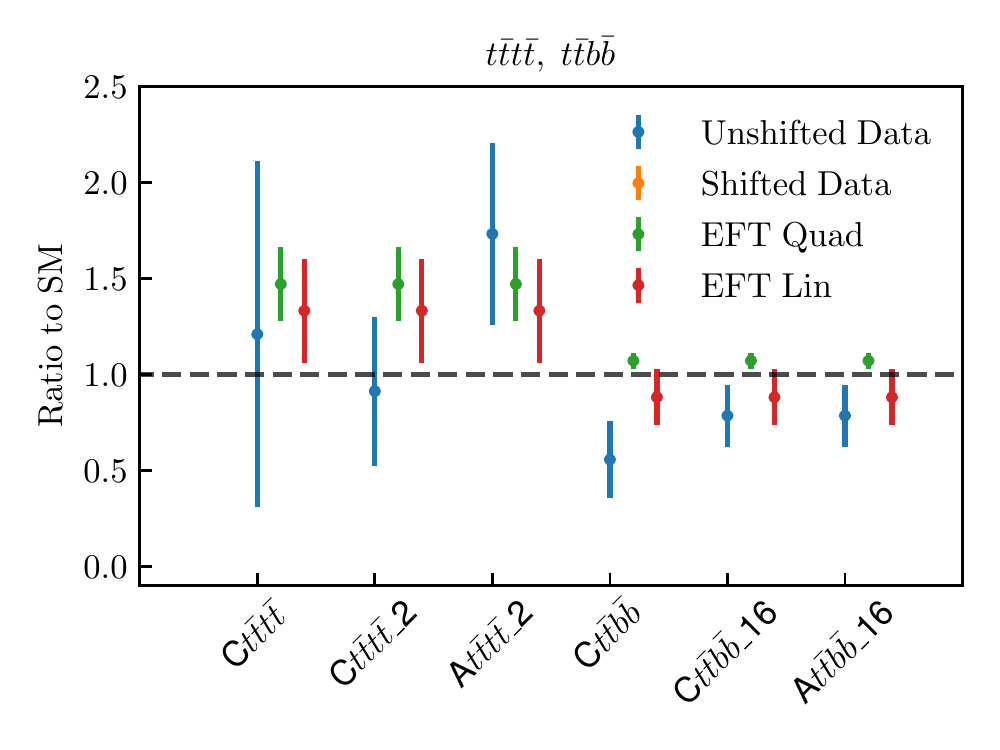}
  \includegraphics[width=0.325\linewidth]{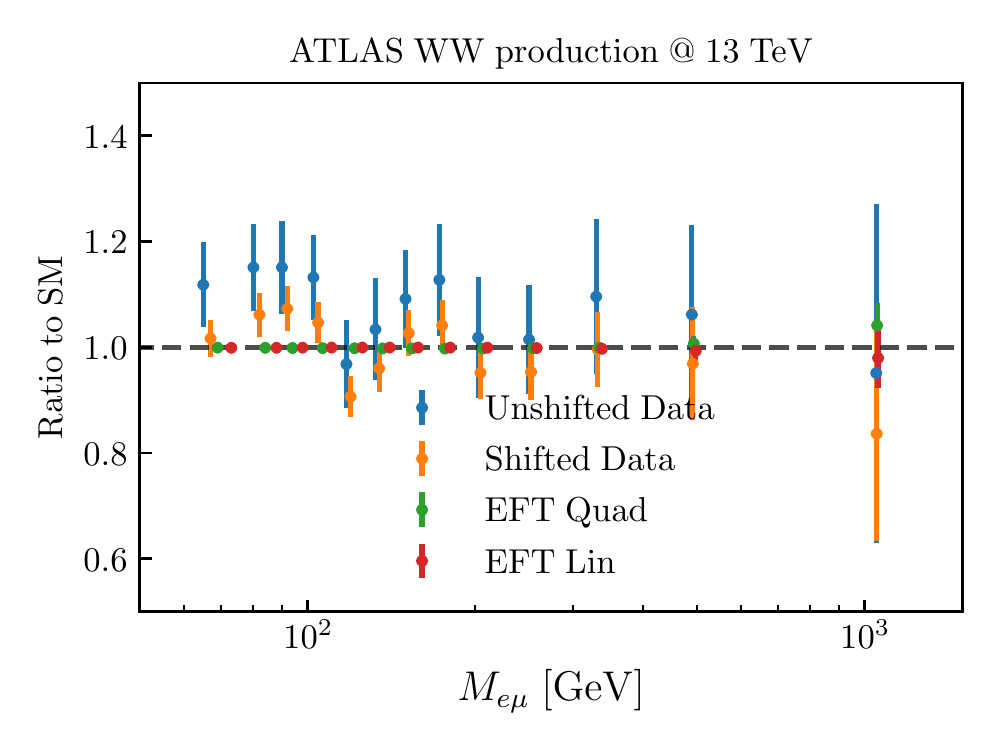}
  \includegraphics[width=0.325\linewidth]{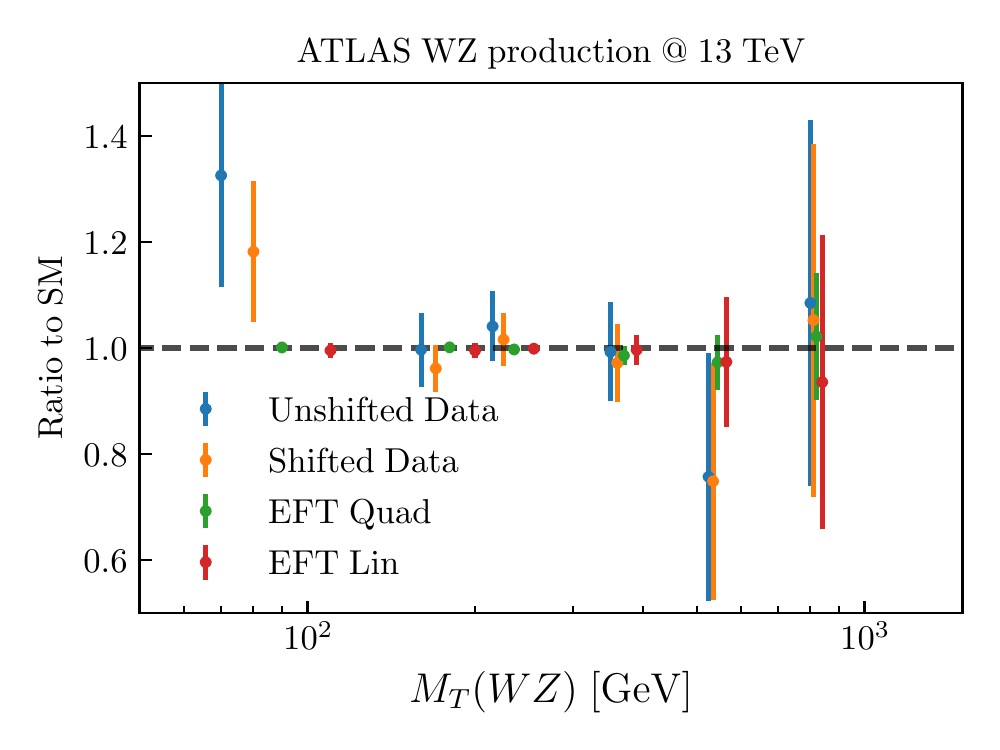}
  \includegraphics[width=0.325\linewidth]{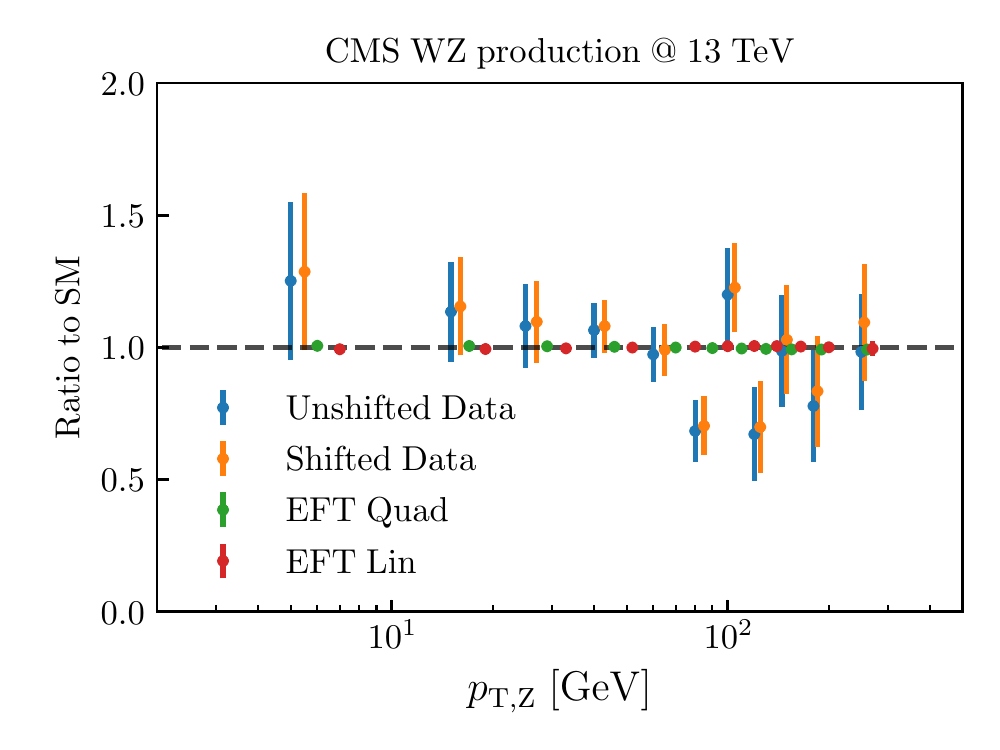}
  \includegraphics[width=0.325\linewidth]{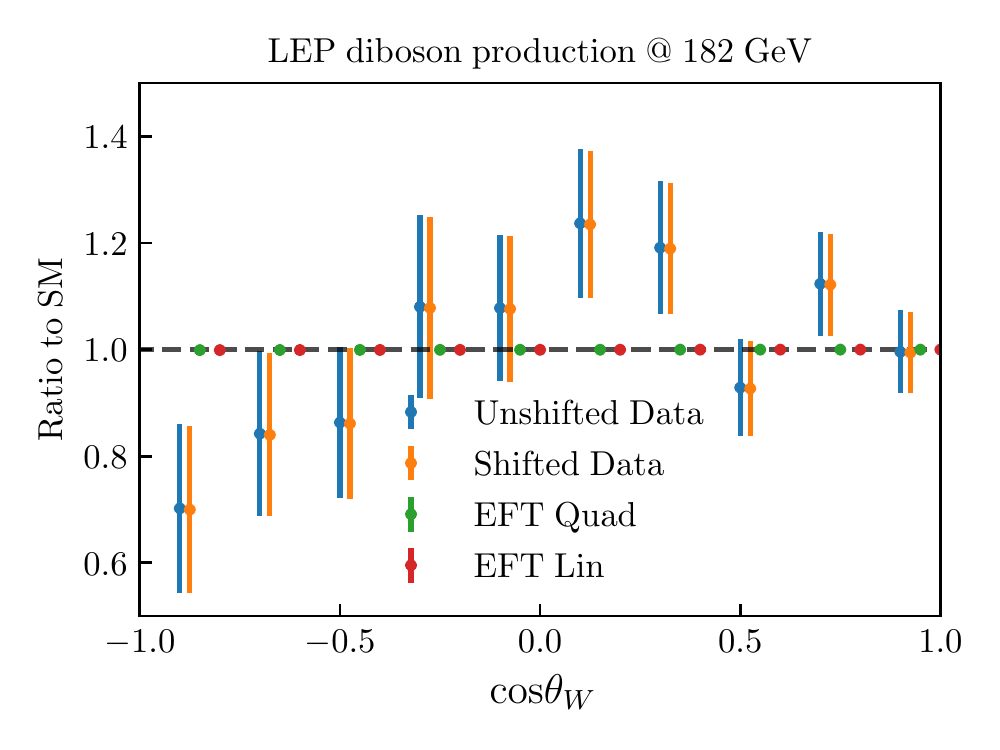}
  \includegraphics[width=0.325\linewidth]{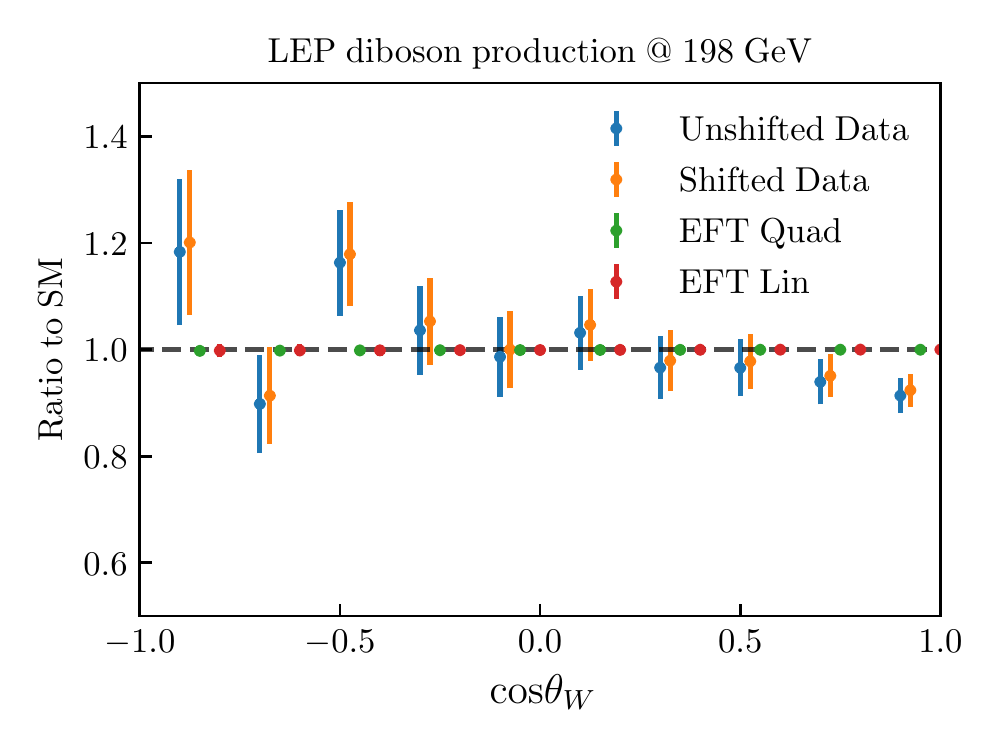}
  \includegraphics[width=0.325\linewidth]{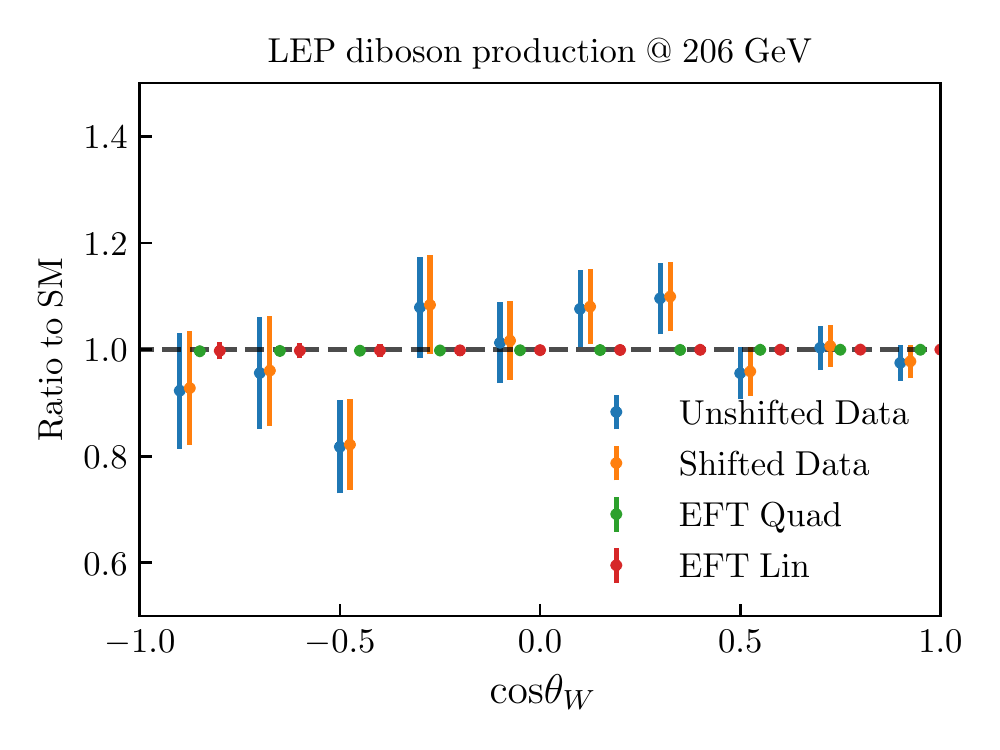}
  \caption{\small Same as Fig.~\ref{fig:data_vs_theory_top_diff} now for the $W$ helicity fractions,
    the single-top $s$-channel and $tV$ total cross-sections, the four-heavy quark fiducial
    measurements, the LHC diboson differential distributions at 13 TeV, and the LEP diboson cross-sections
    at different center of mass energies.
 \label{fig:data_vs_theory_v2} }
 \end{center}
\end{figure}

Then Fig.~\ref{fig:data_vs_theory_v2} displays the same comparison between data and the SM and EFT
predictions as in Fig.~\ref{fig:data_vs_theory_top_diff} now for the $W$ helicity fractions,
the single-top $s$-channel and $tV$ total cross-sections, the four-heavy-quark fiducial
measurements, the LHC diboson differential distributions at 13 TeV, and the LEP diboson cross-sections
at different center-of-mass energies.
Note that contrary to the rest of the datasets, the comparison for the $W$ helicity fraction is carried
out at the absolute rather than at the normalised level.
Concerning the single top measurements, the best-fit EFT predictions tend
to move towards the experimental data, which in most cases is somewhat
higher than the SM prediction.
For some datasets, such as single-top $s$-channel cross-section at 8 TeV and the $tW$ cross-sections
at 13 TeV, the agreement between ATLAS and CMS is at best marginal and thus
the EFT fit interpolates between the two measurements.
A similar behaviour is observed for the $t\bar{t}t\bar{t}$ cross-sections at 13 TeV.
Furthermore, as was the case for the processes
considered in Fig.~\ref{fig:data_vs_theory_top_diff}, the EFT fit uncertainties
appear to be reduced in the quadratic case.

\begin{figure}[t]
  \begin{center}
    \includegraphics[width=0.49\linewidth]{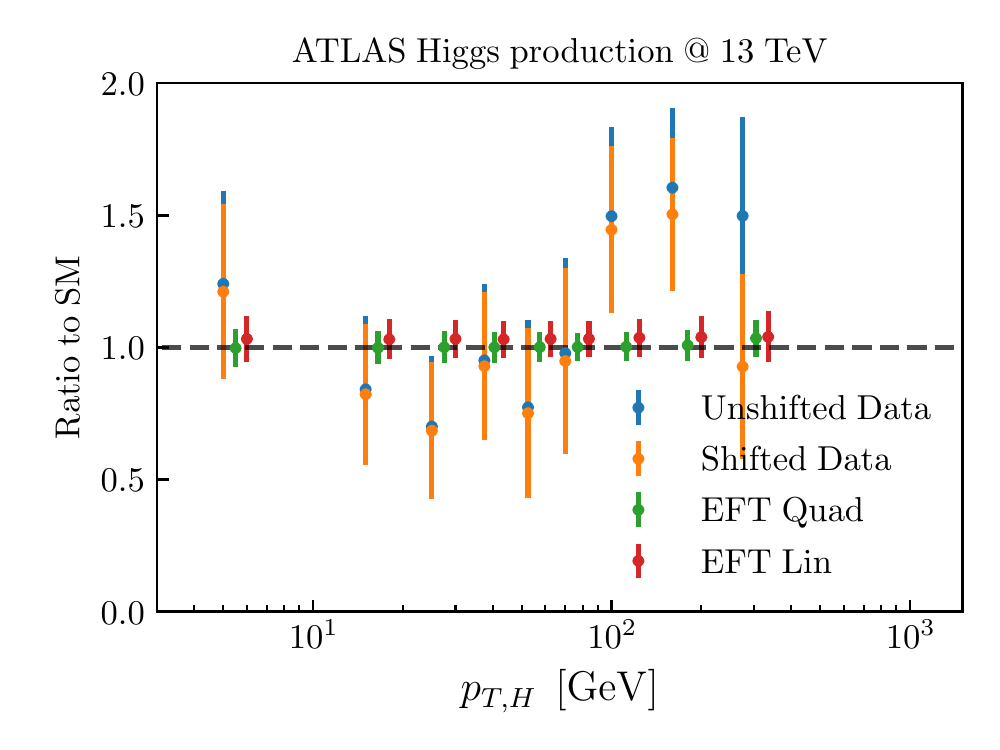}
    \includegraphics[width=0.49\linewidth]{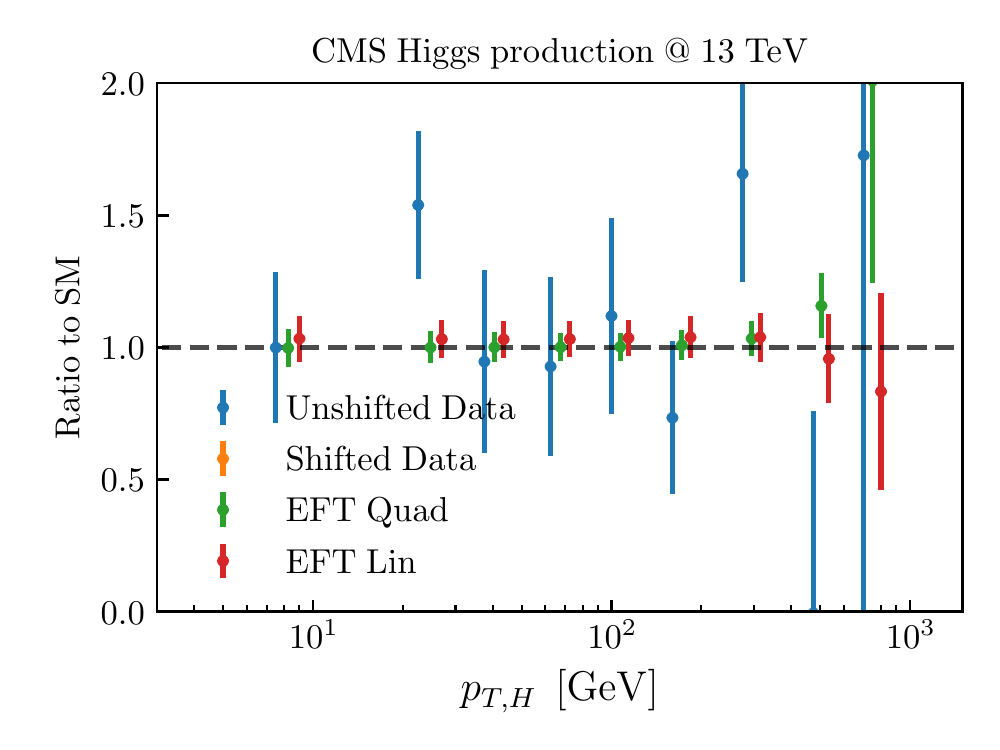}
  \includegraphics[width=0.49\linewidth]{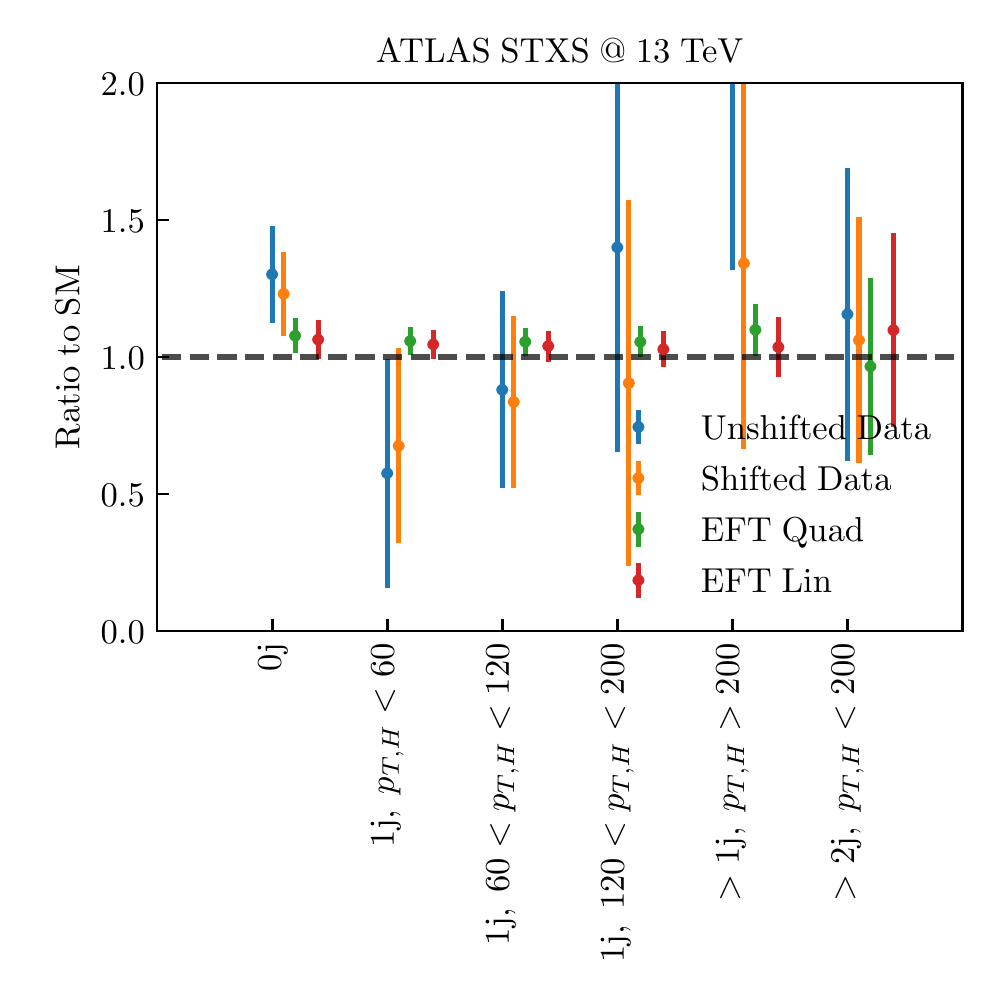}
  \includegraphics[width=0.49\linewidth]{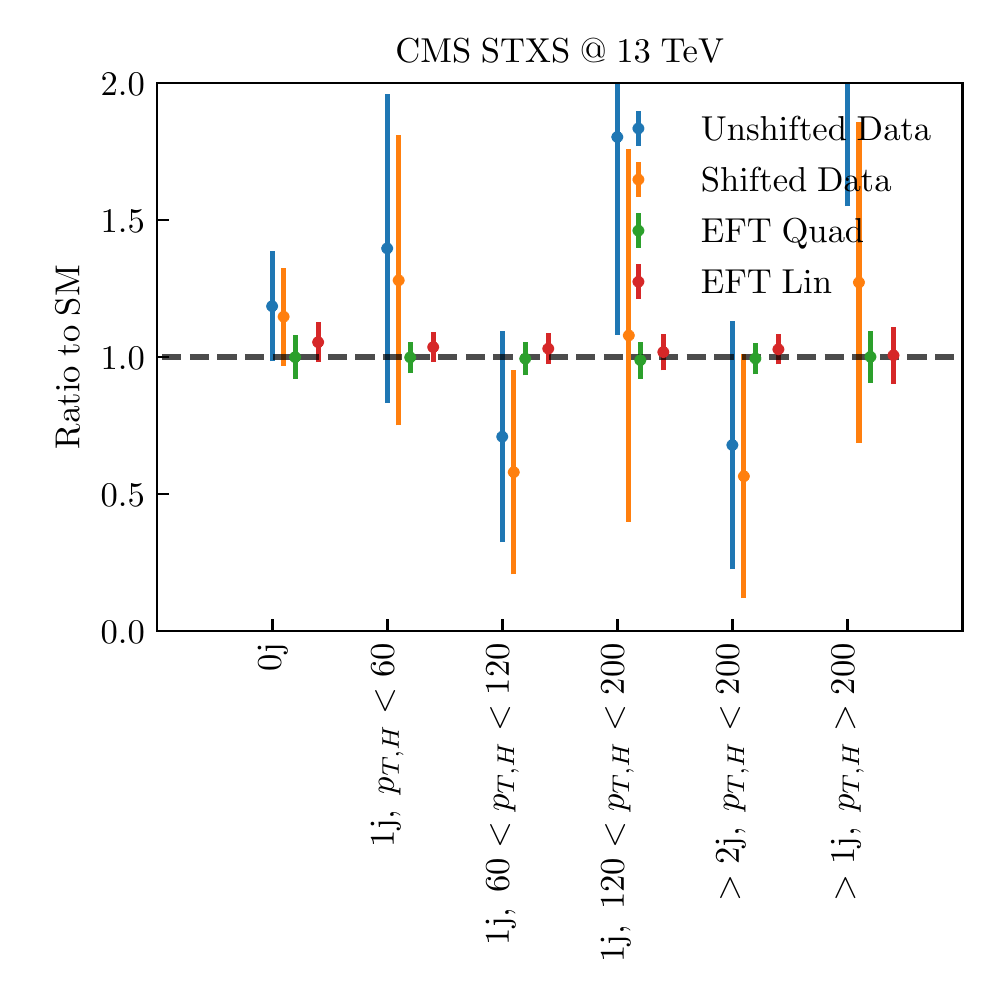}
  \vspace{-0.4cm}
  \caption{\small Same as Fig.~\ref{fig:data_vs_theory_top_diff} for
   representative Higgs  measurements from ATLAS and CMS at $\sqrt{s}=13$ TeV,
    namely the $p_{T,H}$ distributions summing over all production modes and final states (upper panels),
    and the Simplified Template Cross-Section measurements (bottom panels)
    corresponding
    to the $ZZ$ (left) and the $\gamma\gamma$ final states (right panel).
 \label{fig:data_vs_theory_v4} }
 \end{center}
\end{figure}

\begin{figure}[htbp]
  \begin{center}
  \includegraphics[width=0.86\linewidth]{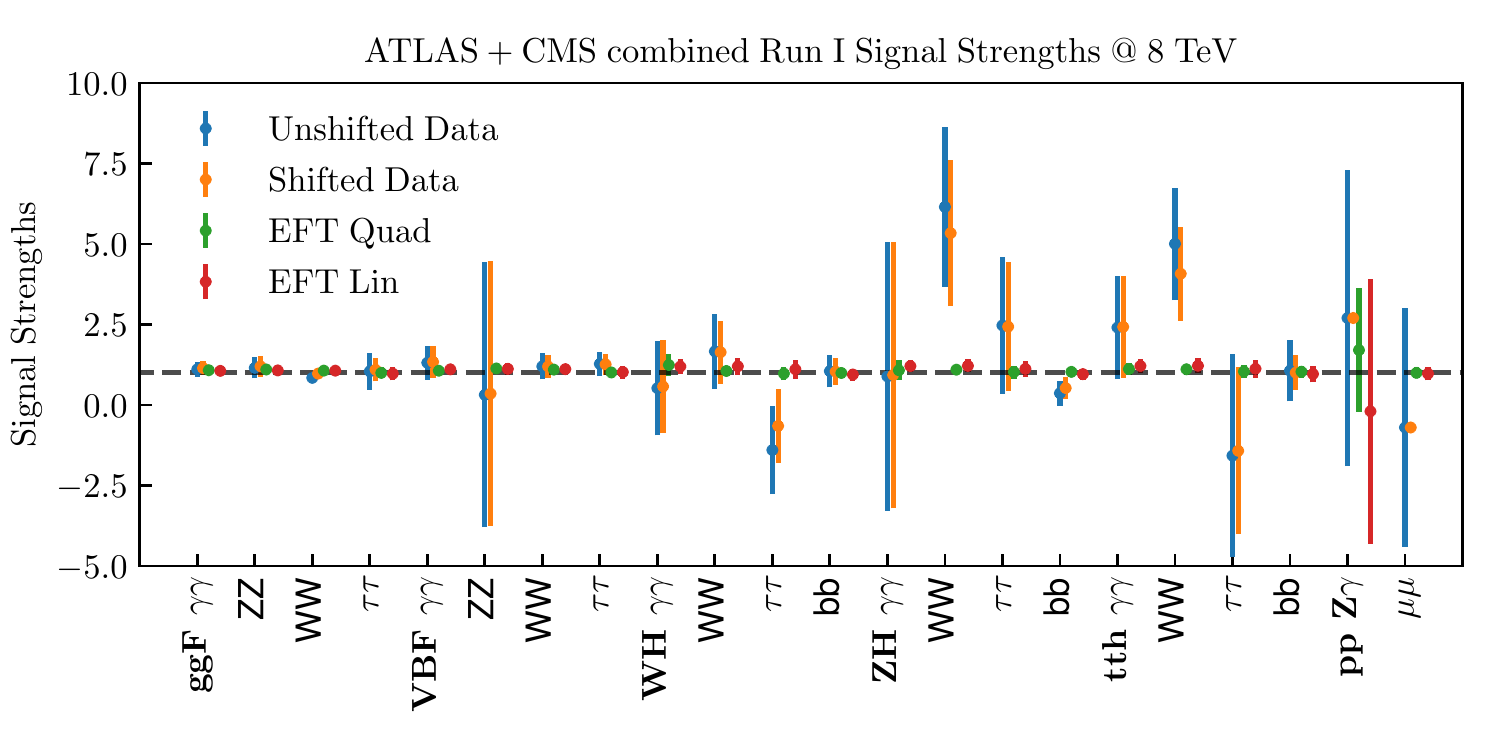}
  \includegraphics[width=0.86\linewidth]{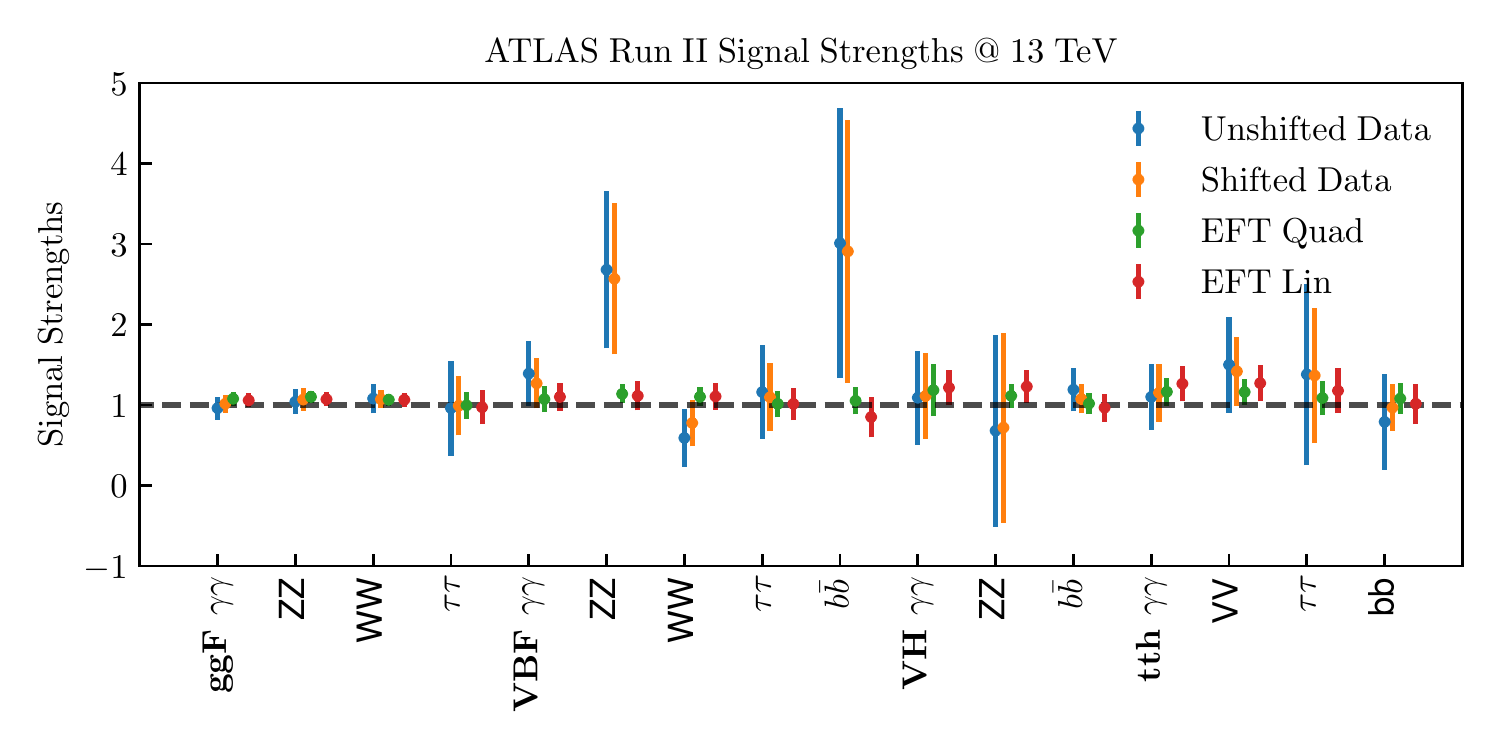}
  \includegraphics[width=0.86\linewidth]{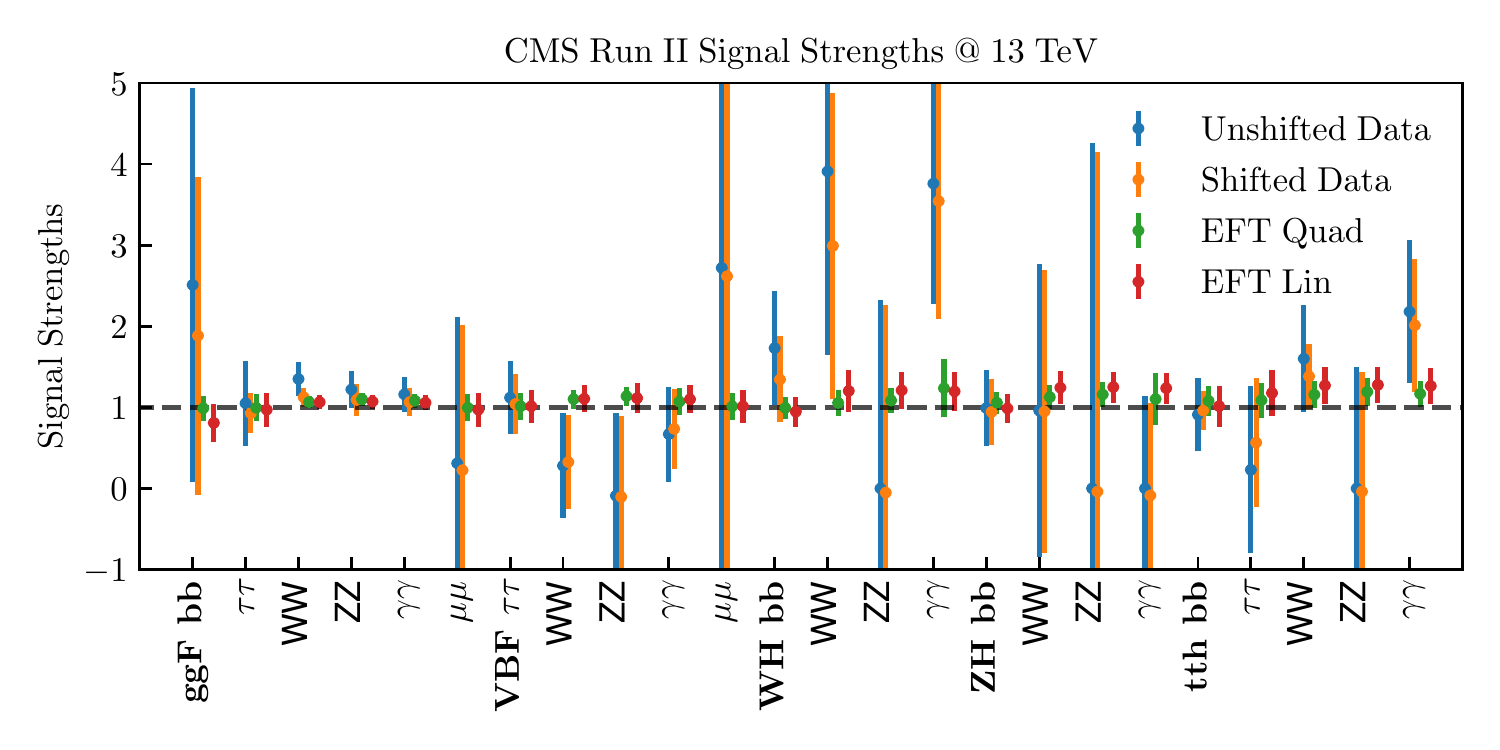}
  \caption{\small Same as Fig.~\ref{fig:data_vs_theory_top_diff} for the Higgs boson
    signal strengths corresponding to different production mechanisms and decay channels.
    From top to bottom we show the ATLAS+CMS Run I combination and the ATLAS and CMS Run II measurements
    at 13 TeV.
    Note that by the definition of the signal strengths the SM predictions correspond to $\mu_i^{(f)}=1$
    in all cases, see also App.~\ref{sec:signalstrenghts}.
 \label{fig:data_vs_theory_v3} }
 \end{center}
\end{figure}

\begin{figure}[htbp]
  \begin{center}
  \includegraphics[width=0.49\linewidth]{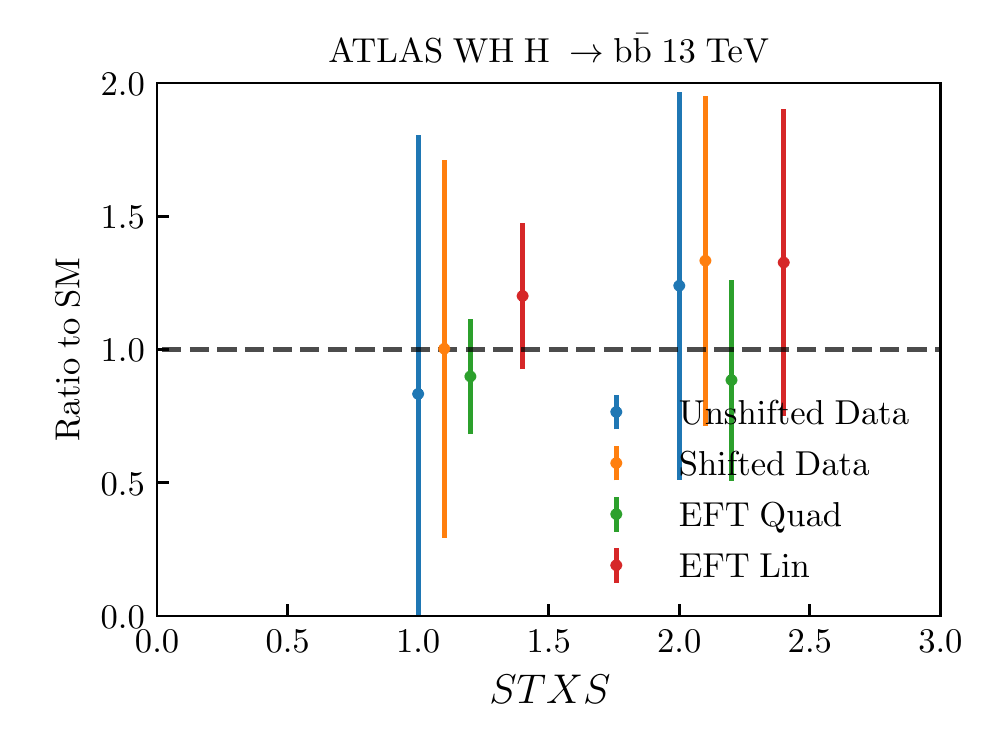}
  \includegraphics[width=0.49\linewidth]{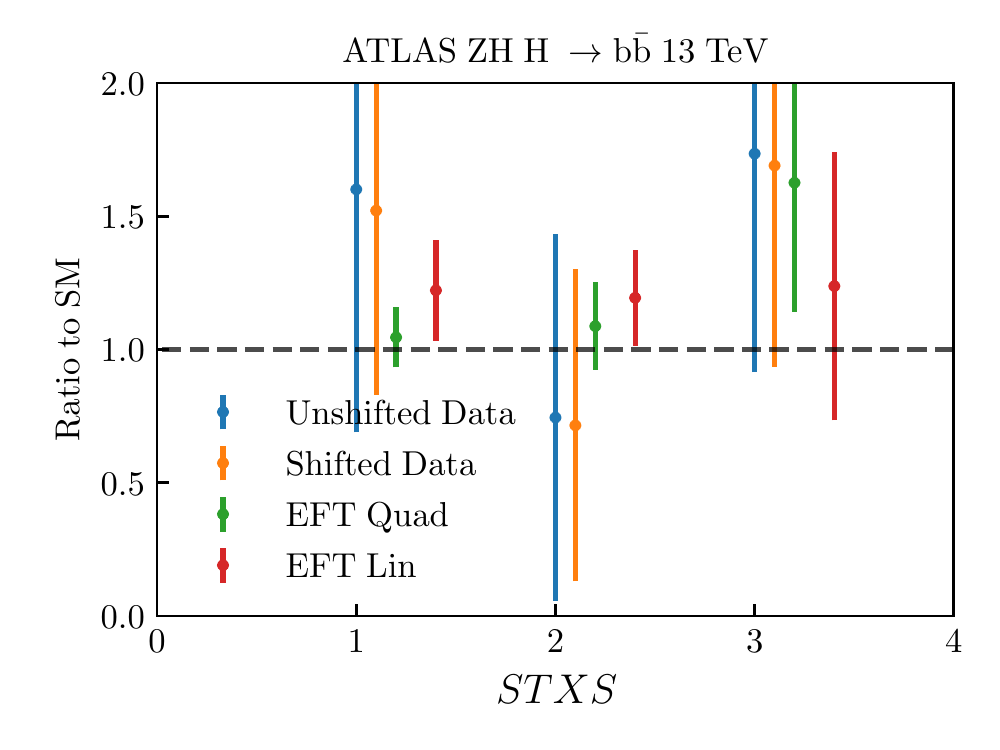}
  \caption{\small Same as Fig.~\ref{fig:data_vs_theory_top_diff} for the Higgs boson
    associated production STXS from the ATLAS $VH$ measurement at 13 TeV.
 \label{fig:data_vs_theory_v5} }
 \end{center}
\end{figure}

Moving to the LEP and LHC diboson datasets, one finds that for electron-positron collisions
the EFT fit result is very close to the SM cross-section with a vanishing uncertainty.
This result is likely to be related to the constraints imposed by the EWPOs as well as by the
LHC diboson data.
Nevertheless, the SM predictions are in good agreement with the LEP data for all
four center-of-mass energies considered to begin with.
In the case of the LHC measurements, for the ATLAS $m_{e\mu}$ and $m_T^{WZ}$
distributions in the $W^+W^-$ and $WZ$ final states, respectively, the data is in good
agreement with the SM and the net effect of the EFT corrections is small, except
perhaps for the highest-energy bin of the $m_{e\mu}$ and $m_T^{WZ}$ distribution.
Similar considerations apply for the CMS 13 TeV $WZ$ dataset, where we observe good
agreement between theory and data also for the high $p_{T}^Z$ region.

Concerning the comparison between experimental data and theory calculations for the Higgs production
and decay measurements,  Fig.~\ref{fig:data_vs_theory_v4} displays
representative Higgs  measurements from ATLAS and CMS at $\sqrt{s}=13$ TeV,
namely the $p_{T,H}$ distributions inclusive over all production modes and final states,
and the Simplified Template Cross-Section measurements 
corresponding
to the $ZZ$  and the $\gamma\gamma$ final states for ATLAS and CMS
respectively.
Then Fig.~\ref{fig:data_vs_theory_v3} summarizes
the results corresponding to Higgs boson signal strengths for
different production mechanisms and decay channels.
Specifically, we show  the ATLAS+CMS Run I combination and the ATLAS and CMS Run II measurements at 13 TeV.
Note that, by construction, in the signal strengths the SM predictions correspond to $\mu_i^{(f)}=1$,
see the discussion of App.~\ref{sec:signalstrenghts} for more details.

In the case of the differential Higgs distributions, we can observe the good agreement
both the SM and the EFT predictions within the relatively large
experimental uncertainties.
For these distributions, the EFT effects can reach a magnitude of up to a few percent
in the global fit.
For instance, for the CMS $p_{T,H}$ distribution in the quadratic fit,
the best-fit results are $\simeq 15\%$ higher than the SM for the $p_{T,H}=1$ TeV bin.
For the case of the signal strengths, also a fair agreement is found, though for some combinations
of production channel and decay mode the experimental uncertainties still remain
rather large.


%% file: app-Higgs-Signal-Strenghts.tex
\section{Implementation of Higgs signal strengths}
\label{sec:signalstrenghts}

In this appendix, we describe how the Higgs signal strengths
have been implemented in the present analysis.
For a generic Higgs production and decay cross-section, denoted as $\sigma(pp \to h \to X)$,
the experimentally measured signal strength is defined as the product of the production
cross-section times the corresponding branching ratio, normalised
to the Standard Model predictions:
\be
\mu_{pp\to h \to X}^{\rm (exp)} = \frac{\sigma^{(\rm exp)}(pp \to h \to X)}{
  \sigma^{(\rm sm)}(pp \to h \to X)} = \frac{\sigma^{(\rm exp)}(pp \to h \to X)}{
\sigma^{(\rm sm)}(pp \to h) \times {\rm BR}^{(\rm sm)}(h\to X) }
 \, ,
 \ee
 which in turn can be expressed as
 \be
 \label{eq:signalstrenghts1}
\mu_{pp\to h \to X}^{\rm (exp)} = \frac{\sigma^{(\rm exp)}(pp \to h \to X)}{
  \sigma^{(\rm sm)}(pp \to h \to X)} = \frac{\sigma^{(\rm exp)}(pp \to h \to X)}{
\sigma^{(\rm sm)}(pp \to h) \times \lp  \Gamma^{(\rm sm)}_{X} / \Gamma^{(\rm sm)}_{{\rm tot}}\rp  }
 \, ,
 \ee
where $\Gamma_{{\rm tot}}$ indicates the total Higgs width
and $\Gamma_{ X}$ is the partial width for
the decay into the specific final state $X$.
In this work, we assume that the Higgs boson decays only to known particles and hence set to zero
its branching ratio to invisible final states.
The theoretical prediction corresponding to the measurement
of the signal strengths Eq.~(\ref{eq:signalstrenghts1}) in the SMEFT is given by
\be
\label{eq:signalstrenghts2}
\mu_{pp\to h \to X}^{\rm (th)}({\boldsymbol c}) =  \lp \frac{\sigma^{(\rm EFT)}(pp \to h)({\boldsymbol c}) \times \lp  \Gamma^{(\rm EFT)}_{X}({\boldsymbol c}) / \Gamma^{(\rm EFT)}_{{\rm tot}}({\boldsymbol c})\rp  }{
\sigma^{(\rm sm)}(pp \to h) \times \lp  \Gamma^{(\rm sm)}_{X} / \Gamma^{(\rm sm)}_{{\rm tot}}\rp  }\rp 
 \, ,
 \ee
 At any order in the EFT expansion, we can express Eq.~(\ref{eq:signalstrenghts2}) as
 \bea
 \label{eq:smeft_signal_1}
 \mu_{pp\to h \to X}^{\rm (th)}({\boldsymbol c}) =  \lp
 \frac{\sigma^{(\rm EFT)}(pp \to h)({\boldsymbol c}) }{\sigma^{(\rm sm)}(pp \to h)}\rp \times
 \lp \frac{  \Gamma^{(\rm EFT)}_{X}({\boldsymbol c}) }{ \Gamma^{(\rm sm)}_{X} }
 \rp \times
 \lp \frac{  \Gamma^{(\rm EFT)}_{\rm tot}({\boldsymbol c}) }{ \Gamma^{(\rm sm)}_{\rm tot} }
 \rp^{-1} \, ,
 \eea
 where the total width is evaluated as the sum of all relevant partial decay widths,
 \bea
 \label{eq:smeft_signal_10}
  \mu_{pp\to h \to X}^{\rm (th)}({\boldsymbol c}) =  \lp
 \frac{\sigma^{(\rm EFT)}(pp \to h)({\boldsymbol c}) }{\sigma^{(\rm sm)}(pp \to h)}\rp \times
 \lp \frac{  \Gamma^{(\rm EFT)}_{X}({\boldsymbol c}) }{ \Gamma^{(\rm sm)}_{X} }
 \rp \times
 \lp \frac{  \sum_Y \Gamma^{(\rm EFT)}_{Y}({\boldsymbol c}) }{ \sum_{Z}\Gamma^{(\rm sm)}_{Z} }
 \rp^{-1} \, ,
 \eea
 where $X$, $Y$, and $Z$ indicate possible (SM) final states in which the Higgs boson can decay.
 Assuming now that $\Lambda=1$ TeV and working at linear order in the EFT expansion one has
 \be
 \sigma^{(\rm EFT)}(pp \to h)({\boldsymbol c}) = \sigma^{(\rm sm)}(pp \to h) + \sum_{i=1}^{n} c_i \kappa_{\sigma,i} \, ,
 \ee
 \be
 \Gamma^{(\rm EFT)}_{X}({\boldsymbol c})   = \Gamma^{(\rm sm)}_{X}
+ \sum_{i=1}^{n} c_i \kappa_{\gamma_x,i} \, ,
 \ee
 where $n$ is the number of independent dimension-6 operators in the fitting basis
 and $\{ \kappa_{\sigma,i}\}$ and $\{ \kappa_{\gamma_x,i}\}$, 
 are the (absolute) EFT corrections associated to the production
 cross section and partial width, respectively, corresponding to the operator $c_i$.

 Inserting now the linear EFT expansion into Eq.~(\ref{eq:smeft_signal_10}), one gets
  \bea
  \mu_{pp\to h \to X}^{\rm (th)}({\boldsymbol c}) &=&  \lp
 \frac{\sigma^{(\rm sm)} + \sum_{i=1}^{n} c_i \kappa_{\sigma,i} }{\sigma^{(\rm sm)}}\rp \times
 \lp \frac{  \Gamma^{(\rm sm)}_{X}
+ \sum_{j=1}^{n} c_j \kappa_{\gamma_x,i} }{ \Gamma^{(\rm sm)}_{X} }
 \rp  \nonumber \\ & \times&   \label{eq:smeft_signal_11}
 \lp \frac{  \sum_Y \lp   \Gamma^{(\rm sm)}_{Y}
+ \sum_{k=1}^{n} c_k \kappa_{\gamma_y,i}\rp }{ \sum_{Z}\Gamma^{(\rm sm)}_{Z} }
 \rp^{-1}  \\
 &=&  \lp
 1+ \sum_{i=1}^{n} c_i \frac{\kappa_{\sigma,i}}{\sigma^{(\rm sm)} }\rp \times
 \lp
   1
+ \sum_{j=1}^{n} c_j \frac{\kappa_{\gamma_x,j}}{\Gamma^{(\rm sm)}_{X}} 
 \rp  \nonumber 
 \lp \sum_Y \lp   \frac{\Gamma^{(\rm sm)}_{Y}}{\Gamma^{(\rm sm)}_{\rm tot}}
+ \sum_{k=1}^{n} c_k \frac{\kappa_{\gamma_y,k}}{\Gamma^{(\rm sm)}_{\rm tot}}\rp 
 \rp^{-1} \nonumber
 \, ,
 \eea
 which can be further simplified to
 \be
 \mu_{pp\to h \to X}^{\rm (th)}({\boldsymbol c}) =
\lp 1+ \sum_{i=1}^{n} c_i \frac{\kappa_{\sigma,i}}{\sigma^{(\rm sm)} }\rp \times
 \lp
   1
+ \sum_{j=1}^{n} c_j \frac{\kappa_{\gamma_x,j}}{\Gamma^{(\rm sm)}_{X}} 
 \rp  \times   
 \lp  1
+ \sum_Y \lp \sum_{k=1}^{n} c_k \frac{\kappa_{\gamma_y,k}}{\Gamma^{(\rm sm)}_{\rm tot}}\rp 
 \rp^{-1} \, .
 \ee
 If we Taylor expand the last term at $\mathcal{O}\lp \Lambda^{-2}\rp$,
 we obtain the required expression for the theoretical prediction of the
 Higgs signal strengths at linear order in the EFT,
 \be
 \label{app:eqb11}
 \mu_{pp\to h \to X}^{\rm (th)}({\boldsymbol c}) =
 1+ \sum_{i=1}^{n} c_i \lc \frac{\kappa_{\sigma,i}}{\sigma^{(\rm sm)} }
 +
 \frac{\kappa_{\gamma_x,i}}{\Gamma^{(\rm sm)}_{X}} -
 \sum_Y \lp \frac{\kappa_{\gamma_y,i}}{\Gamma^{(\rm sm)}_{\rm tot}}\rp 
 \rc \, .
 \ee
 For simplicity, we can replace the total Higgs decay width in the SM
with the corresponding branching fractions,
 \be
    {\rm BR}^{(\rm sm)}_X \equiv \frac{\Gamma^{(\rm sm)}_{X}}{
      \Gamma^{(\rm sm)}_{\rm tot}} \quad \to \quad
    \frac{1}{
      \Gamma^{(\rm sm)}_{\rm tot}} = \frac{{\rm BR}^{(\rm sm)}_X }{\Gamma^{(\rm sm)}_{X}} \, ,
    \ee
   which allows us to express Eq.~(\ref{app:eqb11}) as
    \be
 \mu_{pp\to h \to X}^{\rm (th)}({\boldsymbol c}) =
 1+ \sum_{i=1}^{n} c_i \lc \frac{\kappa_{\sigma,i}}{\sigma^{(\rm sm)} }
 +
 \frac{\kappa_{\gamma_x,i}}{\Gamma^{(\rm sm)}_{X}} -
 \sum_Y \lp \frac{\kappa_{\gamma_y,i}}{\Gamma^{(\rm sm)}_{Y}}{\rm BR}^{(\rm sm)}_Y \rp 
 \rc \, .
 \ee
 Hence we find that we can evaluate the Higgs signal strengths in the EFT as
   \be
 \mu_{pp\to h \to X}^{\rm (th)}({\boldsymbol c}) =
 1+ \sum_{i=1}^{n} c_i \beta_{pp\to h \to X,i}  \, ,
 \ee
 where we have defined
 \be
\beta_{pp\to h \to X,i} \equiv \frac{\kappa_{\sigma,i}}{\sigma^{(\rm sm)} }
 +
 \frac{\kappa_{\gamma_x,i}}{\Gamma^{(\rm sm)}_{X}} -
 \sum_Y \lp \frac{\kappa_{\gamma_y,i}}{\Gamma^{(\rm sm)}_{Y}}{\rm BR}^{(\rm sm)}_Y \rp \, .
 \ee
 Note that in this notation we use $\beta$ to indicate relative
 EFT corrections while the $\kappa$ always indicate instead
  absolute corrections.
  A similar expression, although somewhat more cumbersome, can be derived to account
  for the quadratic EFT contributions to the Higgs signal strengths.

%% file: app-covariance-mat.tex
\section{Correlation matrices in EFT space}
\label{sec:fullcovmat}

\begin{figure}[htbp]
  \begin{center}
  \includegraphics[width=0.495\linewidth]{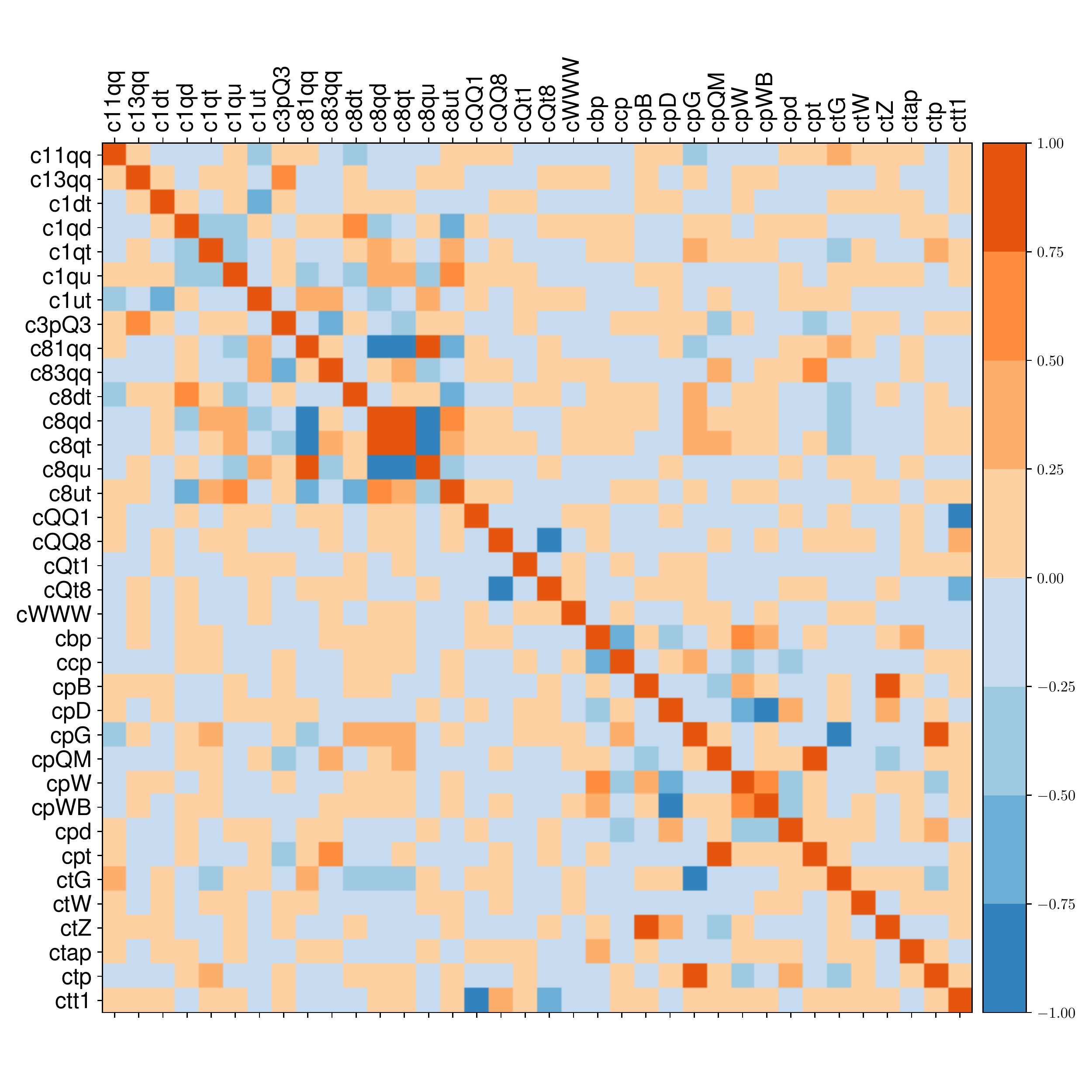}
  \includegraphics[width=0.495\linewidth]{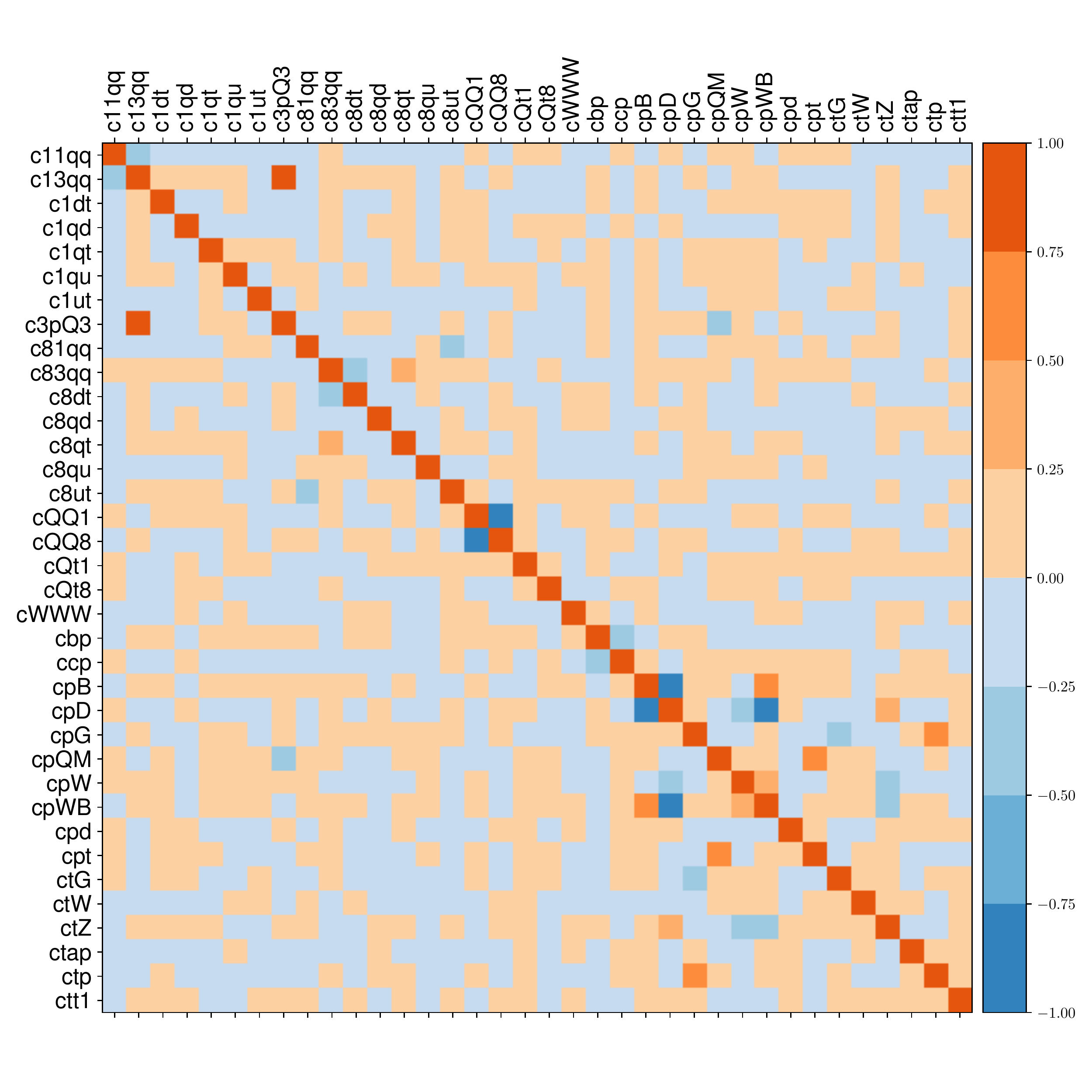}
   \includegraphics[width=0.495\linewidth]{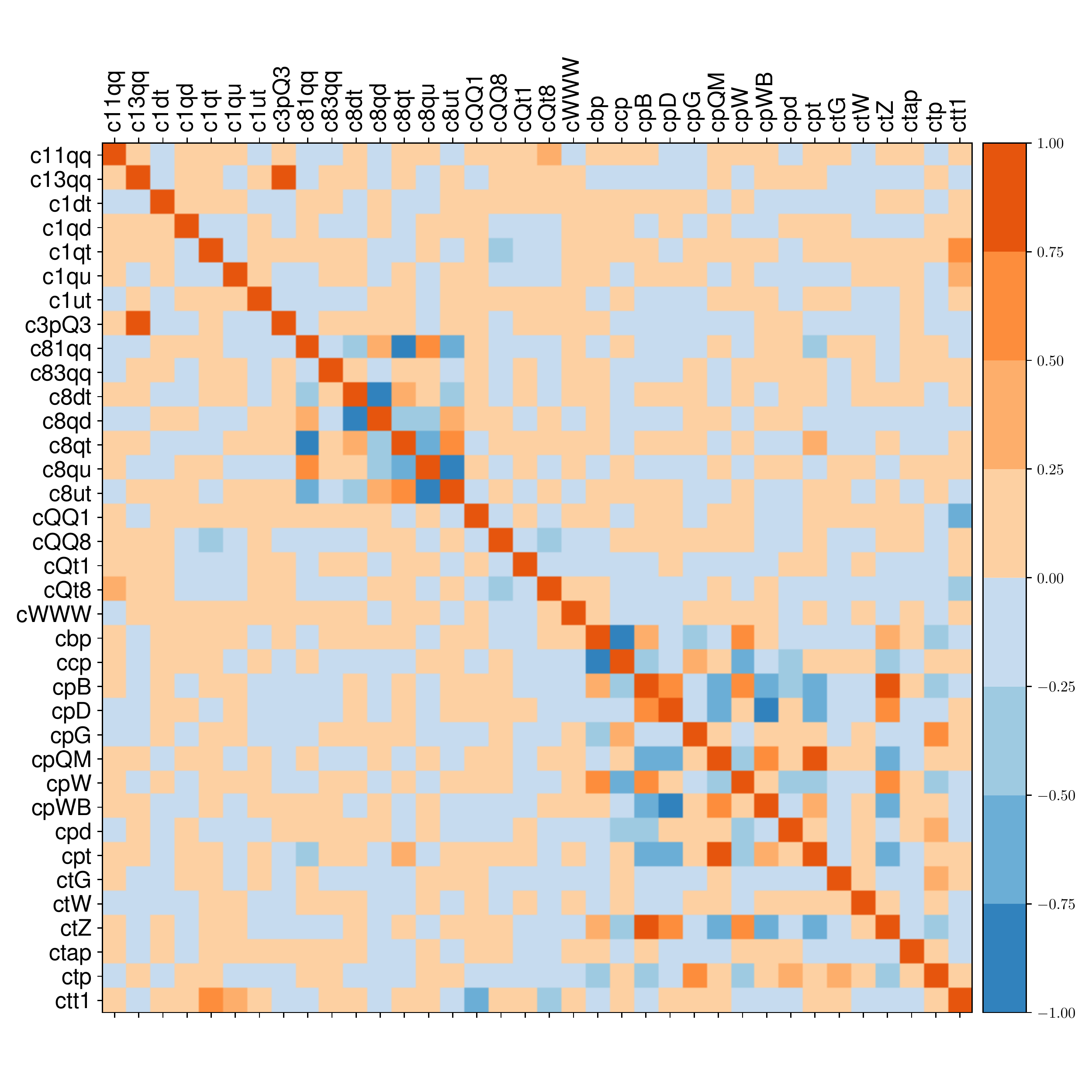}
  \includegraphics[width=0.495\linewidth]{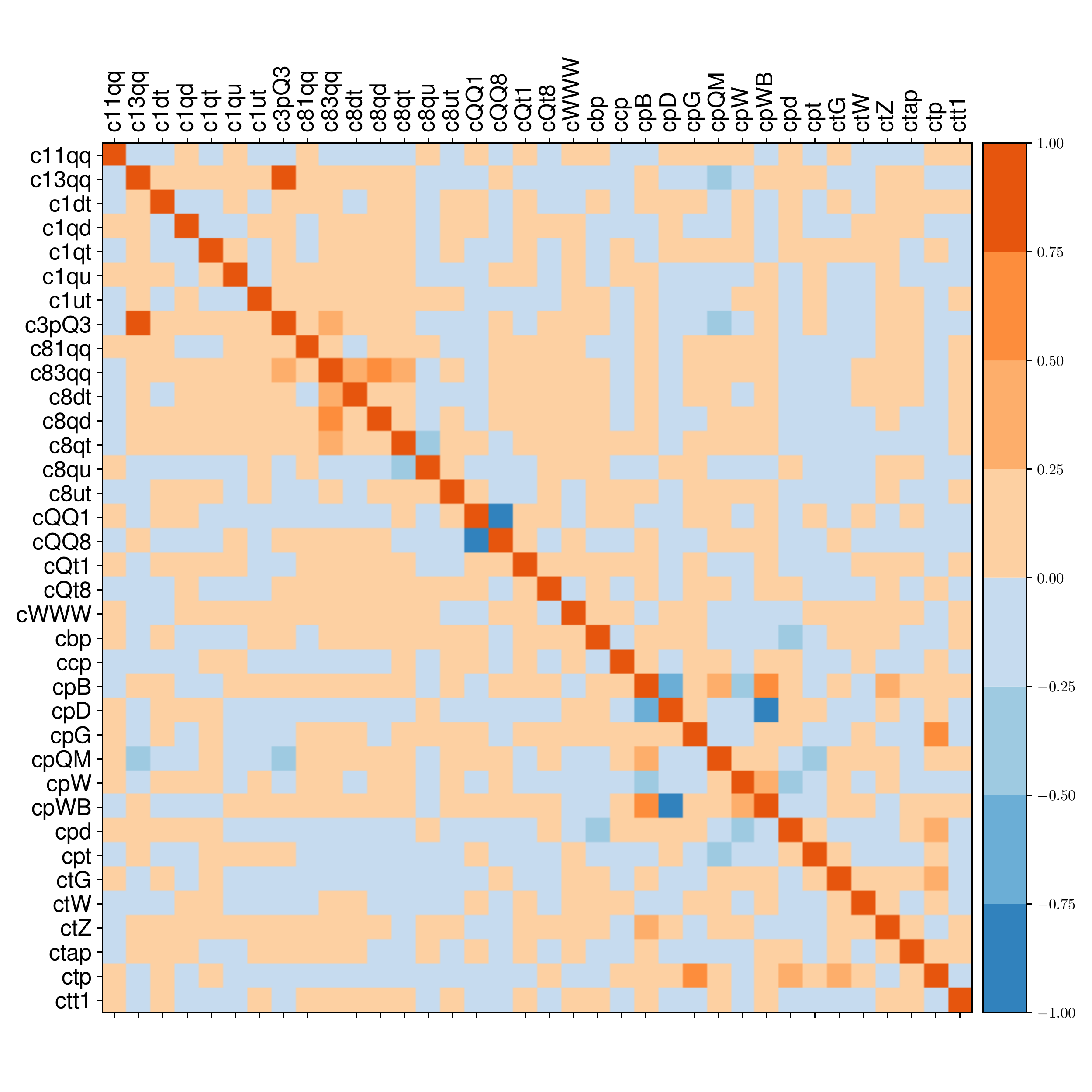}
  \caption{\small  
    The correlation coefficients $\rho\lp c_i,c_j\rp$
    for the full set of EFT coefficients
    that enters the present global analysis.
    We show the results corresponding to the linear
    (left panels) and quadratic (right panels) fits,
    both for the case where the EFT cross-sections include
    NLO QCD corrections (top panels) and where they do not
    (bottom panels).
      \label{fig:globalfitfull-correlations-NLO} }
 \end{center}
\end{figure}

{ 
  In Sect.~\ref{sec:results} we have presented results for
  the correlation coefficients associated to a specific subset of operators
  that define our EFT parameter space, in particular those
  pairs exhibiting large absolute correlations, $|\rho(c_i,c_j)|\ge 0.5$.
  For completeness, we display here the full correlations
  matrices in the EFT parameter space.
  Fig.~\ref{fig:globalfitfull-correlations-NLO} shows the
  correlation coefficients $\rho\lp c_i,c_j\rp$
    for the complete set of EFT coefficients
    that enters the present global analysis.
    We present the results corresponding to both the linear
    (left panels) and quadratic (right panels) fits,
    for the case where the EFT cross-sections include
    NLO QCD corrections (top panels) and where they do not
    (bottom panels).

    Inspection of Fig.~\ref{fig:globalfitfull-correlations-NLO} confirms
    two main findings of Sect.~\ref{sec:results}
    concerning the correlation patterns of the global EFT fit.
    First of all, how quadratic corrections decrease the
    correlation between fit parameters.
    Second, though perhaps a less marked effect, how NLO QCD corrections
    also lead to a decrease in the absolute value
    of these correlations coefficients.
    In both cases, as previously mentioned, the decrease
    in the values of $|\rho(c_i,c_j)|$ arises from
    the additional sensitivity to new directions
    in the SMEFT parameter space introduced by the quadratic
    corrections and by the NLO QCD corrections to the EFT
    hard-scattering cross-sections.
}

%% file: app-delivery.tex
\section{Usage of SMEFiT results}
\label{sec:delivery}

The results of this work are made available via
the website of the {\tt SMEFiT} project:
\begin{center}
\url{https://lhcfitnikhef.github.io/SMEFT/}
\end{center}
as well as via the corresponding public {\tt GitHub} repository:
\begin{center}
\url{https://github.com/LHCfitNikhef/SMEFiT/}
\end{center}
Specifically, we provide the full set of  $N_{\rm spl}$ samples corresponding
to the NS fits presented in this work for the $n_{\rm op}=50$ Wilson coefficients.
We recall that not all of these coefficients are associated to independent degrees
of freedom, and that 14 of them are constrained by the EWPO relations
as discussed in Sect.~\ref{sec:smefttheory}.

These $N_{\rm spl}$ samples provide a representation of the probability
density associated to the EFT coefficients.
From these samples, it is easy to evaluate statistical estimators such as
means, standard deviations, and correlations, e.g.,
\begin{align}
  \la c_i\ra &= \frac{1}{N_{\rm spl}} \sum_{k=1}^{N_{\rm spl}}c_i^{(k)} \, ,\quad i=1,\ldots,n_{\rm op} \, , \\
  \sigma_{c_i} &= \lp \frac{1}{N_{\rm spl}-1} \sum_{k=1}^{N_{\rm spl}} \lp c_i^{(k)}- \la c_i\ra
  \rp^2  \rp^{1/2} \, ,\quad i=1,\ldots,n_{\rm op} \, , \\
  \rho(c_i,c_j) &=  \lp \frac{1}{N_{\rm spl}} \sum_{k=1}^{N_{\rm spl}}c_i^{(k)}c_j^{(k)} -
  \la c_i\ra\la c_j\ra \rp \Bigg/ \sigma_{c_i} \sigma_{c_k}
  \, ,\quad i,j=1,\ldots,n_{\rm op} \, , 
\end{align}
as well as other estimators such as confidence level intervals and higher moments beyond
the quadratic approximation.
One should emphasize that, as discussed in Sect.~\ref{sec:results}, the Gaussian approximation
is in general not reliable for most of the EFT parameters
in the case that  the quadratic
$\mathcal{O}\lp \Lambda^{-4}\rp$ corrections are accounted for.

These samples spanning the probability density in the space of
EFT Wilson coefficients could also be used to quantify {\it a posteriori} the impact
in the fit of new measurements by means of the Bayesian reweighting method
presented in~\cite{vanBeek:2019evb}.

To facilitate the usage of the results presented in this work, we also make available
via the {\tt GitHub} repository
a {\tt Python} analysis code that takes as input the samples corresponding
to a specific EFT fit and then evaluates means, correlations, and 95\% CL intervals,
as well as produces most of the plots and statistical estimators
displayed in Sect.~\ref{sec:results}.
In a future installment of our global EFT analysis,
we plan to release the full {\tt smefit} fitting code and the associated
theoretical and experimental inputs, together with a complete documentation
and user-friendly examples.